\documentclass[a4, useAMS,usenatbib]{mn2e}
\usepackage{graphicx}
\usepackage{bmpsize}
\usepackage{caption}
\usepackage{booktabs}
\usepackage{hyperref}
\usepackage{deluxetable}
\usepackage{multicol}
\usepackage{multirow}
\usepackage{amssymb}
\usepackage[fleqn]{amsmath}
\usepackage{pdflscape}

\usepackage[dvipsnames]{xcolor}
\definecolor{myblue}{HTML}{268BD2}
\definecolor{mygreen}{HTML}{859900}
\definecolor{myred}{HTML}{DC322F}
\definecolor{mymagenta}{HTML}{D33682}

\hypersetup{
bookmarks=true,         
unicode=true,          
colorlinks=true,       
linkcolor=myred,          
citecolor=myblue,        
filecolor=magenta,      
urlcolor=mygreen           
}
\newcommand{\galex}{\textit{GALEX}}
\newcommand{\hst}{\textit{HST}}
\newcommand{\swift}{\textit{Swift}}
\newcommand{\spitzer}{\textit{Spitzer}}

\newcommand{\wise}{\textit{WISE}}

\newcommand{\aap}{A\&A}
\newcommand{\aaps}{A\&AS}
\newcommand{\aj}{AJ}
\newcommand{\apj}{ApJ}
\newcommand{\apjs}{ApJS}
\newcommand{\apss}{Ap\&SS}
\newcommand{\apjl}{ApJ}
\newcommand{\araa}{ARA\&A}
\newcommand{\atel}{ATel}
\newcommand{\cbet}{CBET}
\newcommand{\iaucirc}{IAU Circ.}
\newcommand{\mnras}{MNRAS}
\newcommand{\nat}{Nature}
\newcommand{\pasp}{PASP}

\newcommand{\pasj}{PASJ}
\newcommand{\ssr}{SSR}
\newcommand{\sovast}{Soviet Ast.}
\newcommand{\procspie}{Proc. SPIE}

\newcommand{\ion}[2]{#1 \textsc{#2}}

\title[SLSN host galaxies throughout cosmic time]{Cosmic evolution and metal aversion in super-luminous supernova host galaxies}
\author[S. Schulze et al.]{S. Schulze$^{1,2,3}$\thanks{E-mail:
steve.schulze@weizmann.ac.il},
T. Kr\"uhler$^{4}$,
G. Leloudas$^{1,5}$,
J.~Gorosabel$^{6,7,8}\dagger$,
A.~Mehner$^{9}$,
\newauthor
J.~Buchner$^{2,3}$,
S.~Kim$^{2}$,
E.~Ibar$^{10}$,
R.~Amor\'in$^{11,12}$,
R.~Herrero-Illana$^{8,9}$,
\newauthor
J.~P.~Anderson$^{9}$,
F.~E.~Bauer$^{2,3,13}$,
L.~Christensen$^{5}$,
M.~de~Pasquale$^{14}$
\newauthor
A.~de Ugarte Postigo$^{8}$,
A.~Gallazzi$^{15}$,
J.~Hjorth$^{5}$,
N.~Morrell$^{16}$,
D.~Malesani$^{5}$,
\newauthor
M.~Sparre$^{17}$,
B.~Stalder$^{18}$,
A.~A.~Stark$^{19}$,
C.~C. Th\"one$^{8}$,
J.~C.~Wheeler$^{20}$
\\
Affiliations are listed at the end of the paper
}
\begin{document}

\date{Accepted 8 September 2017. Received 18 December 2016}

\pagerange{\pageref{firstpage}--\pageref{lastpage}} \pubyear{XXX}

\maketitle

\label{firstpage}

\begin{abstract}
The SUperluminous Supernova Host galaxIES (SUSHIES) survey aims to provide strong new constraints
on the progenitors of superluminous supernovae (SLSNe) by understanding the relationship to their
host galaxies. We present the photometric properties of 53 H-poor and 16 H-rich
SLSN host galaxies out to $z\sim4$. We model their spectral energy distributions to
derive physical properties, which we compare with other galaxy populations. At low
redshift, H-poor SLSNe are preferentially found in very blue, low-mass galaxies with high average specific
star-formation rates. As redshift increases, the host population follows the general evolution
of star-forming galaxies towards more luminous galaxies. After accounting for secular evolution,
we find evidence for differential evolution in galaxy mass, but not in the $B$-band and the far
UV luminosity ($3\sigma$ confidence). Most remarkable is the scarcity of hosts with stellar masses above
$10^{10}~M_\odot$ for both classes of SLSNe. In the case of H-poor SLSNe, we attribute this to a
stifled production efficiency above $\sim0.4$ solar metallicity. However, we argue that, in addition
to low metallicity, a short-lived stellar population is also required to regulate the SLSN
production. H-rich SLSNe are found in a very diverse population of star-forming galaxies. Still, the scarcity
of massive hosts suggests a stifled production efficiency above $\sim0.8$ solar metallicity. The
large dispersion of the H-rich SLSNe
host properties is in stark contrast to those of gamma-ray burst, regular core-collapse SN, and
H-poor SLSNe host galaxies. We propose that multiple progenitor channels give rise to this sub-class.
\end{abstract}


\section{Introduction}

In the past decade, untargeted supernova (SN) surveys, e.g.,
the Texas SN Search \citep{Quimby2005a}, the ROTSE SN Verification Project \citep{Yuan2007a},
the Palomar Transient Factory \citep[PTF;][]{Law2009a}, and Pan-STARRS \citep[PS;][]{Tonry2012a}, discovered a new class of SNe with peak magnitudes exceeding $M_V=-21$~mag \citep{GalYam2012a}.
These so-called super-luminous supernovae have been a focus of SN science ever since,
because of the opportunity they provide to study new explosion channels of very massive stars
in the distant Universe \citep{Howell2013a,Cooke2012a},
the interstellar medium (ISM) in distant galaxies \citep{Berger2012a,Vreeswijk2014a} and their
potential use for cosmology \citep{Inserra2014a, Scovacricchi2016a}.
In addition, SLSNe provide a new opportunity
to pinpoint star-forming galaxies independently of galaxy properties, which can ultimately
lead to a better understanding of galaxy evolution at the faint-end of luminosity and
mass \citep{Lunnan2014a, Leloudas2015a, Angus2016a, Chen2016a, Perley2016a}. Despite these prospects, SLSNe
are very rare. At $z\sim0.2$, one H-poor SLSN is expected to be produced for every 1000--20000 core-collapse SNe
(hydrogen-rich SLSNe have a higher rate; \citealt{Quimby2013a}).

Phenomenologically, SLSNe can be classified by their hydrogen content into
H-poor and H-rich SLSNe. The light curves of H-poor SLSNe (SLSNe-I), identified as a new class of
transients by \citet{Quimby2011a}, are $\sim3.5$~mag brighter and three-times broader than
regular stripped-envelope SNe, but the shapes of their light-curves are similar
\citep[e.g.,][]{Quimby2011a,
Inserra2013a, Nicholl2015b}. Early
spectra of H-poor SLSNe show a characteristic w-shaped absorption feature at $\sim4200$~\AA\, due to oxygen in
the ejecta \citep{Quimby2011a} that is usually not seen in Type Ibc SNe \citep[e.g.,][]{Modjaz2009a}.
About a month after maximum light, the ejecta cool down to temperatures typical
of regular Type Ibc SNe at maximum light. At that point, SLSN spectra also exhibit
absorption features similar to Type Ibc SNe \citep[e.g.,][]{Pastorello2010a, Inserra2013a,
Nicholl2014a}.

A subgroup of H-poor SLSNe shows exceptionally slowly-rising and slowly-declining light curves ($\tau_{\rm rise}>25$~days,
$\tau_{\rm decay}>50$~days; \citealt{Nicholl2015b}), hereafter called slow-declining SLSN-I.
In some cases the decay slope is comparable to
that of the radioactive decay of $^{56}$Ni. \citet{GalYam2009a}
argued that in the case of SN2007bi, the supernova was powered by the radioactive decay
of several solar masses of $^{56}$Ni \citep{GalYam2012a}, which were synthesised
during a pair-instability SN (PISN) of a star with a zero-age-main-sequence (ZAMS) mass of $M_{\rm ZAMS}\sim200\,M_\odot$
\citep[e.g.,][]{Fowler1964a, Barkat1967a, Bisnovatyi1967a, Rakavy1967a, Fraley1968a, Heger2003a,
Woosley2007a}. However, the SN was discovered only shortly before it reached
maximum light. Information about the rise time was not available, which is critical to distinguish
between SN models. The well-sampled SLSNe PTF12dam and PS1-11ap, which were spectroscopically similar
to SN2007bi at late times, had rise times that were incompatible with PISN models \citep{Nicholl2013a}.
This also cast doubt on the PISN interpretation of SN2007bi. However, recent findings by
\citet{Kozyreva2016a} showed that PISN models can predict short rise times similar
to that of PTF12dam. Models of PISN spectra, on the other hand, are incompatible with the spectra of PTF12dam
and SN2007bi \citep{Dessart2013a, Chatzopoulos2015a, Jerkstrand2016a}.

The energy source powering H-poor SLSNe is highly debated. The most discussed models include
magnetars formed during the collapse of massive stars \citep[e.g.,][]{Kasen2010a,
Inserra2013a}, the interaction of the SN ejecta with dense H-deficient circumstellar
material (CSM) expelled by the progenitor prior to the
explosion \citep{Woosley2007a, Blinnikov2010a, Chevalier2011a, Chatzopoulos2012a, Quataert2012a, Sorokina2016a}, PISNe,
and pulsational PISNe \citep[e.g.,][]{Woosley2007a, Yan2015a}.

Hydrogen-rich SLSNe are  characterised by an initial blue continuum and narrow Balmer lines,
similar to classical Type IIn SNe \citep{Schlegel1990a, Filippenko1997a, Kiewe2012a} which are powered by the interaction of the
supernova with its circumstellar material \citep[e.g.,][]{Chevalier2011a}. Recent observations
suggest a richer phenomenology. Spectra of the SNe 2008es and 2013hx showed
broad H$\alpha$ emission components and their light curves showed a linear decline after maximum, similar to normal IIL
SNe \citep{Gezari2009a, Miller2009a, Inserra2016a}. Another intriguing object
is CSS121015:004244+132827 (hereafter called CSS121015). It firstly evolved as a H-poor SN but at 49 days
after the maximum, its spectrum showed broad and narrow H$\alpha$ emission lines \citep{Benetti2014a}.
These properties are different from superluminous type IIn SNe. Because
of the similarities to Type II SNe, we label this subclass SLSN-II.

The possible diversity of SLSN progenitors suggests ZAMS masses up to a few
hundred solar masses. Given the characteristic distance scale of SLSNe, a direct search for
their progenitors is unfeasible. Alternatively, host observations have the potential to
indirectly provide constraints on the progenitor population. The first systematic study of a sample of 17 H-poor and -rich SLSNe
by \citet{Neill2011a} suggested that the hosts are low-mass galaxies with high specific
star-formation rates between $10^{-8}$ and $10^{-9}\,{\rm yr}^{-1}$. However, these measurements
are very uncertain because of the limited available wavelength coverage.
This initial finding was supported by studies of the hosts of SN2010gx \citep{Chen2013a} and PS1-10bzj \citep{Lunnan2013a}.
Their spectroscopic observations also showed that both events occurred in low-metallicity galaxies with $Z<0.4\,Z_\odot$.

A survey of 31 H-poor SLSN host galaxies by \citet{Lunnan2014a}
consolidated the picture of H-poor SLSNe exploding in sub-luminous low-mass dwarf galaxies with
median specific star-formation rates of $2\times10^{-9}~{\rm yr}^{-1}$. Furthermore, the
preference for galaxies with a median metallicity of $Z\sim0.5\,Z_\odot$ hinted at a stifled production
efficiency at high metallicity \citep[see also][]{Leloudas2015a}. \citet{Perley2016a} confirmed this trend by modelling
the mass function of 18 SLSN-I hosts at $z<0.5$ from the PTF survey \citep[see also][]{Chen2016a}.
\emph{Hubble Space Telescope}
observations of 16 hosts of H-poor SLSNe by \citet{Lunnan2015a} revealed that the locations of H-poor SLSNe
are correlated with the UV light distribution  within their host galaxies. Yet, they are not as
strongly clustered on the UV-brightest regions of their hosts than long-duration gamma-ray bursts \citep[GRBs; see also][]{Angus2016a, Blanchard2016a},
which are also connected with the death of massive stars \citep[e.g.,][]{Woosley2012a}.
Furthermore, on average, the interstellar medium of SLSN-I host galaxies is characterised by
significantly weaker absorption lines than GRBs  \citep{Vreeswijk2014a}.

In 2012, we initiated the SUperluminous Supernova Host galaxIES (SUSHIES) survey \citep{Leloudas2015a} to characterise a large
set of host galaxies of H-poor and H-rich SLSNe over a large redshift range. The goals of
this survey are to study SLSN host galaxies in context of other star-forming galaxies
and to place constraints on the nature of their progenitors. To achieve this, our survey
has spectroscopic and imaging components to characterise the integrated host properties, such as mass,
metallicity, star-formation rate, age of the stellar populations and dust attenuation.

In the first SUSHIES sample paper, \citet{Leloudas2015a} discussed the spectroscopic properties of 17
H-poor and 8 H-rich SLSN host galaxies. We showed that the host galaxies of H-poor SLSNe are characterised
by hard ionisation fields, low metallicity and very high specific star-formation rates. A
high number ($\sim50\%$) of H-poor SLSNe at $z < 0.5$ occurred in extreme emission-line galaxies
\citep[e.g.,][]{Atek2011a, Amorin2014a, Amorin2015a},
which represent a short-lived phase in galaxy evolution following an intense starburst.
Moreover, in \citet{Thoene2015a} we
performed spatially resolved spectroscopy of the host of PTF12dam, the most extreme host galaxy in the sample with high signal to noise,
and found strong evidence for a very
young stellar population at the explosion site with an age of $\sim3$~Myr. These findings
let us conclude in \citet{Leloudas2015a} that the progenitors of SLSNs are possibly the very
first stars to explode in a starburst, at an earlier evolutionary stage than GRB progenitors.
Therefore, not only metallicity but also age is likely a critical condition for the production
of SLSN progenitors. \citet{Chen2016a} and \citet{Perley2016a} questioned the importance of
the age and proposed that metallicity is the primary factor for SLSN-I progenitors.

While H-poor SLSNe are preferentially found in rather extreme environments, the
findings by \citet{Leloudas2015a} and \citet{Perley2016a} point to a weaker dependence on
environment properties for H-rich SLSNe, e.g., higher average metallicities and softer ionisation
states.

In this second sample paper of the SUSHIES survey, we present photometric data of a
sample of 53 H-poor and 16 H-rich SLSN host galaxies out to $z\sim4$, including almost
every SLSN reported in the literature and detected before 2015.
The scope of this paper is to provide distribution functions of physical properties, such
as luminosities, masses of the stellar populations and star-formation rates, to investigate their
redshift evolution and to compare these results to other samples of starburst galaxies.

Throughout the paper, we adopt a flat $\Lambda$CDM cosmology with $\Omega_m = 0.315$, $\Omega_\Lambda = 0.685$,
$H_0 = 67.3~{\rm km~s}^{-1}~{\rm Mpc}^{-1}$ \citep{Planck2014a}. Uncertainties and dispersions
are quoted at $1\sigma$ confidence. We refer to the solar abundance compiled in \citet{Asplund2009a}.

\begin{table*}
\caption{Properties of the super-luminous supernovae in our sample}
\centering
\begin{tabular}{lccccccc}
\toprule
\multicolumn{1}{l}{\multirow{2}*{Object}}	& R.~A.		& Dec.		& \multicolumn{1}{c}{\multirow{2}*{Redshift}}	& \multicolumn{1}{c}{\multirow{2}*{Type}}	& $E(B-V)_{\rm MW}$	& Decline time			& \multicolumn{1}{c}{\multirow{2}*{Reference}}\\
						& (J2000)	& (J2000)	& 						&						& (mag)			& scale $\tau_{\rm dec}$~(days)	& \\
\midrule
\multicolumn{8}{c}{\textbf{Spectroscopic sample} (23)}\\
\midrule
PS1-10bzj			& 03:31:39.83	& $-$27:47:42.2	& 0.649	& SLSN-I 	& 0.01	& 37.3(fast)	& [1, 2]		\\
PS1-11ap			& 10:48:27.73	& $+$57:09:09.2	& 0.524	& SLSN-I 	& 0.01	& 87.9 (slow)	& [2, 3]		\\
PTF09cnd			& 16:12:08.94	& $+$51:29:16.1	& 0.258	& SLSN-I 	& 0.02	& 75.3 (slow)	& [2, 4]		\\
PTF10heh			& 12:48:52.04	& $+$13:26:24.5	& 0.338	& SLSN-IIn	& 0.02	& \nodata		& [5]			\\
PTF10hgi			& 16:37:47.04	& $+$06:12:32.3	& 0.099	& SLSN-I	& 0.07	& 35.6 (fast)	& [2, 6, 7]		\\
PTF10qaf			& 23:35:42.89	& $+$10:46:32.9	& 0.284	& SLSN-IIn	& 0.07	& \nodata		& [8]			\\
PTF10vqv			& 03:03:06.84	& $-$01:32:34.9	& 0.452	& SLSN-I	& 0.06	& \nodata		& [9]			\\
PTF11dsf			& 16:11:33.55	& $+$40:18:03.5	& 0.385	& SLSN-IIn	& 0.01	& \nodata		& [10]			\\
PTF12dam			& 14:24:46.20	& $+$46:13:48.3	& 0.107	& SLSN-I 	& 0.01	& 72.5 (slow)	& [2, 11]		\\
SN1999as			& 09:16:30.86	& $+$13:39:02.2	& 0.127	& SLSN-I	& 0.03	& \nodata		& [8, 12]		\\
SN1999bd			& 09:30:29.17	& $+$16:26:07.8	& 0.151	& SLSN-IIn	& 0.03	& \nodata		& [8, 13]		\\
SN2006oz			& 22:08:53.56	& $+$00:53:50.4	& 0.396	& SLSN-I	& 0.04	& \nodata		& [14]			\\
SN2006tf\tablenotemark{1}	& 12:46:15.82	& $+$11:25:56.3	& 0.074	& SLSN-IIn	& 0.02	& \nodata		& [15]			\\
SN2007bi\tablenotemark{2}	& 13:19:20.00	& $+$08:55:44.0	& 0.128	& SLSN-I	& 0.02	& 84.5 (slow)	& [2, 16, 17]		\\
SN2008am					& 12:28:36.25	& $+$15:35:49.1	& 0.233	& SLSN-IIn	& 0.02	& \nodata		& [18]			\\
SN2009jh\tablenotemark{3}	& 14:49:10.08	& $+$29:25:11.4	& 0.349	& SLSN-I	& 0.01	& 60.6  (slow)	& [2, 4]		\\
SN2010gx\tablenotemark{4}	& 11:25:46.71	& $-$08:49:41.4	& 0.230	& SLSN-I	& 0.03	& 29.1  (fast)	& [2, 4, 19]		\\
SN2010kd					& 12:08:01.11	& $+$49:13:31.1	& 0.101	& SLSN-I	& 0.03	& \nodata		& [20, 21]		\\
SN2011ke\tablenotemark{5}	& 13:50:57.77	& $+$26:16:42.8	& 0.143	& SLSN-I	& 0.01	& 25.7  (fast)	& [2, 6]		\\
SN2011kf\tablenotemark{6}	& 14:36:57.53	& $+$16:30:56.6	& 0.245	& SLSN-I	& 0.02	& 28.5  (fast)	& [2, 6]		\\
SN2012il\tablenotemark{7}	& 09:46:12.91	& $+$19:50:28.7	& 0.175	& SLSN-I	& 0.02	& 23.2  (fast)	& [2, 6]		\\
SNLS06D4eu					& 22:15:54.29	& $-$18:10:45.6	& 1.588	& SLSN-I	& 0.02	& \nodata		& [22]			\\
SSS120810\tablenotemark{8}	& 23:18:01.82	& $-$56:09:25.7	& 0.156	& SLSN-I	& 0.02	& 30.2 (fast)	& [2, 23]		\\
\midrule
\multicolumn{8}{c}{\textbf{Non-spectroscopic sample} (46)}\\
\midrule
CSS100217\tablenotemark{9}	& 10:29:12.56	& $+$40:42:20.0	& 0.147	& SLSN-IIn	& 0.01	& \nodata		& [24]			\\
CSS121015\tablenotemark{10}	& 00:42:44.34	& $+$13:28:26.5	& 0.286	& SLSN-II	& 0.07	& 37.8  (fast)	& [2, 25]		\\
CSS140925\tablenotemark{11}	& 00:58:54.11 	& $+$18:13:22.2	& 0.460	& SLSN-I	& 0.06	& \nodata		& [26]			\\
DES14S2qri			& 02:43:32.14	& $-$01:07:34.2	& 1.500	& SLSN-I	& 0.03  & \nodata		& [27]			\\
DES14X2byo			& 02:23:46.93	& $-$06:08:12.3	& 0.869	& SLSN-I	& 0.03	& \nodata		& [28]			\\
DES14X3taz			& 02:28:04.46	& $-$04:05:12.7 & 0.608	& SLSN-I	& 0.02	& \nodata		& [29]			\\
iPTF13ajg			& 16:39:03.95	& $+$37:01:38.4	& 0.740	& SLSN-I	& 0.01	& 62.0  (slow)	& [2, 30]		\\
LSQ12dlf\tablenotemark{12}	& 01:50:29.80	& $-$21:48:45.4	& 0.255	& SLSN-I	& 0.01	& 35.4  (fast)	& [2, 23]		\\
LSQ14an				& 12:53:47.83	& $-$29:31:27.2	& 0.163	& SLSN-I	& 0.07	& \nodata 		& [31]			\\
LSQ14mo				& 10:22:41.53	& $-$16:55:14.4	& 0.2561& SLSN-I	& 0.06	& 27.3  (fast)	& [2, 32]		\\
LSQ14bdq			& 10:01:41.60	& $-$12:22:13.4	& 0.345	& SLSN-I	& 0.06	& 71.2  (slow)	& [2, 33]		\\
LSQ14fxj			& 02:39:12.61	& $+$03:19:29.6	& 0.360	& SLSN-I	& 0.03	& \nodata 		& [34]			\\
MLS121104\tablenotemark{13}	& 02:16:42.51 	& $+$20:40:08.5	& 0.303	& SLSN-I	& 0.15	& \nodata 		& [35, 36]		\\
PS1-10ky			& 22:13:37.85 	& $+$01:14:23.6	& 0.956	& SLSN-I	& 0.03	& 32.5  (fast)	& [2, 37]		\\
PS1-10pm			& 12:12:42.20	& $+$46:59:29.5	& 1.206	& SLSN-I	& 0.02	& \nodata		& [38]			\\
PS1-10ahf			& 23:32:28.30	& $-$00:21:43.6	& 1.100	& SLSN-I	& 0.03	& \nodata		& [38]			\\
PS1-10awh			& 22:14:29.83 	& $-$00:04:03.6	& 0.909	& SLSN-I	& 0.07	& \nodata		& [37]			\\
PS1-11tt			& 16:12:45.78 	& $+$54:04:17.0	& 1.283	& SLSN-I	& 0.01	& \nodata		& [39]			\\
PS1-11afv			& 12:15:37.77 	& $+$48:10:48.6	& 1.407	& SLSN-I	& 0.01	& \nodata		& [39]			\\
PS1-11aib			& 22:18:12.22 	& $+$01:33:32.0	& 0.997	& SLSN-I	& 0.04	& \nodata		& [39]			\\
PS1-11bam			& 08:41:14.19 	& $+$44:01:57.0	& 1.565	& SLSN-I	& 0.02	& \nodata		& [40]			\\
PS1-11bdn			& 02:25:46.29 	& $-$05:06:56.6	& 0.738	& SLSN-I	& 0.02	& \nodata		& [39]			\\
PS1-12zn			& 09:59:49.62	& $+$02:51:31.9	& 0.674	& SLSN-I	& 0.02	& \nodata		& [39]			\\
PS1-12bmy			& 03:34:13.12 	& $-$26:31:17.2	& 1.566	& SLSN-I	& 0.01	& \nodata		& [39]			\\
PS1-12bqf			& 02:24:54.62 	& $-$04:50:22.7	& 0.522	& SLSN-I	& 0.02	& \nodata		& [39]			\\
PS1-13gt			& 12:18:02.03 	& $+$47:34:46.0	& 0.884	& SLSN-I	& 0.02	& \nodata		& [39]			\\
PTF09atu			& 16:30:24.55	& $+$23:38:25.0	& 0.501	& SLSN-I	& 0.04	& \nodata		& [4]			\\
PTF11rks			& 01:39:45.51	& $+$29:55:27.0	& 0.190	& SLSN-I	& 0.04	& 22.3  (fast)	& [2, 6, 41]		\\
SCP06F6				& 14:32:27.40	& $+$33:32:24.8	& 1.189	& SLSN-I	& 0.01	& 39.8  (fast)	& [2, 42]		\\
SN2003ma 			& 05:31:01.88	& $-$70:04:15.9	& 0.289 & SLSN-IIn	& 0.31	& \nodata		& [43]			\\
SN2005ap			& 13:01:14.83 	& $+$27:43:32.3	& 0.283	& SLSN-I	& 0.01	& 28.8  (fast)	& [2, 44]		\\
SN2006gy			& 03:17:27.06	& $+$41:24:19.5	& 0.019 & SLSN-IIn	& 0.14	& \nodata		& [45]			\\
SN2007bw\tablenotemark{14}	& 17:11:01.99	& $+$24:30:36.4	& 0.140	& SLSN-IIn	& 0.04	& \nodata		& [46]			\\
SN2008es\tablenotemark{15}	& 11:56:49.13	& $+$54:27:25.7	& 0.205	& SLSN-II	& 0.01	& 38.0 (fast)	& [2, 47, 48]		\\
SN2008fz\tablenotemark{16}	& 23:16:16.60	& $+$11:42:47.5	& 0.133	& SLSN-IIn	& 0.04	& \nodata		& [49]			\\
SN2009de\tablenotemark{17}	& 13:00:37.49	& $+$17:50:57.0	& 0.311	& SLSN-I	& 0.04	& \nodata		& [50, 51, 52]		\\
SN2009nm\tablenotemark{18}	& 10:05:24.54	& $+$51:16:38.7	& 0.210	& SLSN-IIn	& 0.01	& \nodata		& [53, 54]		\\
\bottomrule
\end{tabular}
\label{tab:prop_gen}
\end{table*}

\begin{table*}
\contcaption{Properties of the super-luminous supernovae in our sample}
\centering
\begin{tabular}{lccccccc}
\toprule
\multicolumn{1}{l}{\multirow{2}*{Object}}	& R.~A.		& Dec.		& \multicolumn{1}{c}{\multirow{2}*{Redshift}}	& \multicolumn{1}{c}{\multirow{2}*{Type}}	& $E(B-V)$	& Decline time			& \multicolumn{1}{c}{\multirow{2}*{Reference}}\\
						& (J2000)	& (J2000)	& 						&						& (mag)		& $\tau_{\rm dec}$~(days)	& \\
\midrule
SN2011cp\tablenotemark{19}	& 07:52:32.61	& $+$21:53:29.7	& 0.380	& SLSN-IIn	& 0.05 	& \nodata		& [55]			\\
SN2011ep\tablenotemark{20}	& 17:03:41.78	& $+$32:45:52.6	& 0.280	& SLSN-I	& 0.02	& \nodata		& [56]			\\
SN2013dg\tablenotemark{21}	& 13:18:41.38	& $-$07:04:43.1	& 0.265	& SLSN-I	& 0.04	& 30.7 (fast)	& [2, 23]		\\
SN2013hx\tablenotemark{22}	& 01:35:32.83	& $-$57:57:50.6	& 0.130	& SLSN-II	& 0.02	& 33.6  (fast)	& [2, 57]		\\
SN2013hy\tablenotemark{23}	& 02:42:32.82	& $-$01:21:30.1	& 0.663	& SLSN-I	& 0.03	& \nodata		& [58]			\\
SN2015bn\tablenotemark{24}	& 11:33:41.57	& $+$00:43:32.2	& 0.110	& SLSN-I	& 0.02	& \nodata		& [59]			\\
SN1000+0216$^\dagger$		& 10:00:05.87	& $+$02:16:23.6	& 3.899	& SLSN-I	& 0.02	& \nodata		& [60]			\\
SN2213-1745$^\dagger$		& 22:13.39.97	& $-$17:45:24.5	& 2.046	& SLSN-I	& 0.02	& \nodata		& [60]			\\
SNLS07D2bv			& 10:00:06.62	& $+$02:38:35.8	& $\sim1.5$& SLSN-I & 0.02	& \nodata		& [22]			\\
\bottomrule
\end{tabular}
\tablecomments{
The coordinates refer to the positions of the supernovae. The Galactic extinction measurements
are taken from \citet{Schlafly2011a}. We divide the sample
into the spectroscopic sample (23 objects) presented in \citet{Leloudas2015a} and in a non-spectroscopic sample (46 objects).
The decay-time scale $\tau_{\rm dec}$ is defined as the time when the luminosity of the
pseudo-bolometric $g'r'i'z'$ light curve dropped to $L_{\rm max}/e$. We divide the sample into
fast and slow decliners if $\tau_{\rm dec}<50$ and $>50$ days, respectively.
\newline
$^\dagger$ The classifications of SN1000+0213 and SN2213-1745 are based on photometry. The
light curve of SN1000+0213 shows a bump before the main emission similar to H-poor SLSNe SN2006oz
and LSQ14bdq \citep[for details see][] {Leloudas2012a, Nicholl2015b}.\newline
Alternative SN names:
$^1$\,CSS070320:124616+112555;
$^2$\,SNF20070406-008;
$^3$\,CSS090802:144910+292510, PTF09cwl;
$^4$\,CSS100313:112547-084941, PTF10cwr;
$^5$\,CSS110406:135058+261642, PTF11dij, PS1-11xk;
$^6$\,CSS111230:143658+163057;
$^7$\,CSS120121:094613+195028, PS1-12fo;
$^8$\,SSS120810:231802-560926;
$^9$\,CSS100217:102913+404220;
$^{10}$\,CSS121015:004244+132827;
$^{11}$\,CSS140925:005854+181322;
$^{12}$\,SSS120907:015030-214847;
$^{13}$\,MLS121104:021643+204009, LSQ12fzb;
$^{14}$\,SNF20070418-020;
$^{15}$\,ROTSE3 J115649.1+542725;
$^{16}$\,CSS080922:231617+114248;
$^{17}$\,CSS090102:130037+175057, PSN K0901-1;
$^{18}$\,CSS091120:100525+511639;
$^{19}$\,MLS110426:075233+215330, PSN J07523261+2153297;
$^{20}$\,CSS110414:170342+324553;
$^{21}$\,CSS130530:131841-070443, MLS130517:131841-070443;
$^{22}$\,SMT J013533283-5757506;
$^{23}$\,DES13S2cmm;
$^{24}$\,CSS141223:113342+004332, MLS150211:113342+004333, PS15ae
}
\tablerefs{
[1]:	\citet{Lunnan2013a};
[2]:	\citet{Nicholl2015b};
[3]:	\citet{McCrum2014a};
[4]: 	\citet{Quimby2011a};
[5]:	\citet{Quimby2010c};
[6]: 	\citet{Inserra2013a};
[7]: 	\citet{Leloudas2015a};
[8]: 	\citet{GalYam2012a};
[9]:	\citet{Quimby2010d};
[10]:	\citet{Quimby2011c};
[11]:	\citet{Nicholl2013a};
[12]:	\citet{Knop1999a};
[13]:	\citet{Nugent1999a};
[14]:	\citet{Leloudas2012a};
[15]:	\citet{Smith2008a};
[16]:	\citet{GalYam2009a};
[17]:	\citet{Young2010a};
[18]:	\citet{Chatzopoulos2011a};
[19]:	\citet{Pastorello2010a};
[20]:	\citet{Vinko2012a};
[21]:	\citet{Quimby2013b};
[22]: 	\citet{Howell2013a};
[23]:	\citet{Nicholl2014a};
[24]:	\citet{Drake2011a};
[25]: 	\citet{Benetti2014a};
[26]:	\citet{Campbell2014b};
[27]:	\citet{Castander2015a};
[28]:	\citet{Graham2014a};
[29]:	\citet{Smith2016a};
[30]:	\citet{Vreeswijk2014a};
[31]:	\citet{Leget2014a};
[32]:	\citet{Leloudas2015b};
[33]:	\citet{Nicholl2015a};
[34]:	\citet{Smith2014a};
[35]: 	\citet{Drake2012a};
[36]: 	\citet{Fatkhullin2012a};
[37]:	\citet{Chomiuk2011a};
[38]:	\citet{McCrum2015a};
[39]: 	\citet{Lunnan2014a};
[40]:	\citet{Berger2012a};
[41]: 	\citet{Quimby2011b};
[42]:	\citet{Barbary2009a};
[43]:	\citet{Rest2011a};
[44]:	\citet{Quimby2007a};
[45]:	\citet{Smith2007a};
[46]:	\citet{Agnoletto2010a};
[47]: 	\citet{Gezari2009a};
[48]: 	\citet{Miller2009a};
[49]: 	\citet{Drake2010a};
[50]:	\citet{Drake2009a};
[51]:	\citet{Drake2009d};
[52]:	\citet{Moskvitin2010a};
[53]:	\citet{Drake2009b};
[54]:	\citet{Christensen2009a};
[55]:	\citet{Drake2011d};
[56]:	\citet{Graham2011a};
[57]:	\citet{Inserra2016a};
[58]:	\citet{Papadopoulos2015a};
[59]:	\citet{Nicholl2016a}
[60]: 	\citet{Cooke2012a};
}
\end{table*}

\section{Sample definition, observations and data reduction}\label{sec:sample}

\subsection{Sample definition}

Among all SLSNe reported in the literature ($\sim120$), we selected those that were
discovered before the end of 2014 and announced before April 2015.
Therefore, many of the SLSNe published recently by \cite{Perley2016a} are not included in this paper.
In addition, we screened the Asiago Supernova catalogue \citep{Barbon2010a} for objects with
an absolute magnitude of significantly brighter than $M=-21$~mag and spectroscopic information.
This revealed two
additional H-poor SLSNe, SNe 2009de and 2011ep \citep{Drake2009a, Moskvitin2010a, Graham2011a}, and
two H-rich SLSNe, SNe 2009nm and SN2011cp \citep{Drake2009b, Christensen2009a, Drake2011b, Drake2011c, Drake2011e}.
The SN properties are summarised in Table \ref{tab:prop_gen}.

\begin{figure}
\includegraphics[width=1\columnwidth]{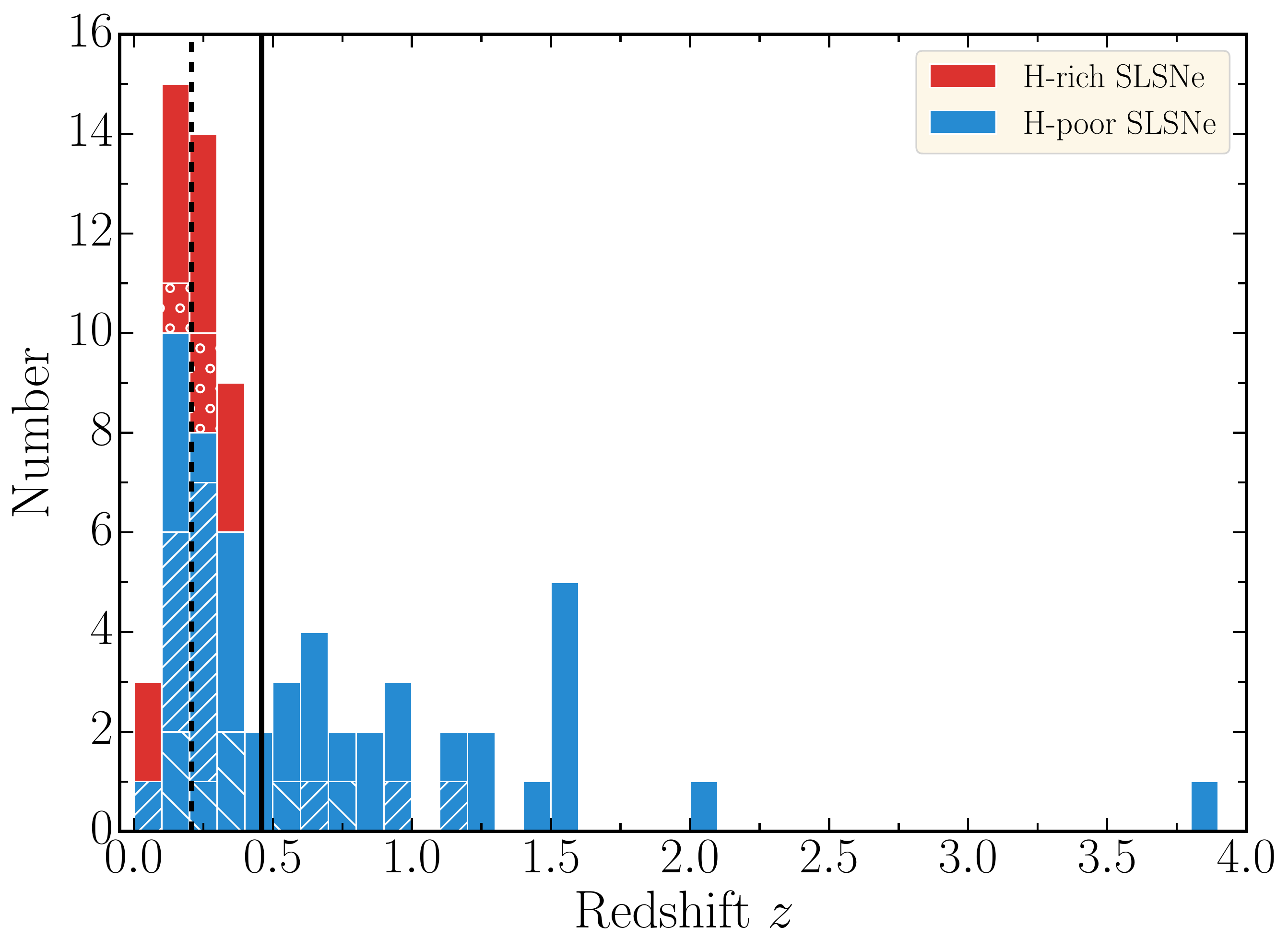}
\caption{
The redshift distribution of the SUSHIES survey.
For 21 H-poor SLSNe, information about the decline time-scale are available. The region
hatched by `//' displays the redshift distribution of the fast-decliners and the region highlighted
by `$\backslash\backslash$' signifies the redshift distribution of the slow-decliners. The redshift
distribution of the three SLSNe-II, CSS121015, SN2008es and SN2013hx, are highlighted by `o'.
The median redshifts of the H-poor and H-rich sample are $\tilde{z}=0.46$ (solid vertical line)
and $\tilde{z}=0.21$ (dashed vertical line), respectively.
}
\label{fig:z_distrib}
\end{figure}

Our final sample comprises of 53 H-poor and 16 H-rich SLSNe. The H-poor sample includes 7 slow-declining
H-poor SLSNe, while the H-rich sample includes the SLSNe-II CSS121015, SN2008es and SN2013hx.
The size of the final sample is not only a factor of $>2$ larger than the SLSN host sample presented
in \citet{Perley2016a} but includes a large population of hosts at $z>0.5$ (which is the highest redshift in
\citealt{Perley2016a}).
Figure \ref{fig:z_distrib} displays the redshift distribution of our sample. It covers a
redshift interval from $z\sim0.1$ to $z\sim2$ with a singular object at $z\sim4$ (SN1000+0216; \citealt{Cooke2012a}).
The redshift distribution of the
H-poor sample covers the full range and has a median of $\tilde{z}=0.46$.
The H-rich sample only extends to $z\sim0.4$ and has a median of $\tilde{z}=0.21$.

\subsection{Observations}

A fundamental goal of our survey is to secure multi-band data from the rest-frame UV to
NIR, to model the spectral energy distributions of the host galaxies. To ensure a sufficient wavelength
coverage and data quality, we aimed to have at least one observation
of the rest-frame UV and of the NIR and two observations of the rest-frame optical, if a
galaxy was brighter than $r'=24$~mag.

To optimise the observing campaign, we queried the \texttt{VizieR} database \citep{Ochsenbein2000a} and public archives for
available catalogues and data, such as the ESO, Gemini and Subaru archives. Our primary source catalogues are from
the Canada-France-Hawaii Telescope Legacy Survey (CFHTLS; \citealt{Hudelot2012a}),
the Cosmological Evolution Survey (COSMOS; \citealt{Scoville2007a}),
the \textit{Galaxy Evolution Explorer} \citep[{\galex};][]{Martin2005a},
the Sloan Digital sky survey (SDSS; \citealt{York2000a}),
the UKIRT Infrared Deep Sky Survey (UKIDSS; \citealt{Lawrence2007a}) and
the \textit{Wide-field Infrared Survey Explorer} \citep[{\wise};][]{Wright2010a}.\footnote{We included \wise\ data of only a few hosts.} These
catalogues were complemented by
the Coma Cluster catalogue \citep{Adami2006a},
the UltraVISTA catalogue \citep{McCracken2012a},
the VISTA Deep Extragalactic Observations survey (VIDEO; \citealt{Jarvis2013a}) and
the VIRMOS deep imaging survey \citep[VIRMOS;][]{LeFevre2004a}.
Furthermore, we
incorporated measurements previously reported in \citet{Inserra2013a}, \citet{Lunnan2014a},
\citet{Nicholl2014a}, \citet{Vreeswijk2014a} and \citet{Angus2016a}.

Between 2012 and 2016, we used observing proposals at the 6.5-m Magellan/Baade Telescope (PI: Schulze, Kim),\footnote{Programme IDs:
CN2013A-195, CN2013B-70, CN2014A-114, CN2014B-127, CN2014B-102, CN-2015A-129, CN2015A-143,
CN-2015B-87, CN2015B-99, CN2016A-108, and CN2016B-98}
ESO's 8.2-m Very Large Telescope (VLT; PI: Leloudas, Kr\"uhler),\footnote{Programme IDs: 089.D-0902, 091.A-0703, 091.D-0734,
and 290.D-5139} the 10.4-m GTC and 3.5-m CAHA telescope (PI: Gorosabel) and
the 0.3-m UV/Optical Telescope \citep[UVOT; ][]{Roming2005a} onboard the \swift\ satellite
\citep[][PI: Leloudas]{Gehrels2004a} to obtain rest-frame UV, optical and NIR
data. In the subsequent sections, we briefly summarise each campaign.

Our Magellan campaign was performed between 2012 and 2016 with the 6.5-m Baade telescope equipped with
the optical wide-field Inamori-Magellan Areal Camera and Spectrograph (IMACS;
\citealt{Dressler2011a}), the Parallel Imager for Southern Cosmological Observations \citep[PISCO;][]{Stalder2014a},
and the near-infrared (NIR) camera FourStar \citep{Persson2013a}. The optical data
were secured in $g'r'i'z'$, primarily with the IMACS f/2 camera, but also with the
IMACS f/4 camera and PISCO. The near infrared observations were performed in $J$ and
$K_{\rm s}$.

The ESO VLT observations were taken in visitor and service mode. The visitor run took place
between 29 May and 2 June 2013. We used the FOcal Reducer and Spectrograph
2 instrument (FORS2; \citealt{Appenzeller1998a}), equipped with the red-sensitive CCD
to secure data in $uBgVRIz$. In addition, we obtained $J$ and $K$ band imaging
with the High Acuity Wide field K-band Imager (HAWK-I; \citealt{Pirard2004a,Casali2006a,Kissler2008a}).
Additional optical and NIR data were obtained with FORS2, the Infrared Spectrometer
And Array Camera (ISAAC; \citealt{Moorwood1998a}) and HAWK-I in queue mode.

The CAHA and GTC campaigns primarily focused on targets on the northern hemisphere. The CAHA
observing programme was carried out with the 4-channel Bonn University Simultaneous CAmera
\citep[BUSCA;][]{Reif1999a} in $u'g'r'i'$ at the 3.5-m CAHA telescope in 2012.
We also used the infrared wide-field camera Omega2000 \citep{Kovacs2004a} to secure $J$ and $K$ band observations
between 2013 and 2015 and also in $Y$ and $H$ band for a few targets. The objective of the campaign at the
10.4-m GTC telescope was to secure deep imaging of SNe 2008es and 2009jh with the Optical System
for Imaging and low-Intermediate-Resolution Integrated Spectroscopy \citep[OSIRIS;][]{Cepa2000a}
camera.

Rest-frame UV data are critical to break degeneracies in the SED modelling. For objects at
$z<0.4$, observations in $U$ or bluer filters are needed to probe the UV.
\galex\ provided critical rest-frame UV data for most objects. In addition, we secured UV photometry
of five fields with the UV/optical telescope UVOT on board the \swift\ satellite in
2014 and incorporated archival UVOT data of a further SLSN.

These core observing campaigns were complemented by smaller observing programmes that targeted
selected host galaxies. We observed the field of SN2005ap with the Andalucia Faint Object
Spectrograph and Camera (ALFOSC) at the 2.54-m Nordic Optical Telescope and the field of
SN2007bi with ALFOSC and the 7-channel imager
Gamma-Ray Burst Optical/Near-Infrared Detector \citep[GROND; ][]{Greiner2008a} at
the 2.2-m Max-Planck-Gesellschaft telescope.

To place limits on the total star-formation rate, we used 1.4 GHz data from the VLA Faint
Images of the Radio Sky at Twenty-Centimeters survey (FIRST; \citealt{Becker1995a}), the NRAO VLA
Sky Survey (NVSS, $\nu=1.4$~GHz; \citealt{Condon1998a}), and 843 MHz data from the
Sydney University Molonglo Sky Survey (SUMSS; \citealt{Bock1999a}). In addition, we secured
continuum observations of MLS121104, SN2005ap and SN2008fz with the Karl
Jansky Very Large Array (JVLA; PI: Ibar).\footnote{Programme ID: 15A-224}
The continuum observations were performed in L
band in the most extended A-configuration in July and September 2015. The frequency was
centred at 1.5 GHz with a
total synthesised bandwidth of 1 GHz. We used the standard flux and bandwidth calibrator
3C48 for all the sources except SN2005ap, for which we used 3C286 instead. For phase
calibration purposes we used bright nearby point-like sources from the VLA calibrator list
(MLS121104: J0238+1636; SN2005ap: J1310+3220; and SN2008fz: J2330+1100).
The key properties of each observation is reported in Tables \ref{tab:data}.

\subsection{Data reduction}

We reduced all data in a consistent
way with standard routines in IRAF \citep{Tody1986a}. The typical steps are \textit{i}) bias/overscan
subtraction, \textit{ii}) flat-fielding, \textit{iii}) fringe correction, \textit{iv}) stacking of individual images
and \textit{v}) astrometric calibration.
For a few instruments we used instrument
specific software packages:
the {\tt GEMINI IRAF} package, the GROND pipeline \citep{Yoldas2008a, Kruehler2008a},
{\tt PHOTPIPE} for PISCO data \citep{Bleem2015a},
{\tt SDFRED1} and {\tt SDFRED2} for Subaru Suprime-Cam data \citep{Yagi2002a, Ouchi2004a},
\texttt{THELI} version 2.10.0 \citep{Erben2005a, Schirmer2013a} for the FourStar data,
VLT instrument pipelines for HAWK-I (version 1.8.18) and ISAAC
(version 6.1.3) data,\footnote{\href{http://www.eso.org/sci/software/cpl/esorex.html}{http://www.eso.org/sci/software/cpl/esorex.html}}
and a customised pipeline for the Magellan/IMACS data.
The world-coordinate systems were calibrated with \texttt{astrometry.net} version 0.5 \citep{Lang2010a}.

UVOT data were retrieved from the \swift\ Data Archive.\footnote{\href{http://www.swift.ac.uk/swift_portal/}{http://www.swift.ac.uk/swift\_portal/}}
We used the standard UVOT data analysis software distributed with \texttt{HEAsoft}
version 6.12, along with the standard calibration data.\footnote{\href{http://heasarc.nasa.gov/lheasoft/}{http://heasarc.nasa.gov/lheasoft/}}

The JVLA data were reduced using the Common Astronomy Software Applications package
\citep[CASA;][]{McMullin2007a} and consisted of careful data flagging and standard flux,
bandwidth and phase calibration. No self-calibration was performed to the data. The obtained
flux density root mean squares (r.m.s.) of the images are summarised in Table \ref{tab:data_radio}.

\section{Methods}\label{sec:methods}

\subsection{Host identification}

We aligned our host-only images with the original SN images that we retrieved from archives with
\texttt{Gaia} version 4.4.6.\footnote{\href{http://starlink.eao.hawaii.edu/starlink/2015ADownload}{http://starlink.eao.hawaii.edu/starlink/2015ADownload}}
The average alignment accuracy was $\sim0\farcs17$.
We neither found (suitable) public data for
13 SNe from PanSTARSS, nor for SNe 2006tf, 2009de, 2009nm and 2011cp
(in total 17/69 objects). For those objects we relied on the reported SN positions. Although
this added an uncertainty to the host identification, the SN positions always coincided with
a galaxy, which we assume is the host galaxy.

\subsection{Photometry}

We developed a Python programme that is based on \texttt{Source Extractor} version 2.19.5 \citep{Bertin1996a}
to perform seeing matched aperture photometry. To measure the total
flux of the given object, the source radius was typically 2--4 times
the full-width at half maximum (FWHM) of the stellar PSF. In case another object was close to the SN position or if the host had
a large angular diameter, we adjusted the extraction radius accordingly. If a host evaded
detection in all bands, we measured the flux and its uncertainty at the SN position
using an aperture with a radius of $4\times{\rm FWHM}$. Those measurements have very large
uncertainties but they can be easily included in the SED modelling in contrast to upper limits.

Once an instrumental magnitude was established, it was photometrically calibrated
against the brightness of several standard stars measured in a similar manner
or tied to the SDSS DR8 \citep{Aihara2011a} and the AAVSO (American Association of
Variable Star Observers) Photometric All-Sky Survey (APASS) DR9
\citep{Henden2016a} catalogues. For Bessell/Johnson/Cousins filters, we converted
the photometry of stars in the SDSS catalogue from SDSS using the Lupton colour equations.\footnote{\href{http://www.sdss.org/dr5/algorithms/sdssUBVRITransform.html}{http://www.sdss.org/dr5/algorithms/sdssUBVRITransform.html}}
In the NIR ($JHK_{\rm s}$), the photometry was tied to 2MASS. The UVOT photometry was performed with the
programme \texttt{uvotsource}. UVOT zeropoints are defined for an aperture with
a diameter of $5\arcsec$. We translated these zeropoints into those of our requested apertures
by applying simple aperture correction methods for stars.

Finally, the measurements were corrected for Galactic extinction using the extinction
maps by \citet{Schlafly2011a} and transformed into the AB system using \citet{Blanton2007a}
and \citet{Breeveld2011a}.

In total, we measured the brightness (and limits for the non-detections) of 53 of the 69 objects, which also includes the
re-evaluation of 27 individual data sets from the Two Micron All Sky Survey (2MASS), CFHTLS and SDSS, as well as several
archival data sets. In addition, we augmented the photometry of 31 objects by literature values,
such as {\galex}, Pan-STARRS and \wise\ data. Owing to \galex's and \wise's large
point-spread functions, we only included their photometry if a contamination by neighbouring
objects could be excluded. Among the 16 objects whose photometry is entirely based on literature
results, four galaxies are in the footprint of the COSMOS survey: PS1-12zn, PS1-12bqf,
SN1000+0213 and SNLS07D2bv. Their photometry is discussed here for the first time.
Table \ref{tab:data} summarises the photometry of each object.

\subsection{Spectral-energy distribution fitting}\label{sec:sed_fitting}

We modelled the SEDs with \texttt{Le Phare}
\citep{Arnouts1999a, Ilbert2006a},\footnote{\href{http://www.cfht.hawaii.edu/~arnouts/LEPHARE}{http://www.cfht.hawaii.edu/\~{}arnouts/LEPHARE}}
using a grid of galaxy templates based on \citet{Bruzual2003a} stellar population-synthesis
models with a Chabrier IMF \citep{Chabrier2003a}. The star-formation history was approximated by
a declining exponential function of the form $\exp\left(-t/\tau\right)$, where $t$
is the age of the stellar population and $\tau$ the e-folding time-scale of the star-formation
history (varied in eight steps between 0.1 and 15 Gyr). Furthermore, we assumed the Calzetti dust attenuation
curve \citep{Calzetti2000a}. For a description of the galaxy
templates, physical parameters of the galaxy fitting, and their error estimation, we refer
to \citet{Kruehler2011a}.\footnote{The templates used in this paper do not account for
possible binary star evolution, which could substantially alter SEDs (more hard UV
photons; e.g., \citealt{Stanway2016a})}.

As an extension to \citet{Kruehler2011a}, we relaxed the analysis threshold of the galaxy mass
to $10^4~M_\odot$ (which is pushing the definition of a galaxy), because previous studies
showed that SLSNe can occur in very low-mass galaxies \citep{Lunnan2014a, Leloudas2015a,
Angus2016a}. We modified the gas component in \texttt{Le Phare} by
incorporating the observed relationship between line flux and SFR for [\ion{O}{ii}] and
[\ion{O}{iii}] by \citet{Kruehler2015a}. The attenuation of the ionised gas component
was linked to the stellar attenuation via $E(B-V)_{\rm star}=0.44\times E(B-V)_{\rm gas}$ by \citet{Calzetti2000a}.
All attenuation measurements are reported for $E(B-V)_{\rm gas}$.
Finally, we used the high-resolution BC03 templates, which are defined over
6900 wavelength points instead of 1221 wavelength points from $9.1\times10^{-3}$ to 160~$\mu$m.
To account for zeropoint offsets in
the cross-calibration and absolute flux scale, we added a systematic error of 0.05 mag in
quadrature to the uncertainty introduced by photon noise. For {\galex}, UVOT and $K$-band data this
systematic error was increased to 0.1~mag.

The absolute magnitudes were computed directly by convolving the filter response functions with
the best-fit template. To compute the corresponding error $\sigma(M_Q)$ in the rest-frame bandpass
$Q$, we interpolated between the errors of the apparent magnitudes $\sigma\left(m_k\right)$
and $\sigma\left(m_l\right)$ of the observed band-pass $k$ and $l$, respectively, via:
\begin{eqnarray}
\sigma\left(M_Q\right)=\frac{\sigma\left(m_k\right)-\sigma\left(m_l\right)}{\lambda_{{\rm rest}, k}-\lambda_{{\rm rest}, l}}\left(\lambda_{{\rm rest}, Q} - \lambda_{{\rm rest}, l}\right)+\sigma\left(m_l\right)\nonumber
\end{eqnarray}
where $\lambda_{{\rm rest}, k/l} = \lambda_{{\rm obs}, k/l} / (1+z)$ is the central wavelength of
the observer-frame bandpass $k$ and $l$ in the rest-frame of the SLSNe. In case a rest-frame bandpass
lies blueward/redward of the observation in the bluest/reddest filter, we set the
error $\sigma(M_Q)$ to the error of the observation in the bluest/reddest filter.

Our observations were characterised by a large set of different filters, of which
several have similar bandpasses. To simplify the fitting, we homogenised the filter
set. Specifically, we set the filter response function of $F336W$, $u_{\rm PS1}$, $u^*$, $uvu$ to $u'$,
$F475$,  $g_{\rm DES}$, $g_{\rm High}$, $g_{\rm PS1}$, $g+$ to $g'$,
$r_{\rm DES}$, $r_{\rm PS1}$, $r+$ to $r'$,
$F775W$, $i_{\rm DES}$, $i_{\rm PS1}$, $i+$ to $i'$',
$F850LP$, $z_{\rm DES}$, $z_{\rm Gunn}$, $z_{\rm PS1}$, $z^+$ to $z'$,
$F390W$ $U38$ to $U$,
$Bj$ to $B$,
$Vj$ to $V$,
$I_c$, $F814W$ to $I$,
$y_{\rm PS1}$ to $Y$,
$F160W$ to $H$,
$W1$ to \spitzer/$3.6~\mu\rm m$, and
$W2$ to \spitzer/$4.5~\mu\rm m$.
It can be seen from our fits (Figs. \ref{fig:SED}, \ref{fig:SED_app_1} and \ref{fig:SED_app_2}),
and quality of the derived host properties (Table \ref{tab:sed_results}), that the impact of these assumptions is negligible.

Studies of SLSN host galaxies and extreme emission-line
galaxies \citep[e.g.,][]{Amorin2015a} showed that emission lines can significantly affect
the SED fitting. To quantify this effect, we repeated the SED fitting for our spectroscopic
sample (\citealt{Leloudas2015a}; Table \ref{tab:prop_gen}). The contribution of the emission line $i$ on the
photometry in filter $j$ is given by
\begin{eqnarray}
\Delta m_{i,j}&=& -2.5\,\log\left(\frac{f_{\lambda,c}\left(\lambda\right) + f^{i}_{\lambda,l}\left(\lambda\right)}{f_{\lambda,c}\left(\lambda\right)}\right)\nonumber\\
& = & -2.5\,\log\left(1+\frac{\int d \lambda\,f^{i}_{\lambda,l}\left(\lambda\right)\,T_j\left(\lambda\right)}{\int d \lambda\,f_{\lambda,c}\left(\lambda\right)\,T_j\left(\lambda\right)}\right)\nonumber
\end{eqnarray}
where $f^{i}_{\lambda,l}$ is the flux density of the emission line $i$,
$f_{\lambda,c}$ is the flux density of the stellar continuum and $T_j\left(\lambda\right)$ is
the transmission function of the filter $j$. The strength of an emission line can be characterised
by its equivalent width, EW, hence $f^{i}_{\lambda,l} = f_{\lambda,c}\times {\rm EW}_i$. Assuming that all
emission lines are narrow compared to the width of the broad-band filter, the above expression
simplifies to
\begin{eqnarray}
\Delta m_{i,j} &=& -2.5\,\times\,\log\,\left(1+\frac{{\rm EW}_i\,T_j\left(\lambda_i\right)}{\Delta \lambda_{j, \rm eff}}\right)\nonumber
\end{eqnarray}
where $T_j\left(\lambda_i\right)$ is the filter response function of filter $j$ at the wavelength
of the emission line $i$ (in the air reference frame) and $\Delta \lambda_{j, \rm eff}$ is the effective width of the filter.
In contrast to the SED fitting, it was necessary to use the exact filter transmission function of each
instrument.

We subtracted the contribution of H$\alpha$--H$\delta$,
[\ion{O}{ii}], [\ion{O}{iii}], [\ion{N}{ii}], [\ion{Ne}{ii}] and [\ion{S}{ii}]
from the measured brightness in the broadband filter. Afterwards, we explicitly switched off
the contribution from the ionised gas of \ion{H}{ii} regions in \texttt{Le Phare} and repeated the fits with
the emission-line-subtracted SEDs. The result of this experiment is discussed in Sect. \ref{sec:em_lines}.

\subsection{Ensemble statistics}\label{sec:statistics}

To compare observed distributions with distributions of other galaxy samples (parent distributions),
such as extreme emission-line galaxies (hereafter EELGs), GRBs and SNe, we performed an Monte-Carlo (MC) simulation
as follows. Each SLSN host measurement was represented by a normal distribution centred at the observed
value and with a width ($1\sigma$) determined from the asymmetric error or a uniform distribution between the
upper limit and the smallest/faintest value in the sample for those objects with upper limits only. A
two-sided Anderson-Darling (AD) test was performed between the resampled distributions and the parent distributions,
using the \texttt{R} package \texttt{kSamples}. This process was repeated 10\,000 times and a mean AD value
obtained. We rejected the null hypothesis of two distributions being drawn from the same parent
distribution if the corresponding chance probability $p_{\rm ch}$ was smaller than 0.01.

To complement the one-dimensional Anderson-Darling tests, we also performed two-dimensional tests in the
mass-SFR plane. We first computed the mean mass and SFR of the SLSN-host sample. After that, we bootstrapped
10\,000 samples of size $N$ from the other galaxy samples, where $N$ is the number of SLSNe in the given
redshift interval, and computed the mean mass and SFR of each bootstrapped sample. Measurement errors were
propagated through a MC simulation as described above. Finally, we computed the region
that contained 99\% of all realisations using the python package \texttt{corner.py} \citep{ForemanMackey2016a}. If the estimator of the SLSN sample did not fall
in that region, the chance probability $p_{\rm ch}$ is less than 0.01 and we rejected the null
hypothesis of both distributions being statistically similar.

For each statistical test, we also performed a two-sided AD test on the redshift distributions to
minimise systematic errors introduced by cosmic evolution, similar to \citet{Japelj2016a}.

We extract robust estimates of the ensemble distribution functions with a Bayesian approach, which
incorporates the varying and asymmetric measurement uncertainties of individual sources and the
limited sample size. For this we fit to the sample measurements in a quantity (e.g., in $M_\star$ or
SFR) a normal distribution. We constrain its parameters, the mean $\mu$ and standard deviation $\sigma$,
with a likelihood defined as the product of convolutions of that distribution and the measurement
probability distributions. The fit uncertainties were obtained with the
\texttt{MultiNest} package \citep{Feroz2013a} through the python package \texttt{PyMultiNest}
(\citealt{Buchner2014a}). Flat priors were assumed on $\mu$ and $\log \sigma$.

\subsection{Comparison samples}

We built several comparison samples to put SLSN host galaxies in context with the cosmic
star-formation history and to better understand the peculiar conditions that gave rise to
this class of stellar explosion.

\paragraph*{Core-collapse supernova host galaxies:}
Because of the connection between SLSNe and massive stars, we compiled core-collapse supernova
(CCSN) host galaxy samples. As in \citet{Leloudas2015a}, we used SNe from untargeted (with
respect to galaxies) surveys.  At $z<0.3$, we use
objects studied in \citet{Leloudas2011a}, \citet{Sanders2012a} and \citet{Stoll2013a}.
All SNe in these samples have robust spectroscopic classifications. The combined sample
consists of 44 type Ib/c SNe and 46 type II SNe.
These studies provide multi-band data, which are primarily based on SDSS photometry
and also spectroscopy for a number of hosts.
We adopt the SED modelling by \citet{Leloudas2015a} for the \citet{Leloudas2011a} and \citet{Sanders2012a} samples. Note,
the spectral energy distributions in \citet{Stoll2013a} were modelled with the FAST stellar population synthesis code
\citep{Kriek2009a} with the \citet{Bruzual2003a} templates and a Salpeter IMF. We reduced their
SFRs and galaxy masses by a factor of 1.8, to convert from a Salpeter to a Chabrier IMF,
used in this paper \citep{Kennicutt1998a}.

To expand the SN sample to redshifts larger than $z>0.3$, where most of our SLSNe are found, we added the
SN sample from the Great Observatories Origins Deep Survey (GOODS) and Probing Acceleration
Now with Supernovae (PANS) surveys \citep{Riess2004a}. GOODS/PANS were \hst\ surveys to
detect Type Ia SNe at high redshift. This survey also located 58 distant CCSNe between
$z=0.28$ and  $z=1.3$ (the median being $\tilde{z}=0.47$). In contrast to the low-$z$ samples,
their classification relied on photometric data.
The method allowed a distinction between Type Ia and CCSNe, but not a categorisation into
sub-types. Thanks to the overlap with the GOODS field, each SN host has deep
rest-frame UV to NIR data. We adopt the results of the SED modelling by \citet{Svensson2010a}.
Note, these authors modelled the SEDs with their own software that uses observed SEDs of
local galaxies and SEDs produced with various spectral synthesis codes as templates.
Furthermore, they assumed a Salpeter IMF. Similar to \citet{Stoll2013a}, the SFRs and the masses
were reduced by a factor of 1.8 to convert from a Salpeter to a Chabrier IMF.

\paragraph*{GRB host galaxies:}
A member of our team (T. Kr\"uhler) collected multi-band data of long GRBs. These GRBs
are selected to be part of one of the following complete GRB samples: GROND 4-hour sample
\citep{Greiner2011a}, TOUGH survey (The Optically Unbiased GRB Host Galaxy survey; \citealt{Hjorth2012a}),
BAT-6 \citep{Salvaterra2012a} or  SHOALS (\swift\ Gamma-Ray Burst Host Galaxy Legacy Survey;
\citealt{Perley2016a}). The individual measurements are reported in \citet{Kruehler2017a}.
Among all hosts, we selected those at
$z<1$ (52 in total). At these redshifts, it is relatively easy to secure the GRB redshift,
because of the sparsity of dust-obscured bursts at $z<1$, and to build host samples with a high
detection completeness. The SEDs of this sample were analysed in a similar way as our
SLSN host galaxy sample.

\begin{table*}
\caption{Properties of the comparison samples and their selection criteria}
\centering
\begin{tabular}{lcccc}
\toprule
Sample			& Selection criteria					& Number of	& Redshift	& Which properties\\
			& 							& objects	& interval	& used?		\\
\midrule
\multicolumn{5}{c}{\textbf{Core-collapse supernova host galaxies (total number 265)}}\\
\midrule
\citet{Leloudas2011a}	& Ib/c SNe, detected by untargeted surveys		& 12	& $0.02\leq z\leq0.18$	& $M_B$, mass, SFR\tablenotemark{1}	\\
(L11)			& spectroscopic classification				&	& $\tilde{z}=0.04$	& 					\\
\citet{Sanders2012a}	& Ib/c SNe, detected by untargeted surveys		& 31	& $0.01\leq z\leq0.26$	& $M_B$, mass, SFR\tablenotemark{1}	\\
(S12)			& spectroscopic classification				&	& $\tilde{z}=0.03$	& 					\\
\citet{Svensson2010a}	& GOODS SN sample					& 165	& $0.28\leq z\leq1.30$	& $M_B$, mass, SFR			\\
			& photometric SN classification				&	& $\tilde{z}=0.47$	& 					\\
\citet{Stoll2013a}	& first-year PTF CCSN sample				& 58	& $0.01\leq z\leq0.18$	& $M_B$, mass, SFR			\\
(S13)			& primarily Type II SNe					&	& $\tilde{z}=0.04$	&					\\
\midrule
\multicolumn{5}{c}{\textbf{Extreme emission-line galaxies (total number 227)}}\\
\midrule
\citet{Amorin2014a}	& VUDS survey \citep{LeFevre2015a}, 			& 31	& $0.21\leq z\leq 0.86$	& colour, $m_R$, $M_B$, 		\\
			& $23~{\rm mag}<I({\rm AB})<25$~mag			&	& $\tilde{z}=0.57$	& mass, SFR				\\
\citet{Amorin2015a}	& zCOSMOS survey, $I({\rm AB})\leq22.5$	~mag		& 165	& $0.11<z<0.92$		& colour, $m_R$, $M_B$,			\\
			& ${\rm EW}_{\rm rest}\left({\left[{\rm \ion{O}{iii}}\right]\lambda5007}\right)>100~\rm \AA$	& 	& $\tilde{z}=0.48$	&  mass, SFR \\
\citet{Atek2011a}	& WISPS survey \citep{Atek2010a}, $0.5<z<2.3$		& 9	& $0.9\leq z\leq2.04$	& mass, SFR				\\
			& ${\rm EW}_{\rm rest}\left({\left[{\rm \ion{O}{iii}}\right]\lambda5007}\right)>200~\rm \AA$	& 	& $\tilde{z}=1.36$	&\\
\citet{Maseda2014a}	& 3D-HST survey \citep{Brammer2012a}, colour selection	& 22	& $1.3\leq z\leq2.3$	& mass, SFR				\\
			& emission-lines do not fall in the NIR band-gaps	& 	& $\tilde{z}=1.65$	& 					\\
\midrule
\multicolumn{5}{c}{\textbf{Field galaxies (total number 150\,900)}}\\
\midrule
\cite{Muzzin2013a}	& $K$-band selected COSMOS/UltraVISTA survey				& 150\,900& $0.01 \leq z \leq 3.96$& colour, $m_R$, mass, \\
			& ${\rm SFR}>10^{-3}~M_\odot\,{\rm yr}^{-1}$, ${\rm USE} = 1$, $z<4$	& 	& $\tilde{z}=0.97$	& SFR			\\
			& $10^{-13}~{\rm yr}^{-1}<{\rm sSFR}<10^{-7.5}~{\rm yr}^{-1}$		&    	& 			& 			\\
\midrule
\multicolumn{5}{c}{\textbf{Long GRB host galaxies (total number 52)}}\\
\midrule
\citet{Kruehler2017a}	& $z<1$, long-duration \swift~GRBs detected before	& 52	& $0.06\leq z\leq 0.98$	& colour, $m_R$, $M_B$,	\\
			& May 2014, part of the GROND 4-hour, TOUGH, SHOALS	&	& $\tilde{z}=0.67$	& mass, SFR 	\\
			& BAT-6 samples						&	&			& 		\\
\bottomrule
\end{tabular}
\tablecomments{The selection criteria consist of the criteria from each individual survey and those we imposed to build the final samples.
All samples were cleaned from duplicates.\\$^1$ We used the re-computed values in \citet{Leloudas2015a}.}
\label{tab:comp_samples}
\end{table*}

\paragraph*{COSMOS/UltraVISTA survey:}
To compare SLSN host galaxies to field galaxies, we used the ultra-deep NIR survey UltraVISTA
that observed an area of 1.8~deg$^2$ down to $K({\rm AB})_{\rm s}=23.9$~mag ($5\sigma$ confidence).
We chose the $K$-band, i.e., mass, selected catalogue by \citet{Muzzin2013a} that
overlaps with the COSMOS field. This catalogue provides observations in 30 bands from
rest-frame UV to NIR. Among all galaxies, we selected
those at $z<4$ with SFRs of at least $10^{-3}~M_\odot\,{\rm yr}^{-1}$,
specific SFRs between $10^{-13}~{\rm yr}^{-1}$ and $10^{-7.5}~{\rm yr}^{-1}$, and
``USE" flags equal to one. This sample
comprises $\sim151\,000$ galaxies with a median redshift of $\tilde{z}=0.97$. Because of the small
survey area, the number of hosts at $z<0.1$ is small. This does not affect our analysis
because only two SLSNe in our sample are at lower redshifts.

\paragraph*{EELGs:}
\citet{Leloudas2015a} showed that H-poor SLSNe are preferentially found in EELGs. We built
a master sample including results from \citet{Atek2011a}, \citet{Amorin2014a, Amorin2015a}
and \citet{Maseda2014a}. Those samples selected EELGs by applying different brightness cuts,
colour selection criteria, spectroscopy and redshift constraints. The total sample consists
of 227 galaxies with rest-frame [\ion{O}{iii}]$\lambda$5007 equivalent widths of $>100$~\AA\
between $z=0.11$ and $z=2.3$. All surveys reported stellar mass and SFR for each galaxy, but other
properties, such as brightness, colour or $M_B$, were only reported for certain subsamples.

A summary of the individual surveys and which properties are used in this study is presented
in Table \ref{tab:comp_samples}.

\section{Results}\label{sec:results}
\subsection{Spectral-energy distribution modelling}\label{res:sed_fitting}
\subsubsection{Quality of the SED modelling}\label{res:sed_fitting_quality}

\begin{figure*}
\begin{center}
\includegraphics[width=0.24\textwidth]{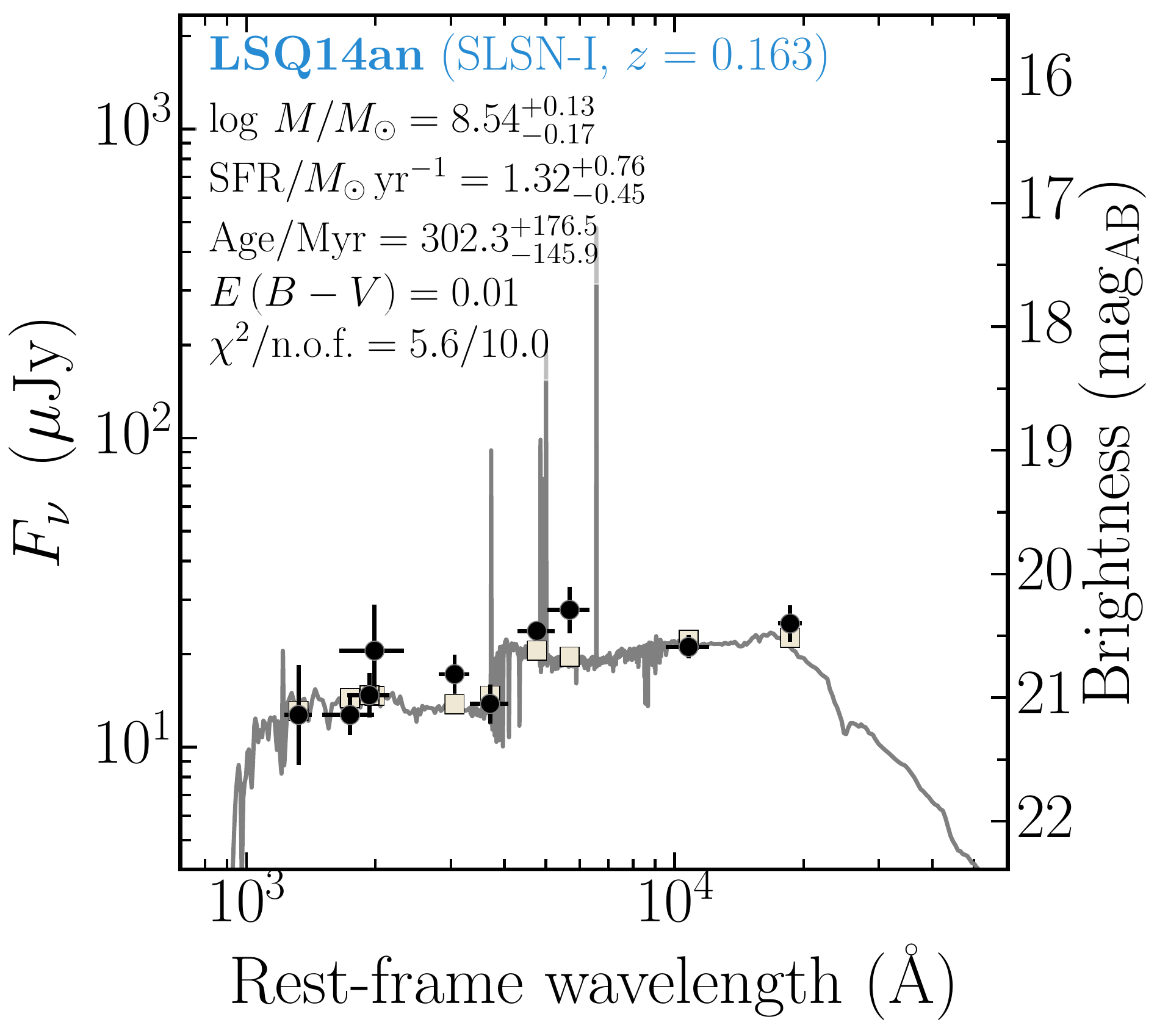}
\includegraphics[width=0.24\textwidth]{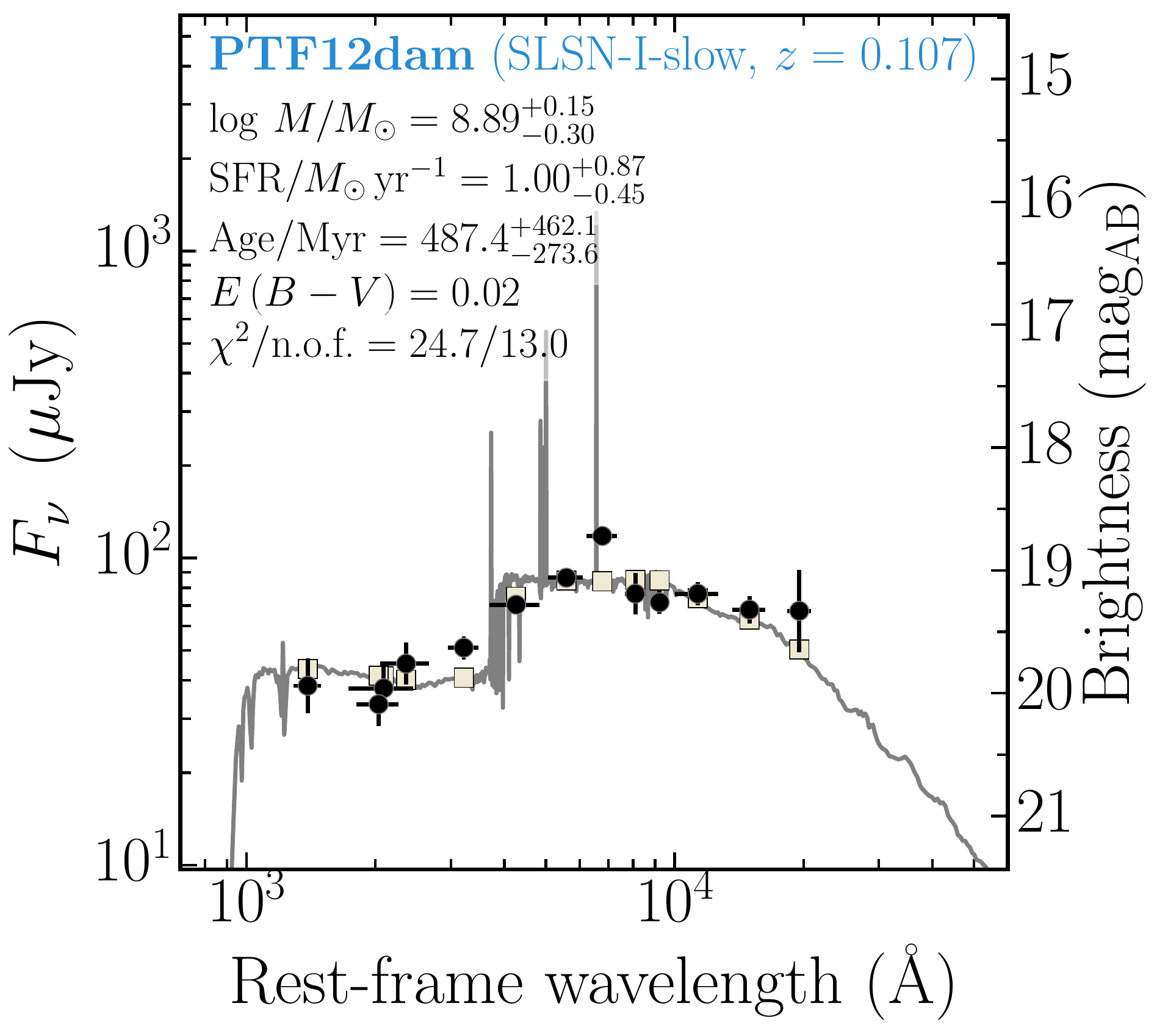}
\includegraphics[width=0.24\textwidth]{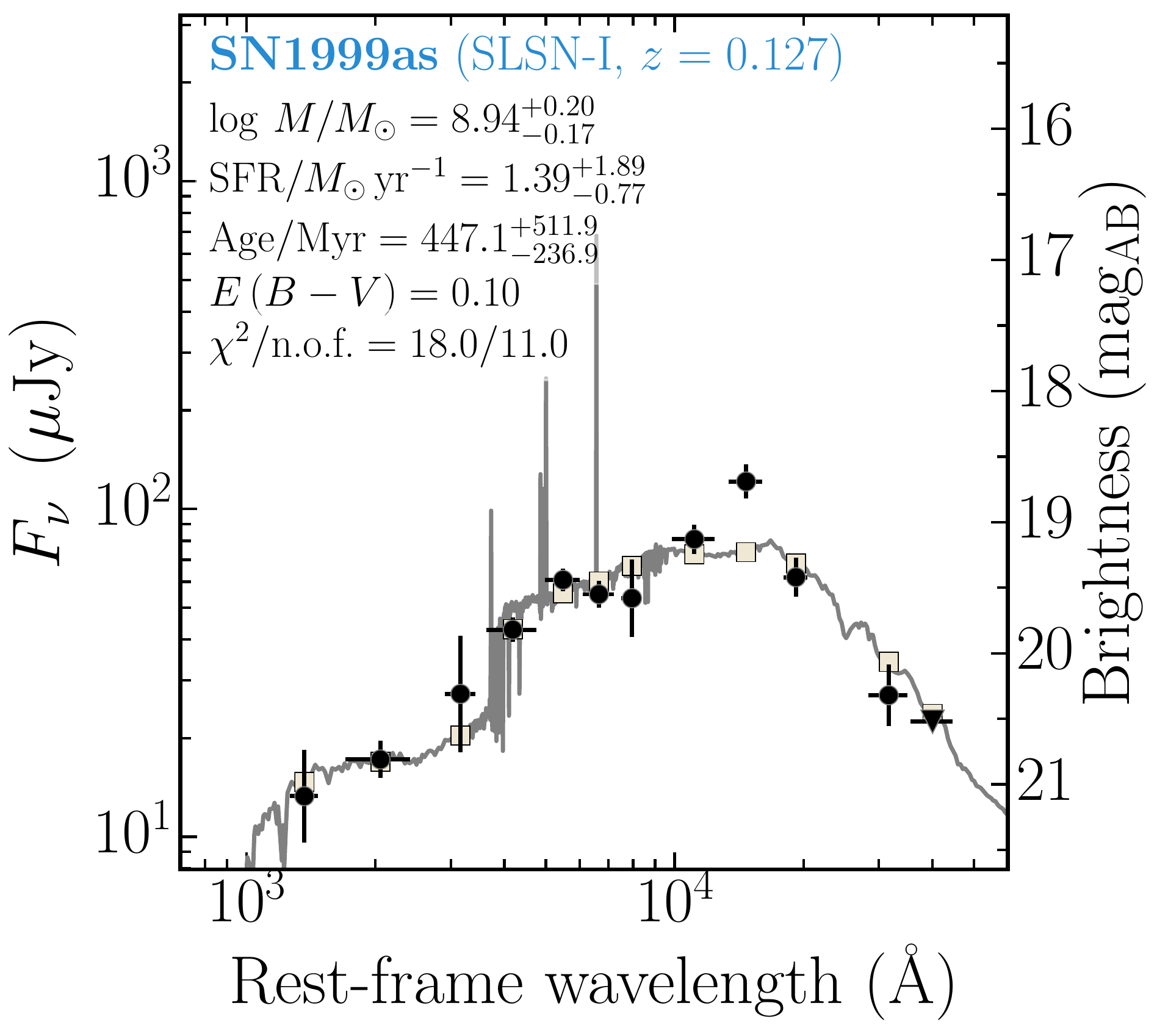}
\includegraphics[width=0.24\textwidth]{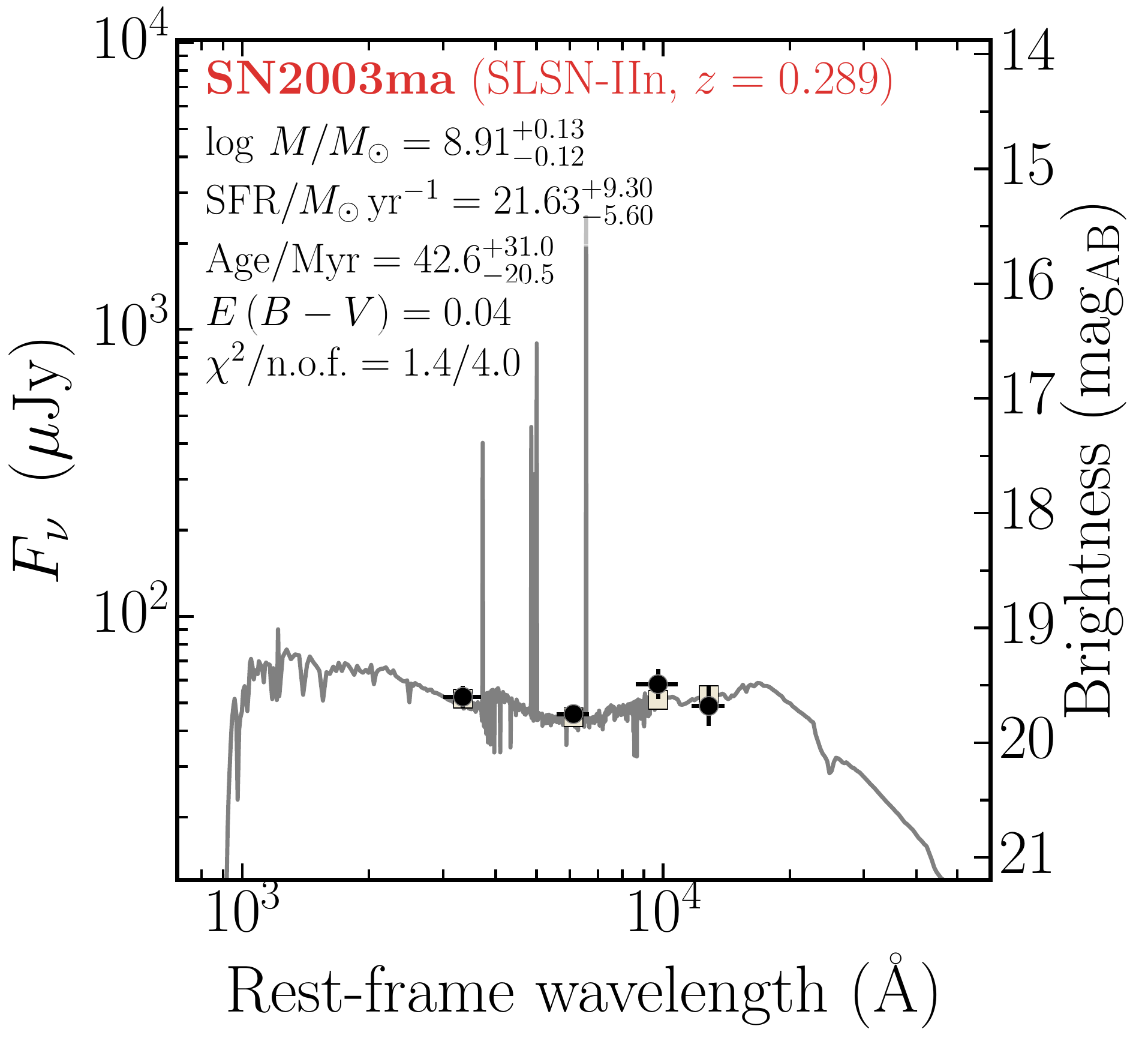}
\includegraphics[width=0.24\textwidth]{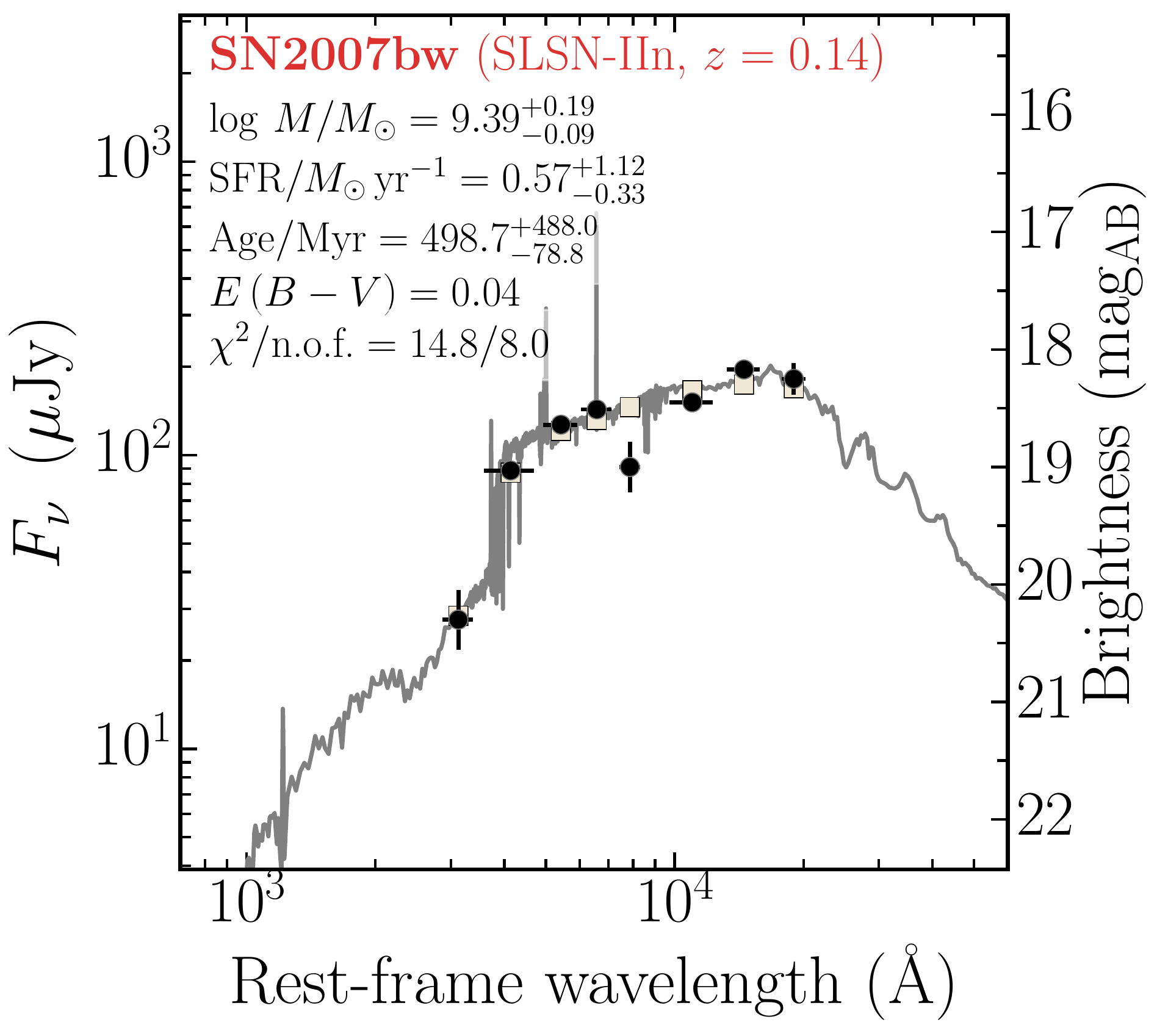}
\includegraphics[width=0.24\textwidth]{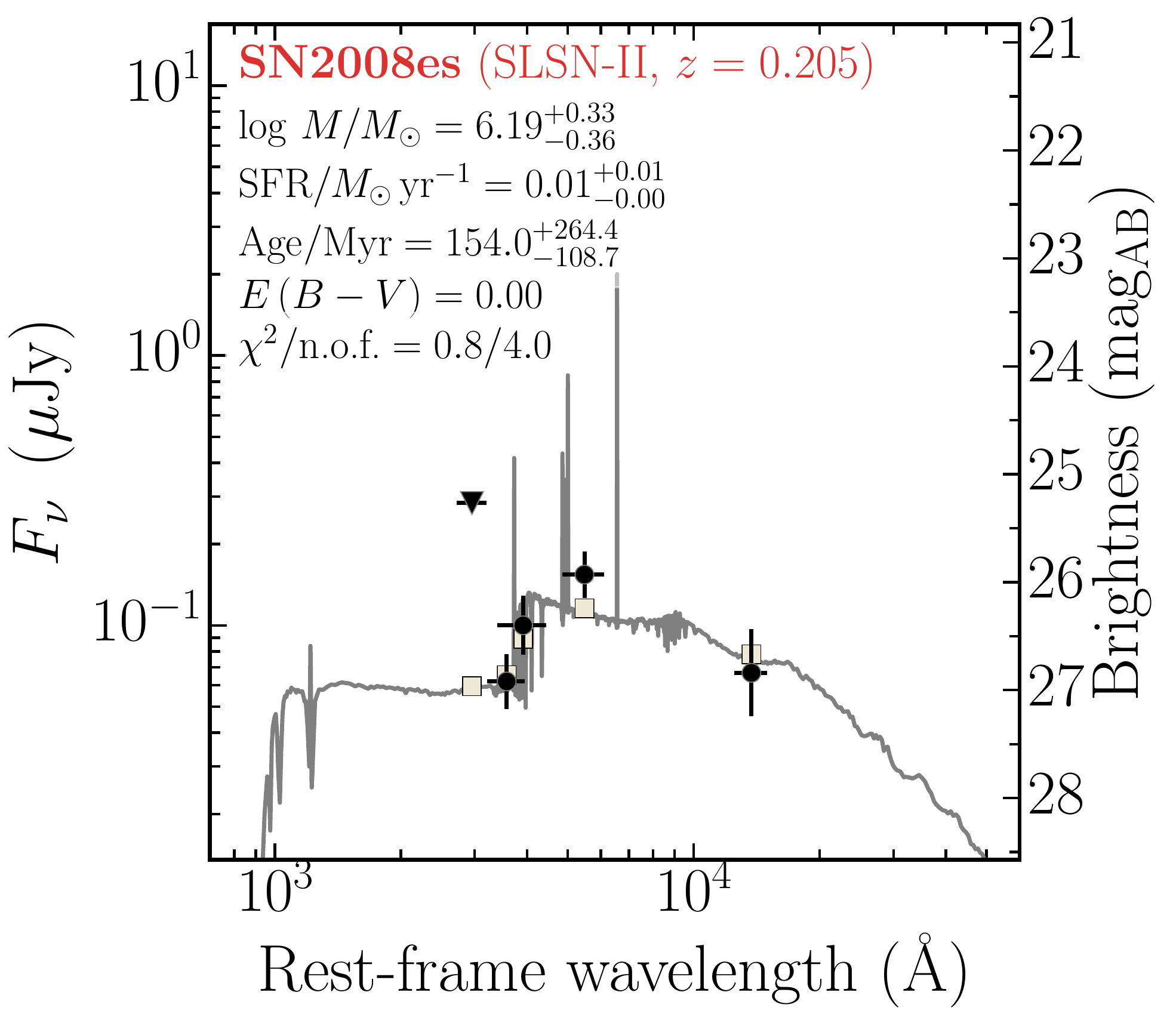}
\includegraphics[width=0.24\textwidth]{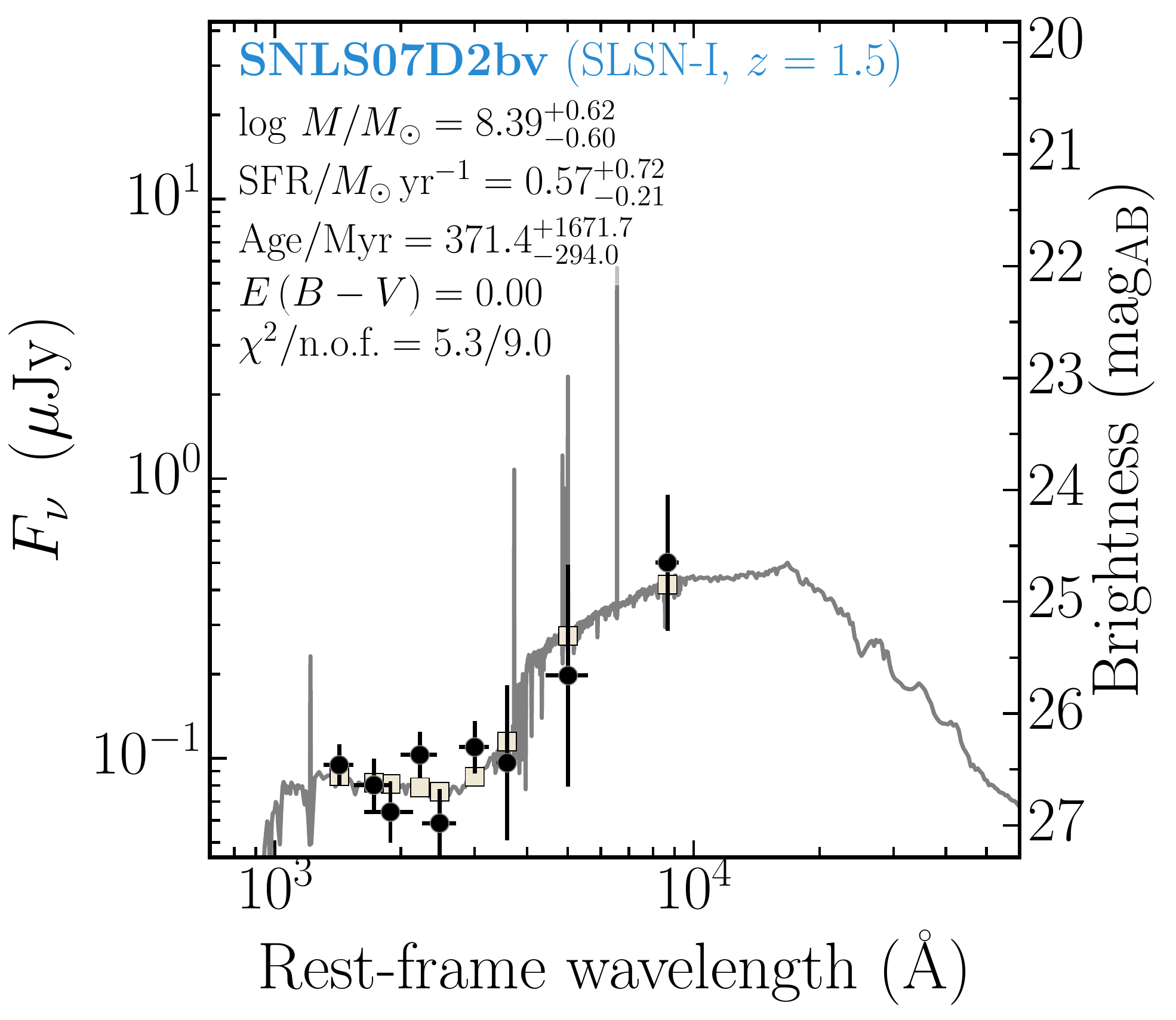}
\includegraphics[width=0.24\textwidth]{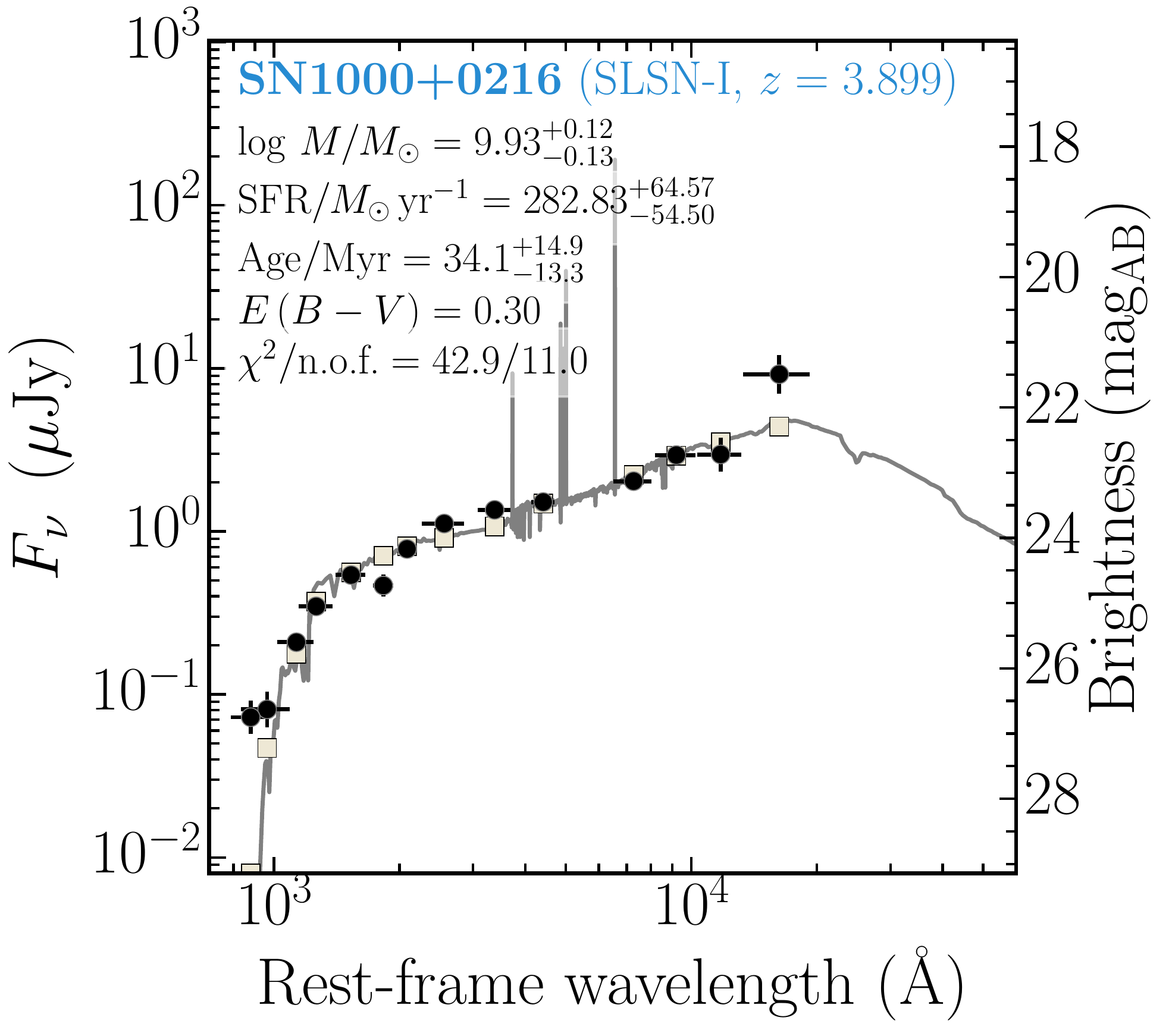}
\end{center}
\caption{Selection of spectral energy distributions of hosts of H-poor and -rich SLSNe from
700 to 60000 {\AA} (detections: $\bullet$; upper limits: $\blacktriangledown$).
The solid line displays the best-fit model of the SED. The squares in a lighter shade are the model predicted magnitudes.
The fitting parameters are displayed for each SED. See Table \ref{tab:sed_results} and Sect.
\ref{sec:sed_fitting} for details. The full collection of SEDs are shown in Figs.
\ref{fig:SED_app_1} and \ref{fig:SED_app_2}.
}
\label{fig:SED}
\end{figure*}

We made two assumptions to model all SEDs in an automatic and self-consistent way: \textit{i})
the SEDs can be described by a stellar component with an exponentially declining star-formation history
and a contribution from the ionised gas of the \ion{H}{ii} regions and \textit{ii}) the number of filters (n.o.f.) can be reduced to the
homogenised filter set in Sect. \ref{sec:sed_fitting}. Over 90\% of our hosts have
good fits with an average $\chi^2/{\rm n.o.f.}$ of 0.5
and derived physical parameters that are comparable to other galaxy samples
(Table \ref{tab:sed_results}, Figs. \ref{fig:SED}, \ref{fig:SED_app_1}, \ref{fig:SED_app_2}).

The fits of only six hosts had $\chi^2/{\rm n.o.f.}$ between 3.9 and 10.4. The fits of PS1-11bdn and SN1000+0216
are of poorer quality ($\chi^2/\rm{n.o.f.}=3.9$ and 6.3, respectively) caused by a few data
points. The host of PS1-10bzj has very strong emission lines that fall in the wings of
the $i'$-band transmission function, which increased the normalised $\chi^2$ to 10.4.
Apart from data points in a few individual filters, the fits are nonetheless very good and can be used without restriction.

The fits of CSS100217, PTF11dsf, SN1999bd and SN2006gy have to be used with more caution. \citet{Drake2011a}
revealed a narrow-line Seyfert in the host galaxy of CSS100217. Furthermore, \citet{Leloudas2015a} reported
on the discovery of broad H$\alpha$ and [\ion{O}{iii}] in the host spectrum of PTF11dsf, which could be due
to an AGN as well.
The hosts of SLSNe-IIn SN1999bd and SN2006gy
are evolved galaxies that experienced a recent starburst. This is demonstrated by the detection of
Balmer lines in both spectra \citep{Smith2007a, Leloudas2015a, Fox2015a}, while the SED cannot be
modelled by an exponentially declining star-formation history.
A reliable modelling of the SEDs of these three hosts requires a detailed modelling of their star-formation histories
and the inclusion of an AGN component, which is beyond the scope of this paper.
\citet{Leloudas2015a} mentioned that the host of PTF11dsf could also harbour an AGN. Similar to the
three aforementioned hosts, we only use the mass and the $B$-band luminosities of PTF11dsf's host
in our discussion, but not the SFR.

\begin{figure*}
\includegraphics[width=2\columnwidth]{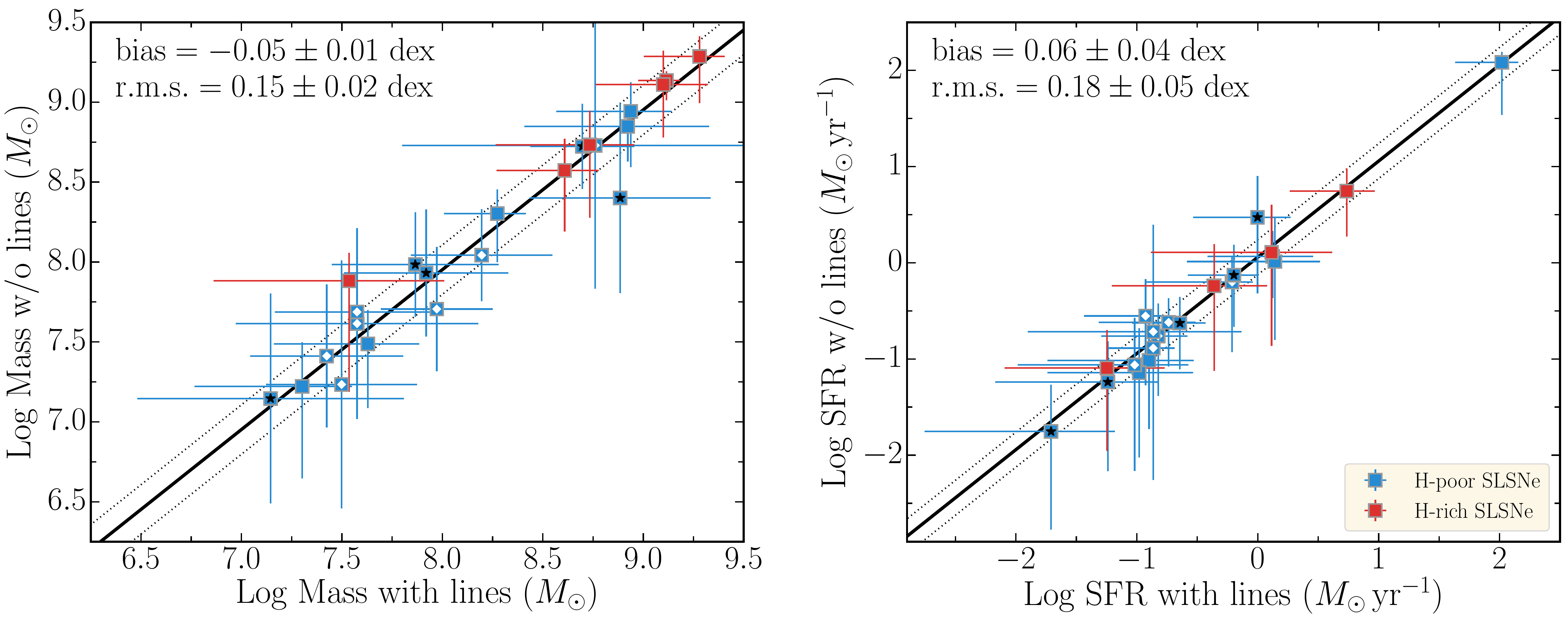}
\caption{
Derived masses (left) and SFRs (right) of galaxies from the spectroscopic sub-sample.
The SEDs are fitted with two different procedures:
\textit{i}) the photometry of the galaxies with the contribution of the emission
lines is fitted with galaxy templates and an emission line component in \texttt{Le Phare};
\textit{ii}) the photometry of the same galaxies is fitted after removal of the emission line
contribution and switching off the ionised gas component in \texttt{Le Phare}.
The values in the upper left corners report the mean bias deviations and the average root square errors
(r.m.s.) between the measurements with and without emission-line contribution and their corresponding errors.
The solid line indicates the bias between both diagnostics and the dotted lines the mean r.m.s.
centred around the bias. The agreement is very good, showing
that we can obtain reliable results with \texttt{Le Phare} also for the galaxies where
spectroscopic information is not available. The hosts of fast and slow-declining H-poor
SLSNe are signified by `$\star$' and `$\diamond$', respectively.
}
\label{fig:sed}
\end{figure*}

\begin{figure}
\includegraphics[width=1\columnwidth]{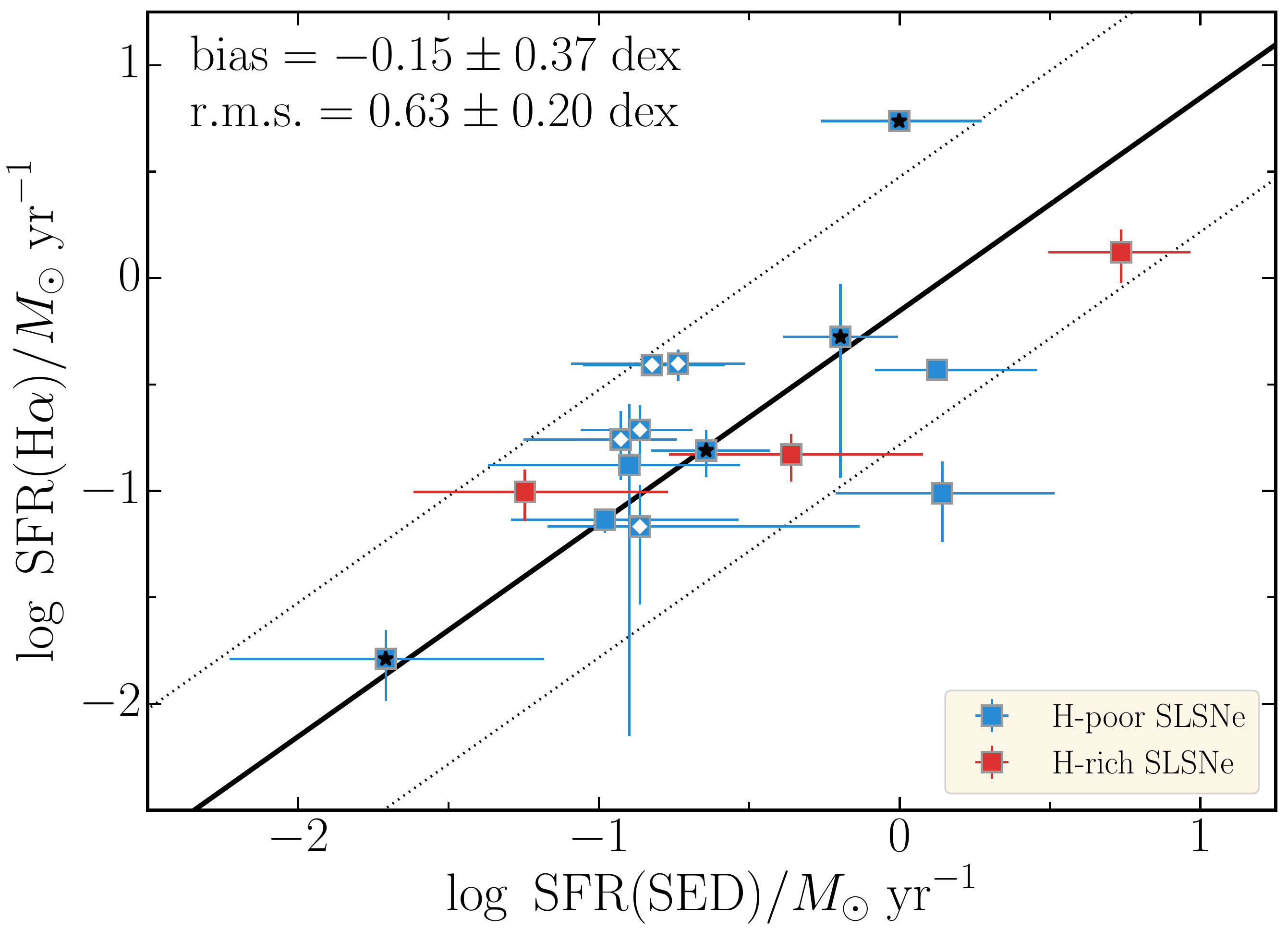}
\caption{
Star-formation rates obtained from SED modelling and from emission lines for the spectroscopic sub-sample. The values
in the upper left corners report the mean bias deviation and the mean r.m.s. between the H$\alpha$ and SED-derived
star-formation rates. The solid line indicates the bias between both diagnostics and the dotted lines the mean r.m.s.
centred around the bias. Symbols are identical to Fig. \ref{fig:sed}.
}
\label{fig:sed2}
\end{figure}

\subsubsection{Contribution of emission lines}\label{sec:em_lines}

Our SED modelling includes the contribution of the \ion{H}{ii} regions. This is of particular importance because previous studies showed that
emission lines can significantly affect the SED fitting \citep[e.g.,][]{Castellano2014a, Lunnan2014a, Chen2015a, Santini2015a}.
This motivated \citet{Lunnan2014a}
to omit filters that were affected by [\ion{O}{iii}]$\lambda$5007, if [\ion{O}{iii}] had a
large equivalent width, and \citet{Chen2015a} to subtract the emission line contribution from
the broad-band photometry. Both approaches are strictly limited to objects with
host spectroscopy.

Thanks to \texttt{Le Phare} capabilities, we quantify the impact of emission-lines on the
SED fitting with a more sophisticated approach. First, we fit the SEDs of the spectroscopic subsample
with templates that include a stellar and a gas component. Then, we subtract the contribution of
the emission lines from the broad-band photometry and fit the new SEDs with a stellar component only,
i.e., the gas component is explicitly switched off in \texttt{Le Phare}.

Figure \ref{fig:sed} shows how the primary diagnostics mass and SFR change if emission lines are
included in the SED fitting. The absolute value of the average mean bias deviation
and the average root mean square error in the mass and SFR estimates are $<0.06$~dex and $<0.18$~dex, respectively,
and smaller than the $1\sigma$ error bars of individual measurements.
The most critical object in this analysis is PTF12dam,
the most extreme SLSN host galaxy known to date. Its deviations between the mass and SFR
estimates with and without lines are
$\Delta {\rm SFR} = \log {\rm SFR}_{\rm w/~lines} - \log {\rm SFR}_{\rm w/o~lines}=-0.47\pm0.45~\rm dex$,
$\Delta M = 0.48\pm0.42~\rm dex$.
Apart from this object, the agreement between the two fits is excellent. This reflects the fact that we have good photometry spanning a large
wavelength interval \textit{and} a good handle on the gas emission in the SED fitting, so that the uncertainty
in the emission-line contribution does not affect our results.

\subsubsection{SED vs. emission-line diagnostics}

By combining results from the spectroscopic observations in \citet{Leloudas2015a} with the results from our SED modelling,
we have two independent estimates on the recent star-formation activity for our spectroscopic
sub-sample. Both diagnostics assume a particular star-formation history and a particular initial mass function.
In addition, different diagnostics average the star-formation activity over different time intervals,
e.g., the H$\alpha$ SFR-indicator is sensitive to the star-formation activity over the past 6~Myr,
whereas the SFR derived from rest-frame UV continuum averages over a time period of 100~Myr \citep[e.g.,][]{Kennicutt2012a, Calzetti2013a}.
Because of the extreme nature of SLSNe, we examine whether we can isolate the differences that occur
due to the time-scales that the H$\alpha$ and SED-inferred SFRs probe.

Assessing these differences requires that the systematic uncertainties in the data are well understood. Spectroscopic
observations with slits are subject to flux losses, because a slit may only cover a part
of a given galaxy. Most SLSN host galaxies are relatively compact \citep{Lunnan2015a} so that the expected
losses are small. To correct these, \citet{Leloudas2015a} convolved the spectrum of a given object with the
filter bandpasses of its imaging data to extract synthetic photometry. In most cases, a simple rescaling
was sufficient to adjust the absolute flux scale, i.e., the extracted spectrum is representative for the
entire galaxy. Only a few objects required low-order polynomials to correct the warping of the spectrum.
In the following, we use the spectroscopic data of a sub-sample of 16 host galaxies with a reliable absolute
flux scale.

Figure \ref{fig:sed2} compares the extinction-corrected SFR's from SED modelling and H$\alpha$ emission
lines of these 16 hosts. Both diagnostics reassuringly show consistency. The mean bias deviation and the mean r.m.s.
between the H$\alpha$ and SED derived SFRs are $-0.16\pm0.37$~dex and $0.63\pm0.20$~dex, respectively. \citet{Conroy2013a}
pointed out that a systematic uncertainty in the SED-based SFRs of a factor of 0.3 dex is expected.
Our observed value is larger than the expected value but consistent within $2\sigma$.

The most interesting object in our sample to identify differences in the SFR indicators is again the host of PTF12dam.
\citet{Thoene2015a} reported that the head of the tadpole
galaxy is characterised by a very young stellar population which is $\sim3$~Myr old. \citet{Calzetti2013a}
showed that in such cases, the UV SFR estimator will be underestimated by a factor of a few. We measure
an excess of $0.74\pm0.27$~dex in the H$\alpha$ inferred SFR. Even in that case, the deviation between the
H$\alpha$- and SED-inferred SFRs only has a significance of $<2.7\sigma$, reassuring us that even in
such an extreme case the SED modelling can provide robust results.

\begin{figure*}
\begin{center}
\includegraphics[width=0.24\textwidth]{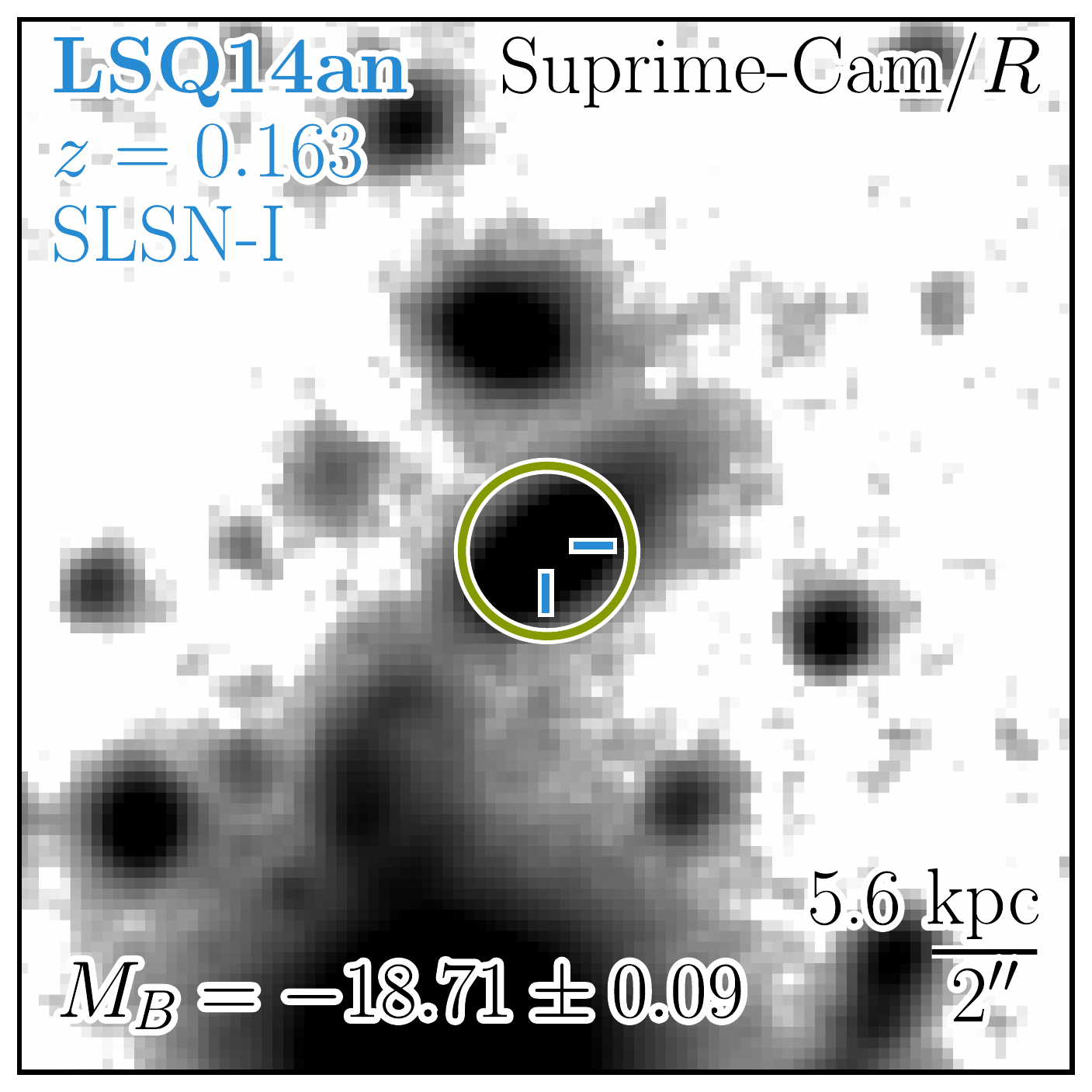}
\includegraphics[width=0.24\textwidth]{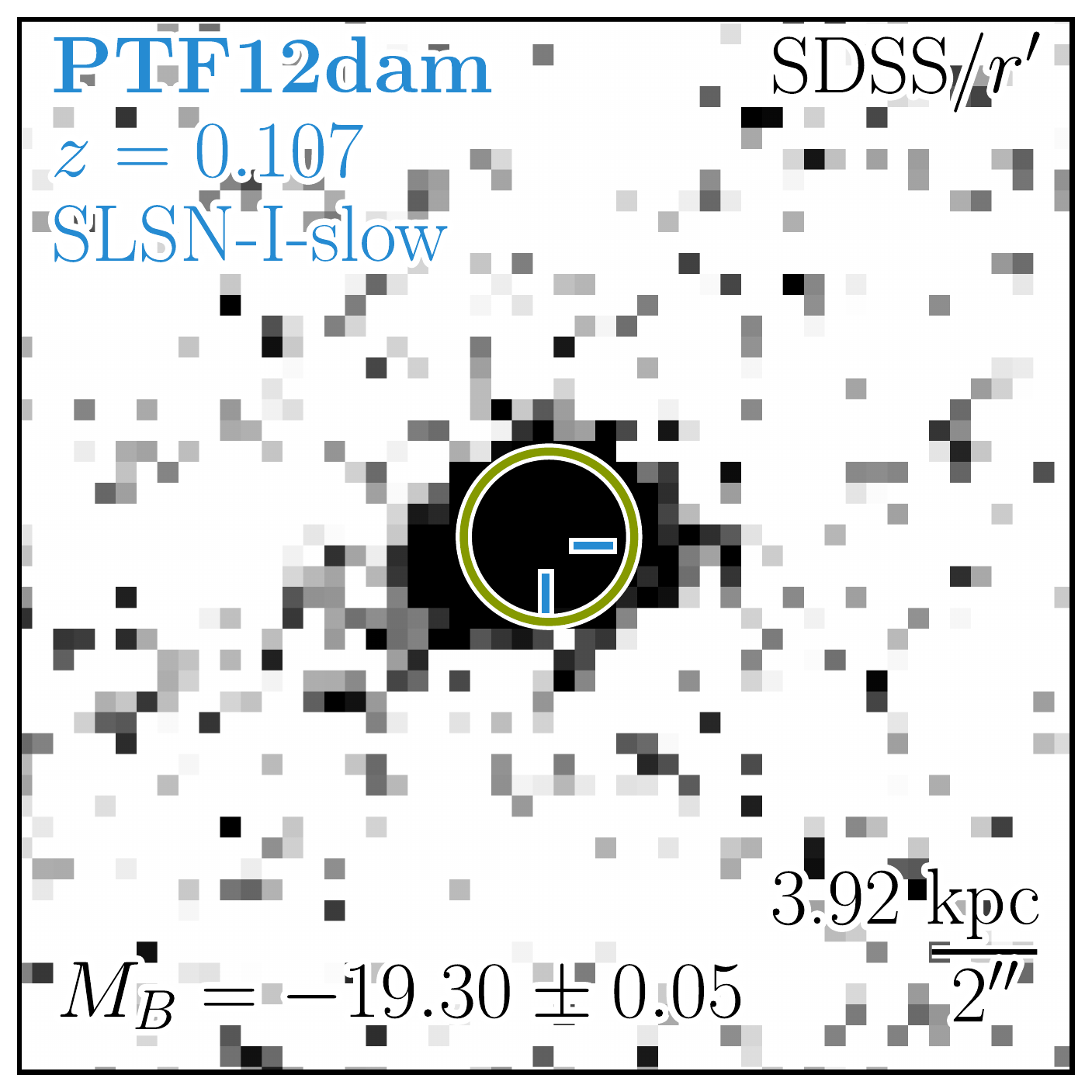}
\includegraphics[width=0.24\textwidth]{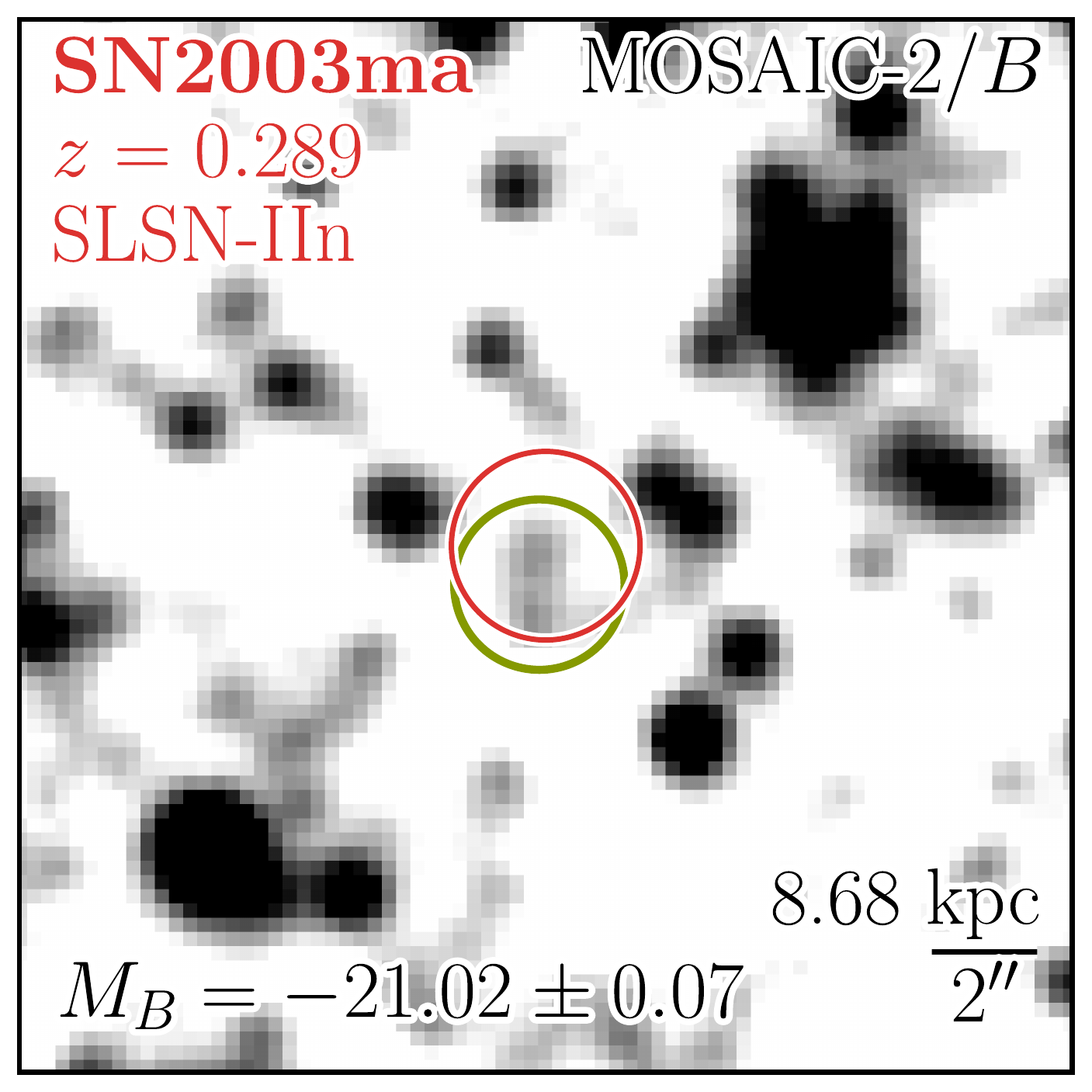}
\includegraphics[width=0.24\textwidth]{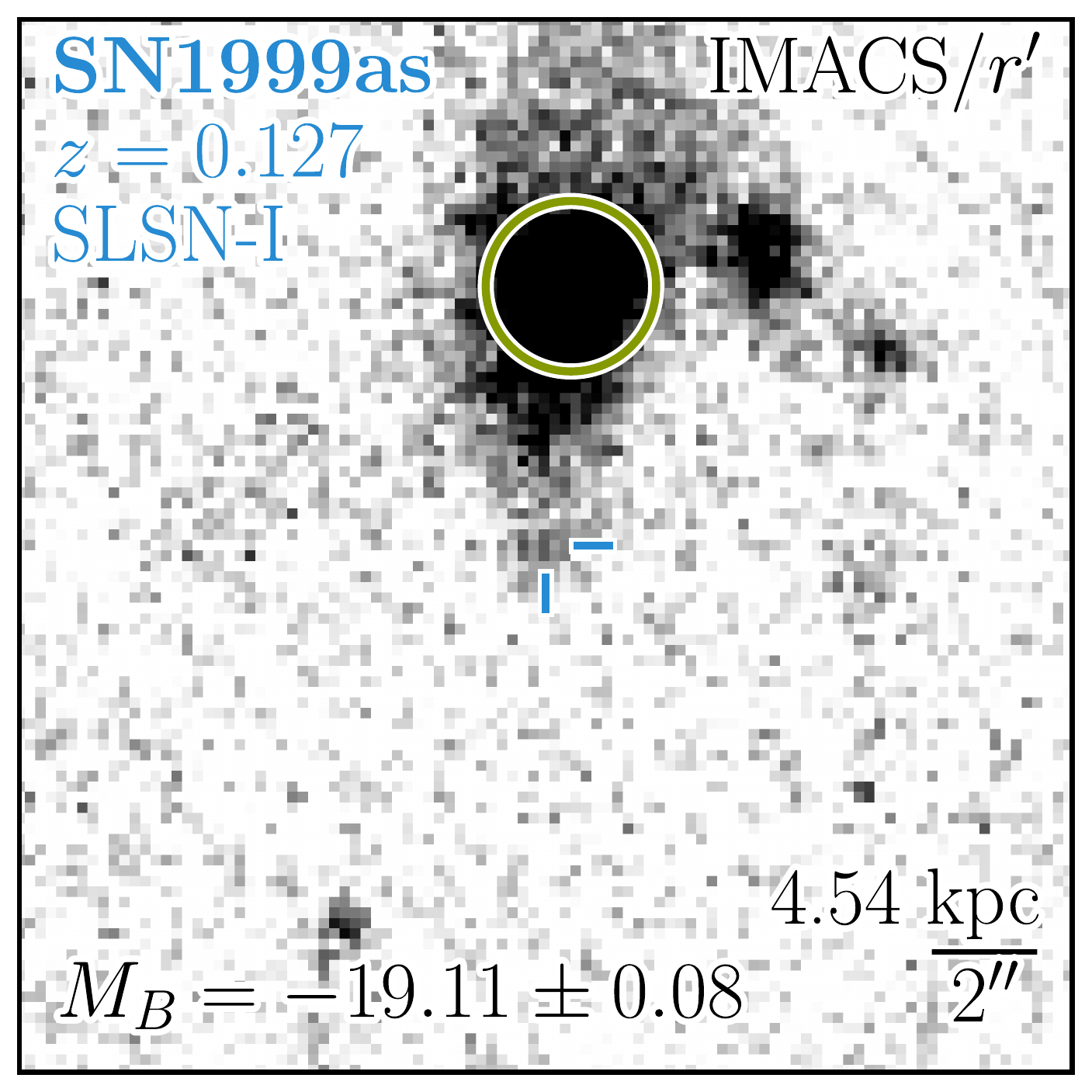}
\includegraphics[width=0.24\textwidth]{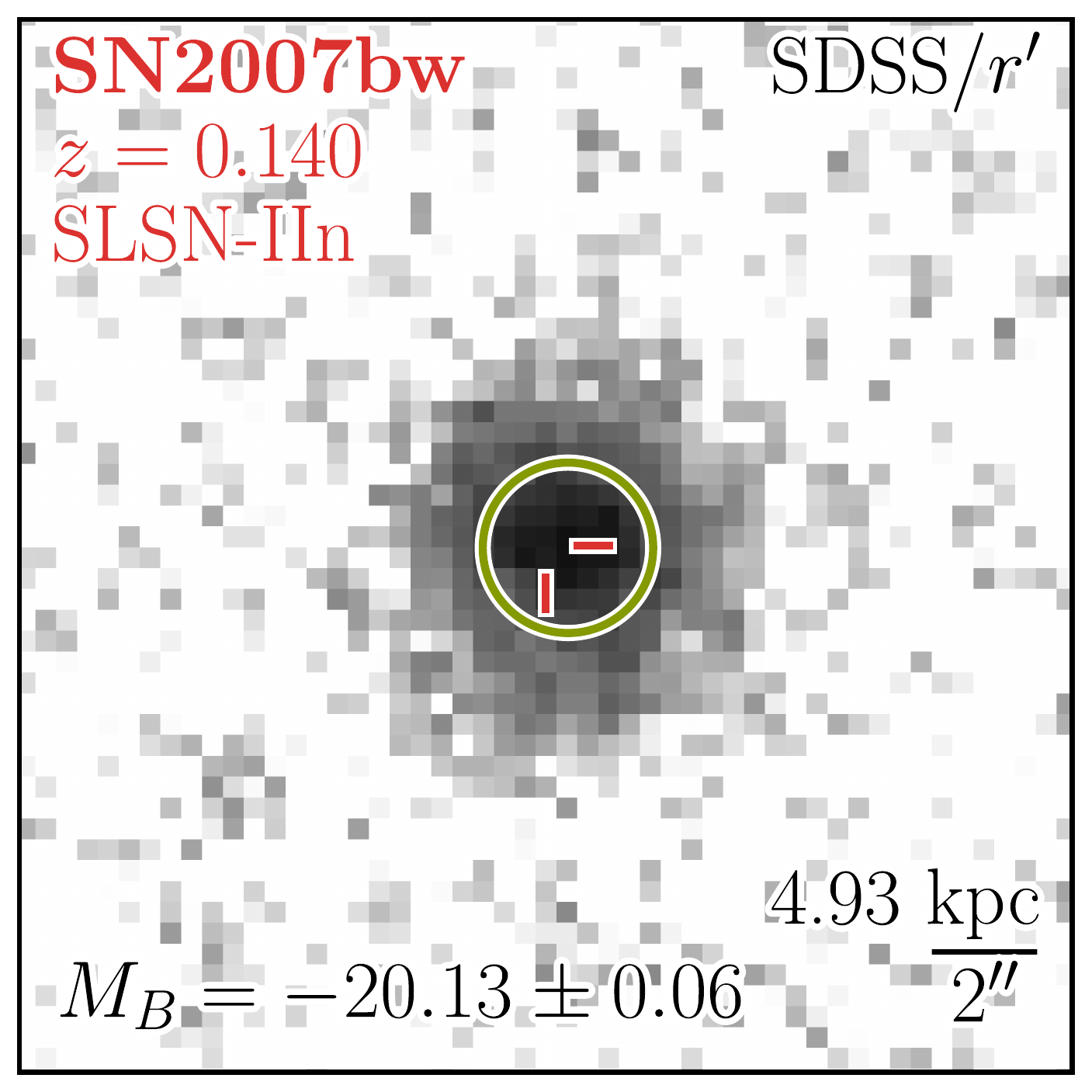}
\includegraphics[width=0.24\textwidth]{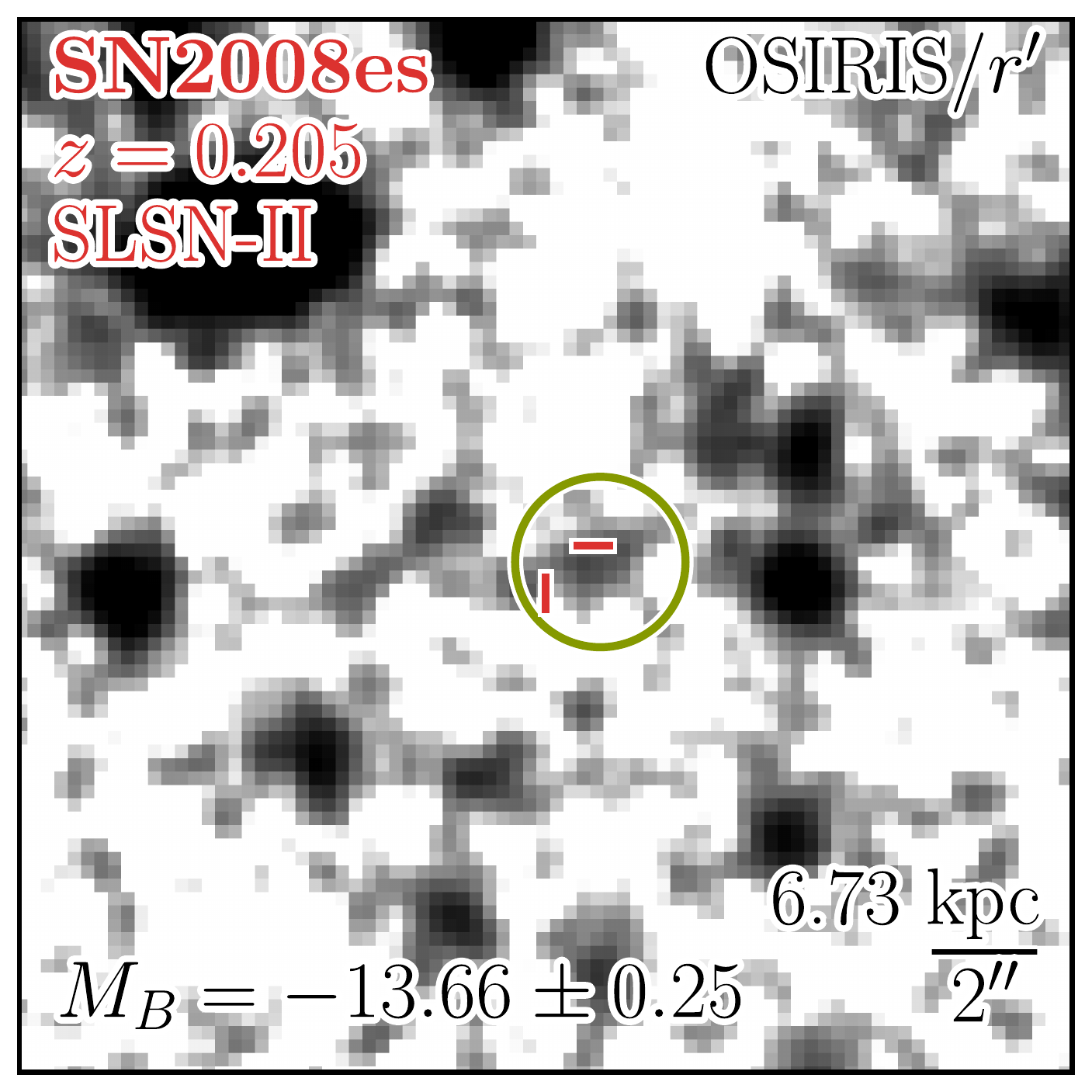}
\includegraphics[width=0.24\textwidth]{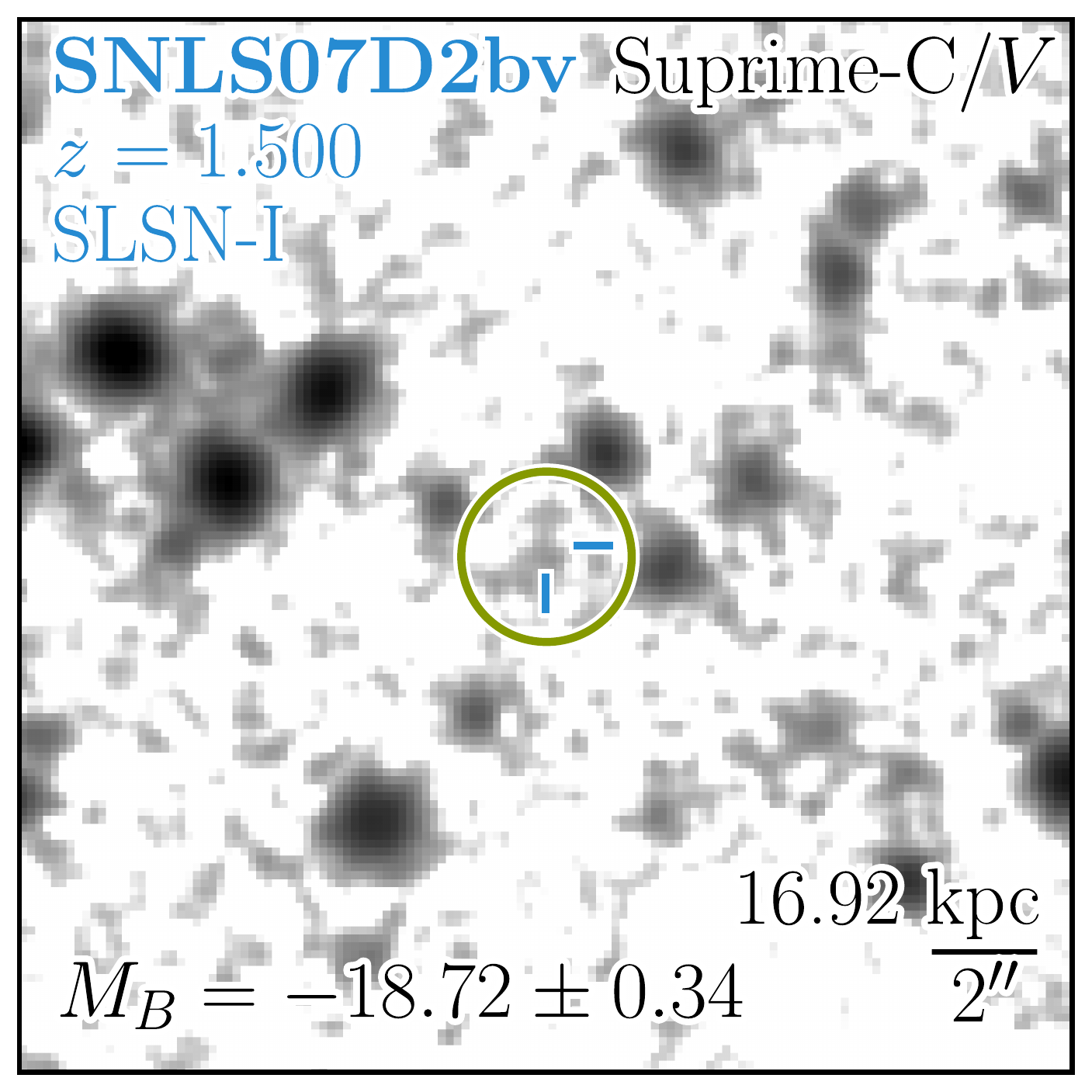}
\includegraphics[width=0.24\textwidth]{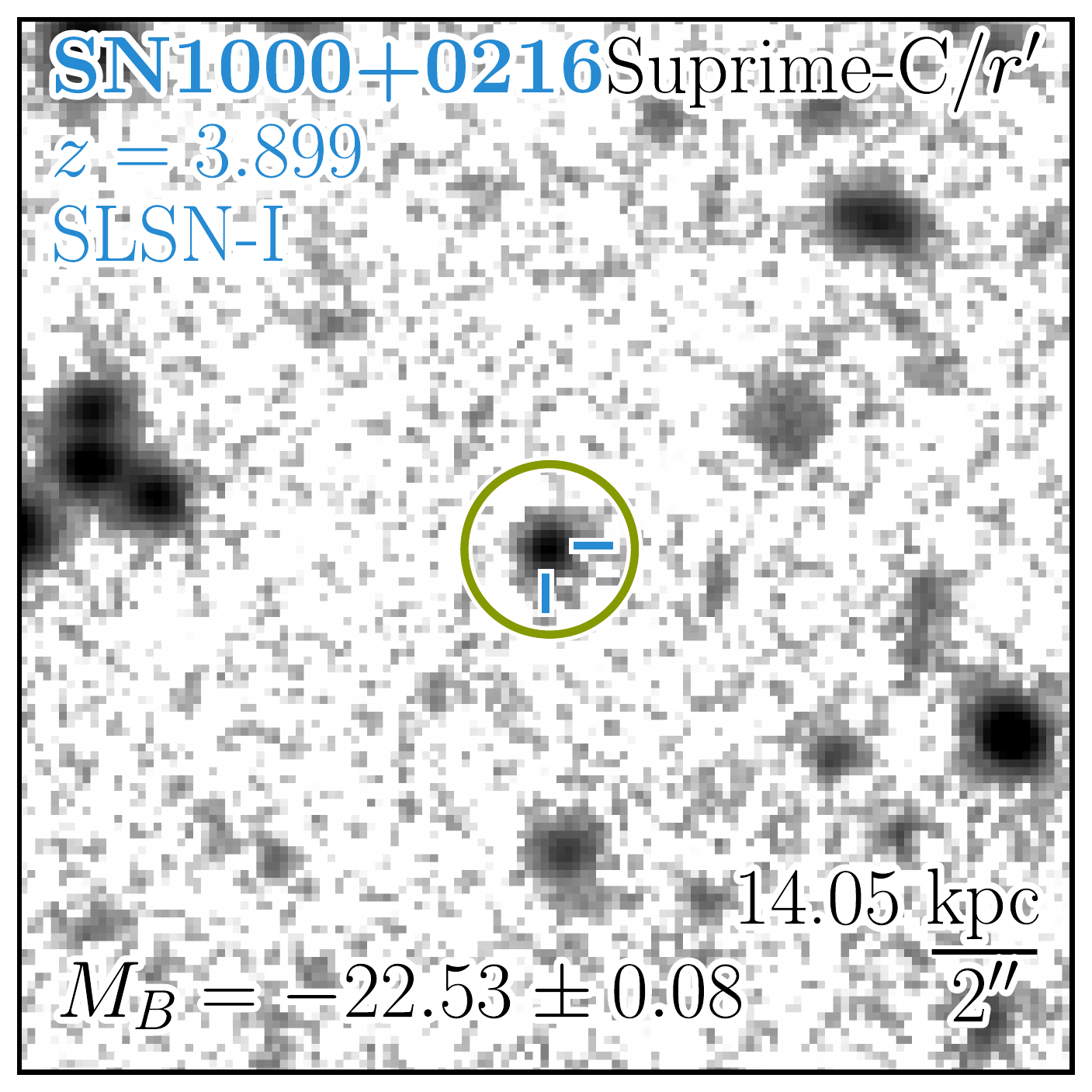}
\end{center}
\caption{Selection of postage stamps of the hosts of H-poor and -rich SLSN host galaxies in our sample.
The images were taken before the SN occurred or after the SN faded.
Each panel has a size of $20\arcsec\times20\arcsec$ where North is up and East is left.
The crosshair marks the position of the SNe after aligning on a SN and a host image (H-poor SLSN: blue; H-rich SLSNe: red).
{If no SN image was available, a circle in blue or red (arbitrary radius) is shown instead, indicating
the SN position reported in the literature.} The average alignment error is $0\farcs17$ but it exceeds
$1\farcs0$ in a few cases. The green circle (arbitrary radius) marks the host galaxy.
The observed absolute $B$-band magnitude is displayed in the lower left corner.
The image of SNLS07D2bv is smoothed with a Gaussian kernel (width of 1 px) to
improve the visibility of the field. The complete collection of postage stamps is shown
in Figs. \ref{fig:poststamp_app_1} and \ref{fig:poststamp_app_2}.
}
\label{fig:poststamp}
\end{figure*}

\subsection{Host offsets}\label{res:morphologies}

Figures \ref{fig:poststamp}, \ref{fig:poststamp_app_1} and \ref{fig:poststamp_app_2} show postage stamps of each
field in our sample. The detected host galaxies (detection rate of $\approx90\%$) are marked by green circles.
The SN positions, after astrometrically aligning the SN and the host images, are indicated
by crosshairs. The average uncertainty of 0\farcs17 is dominated by the
different pixel scales of the SN and host images. In a few examples, this uncertainty
exceeds $1''$ because of the coarse spatial resolution of the SN images, the small spatial
overlap of SN and host images, or the low number of reference stars.
We lack SN images for 17 hosts in our sample. Their SN positions are indicated by circles
as reported in the literature.

Thanks to the high host recovery rate (85\% and 100\% for H-poor and H-rich SLSNe, respectively),
we present a relatively complete distribution of the distances between the SN positions and the barycentres
of the host light (predominantly in $r'$ band) of H-poor and H-rich SLSNe. In addition, we incorporate
results on CSS100217 by \citet{Drake2011a}, on SN2003ma by \cite{Rest2011a} and on
Pan-STARRS SLSNe by \citet{Lunnan2015a}. The observed distribution is skewed to small radii
(the expectation value being 1.3~kpc) but has a long tail extending up to 12~kpc. For the smallest offsets,
the measurements are comparable to the errors. In this regime, Gaussian noise superimposed on
a vector with length $\mu$ results in a non-Gaussian probability distribution of the vector
length, i.e., an overestimated host offset \citep{Rice1944a}.
The expected probability distribution function of a host offset measurement $r$ is given by
\begin{equation*}
p\left(r|\mu,\sigma\right)=\frac{r}{\sigma^2}\,I_0\left(\frac{r\,\mu}{\sigma^2}\right)\,\exp\left(-\frac{r^2+\mu^2}{\sigma^2}\right)
\end{equation*}
where $\mu$ is the true offset, $\sigma$ is the dispersion of the distribution, which can be assumed
to be comparable to the measurement error, and $I_0$ is the modified Bessel function of the first kind. By differentiating $p\left(r|\mu,\sigma\right)$
with respect to $r$, a closure relation can be derived between the observed offset, its error and the true
offset \citep{Wardle1974a}:
\begin{equation*}
I_0\left(\frac{r\,\mu}{\sigma^2}\right)\left(1-\frac{r^2}{\sigma^2}\right)+\frac{r\,\mu}{\sigma^2}\,I_1\left(\frac{r\,\mu}{\sigma^2}\right)=0\,.
\end{equation*}

\begin{figure}
\includegraphics[width=1\columnwidth]{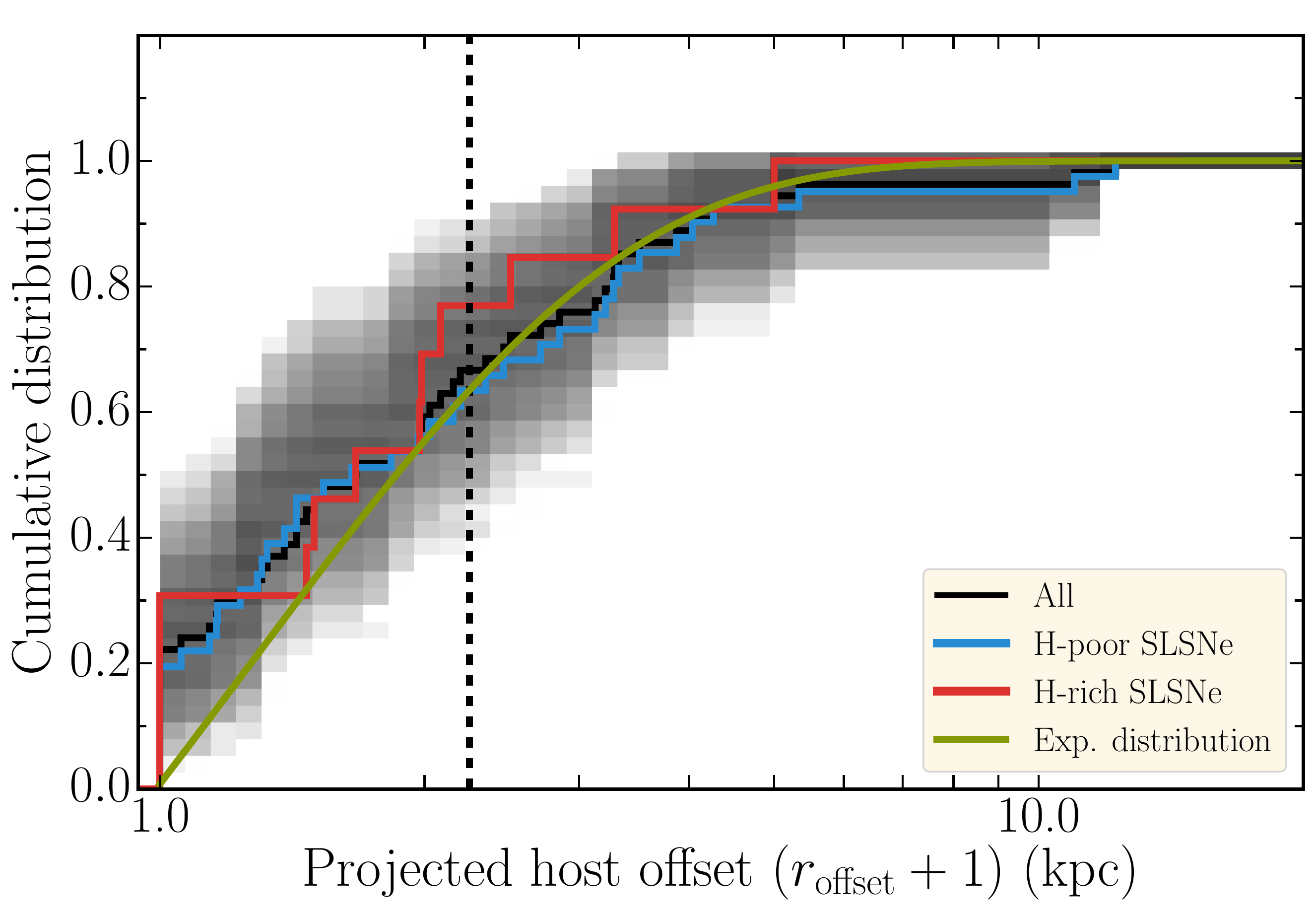}
\caption{
Host offset cumulative distribution for 41 H-poor (blue) and 13 H-rich (red) SLSNe and the total sample (black).
The shaded region displays the expected parameter space after
bootstrapping the sample 30\,000 times. The dotted, vertical line indicates the median offset.
We shifted the distribution by 1~kpc in order to use a logarithmic scaling for presentation purposes.
}
\label{fig:offset}
\end{figure}

We solved this equation numerically to build the intrinsic host offset distribution. The
black curve in Fig. \ref{fig:offset} shows the joint cumulative distribution of H-poor and
-rich SLSNe. The grey-shaded regions display the expected parameter space of our distribution
after bootstrapping the sample 30\,000 times with darker regions, indicating a higher probability.
The distribution is well described by the cumulative distribution function of a negative exponential
distribution $1-\exp\left(-r/r_{\rm mean}\right)$ with a mean offset of $r_{\rm mean}\sim1.3$~kpc.

The fit underpredicts the fraction
of hosts with offsets smaller than $<0.5$~kpc and $>4$~kpc. The discrepancy for small host offsets
can be reconciled with the alignment errors between SN and host image, and intrinsically small
host offsets. As the  alignment error exceeds the offset measurement, the closure relation is only
fulfilled if $\mu=0$. Therefore, the fraction of SLSNe with negligible host offsets
is a strict upper limit. In addition, any inclination will lead to an underestimation
of the true host offset. The blue and red curves in Fig. \ref{fig:offset} show the observed
offset distribution after separating the sample in H-poor and -rich SLSNe, respectively. Both samples
are statistically identical.

The offsets of PTF11rks and SN1999as are $>10$~kpc and therefore they exceed the median of 0.7~kpc by
a large factor. The host of SN1999as is an irregular galaxy interacting with its environment (Fig.
\ref{fig:poststamp}). At the explosion site a faint object is detected in continuum.
The explosion site of PTF11rks is connected by a linear feature with the nucleus \citep{Perley2016a}.
This could point to a spiral galaxy morphology or galaxy interaction whereby the SN exploded in a faint
satellite galaxy. Spectroscopic observation of SN1999as by \citet{Leloudas2015a}
showed that the explosion site is characterised by strong emission lines. In this case,
the true host is a fainter galaxy that is difficult to disentangle from the more massive galaxy.

\subsection{Brightness, colour and luminosity}\label{res:brightness_color}

\begin{figure}
\includegraphics[width=1\columnwidth]{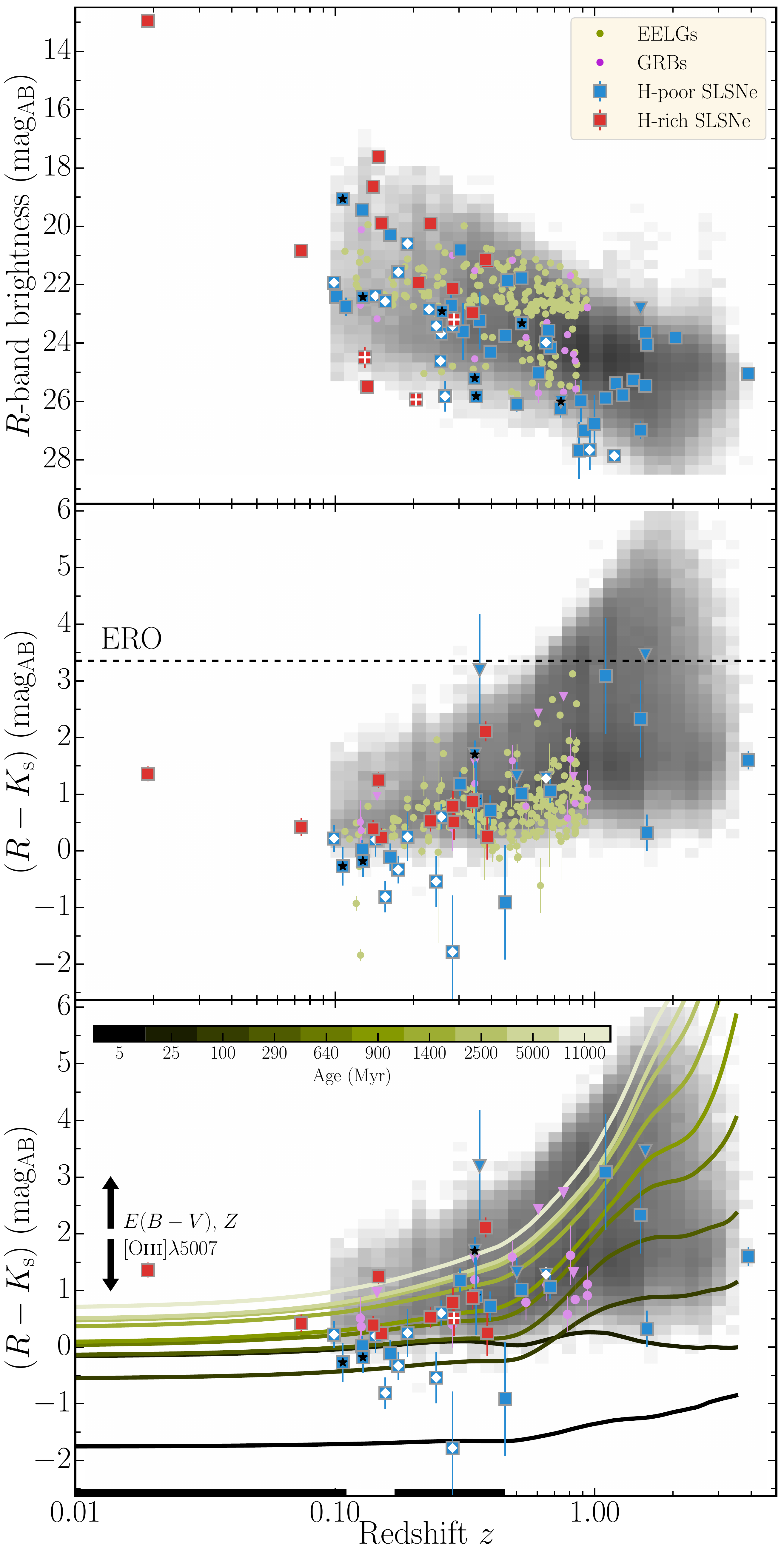}
\caption{\textit{Top}: The observed $R$-band host magnitude as a function of redshift for H-poor (blue) and H-rich
(red) SLSNe. In case of a $R$-band upper limit, the measurement is displayed as a downward pointing triangle.
The hosts of fast and slow-declining H-poor
SLSNe are signified by `$\star$' and `$\diamond$', respectively, and SLSNe-II by `$+$'.
\textit{Middle}: The $R-K_{\rm s}$ colour evolution. The observed $R-K_{\rm s}$ colour
evolution of SUSHIES, GRB host galaxies and star-forming galaxies from the UltraVISTA survey (density
plot). \textit{Bottom}:
The colour evolution of galaxies with a metallicity of 0.2 solar for different stellar
population ages, derived from templates by \citet{Bruzual2003a}.
The tracks are shown up to $z=3.5$ to avoid corrections for Ly$\alpha$ absorption in the host galaxies and in the intergalactic medium.
The vectors on the left indicate how
extinction, metallicity and emission-lines with very large equivalent widths, such as H$\alpha$
and [\ion{O}{iii}]$\lambda$5007, can alter the intrinsic colour. Note, H$\alpha$ and [\ion{O}{iii}]$\lambda$5007
can turn the colour to the blue only at $z\lesssim0.11$ and between $z\sim0.17$ and $z\sim0.45$, respectively
(indicated by the bars at the bottom).
}
\label{fig:colour}
\end{figure}

\subsubsection{Brightness and luminosity}

\begin{table*}
\caption{Statistical properties of H-poor and -rich SLSN host galaxies per redshift bin}
\centering
\begin{tabular}{ccccccccc}
\toprule
\multirow{2}*{Sample}	& \multirow{2}*{Number}	& Mean			& $m_R^{(a)}$	& $(R-K_{\rm s})^{(a)}$	& $M_B$	& \multirow{2}*{$\log M/M_\odot$}	& $\log \rm SFR$ 		&  $\log \rm sSFR$	\\
	& 			& redshift		&(mag)		& (mag)			& (mag)	& 					& $(M_\odot\,\rm{yr}^{-1})	$& $(\rm{yr}^{-1})$	\\
\midrule
\multicolumn{9}{c}{\textbf{$\mathbf{z\leq0.5}$}}\\
\midrule

\multirow{2}*{I-fast}	& \multirow{2}*{11}	& \multirow{2}*{0.21}	&$ 22.96\pm0.48			$&$ -0.10\pm0.24$ (8) 	 &$ -16.71\pm0.37		 $&$ 7.86\pm0.16			$&$ -0.89\pm0.08		 	$&$ -8.70\pm0.11			$\\
	&			&											&$ 1.46^{+0.42}_{-0.33}	$&$ 0.41^{+0.37}_{-0.19}$&$ 1.14^{+0.31}_{-0.24} $&$ 0.45^{+0.14}_{-0.11}	$&$ 0.03^{+0.05}_{-0.02}	$&$ 0.05^{+0.11}_{-0.04}	$\\

\multirow{2}*{I-slow}	& \multirow{2}*{5}	& \multirow{2}*{0.24}	&$ 23.06\pm1.58			$&$ 0.01\pm0.26$ (4)	 &$ -16.76\pm0.96		 $&$ 7.69\pm0.49			$&$ -0.73\pm0.29		 	$&$ -8.55\pm0.33	 		$\\
	&			&											&$ 3.00^{+1.43}_{-0.97}	$&$ 0.07^{+0.20}_{-0.05}$&$ 1.82^{+0.80}_{-0.50} $&$ 0.86^{+0.49}_{-0.31}	$&$ 0.22^{+0.63}_{-0.16}	$&$ 0.15^{+0.52}_{-0.12}	$\\

\multirow{2}*{H-poor}	& \multirow{2}*{27}	& \multirow{2}*{0.24}	&$ 22.68\pm0.34$ 		&$ 0.07\pm0.16$ (16) 	 &$ -17.10\pm0.30		 $&$ 7.94\pm0.13			$&$ -0.61\pm0.11 	 		$&$ -8.59\pm0.10			$\\
	&			&											&$ 1.75^{+0.27}_{-0.24}	$&$ 0.50^{+0.16}_{-0.12}$&$ 1.45^{+0.23}_{-0.20} $&$ 0.62^{+0.12}_{-0.10}	$&$ 0.40^{+0.13}_{-0.10}	$&$ 0.10^{+0.24}_{-0.07}	$\\

\multirow{2}*{II}		& \multirow{2}*{3}	& \multirow{2}*{0.21}	&$ 24.46\pm1.46			$&\nodata				 &$ -15.29\pm1.48		 $&$ 7.22\pm0.93 		 	$&$ -1.27\pm0.72	 		$&$ -8.39\pm0.42			$\\
	&			&											&$ 1.77^{+1.47}_{-0.80}	$&\nodata				 &$ 2.31^{+1.50}_{-0.90} $&$ 1.18^{+0.93}_{-0.52}	$&$ 0.80^{+1.01}_{-0.45}	$&$ 0.08^{+0.26}_{-0.06}	$\\

\multirow{2}*{IIn$^{(b)}$}& \multirow{2}*{13}& \multirow{2}*{0.21}	&$ 20.37\pm0.96 $ (12)	 &$ 0.83\pm0.22	$ (10)	 &$ -18.89\pm0.67		 $&$ 9.08\pm0.35 			$&$ -0.16\pm0.39$ (9)	  	&$ -8.71\pm0.31~(9)		 	$\\
	&			&											&$ 3.25^{+0.82}_{-0.65}	$&$ 0.60^{+0.19}_{-0.14}$&$ 2.30^{+0.56}_{-0.45} $&$ 1.23^{+0.30}_{-0.24}	$&$ 1.03^{+0.36}_{-0.27}	$&$ 0.57^{+0.31}_{-0.20}	$\\

\multirow{2}*{H-rich$^{(b)}$}	& \multirow{2}*{16}	& \multirow{2}*{0.21}	&$ 21.20\pm0.90	$ (15)	 &$ 0.80\pm0.20	$ (11)	 &$ -18.18\pm0.70		 $&$ 8.74\pm0.38		 	$&$ -0.45\pm0.33$ (12)	  	&$ -8.61\pm0.23~(12)		$\\
	&			&											&$ 3.41^{+0.73}_{-0.60}	$&$ 0.57^{+0.17}_{-0.13}$&$ 2.70^{+0.57}_{-0.47} $&$ 1.37^{+0.29}_{-0.24}	$&$ 1.05^{+0.27}_{-0.24}	$&$ 0.46^{+0.32}_{-0.19}	$\\
\midrule
\multicolumn{9}{c}{\textbf{$\mathbf{0.5<z\leq1.0}$}}\\
\midrule
\multirow{2}*{H-poor}	& \multirow{2}*{14}	& \multirow{2}*{0.73}	&$ 25.24\pm0.54	$ (13)	 &$ 1.11\pm0.07	$ (4)	 &$ -17.66\pm0.44		$&$ 8.50\pm0.24				$&$ -0.10\pm0.19	 		$&$ -8.56\pm0.21	 		$\\
	&			&											&$ 1.86^{+0.47}_{-0.37}	$&$ 0.03^{+0.05}_{-0.02}$&$ 1.52^{+0.34}_{-0.28}$&$ 0.71^{+0.22}_{-0.17} 	$&$0.44^{+0.25}_{-0.16} 	$&$ 0.47^{+0.18}_{-0.13}	$\\

\midrule
\multicolumn{9}{c}{\textbf{$\mathbf{1.0<z\leq4.0}$}}\\
\midrule
\multirow{2}*{H-poor}	& \multirow{2}*{12}	& \multirow{2}*{1.67}	&$ 25.38\pm0.43	$ (11)	 &$ 1.59\pm0.60	$ (5) 	 &$ -19.86\pm0.68		$&$ 8.91\pm0.27			 	$&$ 0.70\pm0.30				$&$ -8.00\pm0.23	 		$\\
	&			&											&$ 1.32^{+0.35}_{-0.27}	$&$ 0.75^{+1.00}_{-0.43}$&$ 2.25^{+0.58}_{-0.46}$&$ 0.77^{+0.24}_{-0.18} 	$&$ 0.93^{+0.24}_{-0.19}	$&$ 0.25^{+0.54}_{-0.17}	$\\
\bottomrule
\end{tabular}
\tablecomments{
The first row of each ensemble property shows the mean value and its error and the
second row the standard deviation of the sample. The values of the $R$-band brightness, the $B$-band luminosity and the $R-K_{\rm s}$ colour
are not corrected for host attenuation.
The H-poor and H-rich samples include all SLSNe irrespective of sub-type.\\
$^{(a)}$ The number of objects with measured $R-K_{\rm s}$ colour or with an $F625W/R/r'$-band observation
are given in parenthesis, if they are less than the total number in the sample.\\
$^{(b)}$ SNe 1999bd and 2006gy are not considered in the sSFR and SFR calculations
because their star-formation histories (SFHs) is more complex than assumed in this paper, while CSS100217
and PTF11dsf are excluded because of a possible AGN contamination.
}
\label{tab:statresult}
\end{table*}

More than 87\% of all hosts were detected at $>2\sigma$ confidence in a $R$-band filter. Their
observed distribution, displayed in the upper panel of Fig. \ref{fig:colour}, extends
from $R\sim13.3$~mag (SN2006gy) to $R\sim27.9$~mag (SCP06F6) and shows a clear trend to
fainter galaxies as redshift increases (Table \ref{tab:statresult}).
The average brightness of SLSN-I host galaxies decreases from $m_R\sim22.7$~mag at $z\sim0.5$
to $m_R\sim25.4$~mag at $z>1$, while the dispersion remains at $\sim1.6$~mag at all
redshifts. Compared to a sample of star-forming galaxies from the UltraVISTA survey
(density plot in Fig. \ref{fig:colour}), they are on average fainter and their distributions
become more incompatible as redshift increases.

The class of H-poor SLSNe is comprised of fast- and slow-declining SLSNe, which might have
different progenitors and host environments. Using the gap in the decline time scale
at $\sim50$~days (Table \ref{tab:prop_gen}), we define
a sub-sample of 12 fast and seven slow declining H-poor SLSNe at $z<0.5$ (Table \ref{tab:prop_gen}).
The properties of the two samples appear to be indistinguishable (Table \ref{tab:statresult}).
However, the samples are too small to draw a conclusion yet.

Host galaxies of H-rich SLSNe are on average 1.5~mag brighter than hosts of H-poor SLSNe
at $z<0.5$ (upper panel in Fig. \ref{fig:colour}; Table \ref{tab:statresult}).
Most striking about the SLSN-II/IIn host population is the exceptionally large dispersion of 3.4~mag
that is even a factor of 2--3 larger than that of H-poor SLSNe and the UltraVISTA
sample (Tables \ref{tab:statresult}, \ref{tab:statresult_non_SLSN}; Fig. \ref{tab:statresult_non_SLSN}).
The large dispersion remains after separating out the three SLSNe-II from the H-rich
population (Table \ref{tab:prop_gen}). The distribution is incompatible with the UltraVISTA
sample (chance probability $p_{\rm ch}=7\times10^{-4}$) and with the fainter and narrower distribution of
SLSN-I host galaxies ($p_{\rm ch}=8.4\times10^{-3}$).
Among the hosts of the three SLSNe-II are two of the faintest H-rich SLSN host galaxies in our sample
($R\sim24.6$--26.4; Table \ref{tab:data}). They are more than a hundred times fainter than an
$L^\star_{B}$ galaxy at $z\sim0.2$ \citep{Faber2007a}, and about two magnitudes fainter than the SMC
galaxy at $z\sim0.2$.

Panel A of Figure \ref{fig:sed_result} shows the evolution of the absolute $B$-band
luminosity (not corrected for host reddening) with redshift. The distribution spans a wide range from
$-13$ to $-22$ mag. Compared
with appropriate luminosity functions \citep[e.g.,][tracks in Fig. \ref{fig:sed_result}]{Faber2007a,Ilbert2005a,Marchesini2007a},
the span corresponds to a range from a few thousandths of $L^\star$
to a few $L^\star$. Clear differences are visible between hosts of H-poor and -rich SLSNe.
In their common redshift interval ($z<0.5$), the distribution of the H-poor SLSN hosts is
narrower by $>1$~mag and in addition shifted by $\sim1$~mag towards lower luminosities
(Table \ref{tab:statresult}). Intriguingly, the luminosity distribution shows a rapid evolution
from $0.04~L^\star$ at $z<1$ to $\sim0.2~L^\star$ at $z>1$. We discuss its origin
in Sect. \ref{sec:redshift_evol}.

With the $B$-band luminosity distribution in hand we put SLSN host galaxies into context with
unbiased GRB and regular core-collapse SN host galaxy samples. Between $z=0.3$ and $z=1$,
Type I SLSNe reside in galaxies that are $1.61\pm0.42$~mag less luminous than GRBs. The AD test
gives a chance probability of
$p_{\rm ch}=2\times10^{-4}$ that both distributions are drawn from the same
parent distribution (Fig. \ref{fig:statcomp_1}). This result contradicts \citet{Japelj2016a}, who argued that
previously claimed differences
between the two populations are an artefact of the comparison methodology. We discuss this
finding in Sec. \ref{sec:class_diff_hpoor} in detail. The population of SLSN-I
host galaxies is also incompatible with those of regular core-collapse SNe from
untargeted surveys at all redshifts ($p_{\rm ch}<1\times10^{-5}$; Figs. \ref{fig:statcomp_1}).
In contrast, the SLSN-IIn host population is closer to the GRB host population ($p_{\rm ch}> 0.26$;
Figs. \ref{fig:statcomp_2}).

\subsubsection{$R-K_{\rm s}$ colour}\label{sec:colour}

The middle panel of Fig. \ref{fig:colour} shows the redshift evolution of the $R-K_{\rm s}$ colour of the
25 H-poor and 11 H-rich SLSN hosts with $R$ and $K_{\rm s}$-band observations. The colour varies
between $\sim-2$ and 3~mag, though with large errors. No SLSNe are found in extremely red objects (EROs,
$R-K_{\rm s}\geq3.3~{\rm mag}$).
At $z<0.5$, SLSN-I hosts
are characterised by significantly bluer average colours ($R-K_{\rm s}\sim0.07$~mag; Table \ref{tab:statresult})
than star-forming galaxies from the UltraVISTA survey (grey shaded region; $R-K_{\rm s}\sim1.10$~mag;
Table \ref{tab:statresult_non_SLSN}). The chance of randomly drawing a distribution from the UltraVISTA
sample that is at least as extreme as the SLSN-I is $<10^{-5}$.
The average colour is $>0.45\pm0.19$ mag bluer and statistically incompatible with those
extreme emission galaxies in the VUDS and zCOSMOS surveys ($p_{\rm ch}< 1 \times10^{-2}$).
At $z>1$, the average colour increases to $1.59\pm0.60$~mag, but still remains
below the average colour of UltraVISTA galaxies (2.43~mag; Tables \ref{tab:statresult}, \ref{tab:statresult_non_SLSN}).

The mean colour of hydrogen-rich SLSNe ($R-K_{\rm s}\sim0.80$~mag) is modestly bluer
compared to the general population of star-forming galaxies in the UltraVISTA survey and of GRB host galaxies
(Tables \ref{tab:statresult}, \ref{tab:statresult_non_SLSN}). While the dispersions of the brightness
and luminosity distributions are broader than of other galaxy samples, the colour distribution
has a dispersion comparable to all other samples [$\sigma(R-K_{\rm s})\sim0.57$~mag;
Tables \ref{tab:statresult}, \ref{tab:statresult_non_SLSN}]. Hosts of type II SLSNe tend to
be too faint to obtain meaningful $K_{\rm s}$-band constraints, which prevents contrasting
their properties to the ensemble of type IIn SLSNe.

In the bottom panel of Fig. \ref{fig:colour}, we overlay expected colour-tracks for the
stellar population synthesis templates from \citet{Bruzual2003a} for a metallicity of 0.2 solar
and a wide range of ages. The colour of SLSN-I hosts of $\sim0$~mag at $z<0.5$
points to stellar population ages of several up to a few hundred million years, whereas H-rich SLSNe are found in
galaxies with a redder $R-K_{\rm s}$ colour because of more evolved stellar populations.
However, the exact relation between colour and age is a complicated function of metallicity, extinction,
the equivalent width of emission lines and star-formation histories (for a detailed discussion see \citealt{Conroy2013a}).
The vectors in Fig. \ref{fig:colour} indicate how they can alter the intrinsic colour.

A critical aspect of this analysis is the $R$ and $K_{\rm s}$-band observing completeness. Almost all hosts
were observed in $R$ band, but only $\sim57\%$ were observed in $K_{\rm s}$ band. The colour incompleteness
is a direct consequence of the difficulty to obtain meaningful $K_{\rm s}$-band constraints for hosts fainter
than $K_{\rm s}=23$--24~mag. This is supported by the SED modelling, which always suggests $K_{\rm s}$-band magnitudes
below this detection limit and colours that are comparable to the observed colour distribution. In the
unlikely case that the hosts without $K_{\rm s}$-band observations had $K_{\rm s}=23$--24~mag, the colour distribution
would span a range from 0.3 to 4.7~mag. Such red colours are in stark contrast to the observed
distribution, the SED modelling, and SN observations
\citep[e.g.,][]{Quimby2011a, Inserra2013a, Lunnan2013a, Nicholl2014a}.

\subsection{Physical properties and distribution functions}\label{sec:distrib_function}

\begin{figure}
\includegraphics[width=1\columnwidth]{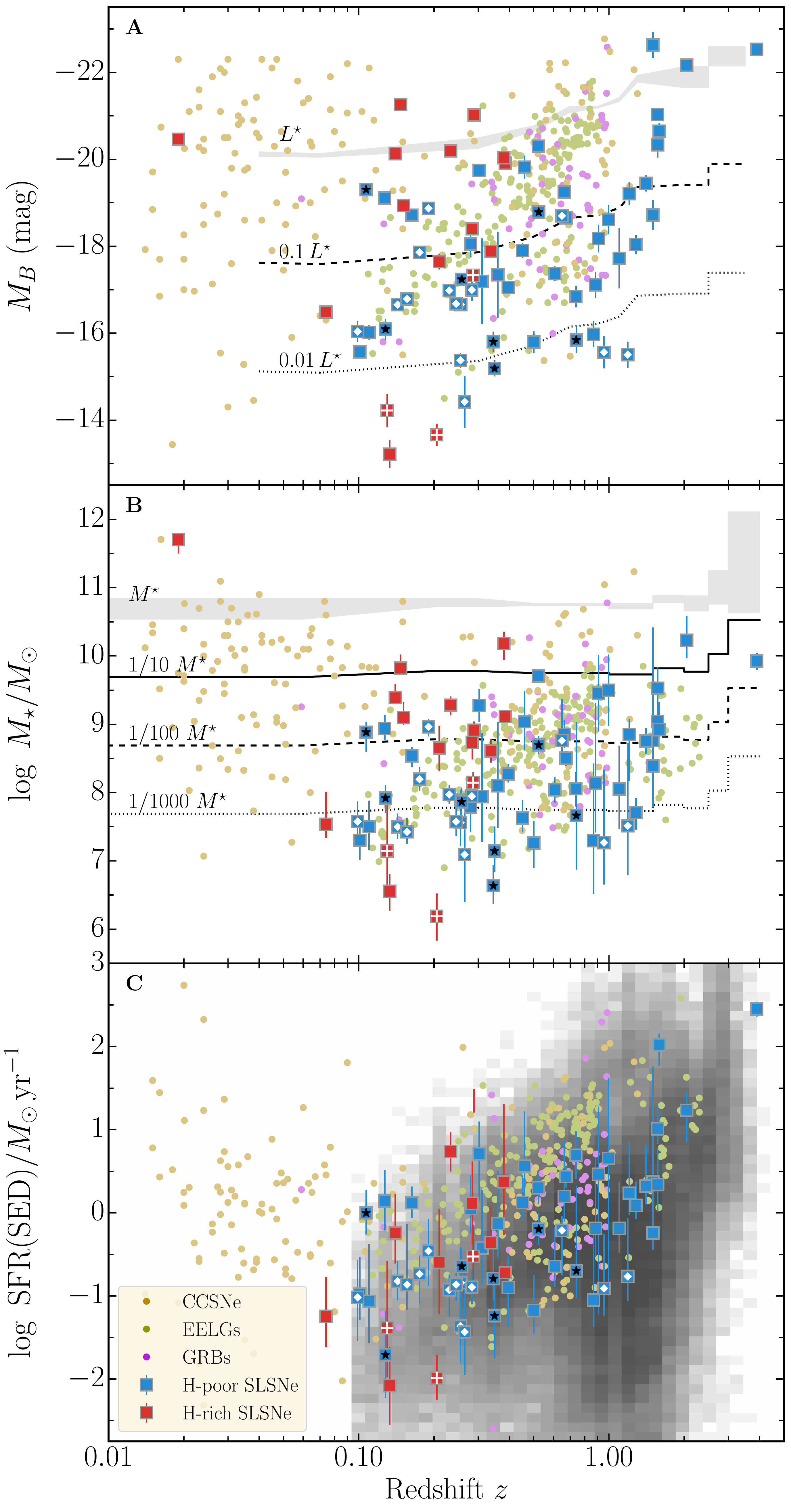}
\caption{
Evolution of the physical properties of SLSN host galaxies and comparison samples with redshift.
Symbols are identical to previous figures. In panel A, we overlay the evolution  of the
characteristic luminosity $L^\star$  of the $B$-band luminosity function of blue galaxies,
reported in \citet{Faber2007a}, \citet{Ilbert2005a} and \citet{Marchesini2007a} in grey, and
several luminosity tracks. In panel B, we overlay the evolution of the characteristic mass $M^\star$ of the mass
function from the GAMA \citep{Baldry2012a} and UltraVISTA surveys in grey, and several mass tracks.
These characteristic masses and luminosities are defined where the power-law form of the
Schechter function cuts off. The parameter space of the UltraVISTA sample is shown as a grey-shaded
density plot in panel C. For clarity, measurement errors are omitted for the comparison
samples. They are comparable to those of the SLSN host galaxies.
}
\label{fig:sed_result}
\end{figure}

\begin{table*}
\caption{Results from the spectral energy distribution modelling}
\begin{tiny}
\centering
\begin{tabular}{lccccccrrrrc}
\toprule
\multirow{2}*{SLSN}		& \multirow{2}*{Redshift}	& \multirow{2}*{$ \chi^2/{\rm n.o.f.}	$}	&$ E(B-V)		$&$ M_{\rm FUV}		$&$ M_{\rm B}	$&$ M_{\rm Ks}	$&\multicolumn{1}{c}{$ \log\,{\rm SFR}			$}& \multicolumn{1}{c}{$ \log\,M	$}& \multicolumn{1}{c}{$ \log\,{\rm sSFR}	$}&$ \log\,{\rm Age}$\\
				&							&											& (mag; host)	& (mag)				& 	(mag)		& (mag)			& \multicolumn{1}{c}{$(M_\odot\,{\rm yr}^{-1})$}& \multicolumn{1}{c}{$(M_\odot)$}	& \multicolumn{1}{c}{ $({\rm yr}^{-1})$}	& (yr)\\
\midrule
\multicolumn{10}{l}{\textbf{SLSN-I host galaxies}}\\
\midrule
CSS140925 	 &0.460 	&$  0.32 /  4 	$&$ 0.50 	$&$ -17.96\pm0.24 	$&$ -19.82\pm0.26 	$&$ <-21.12		 	$&$   0.56^{+0.66}_{-0.34} 	$&$   9.04^{+0.44}_{-0.41} 	$&$  -8.34^{+0.68}_{-0.67} 	$&$   8.37^{+0.57}_{-0.65}$\\
DES14S2qri 	 &1.500 	&$  1.06 /  4 	$&$ 0.07 	$&$ -18.19\pm0.82 	$&$ <-22.63 		$&$ <-24.15	 		$&$   0.37^{+1.39}_{-0.34} 	$&$   8.76^{+1.65}_{-0.87} 	$&$  -8.14^{+0.73}_{-1.07} 	$&$   8.19^{+0.80}_{-0.69}$\\
DES14X2byo 	 &0.869 	&$  0.00 /  2 	$&$ 0.30 	$&$ -14.86\pm0.99 	$&$ <-15.97 		$&$ <-16.71	 		$&$  -1.05^{+0.83}_{-0.32} 	$&$   7.30^{+1.13}_{-0.78} 	$&$  -8.14^{+0.72}_{-1.06} 	$&$   8.19^{+0.79}_{-0.68}$\\
DES14X3taz 	 &0.608 	&$  4.78 /  8 	$&$ 0.00 	$&$ -16.57\pm0.20 	$&$ -17.37\pm0.17 	$&$ -17.13\pm0.19 	$&$  -0.65^{+0.48}_{-0.24} 	$&$   8.04^{+0.19}_{-0.19} 	$&$  -8.64^{+0.46}_{-0.34} 	$&$   8.60^{+0.33}_{-0.39}$\\
iPTF13ajg$^\dagger$ 	 &0.740 	&$  0.00 /  1 	$&$ 0.00 	$&$ -16.68\pm0.21 	$&$ <-15.84		 	$&$ <-15.08	 	 	$&$  -0.70^{+1.02}_{-0.33} 	$&$   7.66^{+1.36}_{-0.79} 	$&$  -8.15^{+0.74}_{-1.15} 	$&$   8.22^{+0.82}_{-0.71}$\\
LSQ12dlf$^\ddagger$ &0.255 	&$  0.54 /  5 	$&$ 0.00 	$&$ -14.72\pm0.25 	$&$ -15.38\pm0.17 	$&$ -15.91\pm0.31 	$&$  -1.36^{+0.54}_{-0.43} 	$&$   7.56^{+0.33}_{-0.34} 	$&$  -8.86^{+0.75}_{-0.85} 	$&$   8.73^{+0.73}_{-0.57}$\\
 LSQ14an 	 &0.163 	&$  5.60 / 10 	$&$ 0.01 	$&$ -18.34\pm0.26 	$&$ -18.71\pm0.09 	$&$ -18.60\pm0.11 	$&$   0.12^{+0.20}_{-0.18} 	$&$   8.54^{+0.13}_{-0.17} 	$&$  -8.42^{+0.27}_{-0.20} 	$&$   8.48^{+0.20}_{-0.29}$\\
 LSQ14mo$^\ddagger$&0.256 	&$  2.37 /  5 	$&$ 0.00 	$&$ -15.92\pm0.08 	$&$ -16.66\pm0.11 	$&$ -16.95\pm0.13 	$&$  -0.84^{+0.42}_{-0.34} 	$&$   7.89^{+0.15}_{-0.19} 	$&$  -8.77^{+0.62}_{-0.43} 	$&$   8.67^{+0.35}_{-0.48}$\\
LSQ14bdq$^\dagger$ 	 &0.345 	&$  5.77 /  5 	$&$ 0.00 	$&$ -16.46\pm0.21 	$&$ -15.80\pm0.23 	$&$ <-14.09	 	 	$&$  -0.79^{+0.39}_{-0.26} 	$&$   6.64^{+0.30}_{-0.27} 	$&$  -7.41^{+0.63}_{-0.52} 	$&$   7.50^{+0.47}_{-0.76}$\\
LSQ14fxj 	 &0.360 	&$  0.27 /  3 	$&$ 0.00 	$&$ -18.40\pm0.99 	$&$ -17.34\pm0.99 	$&$ <-16.03	 	 	$&$  -0.13^{+0.63}_{-0.41} 	$&$   8.10^{+0.94}_{-0.62} 	$&$  -8.10^{+0.71}_{-1.09} 	$&$   8.16^{+0.83}_{-0.67}$\\
MLS121104 	 &0.303 	&$  8.46 /  7 	$&$ 0.20 	$&$ -18.18\pm0.17 	$&$ -19.74\pm0.14 	$&$ -20.57\pm0.13 	$&$   0.71^{+0.39}_{-0.56} 	$&$   9.27^{+0.25}_{-0.24} 	$&$  -8.56^{+0.59}_{-0.82} 	$&$   8.60^{+0.86}_{-0.57}$\\
PS1-10ky 	 &0.956 	&$  0.01 /  4 	$&$ 0.20 	$&$ -15.66\pm0.99 	$&$ -15.56\pm0.37 	$&$ <-13.73	 	 	$&$  -0.91^{+0.65}_{-0.33} 	$&$   7.27^{+1.07}_{-0.61} 	$&$  -8.06^{+0.68}_{-0.99} 	$&$   8.10^{+0.86}_{-0.63}$\\
PS1-10pm 	 &1.206 	&$  0.30 /  4 	$&$ 0.50 	$&$ <-17.61	 		$&$ -19.21\pm0.26 	$&$ -19.68\pm0.09 	$&$   0.24^{+0.62}_{-0.26} 	$&$   8.85^{+0.23}_{-0.69} 	$&$  -8.42^{+0.79}_{-0.57} 	$&$   8.45^{+0.52}_{-0.74}$\\
PS1-10ahf 	 &1.158 	&$  3.98 /  5 	$&$ 0.30 	$&$ -17.10\pm0.25 	$&$ -17.72\pm0.99 	$&$ -17.62\pm0.99 	$&$  -0.19^{+0.56}_{-0.29} 	$&$   8.05^{+0.63}_{-0.59} 	$&$  -8.10^{+0.66}_{-0.75} 	$&$   8.15^{+0.59}_{-0.62}$\\
PS1-10awh 	 &0.909 	&$  0.09 /  4 	$&$ 0.50 	$&$ <-14.11		 	$&$ -18.18\pm0.32 	$&$ -22.01\pm0.28 	$&$   0.46^{+0.82}_{-1.68} 	$&$   9.45^{+0.56}_{-0.56} 	$&$  -9.22^{+0.98}_{-1.36} 	$&$   8.97^{+0.63}_{-0.67}$\\
PS1-10bzj$^\ddagger$ 	 &0.649 	&$ 51.95 /  5 	$&$ 0.00 	$&$ -18.64\pm0.09 	$&$ -18.70\pm0.12 	$&$ -18.18\pm0.17 	$&$  -0.21^{+0.17}_{-0.54} 	$&$   8.76^{+0.61}_{-0.35} 	$&$  -8.95^{+0.49}_{-1.12} 	$&$   8.93^{+0.66}_{-0.41}$\\
PS1-11ap$^\dagger$ 	 &0.524 	&$  1.83 /  5 	$&$ 0.00 	$&$ -18.00\pm0.05 	$&$ -18.79\pm0.11 	$&$ -18.55\pm0.37 	$&$  -0.20^{+0.19}_{-0.19} 	$&$   8.70^{+0.13}_{-0.13} 	$&$  -8.89^{+0.22}_{-0.21} 	$&$   8.71^{+0.28}_{-0.24}$\\
PS1-11tt 	 &1.283 	&$  0.00 /  2 	$&$ 0.00 	$&$ <-18.49		 	$&$ -18.04\pm0.22 	$&$ -17.24\pm0.07 	$&$   0.09^{+0.29}_{-0.17} 	$&$   7.71^{+0.22}_{-0.25} 	$&$  -7.58^{+0.38}_{-0.35} 	$&$   7.65^{+0.34}_{-0.40}$\\
PS1-11afv 	 &1.407 	&$  0.00 /  2 	$&$ 0.10 	$&$ <-18.70	 		$&$ -19.45\pm0.19 	$&$ -19.79\pm0.09 	$&$   0.32^{+0.50}_{-0.22} 	$&$   8.76^{+0.19}_{-0.19} 	$&$  -8.39^{+0.49}_{-0.35} 	$&$   8.42^{+0.29}_{-0.46}$\\
PS1-11aib 	 &0.997 	&$  1.34 /  5 	$&$ 0.20 	$&$ -15.62\pm0.71 	$&$ -18.61\pm0.34 	$&$ -21.11\pm0.32 	$&$   0.65^{+0.97}_{-1.65} 	$&$   9.50^{+0.52}_{-0.52} 	$&$  -9.01^{+1.02}_{-1.52} 	$&$   8.88^{+0.67}_{-0.84}$\\
PS1-11bam 	 &1.565 	&$  0.56 /  5 	$&$ 0.02 	$&$ -20.81\pm0.14 	$&$ -21.03\pm0.15 	$&$ -20.66\pm0.15 	$&$   1.01^{+0.29}_{-0.18} 	$&$   9.04^{+0.37}_{-0.37} 	$&$  -8.01^{+0.56}_{-0.52} 	$&$   8.04^{+0.52}_{-0.51}$\\
PS1-11bdn 	 &0.738 	&$ 31.69 /  5 	$&$ 0.50 	$&$ -15.19\pm0.11 	$&$ -16.84\pm0.25 	$&$ <-16.49		 	$&$   0.69^{+0.29}_{-0.29} 	$&$   8.06^{+0.25}_{-0.26} 	$&$  -7.42^{+0.51}_{-0.48} 	$&$   7.50^{+0.45}_{-0.55}$\\
PS1-12zn 	 &0.674 	&$ 12.73 / 12 	$&$ 0.20 	$&$ -17.76\pm0.14 	$&$ -18.65\pm0.06 	$&$ -19.37\pm0.07 	$&$   0.43^{+0.13}_{-0.13} 	$&$   8.50^{+0.11}_{-0.12} 	$&$  -8.06^{+0.16}_{-0.21} 	$&$   8.14^{+0.23}_{-0.18}$\\
PS1-12bmy 	 &1.566 	&$  3.22 /  6 	$&$ 0.00 	$&$ -18.96\pm0.11 	$&$ -20.33\pm0.29 	$&$ <-19.79	 	 	$&$   0.34^{+0.44}_{-0.33} 	$&$   9.53^{+0.31}_{-0.26} 	$&$  -9.31^{+0.69}_{-0.36} 	$&$   8.92^{+0.58}_{-0.42}$\\
PS1-12bqf 	 &0.522 	&$  9.88 / 15 	$&$ 0.10 	$&$ -18.55\pm0.09 	$&$ -20.30\pm0.05 	$&$ -21.14\pm0.07 	$&$   0.30^{+0.17}_{-0.19} 	$&$   9.71^{+0.04}_{-0.04} 	$&$  -9.40^{+0.16}_{-0.23} 	$&$   8.93^{+0.08}_{-0.05}$\\
PS1-13gt 	 &0.884 	&$  0.00 /  1 	$&$ 0.00 	$&$ <-18.13	 		$&$ <-17.11		 	$&$ <-16.04	 	 	$&$  -0.19^{+0.84}_{-0.35} 	$&$   8.14^{+1.18}_{-0.72} 	$&$  -8.15^{+0.74}_{-1.11} 	$&$   8.22^{+0.82}_{-0.71}$\\
PTF09atu 	 &0.501 	&$  0.93 /  5 	$&$ 0.00 	$&$ <-15.22		 	$&$ -15.80\pm0.25 	$&$ <-14.98 		$&$  -1.18^{+0.42}_{-0.27} 	$&$   7.26^{+0.32}_{-0.36} 	$&$  -8.38^{+0.61}_{-0.55} 	$&$   8.40^{+0.47}_{-0.60}$\\
PTF09cnd$^\dagger$ 	 &0.258 	&$  2.67 /  6 	$&$ 0.00 	$&$ -16.97\pm0.32 	$&$ -17.24\pm0.08 	$&$ -16.87\pm0.46 	$&$  -0.64^{+0.21}_{-0.18} 	$&$   7.87^{+0.20}_{-0.21} 	$&$  -8.49^{+0.32}_{-0.31} 	$&$   8.52^{+0.21}_{-0.31}$\\
PTF10hgi$^\ddagger$ 	 &0.099 	&$  7.40 /  7 	$&$ 0.01 	$&$ -14.36\pm0.24 	$&$ -16.04\pm0.24 	$&$ -16.09\pm0.18 	$&$  -1.02^{+0.44}_{-0.52} 	$&$   7.58^{+0.29}_{-0.31} 	$&$  -8.62^{+0.69}_{-0.71} 	$&$   8.61^{+0.38}_{-0.61}$\\
PTF10vqv 	 &0.452 	&$  0.51 /  6 	$&$ 0.07 	$&$ -18.60\pm0.12 	$&$ -17.90\pm0.17 	$&$ -16.92\pm0.99 	$&$   0.12^{+0.33}_{-0.21} 	$&$   7.63^{+0.26}_{-0.21} 	$&$  -7.48^{+0.46}_{-0.37} 	$&$   7.55^{+0.36}_{-0.54}$\\
PTF11rks$^\ddagger$ 	 &0.190 	&$  8.72 /  9 	$&$ 0.00 	$&$ -17.27\pm0.50 	$&$ -18.87\pm0.07 	$&$ -19.20\pm0.41 	$&$  -0.46^{+0.38}_{-0.43} 	$&$   8.96^{+0.12}_{-0.14} 	$&$  -9.43^{+0.44}_{-0.46} 	$&$   8.98^{+0.36}_{-0.31}$\\
PTF12dam$^\dagger$ 	 &0.107 	&$ 24.66 / 13 	$&$ 0.02 	$&$ -18.65\pm0.19 	$&$ -19.30\pm0.05 	$&$ -18.61\pm0.32 	$&$  -0.00^{+0.27}_{-0.26} 	$&$   8.89^{+0.15}_{-0.30} 	$&$  -8.87^{+0.35}_{-0.19} 	$&$   8.69^{+0.29}_{-0.36}$\\
 SCP06F6$^\ddagger$&1.189 	&$  0.00 /  1 	$&$ 0.00 	$&$ -16.56\pm0.21 	$&$ <-15.50		 	$&$ <-14.19	 	 	$&$  -0.77^{+0.70}_{-0.30} 	$&$   7.51^{+1.21}_{-0.72} 	$&$  -8.10^{+0.71}_{-1.05} 	$&$   8.14^{+0.81}_{-0.66}$\\
SN1999as 	 &0.127 	&$ 17.97 / 11 	$&$ 0.10 	$&$ -17.93\pm0.28 	$&$ -19.11\pm0.08 	$&$ -19.38\pm0.13 	$&$   0.14^{+0.37}_{-0.35} 	$&$   8.94^{+0.20}_{-0.17} 	$&$  -8.83^{+0.48}_{-0.39} 	$&$   8.65^{+0.33}_{-0.33}$\\
SN2005ap$^\ddagger$&0.283 	&$ 10.52 / 10 	$&$ 0.00 	$&$ -16.44\pm0.10 	$&$ -17.00\pm0.25 	$&$ -16.21\pm0.36 	$&$  -0.89^{+0.19}_{-0.21} 	$&$   7.95^{+0.11}_{-0.15} 	$&$  -8.82^{+0.26}_{-0.21} 	$&$   8.66^{+0.25}_{-0.22}$\\
SN2006oz 	 &0.396 	&$ 15.05 /  7 	$&$ 0.15 	$&$ -15.50\pm0.25 	$&$ -17.05\pm0.08 	$&$ -17.53\pm0.22 	$&$  -0.90^{+0.37}_{-0.47} 	$&$   8.27^{+0.14}_{-0.12} 	$&$  -9.18^{+0.38}_{-0.52} 	$&$   8.89^{+0.41}_{-0.34}$\\
SN2007bi$^\dagger$ 	 &0.128 	&$  6.62 /  9 	$&$ 0.04 	$&$ -14.52\pm0.34 	$&$ -16.09\pm0.24 	$&$ -15.73\pm0.24 	$&$  -1.71^{+0.53}_{-0.52} 	$&$   7.92^{+0.20}_{-0.21} 	$&$  -9.68^{+0.65}_{-0.41} 	$&$   8.88^{+0.15}_{-0.27}$\\
SN2009de 	 &0.311 	&$  0.99 /  4 	$&$ 0.30 	$&$ -16.50\pm0.99 	$&$ -17.19\pm0.99 	$&$ <-17.87	 	 	$&$  -0.42^{+0.70}_{-0.45} 	$&$   7.94^{+0.93}_{-0.66} 	$&$  -8.20^{+0.76}_{-1.27} 	$&$   8.26^{+1.04}_{-0.73}$\\
SN2009jh$^\dagger$ 	 &0.349 	&$  5.87 /  5 	$&$ 0.15 	$&$ <-14.85		 	$&$ -15.19\pm0.18 	$&$ -15.70\pm0.16 	$&$  -1.24^{+0.42}_{-0.51} 	$&$   7.15^{+0.36}_{-0.30} 	$&$  -8.36^{+0.69}_{-0.92} 	$&$   8.42^{+0.60}_{-0.69}$\\
SN2010gx$^\ddagger$ 	 &0.230 	&$  2.85 /  5 	$&$ 0.00 	$&$ -16.30\pm0.06 	$&$ -16.98\pm0.06 	$&$ -16.96\pm0.05 	$&$  -0.93^{+0.19}_{-0.32} 	$&$   7.97^{+0.14}_{-0.13} 	$&$  -8.89^{+0.23}_{-0.37} 	$&$   8.87^{+0.13}_{-0.30}$\\
SN2010kd	 	 &0.101 	&$  4.37 /  5 	$&$ 0.15 	$&$ -15.83\pm0.53 	$&$ -15.57\pm0.07 	$&$ -15.22\pm0.07 	$&$  -0.98^{+0.44}_{-0.31} 	$&$   7.30^{+0.25}_{-0.29} 	$&$  -8.25^{+0.61}_{-0.52} 	$&$   8.30^{+0.40}_{-0.59}$\\
SN2011ep 	 &0.280 	&$  0.02 /  4 	$&$ 0.15 	$&$ -18.17\pm0.41 	$&$ <-18.05		 	$&$ <-16.02	 	 	$&$   0.05^{+0.41}_{-0.30} 	$&$   7.79^{+0.42}_{-0.36} 	$&$  -7.71^{+0.51}_{-0.58} 	$&$   7.74^{+0.59}_{-0.49}$\\
SN2011ke$^\ddagger$ 	 &0.143 	&$  2.70 /  6 	$&$ 0.00 	$&$ -16.38\pm0.09 	$&$ -16.66\pm0.06 	$&$ -16.57\pm0.27 	$&$  -0.82^{+0.24}_{-0.23} 	$&$   7.50^{+0.20}_{-0.18} 	$&$  -8.34^{+0.31}_{-0.23} 	$&$   8.40^{+0.25}_{-0.33}$\\
SN2011kf$^\ddagger$ 	 &0.245 	&$  4.60 /  6 	$&$ 0.00 	$&$ -16.41\pm0.08 	$&$ -16.68\pm0.08 	$&$ -15.72\pm0.43 	$&$  -0.86^{+0.18}_{-0.20} 	$&$   7.58^{+0.19}_{-0.22} 	$&$  -8.43^{+0.33}_{-0.32} 	$&$   8.43^{+0.24}_{-0.29}$\\
SN2012il$^\ddagger$ 	 &0.175 	&$ 12.26 / 10 	$&$ 0.02 	$&$ -16.82\pm0.37 	$&$ -17.86\pm0.09 	$&$ -17.19\pm0.21 	$&$  -0.74^{+0.22}_{-0.36} 	$&$   8.20^{+0.18}_{-0.17} 	$&$  -8.92^{+0.27}_{-0.48} 	$&$   8.68^{+0.27}_{-0.19}$\\
SN2013dg$^\ddagger$ 	 &0.265 	&$  0.11 /  3 	$&$ 0.80 	$&$ -11.05\pm0.68 	$&$ -14.42\pm0.60 	$&$ <-17.15	 	 	$&$  -1.43^{+0.80}_{-0.52} 	$&$   7.09^{+0.82}_{-0.70} 	$&$  -8.34^{+0.81}_{-1.32} 	$&$   8.39^{+1.09}_{-0.78}$\\
SN2013hy 	 &0.663 	&$  0.31 /  4 	$&$ 0.01 	$&$ -18.39\pm0.14 	$&$ -19.25\pm0.11 	$&$ -19.06\pm0.18 	$&$   0.20^{+0.68}_{-0.30} 	$&$   8.85^{+0.21}_{-0.19} 	$&$  -8.59^{+0.62}_{-0.42} 	$&$   8.56^{+0.39}_{-0.54}$\\
SN2015bn 	 &0.110 	&$  8.18 /  6 	$&$ 0.30 	$&$ -14.81\pm0.59 	$&$ -16.02\pm0.17 	$&$ -17.27\pm0.41 	$&$  -1.06^{+0.69}_{-0.50} 	$&$   7.50^{+0.38}_{-0.35} 	$&$  -8.51^{+0.72}_{-0.73} 	$&$   8.52^{+0.50}_{-0.66}$\\
SN1000+0216 	 &3.899 	&$ 42.94 / 11 	$&$ 0.30 	$&$ -21.52\pm0.08 	$&$ -22.53\pm0.08 	$&$ -23.65\pm0.28 	$&$   2.45^{+0.09}_{-0.09} 	$&$   9.93^{+0.12}_{-0.13} 	$&$  -7.46^{+0.14}_{-0.18} 	$&$   7.53^{+0.16}_{-0.21}$\\
SN2213-1745 	 &2.046 	&$  0.34 /  6 	$&$ 0.02 	$&$ -21.00\pm0.05 	$&$ -22.16\pm0.13 	$&$ -21.44\pm0.13 	$&$   1.23^{+0.23}_{-0.38} 	$&$  10.23^{+0.36}_{-0.26} 	$&$  -9.15^{+0.55}_{-0.34} 	$&$   8.90^{+0.43}_{-0.41}$\\
SNLS06D4eu 	 &1.588 	&$  2.47 /  7 	$&$ 0.30 	$&$ -20.00\pm0.05 	$&$ -20.65\pm0.18 	$&$ <-19.34		 	$&$   2.02^{+0.14}_{-0.25} 	$&$   8.92^{+0.41}_{-0.11} 	$&$  -6.91^{+0.25}_{-0.65} 	$&$   6.96^{+0.65}_{-0.27}$\\
SNLS07D2bv 	 &1.500 	&$  5.34 /  9 	$&$ 0.00 	$&$ -17.66\pm0.20 	$&$ -18.72\pm0.34 	$&$ -19.23\pm0.60 	$&$  -0.24^{+0.35}_{-0.20} 	$&$   8.39^{+0.62}_{-0.60} 	$&$  -8.58^{+0.76}_{-0.79} 	$&$   8.57^{+0.74}_{-0.68}$\\
SSS120810$^\ddagger$&0.156 	&$  5.20 /  7 	$&$ 0.00 	$&$ -16.61\pm0.18 	$&$ -16.79\pm0.11 	$&$ -15.80\pm0.24 	$&$  -0.86^{+0.73}_{-0.31} 	$&$   7.42^{+0.21}_{-0.17} 	$&$  -8.35^{+1.00}_{-0.31} 	$&$   8.27^{+0.30}_{-0.82}$\\
\midrule
\multicolumn{10}{l}{\textbf{SLSN-IIn host galaxies}}\\
\midrule
CSS100217 	 &0.147 	&$ 78.34 / 11 	$&$ 0.50 	$&$ -19.64\pm0.09 	$&$ -21.26\pm0.05 	$&$ -21.76\pm0.05 	$&$   2.35^{+0.24}_{-0.14} 	$&$   9.82^{+0.20}_{-0.07} 	$&$  -7.46^{+0.10}_{-0.11} 	$&$   7.53^{+0.08}_{-0.05}$\\
PTF10heh 	 &0.338 	&$  2.97 /  7 	$&$ 0.15 	$&$ -15.84\pm0.21 	$&$ -17.88\pm0.09 	$&$ -18.66\pm0.13 	$&$  -0.36^{+0.44}_{-0.41} 	$&$   8.61^{+0.17}_{-0.17} 	$&$  -8.97^{+0.56}_{-0.54} 	$&$   8.87^{+0.60}_{-0.47}$\\
PTF10qaf 	 &0.284 	&$  0.49 /  6 	$&$ 0.00 	$&$ -17.71\pm0.16 	$&$ -18.40\pm0.15 	$&$ -18.98\pm0.19 	$&$   0.11^{+0.50}_{-0.50} 	$&$   8.73^{+0.22}_{-0.25} 	$&$  -8.61^{+0.71}_{-0.73} 	$&$   8.63^{+0.69}_{-0.67}$\\
PTF11dsf 	 &0.385 	&$ 18.56 /  8 	$&$ 0.00 	$&$ -18.84\pm0.30 	$&$ -19.91\pm0.07 	$&$ -19.09\pm0.38 	$&$  -0.72^{+0.12}_{-0.13} 	$&$   9.12^{+0.07}_{-0.07} 	$&$  -9.85^{+0.13}_{-0.16} 	$&$   8.67^{+0.04}_{-0.05}$\\
SN1999bd 	 &0.151 	&$ 52.37 / 11 	$&$ 0.80 	$&$ -16.02\pm0.33 	$&$ -18.95\pm0.07 	$&$ -19.99\pm0.10 	$&$   2.42^{+0.22}_{-0.10} 	$&$   9.10^{+0.22}_{-0.11} 	$&$  -6.70^{+0.07}_{-0.07} 	$&$   6.70^{+0.03}_{-0.03}$\\
SN2003ma 	 &0.289 	&$  1.43 /  4 	$&$ 0.04 	$&$ -21.31\pm0.08 	$&$ -21.02\pm0.07 	$&$ -20.91\pm0.17 	$&$   1.34^{+0.16}_{-0.13} 	$&$   8.91^{+0.13}_{-0.12} 	$&$  -7.55^{+0.21}_{-0.24} 	$&$   7.63^{+0.24}_{-0.29}$\\
SN2006gy 	 &0.019 	&$ 46.32 / 10 	$&$ 0.15 	$&$ -13.54\pm0.10 	$&$ -20.46\pm0.05 	$&$ -22.99\pm0.05 	$&$  -1.12^{+0.08}_{-0.08} 	$&$  11.70^{+0.06}_{-0.21} 	$&$  -6.78^{+0.12}_{-7.17} 	$&$   9.91^{+0.10}_{-3.18}$\\
SN2006tf 	 &0.074 	&$  8.54 / 11 	$&$ 0.07 	$&$ -15.45\pm0.18 	$&$ -16.49\pm0.06 	$&$ -17.05\pm0.10 	$&$  -1.25^{+0.48}_{-0.37} 	$&$   7.54^{+0.47}_{-0.20} 	$&$  -8.88^{+0.35}_{-0.29} 	$&$   8.86^{+0.40}_{-0.35}$\\
SN2007bw 	 &0.140 	&$ 14.85 /  8 	$&$ 0.04 	$&$ -17.72\pm0.25 	$&$ -20.13\pm0.06 	$&$ -20.59\pm0.09 	$&$  -0.24^{+0.47}_{-0.37} 	$&$   9.39^{+0.19}_{-0.09} 	$&$  -9.74^{+0.53}_{-0.25} 	$&$   8.70^{+0.30}_{-0.07}$\\
SN2008am 	 &0.233 	&$  5.63 / 12 	$&$ 0.20 	$&$ -19.02\pm0.11 	$&$ -20.19\pm0.06 	$&$ -20.70\pm0.14 	$&$   0.74^{+0.23}_{-0.24} 	$&$   9.28^{+0.13}_{-0.15} 	$&$  -8.50^{+0.20}_{-0.35} 	$&$   8.57^{+0.33}_{-0.24}$\\
\bottomrule
\end{tabular}
\end{tiny}
\label{tab:sed_results}
\end{table*}

\begin{table*}
\contcaption{Results from the spectral energy distribution modelling}
\begin{tiny}
\centering
\begin{tabular}{lccccccrrrrc}
\toprule
\multirow{2}*{SLSN}		& \multirow{2}*{Redshift}	& \multirow{2}*{$ \chi^2/{\rm n.o.f.}	$}	&$ E(B-V)		$&$ M_{\rm FUV}		$&$ M_{\rm B}	$&$ M_{\rm Ks}	$&\multicolumn{1}{c}{$ \log\,{\rm SFR}			$}& \multicolumn{1}{c}{$ \log\,M	$}& \multicolumn{1}{c}{$ \log\,{\rm sSFR}	$}&$ \log\,{\rm Age}$\\
				&							&											& (mag; host)	& (mag)				& 	(mag)		& (mag)			& \multicolumn{1}{c}{$(M_\odot\,{\rm yr}^{-1})$}& \multicolumn{1}{c}{$(M_\odot)$}	& \multicolumn{1}{c}{ $({\rm yr}^{-1})$}	& (yr)\\
\midrule
\multicolumn{10}{l}{\textbf{SLSN-IIn host galaxies (continued)}}\\
\midrule
SN2008fz 	 &0.133 	&$  1.53 /  6 	$&$ 0.01 	$&$ -12.43\pm0.55 	$&$ -13.22\pm0.32 	$&$ -13.56\pm0.08 	$&$  -2.08^{+0.47}_{-0.48} 	$&$   6.55^{+0.25}_{-0.28} 	$&$  -8.64^{+0.71}_{-0.67} 	$&$   8.62^{+0.41}_{-0.62}$\\
SN2009nm 	 &0.210 	&$  2.39 /  5 	$&$ 0.15 	$&$ -14.61\pm0.21 	$&$ -17.65\pm0.18 	$&$ -17.71\pm0.21 	$&$  -0.60^{+0.65}_{-0.62} 	$&$   8.65^{+0.33}_{-0.34} 	$&$  -9.20^{+0.79}_{-0.83} 	$&$   8.95^{+0.62}_{-0.52}$\\
SN2011cp 	 &0.380 	&$ 10.25 /  9 	$&$ 0.30 	$&$ -16.90\pm0.28 	$&$ -20.04\pm0.14 	$&$ -21.79\pm0.08 	$&$   0.37^{+0.93}_{-0.64} 	$&$  10.18^{+0.17}_{-0.25} 	$&$  -9.88^{+1.28}_{-0.70} 	$&$   9.53^{+0.32}_{-0.89}$\\
\midrule
\multicolumn{10}{l}{\textbf{SLSN-II host galaxies}}\\
\midrule
CSS121015$^\ddagger$&0.287 	&$  0.97 /  6 	$&$ 0.00 	$&$ -16.70\pm0.08 	$&$ -17.33\pm0.07 	$&$ -17.53\pm0.29 	$&$  -0.52^{+0.38}_{-0.29} 	$&$   8.15^{+0.15}_{-0.17} 	$&$  -8.69^{+0.51}_{-0.35} 	$&$   8.65^{+0.33}_{-0.43}$\\
SN2008es$^\ddagger$&0.205 	&$  0.84 /  4 	$&$ 0.00 	$&$ -12.95\pm0.30 	$&$ -13.66\pm0.25 	$&$ -12.79\pm0.40 	$&$  -1.99^{+0.28}_{-0.27} 	$&$   6.19^{+0.33}_{-0.36} 	$&$  -8.15^{+0.57}_{-0.54} 	$&$   8.19^{+0.43}_{-0.53}$\\
SN2013hx$^\ddagger$&0.130 	&$  1.55 /  3 	$&$ 0.50 	$&$ -12.04\pm0.38 	$&$ -14.22\pm0.38 	$&$ -16.43\pm0.33 	$&$  -1.38^{+0.81}_{-0.60} 	$&$   7.14^{+0.71}_{-0.67} 	$&$  -8.33^{+0.79}_{-1.32} 	$&$   8.38^{+1.10}_{-0.77}$\\
\bottomrule
\end{tabular}
\tablecomments{The absolute magnitudes are not corrected for host reddening, to compare those measurements
with luminosity functions from flux-limited surveys.
The star-formation rates are corrected for host reddening. The host attenuation was modelled with the
Calzetti model. The abbreviation `n.o.f.' stands for number of filters.
The age refers to the age of the stellar population.
Objects with measured decline time-scale are marked by a $^\dagger$/$^\ddagger$ if
their decay is slower/faster than 50 days.
For details on the fitting, see Sect. \ref{sec:sed_fitting}.
}
\end{tiny}
\end{table*}

In the following, we take advantage of the full SUSHIES sample and present distribution
functions of the primary diagnostics mass and SFR of H-poor and -rich SLSNe host galaxies.\footnote{We omit
discussing the age of the stellar populations and their attenuation. In particular, the age is notoriously
difficult to measure accurately and precisely.}
Figures \ref{fig:SED}, \ref{fig:SED_app_1} and \ref{fig:SED_app_2} show the best fit of each host galaxy
and the evolution of the galaxy properties are shown in Fig. \ref{fig:sed_result}. Table
\ref{tab:sed_results} lists the model parameters. The ensemble properties in different redshift bins
are summarised in Table \ref{tab:statresult}.

\subsubsection{Stellar mass}\label{sec:mass}

The host masses (panel B in Fig. \ref{fig:sed_result}) span a range between $10^6$
and $10^{10}~M_\odot$ for both classes of SLSNe. This dearth of hosts above $10^{10}~M_\odot$
is remarkable. Assuming that SLSNe populate galaxies according to their star-formation rate, we
would in fact expect $\sim40\%$ of hosts galaxies to have masses above $10^{10}~M_\odot$. However, only one of the
53 SLSNe-I and two of the 16 H-rich SLSNe have such a high stellar mass. The probability of randomly
drawing a sample that is at least as extreme as the SLSN-I sample from UltraVISTA, weighted by the SFR, is $<10^{-5}$
at all redshifts (Fig. \ref{fig:statcomp_1}). For H-rich SLSNe, this scenario cannot be excluded,
however, as we will show below, the H-rich SLSN host sample has also some peculiar properties compared to
the general population of star-forming galaxies. The lack of massive galaxies for both classes strongly
argues for a stifled production efficiency in massive galaxies (see also \citealt{Perley2016a}).
We investigate its origin in detail in Sect. \ref{sec:metallicity_bias}.

Apart from the dearth of massive hosts, we observe clear differences between the host populations of both SLSN
classes. H-poor SLSNe are preferentially found in galaxies with average masses of $\sim10^{7.9}~M_\odot$ at $z=0.5$.
As redshift increases, the average masses gradually increase to $\sim10^{8.9}~M_\odot$ at $z>1$, while
the dispersion remains constant at $\sim0.65$~dex (Figs. \ref{fig:sed_result}, \ref{fig:statplot_app}; Table \ref{tab:statresult}).
Using the parametrisation of the mass function in \citet{Muzzin2013a}, the average masses correspond
to $1/500~M^\star$ and $1/50~M^\star$ at $z\sim0.5$ and $z\sim1$, respectively. Differences between
the hosts of fast and slow-declining SLSNe  are not present in our sample. A two-sided Anderson-Darling test
gives $p$ value of 0.72.

Hydrogen-rich SLSNe, in contrast to SLSNe-I, probe a significantly larger portion of the parameter space of
the general population of star-forming galaxies. Their distribution is not only shifted by 0.8 dex to higher
masses, but the distribution also includes three hosts that are even less massive than the
least massive SLSN-I host. The dispersion is $\sim0.8$ dex broader compared to the H-poor sample and even
$\sim0.5$~dex broader compared to the UltraVISTA survey (Tables \ref{tab:statresult}, \ref{tab:statresult_non_SLSN}).
Despite a larger dispersion, the probability of randomly drawing a distribution that is at least
as extreme as the H-rich population from the UltraVISTA sample is 25\% and hence does not point
to a significant difference to the general population of star-forming galaxies
Even after separating out the three SLSNe-II, of which two occurred in galaxies with masses between
$10^6$ and $10^7~M_\odot$, the dispersion remains unchanged. While this result is noteworthy, the chance
probability to randomly draw the SLSN-IIn sample from the UltraVISTA sample is 21\% (Fig. \ref{fig:statcomp_2}).

\begin{figure*}
\includegraphics[width=0.99\textwidth]{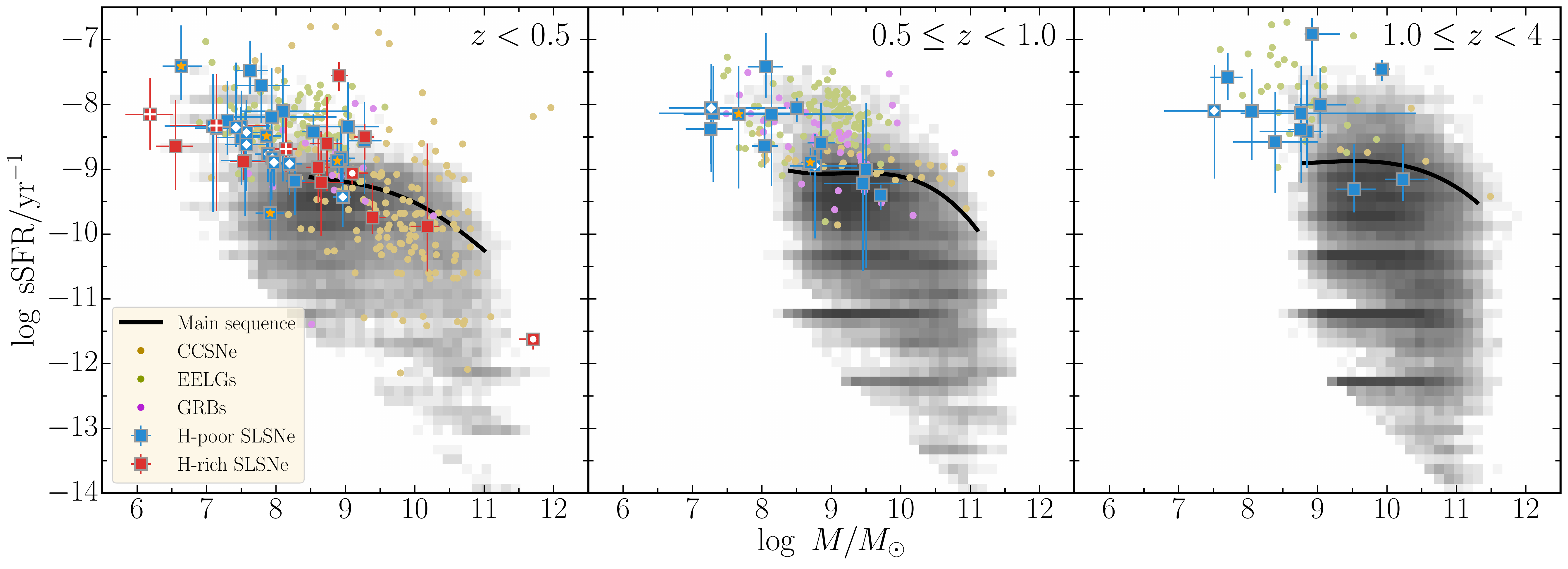}
\caption{
Specific star-formation rate versus stellar mass in three different redshift intervals.
The SUSHIES sample is displayed in red and blue. Similar to previous figures, hosts of
slow- and fast-declining SLSNe-I are signified by ``$\star$'' and ``$\diamond$'', respectively.
In contrast to the other plots, we use the H$\alpha$ and IR luminosity as an SFR indicator for SLSNe-IIn
SN1999bd and SN2006gy, respectively (highlighted by a $\bullet$; measurements taken from \citealt{Smith2007a}
and \citealt{Leloudas2015a}).
Overlaid are the locus of star-forming galaxies from the UltraVISTA survey (grey shaded area)
and of other comparison samples (in colour).
The black curve shows the location of the galaxy main sequence in each redshift bin. The
values were taken from \citet{Whitaker2014a} and \citet{Lee2015a}.
Measurement errors are omitted for comparison samples. They are similar to those of SLSN
host galaxies.
}
\label{fig:ssfr_vs_mass}
\end{figure*}

\subsubsection{Star-formation rate}

Panel C in Fig. \ref{fig:sed_result} displays the evolution of the dust-corrected star-formation rate (SFR).
Hosts of H-poor SLSNe have similar SFRs to the general population of star-forming galaxies (Tables
\ref{tab:statresult}, \ref{tab:statresult_non_SLSN}), but smaller SFRs than host galaxies of GRBs and regular
core-collapse SNe.
The mean SFR rapidly grows with increasing look-back time from $0.25~M_\odot\,{\rm yr}^{-1}‚$ at $z<0.5$ to $5~M_\odot\,{\rm yr}^{-1}$ at $z>1$
(Table \ref{tab:statresult}). In singular cases, the SFR reaches $>100~M_\odot\,\rm{yr}^{-1}$
(SN1000+0216 and SNLS06D4eu; Table \ref{fig:sed_result}). While the mean value evolves with redshift, the
dispersion remains constant at $\sim0.4$~dex. The SFR increases somewhat faster compared to UltraVISTA,
out to $z\sim2$, but statistically both distributions remain similar (Fig. \ref{fig:statcomp_1}).

Host galaxies of H-rich SLSNe exhibit different characteristics. The three H-rich SLSNe
with broad Balmer emission lines exploded in galaxies with low SFRs. Two of the hosts (SNe 2008es and
2013hx) have very low SFRs between 0.01 and $0.1~M_\odot\,{\rm yr}^{-1}$. In contrast, SLSNe-IIn are found
in a more diverse population of star-forming galaxies. Their defining property is again the large dispersion of
$\sim1$~dex (Table \ref{tab:statresult}). Their average SFR is only modestly larger compared to the
galaxy samples discussed in this paper (Fig. \ref{fig:statplot_app}).

Although the SFRs of SLSN-I hosts are similar to the general population of star-forming galaxies, they
are on average less vigorously star-forming than GRB and regular CCSN host galaxies. However, in the
previous section, we revealed that especially the H-poor SLSNe are found in very low mass galaxies.
Likewise, hosts of H-rich SLSNe have higher average SFRs but their mass distribution is skewed to higher masses
and is substantially broader. To better understand how SLSN host galaxies fit in the context of other
galaxy samples, we normalise the SFR by the stellar mass (so-called specific star-formation rate, sSFR).

Figure \ref{fig:ssfr_vs_mass} displays the two classes of SLSNe in the sSFR-mass plane in three different
redshift intervals. Both classes are characterised by high sSFR between $10^{-8.7}~{\rm yr}^{-1}$ and
$10^{-8.0}~{\rm yr}^{-1}$ at all
redshifts. They reside in a part of the parameter space well above the galaxy main-sequence (black curves
in Fig. \ref{fig:ssfr_vs_mass}) that is occupied by starburst galaxies. The most extreme hosts have
sSFR that are two orders of magnitude in excess to the galaxy main sequence, indicating that some
hosts experience very extreme starbursts. In general, SLSN-I hosts are found in the region of the
parameter space that is occupied by extreme emission-line galaxies and more extreme than of GRBs and
of regular CCSNe, which trace the bulk of the population of star-forming galaxies. Host galaxies of
H-rich SLSNe have high sSFR as well but because of their high stellar masses, their parameter space
is more extended.

\subsection{A radio perspective on SLSN host galaxies}\label{sec:radio}

Radio emission from star-forming galaxies is an excellent tracer of the total SFR
\citep{Condon1992a, Schmitt2006a, Murphy2011a, Calzetti2013a}. In contrast to SED modelling
and emission-line diagnostics, e.g., Balmer lines, it is independent of any extinction correction,
although radio SFRs do suffer from time-delay for SNe to explode and create sufficient
cosmic rays.

Almost all SLSN hosts lie
in the footprints of wide-field radio surveys, such as FIRST, NVSS and SUMSS. All hosts
evaded detection in individual images down to the nominal r.m.s. levels of the surveys: FIRST $\sim0.15$ mJy/beam,
NVSS $\sim0.45$ mJy/beam and SUMSS $\sim1.3$ mJy/beam (see Table \ref{tab:data_radio} for individual measurements).
To place tighter constraints on the average radio brightness of the host populations, we
stack the data of the 51 fields with VLA FIRST data. We first divide the sample into three
redshift bins ($z\leq0.5$, $0.5<z\leq1.0$ and $z>1$) and according to the SN type. Afterwards,
we centre the images on the supernova positions and median-combine them. Also, in the stacks
no host population is detected down to an r.m.s. of 32--60 $\mu$Jy/beam at all redshifts (Table \ref{tab:radio}).

Following the method in \citet{Michalowski2009a}, we translate the flux density into SFR limits.\footnote{
This method is based on \citet{Bell2003a} and assumes a power-law
shaped radio continuum with a spectral index of $\alpha=-0.75$ \citep[$F_\nu\propto\nu^\alpha$;][]{Condon1992a, Ibar2010a}.}
The non-detections correspond to $4\sigma$ SFR limits between $8.0~M_\odot\,{\rm yr}^{-1}$ at $z\sim0.23$ to
$326~M_\odot\,{\rm yr}^{-1}$ at $z\sim1.41$, and exceed the SED-derived SFRs by
factors 21 to 120 (Table \ref{tab:radio}). This allows ruling out truly extreme obscured star
formation, in agreement with the observed $R-K_{\rm s}$ colours and the absence of reddened
SLSNe in our sample.

In addition to the survey data, the hosts of MLS121104, SN2005ap and SN2008fz were targets
of our JVLA campaign. All three hosts evaded detection down to nominal r.m.s. values of
15, 25 and 15~$\mu$Jy/beam for MLS121104, SN2005ap and SN2008fz, respectively. Those limits
correspond to $4\sigma$ SFR limits of 6.2, 9.0 and $1.6\,M_\odot\,{\rm yr}^{-1}$, respectively.
The limit on MLS121104 is of particular
interest. It is the only known host with a super-solar metal abundance. The SED modelling
revealed a dust-corrected SFR of $5.13^{+7.46}_{-3.72}\,M_\odot\,{\rm yr}^{-1}$ (Table \ref{tab:sed_results}),
which is comparable to the radio limit within errors, implying that the optical diagnostics
probed the total star formation activity in the galaxy.
The high upper limits on the hosts of SNe 2005ap and 2008fz exceed the SED-SFRs by at least a factor
of 50 and, hence, do not have much meaning (Table \ref{tab:sed_results}).

\begin{table}
\caption{Properties of the stacked FIRST data}
\centering
\begin{scriptsize}
\begin{tabular}{ccccc}
\toprule
Redshift	& \multirow{2}*{Number}	&  r.m.s.		& $\log \rm SFR$(tot.)		& $\log \rm SFR$(SED)	\\
interval	&			& ($\mu$Jy/beam)	& $(M_\odot\rm{yr}^{-1})$	& $(M_\odot\rm{yr}^{-1})$	\\
\midrule
\multicolumn{4}{l}{\textbf{H-poor SLSN host galaxies}}\\
\midrule
$z\leq0.5$								& 17	& 42.5	& $<1.11$	&$ -0.61\pm0.12 $\\
($\left\langle z\right\rangle=0.26$)\\
$0.5<z\leq1.0$							& 12	& 44.2	& $<1.96$	&$ -0.10\pm0.19 $\\
($\left\langle z\right\rangle=0.74$)\\
$1.0<z\leq4.0$							& 9		& 56.3	& $<2.51$	&$  0.68\pm0.30 $\\
($\left\langle z\right\rangle=1.41$)\\
\midrule
\multicolumn{4}{l}{\textbf{H-rich SLSN host galaxies}}\\
\midrule
$z\leq0.5$								& 13	& 49.4	&$<1.00$	&$ -0.44\pm0.36 $\\
($\left\langle z\right\rangle=0.21$)\\
\midrule
\multicolumn{4}{l}{\textbf{H-poor and H-rich SLSN host galaxies}}\\
\midrule
$z\leq0.5$								& 30	& 32.2	&$<0.90$	&$ -0.42\pm0.17 $\\
($\left\langle z\right\rangle=0.23$)\\
\bottomrule
\end{tabular}
\end{scriptsize}
\begin{minipage}{1\columnwidth}
\vspace{3ex}
\hskip1em\rmfamily
Note. ---  The r.m.s. level is calculated from the stacked FIRST image and converted
into a $4\sigma$ limit on the total unobscured star-formation rate at the median redshift
of each sample. The weighted means of
the SED-derived SFR is reported for comparison. For details, see Sect. \ref{sec:radio}.
The second value in the redshift column reports the mean redshift of each redshift interval.
\end{minipage}
\label{tab:radio}
\end{table}

\section{Discussion}\label{sec:discussion}

\subsection{Evolution of SLSN-I host galaxies}\label{sec:redshift_evol}

\begin{figure}
\includegraphics[width=1\columnwidth]{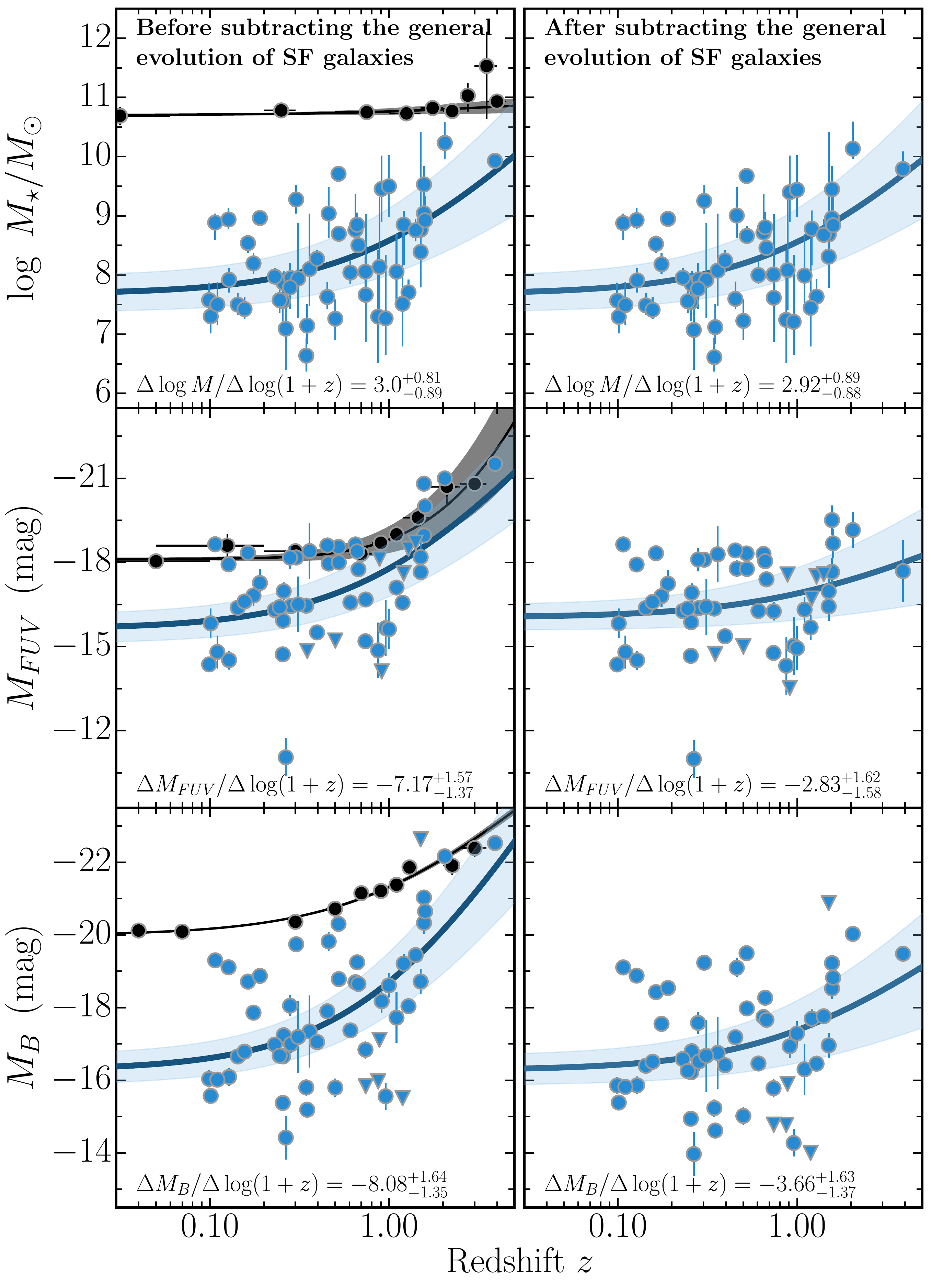}
\caption{Mass, FUV luminosity at 1500 \AA\ (as proxy of the observed SFR) and $B$-band
luminosity plotted vs. redshift (detections: $\bullet$; non-detections: $\blacktriangledown$). The observed
evolution (left panels) is the sum of the differential evolution of SLSN-I host galaxies and
the general cosmic evolution of star-forming galaxies. This general cosmic evolution is
indicated by the evolution of the characteristic luminosity and mass of appropriate
luminosity and mass functions (black data points; x-errors indicate the redshift intervals
of the luminosity and mass functions). The right panels display the differential evolution of
SLSN-I host galaxies after detrending. Each data set was fitted with the linear model
$Y=A+B\,\log\,\left(1+z\right)$. The curves represent the best fit and the shaded regions
the $1\sigma$ error contour. The slopes of the best fits are displayed at the bottom of
the panels. Note the significant change in the redshift evolution of the FUV and $B$-band
luminosity after detrending, while the evolution of the galaxy mass remains unchanged.
}
\label{fig:redshift_evol}
\end{figure}

In the previous sections, we revealed a rapid evolution of $B$-band luminosity and the SFR of
SLSN-I host galaxies. In the following, we quantify  how mass, FUV luminosity (as a tracer of the
SFR) and the $B$-band luminosity of the SLSN-I host population evolve throughout cosmic time. The
redshift evolution of these diagnostics is displayed in Fig. \ref{fig:redshift_evol} (left panels).
We fit these data with the linear model $Y=A+B\,\log\,\left(1+z\right)$
and propagate errors through an MC simulation and bootstrapping, as described in Sect. \ref{sec:statistics}.

The left panels in Fig. \ref{fig:redshift_evol} show the best fits and their $1\sigma$ error contours.
The mass, FUV and the $B$-band luminosity of SLSN-I hosts show a moderate to strong
redshift dependence with a linear correlation coefficient between $\left|r\right|=0.5$ and
$\left|r\right|=0.6$ (Table \ref{tab:redshift_evol}). The probability
of generating each of these linear correlations by chance is between $4\times10^{-5}$ and
$3.5\times10^{-6}$, respectively ($\sim4.0$--$4.5\sigma$; Table \ref{tab:redshift_evol}).

To isolate the differential evolution of SLSN host galaxies from known global trends, we
repeat the analysis after subtracting the evolution of the mass function, and the FUV
and $B$-band luminosity functions of star-forming galaxies.
As tracers for the secular evolution, we use the characteristic luminosities
and masses of the luminosity and mass functions: FUV: \citet{Wyder2005a} and \citet{Cucciati2012a};
$B$ band: \citet{Madgwick2002a}, \citet{Faber2007a} and \citet{Marchesini2007a}; and
mass: \citet{Baldry2012a}, \citet{Muzzin2013a} and \citet{Grazian2015a}.

The right panels in Fig. \ref{fig:redshift_evol} show the redshift evolution of the host properties
after detrending. The strong redshift evolution in the $B$ band and the FUV is consistent with the general
cosmic evolution of star-forming galaxies.
After detrending the data, the differential
evolution in the FUV and $B$ band is consistent with no evolution. The chance probability
increases from $<4\times10^{-5}$ to $>2\times10^{-2}$ (i.e., $<3\sigma$; Table \ref{tab:redshift_evol}).
The galaxy mass, on the other hand, still shows a moderate redshift dependence [$\Delta M/\Delta \log (1+z) =2.92^{+0.89}_{-0.88}$],
though with a significantly higher chance probability of $1.1\times10^{-4}$ (equivalent to $3.9\sigma$; Table \ref{tab:redshift_evol}).

Intriguingly, the rate with which the stellar mass of SLSN-I host galaxies increases with
redshift before and after detrending is close to the redshift dependence of the
characteristic mass in the mass-metallicity relation [$\Delta M/\Delta \log (1+z) \sim2.64$;
\citealt{Zahid2014a}]. This suggests that metallicity could be a regulating factor in the SLSN
production (as argued by \citealt{Chen2016a} and \citealt{Perley2016a}). In the following
section, we investigate this relationship in detail.

Due to the small redshift range probed by our H-rich SLSN sample, the redshift dependence
of their physical properties is inconclusive.

\begin{table}
\caption{Redshift evolution of SLSN-I host galaxies}
\centering
\begin{tabular}{lrrrr}
\toprule
\multirow{2}*{Property}	& \multicolumn{2}{c}{Linear correlation}						& \multicolumn{2}{c}{Linear model} \\
				& \multicolumn{1}{c}{$r$}		& \multicolumn{1}{c}{$p_{\rm ch}$}		& \multicolumn{1}{c}{slope}		& \multicolumn{1}{c}{intercept}\\
\midrule
\multicolumn{5}{c}{\textbf{Before removing the cosmic evolution of SF galaxies}}\\
\midrule
Mass			&$ 0.52^{+0.13}_{-0.18}	$&$ 7.7\times10^{-5}	$&$  3.00^{+0.81}_{-0.89}	$&$  7.68^{+0.30}_{-0.31}	$\\
$M_{\rm FUV}$	&$-0.53^{+0.13}_{-0.10}	$&$ 4.0\times10^{-5}	$&$ -7.17^{+1.57}_{-1.37}	$&$ -15.63^{+0.53}_{-0.50}	$\\
$M_B$			&$-0.59^{+0.13}_{-0.10}	$&$ 3.5\times10^{-6}	$&$ -8.08^{+1.64}_{-1.35}	$&$ -16.28^{+0.41}_{-0.40}	$\\
\midrule
\multicolumn{5}{c}{\textbf{After removing the cosmic evolution of SF galaxies}}\\
\midrule
Mass			&$  0.51^{+0.14}_{-0.18}	$&$ 1.1\times10^{-4}	$&$  2.92^{+0.89}_{-0.88}	$&$   7.68^{+0.29}_{-0.31}	$\\
$M_{\rm FUV}$	&$ -0.24^{+0.14}_{-0.13}	$&$ 7.7\times10^{-2}	$&$ -2.83^{+1.62}_{-1.58}	$&$ -16.04^{+0.46}_{-0.44}	$\\
$M_B$			&$ -0.32^{+0.15}_{-0.13}	$&$ 2.1\times10^{-2}	$&$ -3.66^{+1.63}_{-1.37}	$&$ -16.28^{+0.41}_{-0.40}	$\\
\bottomrule
\end{tabular}
\begin{minipage}{1\columnwidth}
\vspace{3ex}
\hskip1em\rmfamily
Note. ---  The two sets of fits show the redshift evolution before and after correction for global
trends of star-forming (SF) galaxies.
The columns of the linear correlation analysis display the linear correlation coefficient
$r$, and the corresponding chance probability $p_{\rm ch}$. The redshift evolution is
parametrised with the linear model $Y = A+B\,\log\,\left(1+z\right)$.
\end{minipage}
\label{tab:redshift_evol}
\end{table}

\subsection{Metallicity bias}\label{sec:metallicity_bias}
\subsubsection{Dependence of SLSN formation on host galaxy mass}

To quantify the effect of the physical parameters of SLSN host galaxies on SLSN
formation, we contrast the galactic environments of SLSN explosions to those of star-forming
galaxies in general. In addition to our SLSN host data, we hence require a census of
cosmic star-formation in the respective redshift range as complete as possible. Fortunately,
numerous deep-field photometric galaxy surveys compiled in recent years provide a good
match to our SLSN imaging data.

The deepest surveys that probe a sufficient cosmic volume are COSMOS \citep{Scoville2007a}
and CANDELS \citep{Grogin2011a, Koekemoer2011a}; both have high completeness levels for galaxies above
stellar masses of $M_\star \gtrsim 10^8 M_\odot$ at $z\sim0.5$ \citep[e.g.,][]{Tomczak2014a}.
However, this is still two orders of magnitude higher than our least massive SLSN hosts
(Table~\ref{tab:sed_results}). Nonetheless, we extrapolate the mass functions to the lowest
observed galaxy masses ($M\sim10^6~M_\odot$). This extrapolation will add some uncertainty, but
mass and luminosity functions of star-forming galaxies are rather well constrained and show no hints
for plunging at the faint-end.

The primary parameter that we are interested in is galaxy stellar mass $M_\star$, because
it is known to correlate well with the average galaxy metallicity. Metallicity, in turn,
has a strong effect on the evolution of massive stars through line-driven stellar winds. Similar
considerations have previously been applied to GRB hosts, where after a long debate,
the impact of metallicity on long GRB-selected galaxies is now relatively robustly
established \citep[e.g.,][]{Kruehler2015a, Schulze2015a, Vergani2015a, Perley2016c}.

In addition to galaxies from wide-field surveys, we also compare the mass distribution of
our SLSN hosts to those of star-forming galaxies selected through GRBs \citep{Hjorth2012a, Perley2016b} and
low-redshift core-collapse supernovae from untargeted surveys \citep{Stoll2013a}. The latter is a particularly suitable control sample,
as normal CCSNe are thought to trace all star-forming environments in a relatively direct
and unbiased way \citep{Stoll2013a}. For simplicity and the sake of clarity, we do not
differentiate between CCSNe {sub-types}.

\begin{figure}
\includegraphics[width=1\columnwidth]{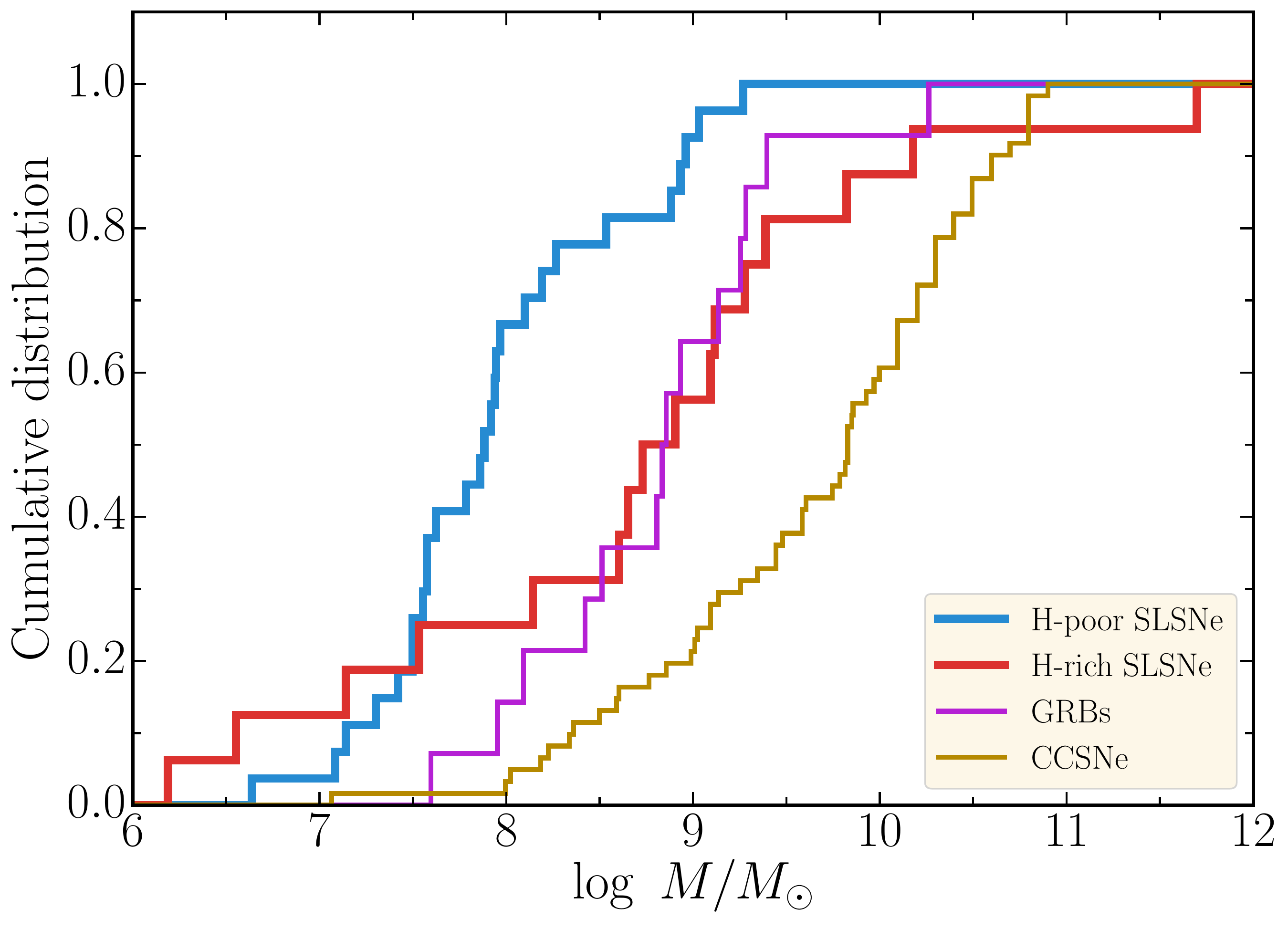}
\caption{
Cumulative histograms of the stellar-mass distributions of various galaxy samples at $z<0.5$. SLSNe-I
show a strong preference for the least massive hosts, even compared to GRBs. The mass distribution
of H-rich SLSNe and GRBs is similar and skewed by 0.6 dex to higher masses than the SLSN-I sample.
The SN sample was taken from \citet{Stoll2013a}.
}
\label{fig:mass_hist}
\end{figure}

Figure~\ref{fig:mass_hist} shows the cumulative histograms of stellar masses for the four kinds of
transients at $z<0.5$. Clearly, SLSNe-I trace the least massive systems.
The median stellar mass increases towards GRB hosts and galaxies selected by more
frequent regular CCSNe (Fig. \ref{fig:mass_hist}). An Anderson-Darling test between GRB and SLSN-I host
galaxies at $z<0.5$ rejects the notion that long GRBs and SLSN-I
have similar host mass distributions ($p_{\rm ch}< 8\times10^{-4}$).
Moreover, at $z<1.0$, none of the SLSN-I hosts in our sample of 41 events has a stellar mass
above $10^{10}\,M_\odot$, whereas $\sim 40\%$ of CCSNe form in such massive galaxies.
Thus, it is immediately obvious that a strong effect prevents SLSN-I from forming in galaxies of
high stellar mass.

SLSN-IIn hosts are 0.8 dex more massive than SLSN-I hosts, as noted previously in \citet{Leloudas2015a} and \citet{Perley2016a}.
Their mass distribution is comparable to the GRB hosts (within the limited number
statistics). Here, we also find a lack of massive hosts above
$10^{10}~M_\odot$, though the metallicity dependence is weaker.

\subsubsection{SLSNe are biased tracers of SFR}\label{sec:slsn_vs_sfr}

Under the working hypothesis that massive stars are the progenitors of SLSNe, they should
also trace star formation in a particular way. However, previous experience with GRB
hosts has illustrated that environmental factors, most commonly attributed to a
low progenitor metallicity, can have a significant effect \citep[e.g.,][]{Graham2013a,
Schulze2015a, Perley2016a}. This effect is presumably even stronger in SLSN-selected galaxies,
considering their mass distributions (Fig. \ref{fig:mass_hist}).

To better illustrate the efficiency of SLSN production with host stellar mass (or metallicity),
we need to normalise the number of SLSN-selected galaxies by the contribution of similar
massive systems to the cosmic star-formation at the given redshifts. We derive this by
starting with the stellar mass function $\Phi(M)dM$ of star-forming galaxies from CANDELS.
This yields the number density of galaxies per stellar mass bin. We use the parametrisation
of $\Phi$ for star-forming galaxies from Table 2 of \citet{Tomczak2014a} and note that the
stellar-mass functions from \citet{Ilbert2013a} or \citet{Muzzin2013a} are similar and do not
alter our conclusions significantly.

\begin{figure}
\includegraphics[angle=0, width=0.99\columnwidth]{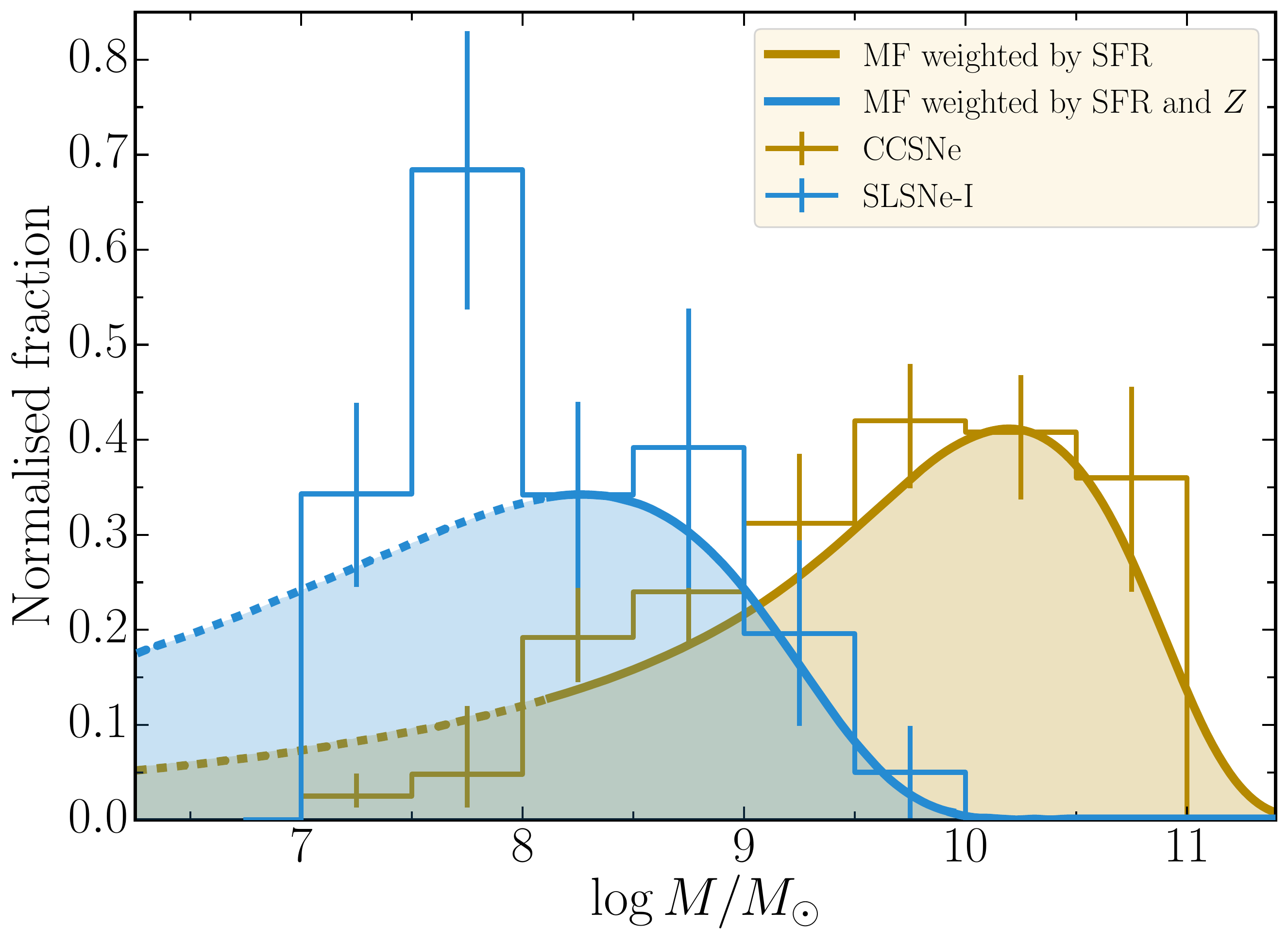}
\caption{Histogram of the mass distribution of SLSN-I host galaxies and hosts of CCSNe
from the \citet{Stoll2013a} sample at $z<1$. The area of each histogram is normalised to unity.
The yellow curve shows the SFR-weighted CANDELS mass function. This model describes the observed
distribution for CCSNe reasonably well. To match the distribution of H-poor SLSN host galaxies,
a further weighting is required that stifles the SLSN production in high-mass galaxies.
This mass-dependent (i.e., metallicity dependent) production efficiency can be modelled by
an exponential metallicity cut-off at $12+\log {\rm O/H}=8.31^{+0.16} _{-0.26}$ (blue curve).
The dashed lines of the model fits indicate the mass regime where the CANDELS mass function (MF)
had to be extrapolated.
}
\label{fig:slsnhist}
\end{figure}

Then, we sum the star-formation rate of all contributing galaxies by integrating over the
scatter of all galaxies in the galaxy main sequence at a given stellar mass \citep[e.g.,][]{Whitaker2012a,
Sobral2014a, Speagle2014a, Tasca2015a}. The SFR-weighted mass histogram, shown in Fig.~\ref{fig:slsnhist}
in yellow, peaks at around $10^{9.5-10.5}~M_\odot$, and provides a good match to the sample
of host galaxies of CCSN selected from untargeted surveys.
In contrast, the mass histogram of SLSN-hosting galaxies peaks two orders of magnitudes
lower, which is clearly inconsistent with the typical environments where the bulk of the stars are
produced at $z\sim0.5$.

\subsubsection{SLSNe production efficiency}\label{sec:cutoff}

We modelled the SLSN-I host stellar mass histogram by applying a function that describes an
efficiency $\rho(M)$ of producing SLSNe from star-formation. We chose $\rho(M)$ as an exponential
function in the form of $\rho = \exp(-\beta\,M/M_0)$,
where  $M_0$ is a characteristic cut-off mass, where the production efficiency dropped to
$1/e$, and $\beta$ a cut-off strength. This essentially shuts off SLSN production in galaxies
of high stellar mass. Physically, this can be interpreted as a decrease in the probability of
creating SLSNe-I from massive stars above a characteristic cut-off metallicity, where we
assume that stellar mass at a given star-formation rate relates to host metallicity at
stellar masses below $\sim10^{10}\, M_\odot$\citep[e.g.,][]{Maiolino2008a, Yates2012a}.

\begin{figure}
\includegraphics[angle=0, width=0.99\columnwidth]{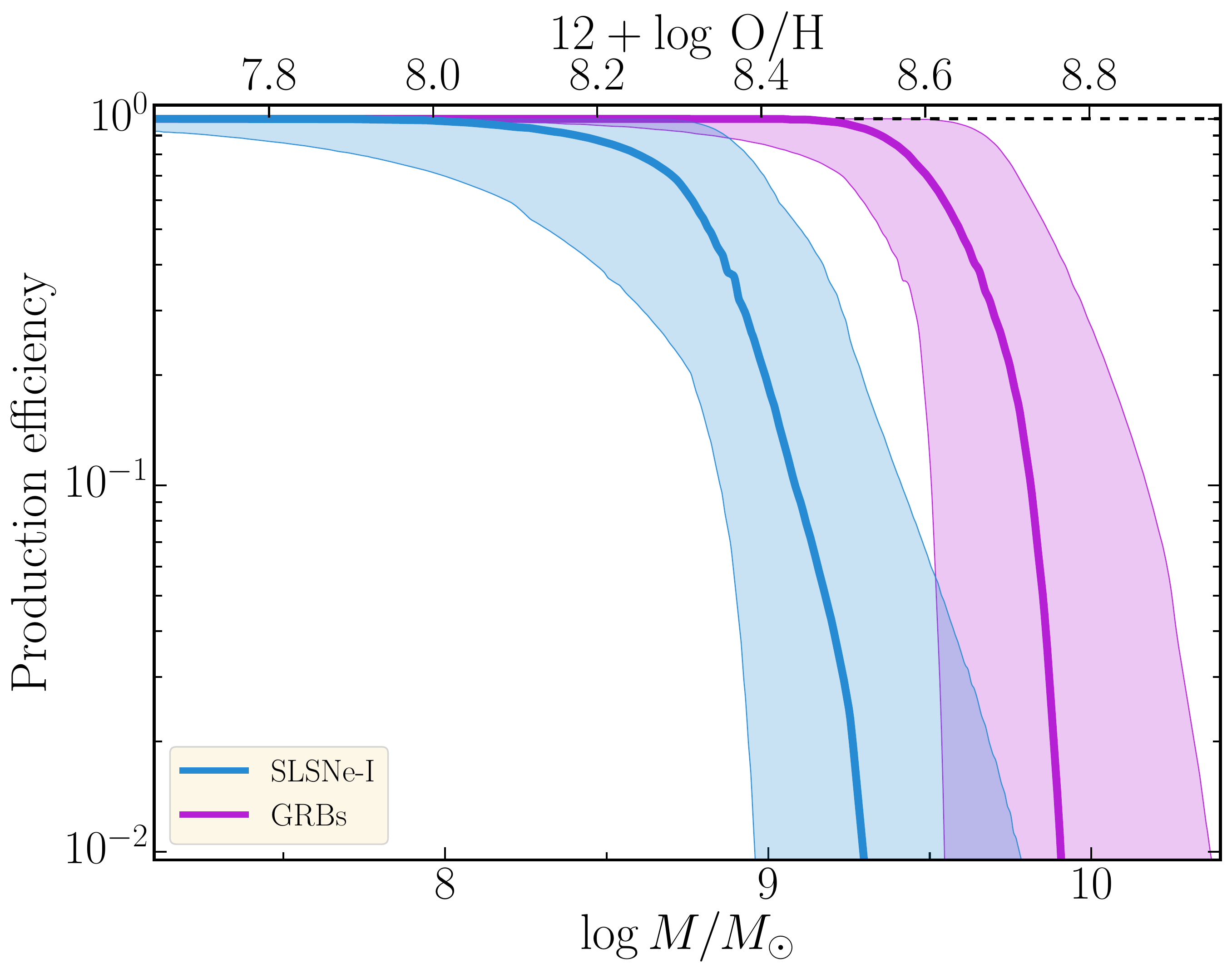}
\caption{Production efficiency of H-poor SLSNe in galaxies with stellar mass $M$.
Applying the mass-metallicity relation in \citet{Mannucci2011} maps a given mass of a galaxy with a given
metallicity. The shaded regions show the $1\sigma$ uncertainty. For comparison, the
GRB production efficiency is displayed. The production of SLSN progenitors must be stifled in
galaxies with metallicity above $12+\log {\rm O/H}=8.31^{+0.16} _{-0.26}$, 0.3 dex lower
than for GRBs, indicating that SLSN progenitors are on average less metal-enriched than GRBs.
}
\label{fig:slsneff}
\end{figure}

We minimise the deviation between model and data by varying $M_0$ and $\beta$ using an
MC method on $10^{5}$ bootstrapped distributions of SLSN-I host masses derived
from our parent sample. Statistical errors on host masses are included in the procedure by
varying them according to the uncertainties in Table~\ref{tab:sed_results} within each
trial. The best-fit model is obtained at $M_0$ corresponding to $12+\log(\mathrm{O/H})_0 = 8.31^{+0.16}_{-0.26}$
and $\beta = 2.1$. While our procedure can constrain $12+\log(\mathrm{O/H})_0$ relatively
accurately, the cut-off shape is not yet well measured. Acceptable fits are obtained in a range
between $\beta=1$ and $\beta>30$, where the latter illustrates an infinitively sharp cutoff
at $12+\log(\mathrm{O/H})_0 = 8.4$.
Of course, the parameters $M_0$ and $\beta$ are not fully independent. The higher the
cut-off mass, the sharper the cutoff. Figure~\ref{fig:slsneff} shows the best-fit
and a region which contains 68\% of all MC trials.

For comparison,
we modelled the mass distribution of our GRB host galaxy sample with the same model (purple
curve in Fig. \ref{fig:slsneff}). Its mass distribution points to a higher metallicity
cut-off at $12+\log(\mathrm{O/H})_0 \sim 8.6\pm0.10$ (i.e., a 0.3 dex larger oxygen abundance
than SLSN-I host galaxies), in agreement with \citet{Kruehler2015a} and marginally lower than
\citet{Perley2016a}.

For SLSNe-II, number statistics are still too low to derive robust
constraints, but the host mass distribution indicates a behaviour similar to that observed
for GRB hosts.

\subsection{On the factors behind forming H-poor SLSNe}\label{ref:other}

In the first paper of our series \citep{Leloudas2015a}, we showed that the metallicities
(directly measured from spectra) of SLSN-I hosts were low (median value being 0.27 solar metallicity).
They were modestly lower than those of GRB hosts, although the difference was
statistically insignificant. What is even more striking in the case of SLSNe-I is that their
host spectra exhibit emission lines with very large rest-frame equivalent widths. In
$\approx50\%$ of the cases, we observed rest-frame equivalent widths exceeding 100 \AA\ and in some extreme cases reaching up to
500--800 \AA.

The presence of EELGs in our sample is extremely unusual \citep[only
1\% of galaxies in the zCOSMOS survey have rest-frame EW $> 100$ \AA;][]{Amorin2015a}, and
we determined that their frequency could not be a chance coincidence ($p_{\rm ch}\sim 10^{-12}$;
\citealt{Leloudas2015a}). On average, even GRB hosts do not show such
strong emission lines. The difference to the distribution of a complete sample of
GRBs at $z<1$ \citep{Hjorth2012a} was found to be statistically significant, although
the strongest emitters in our sample were mostly found at $z < 0.3$.
The difference was even more pronounced in [\ion{O}{iii}]$\lambda$5007 than in H$\alpha$,
pointing to a higher ionisation fraction in the gas around SLSNe.

These extreme properties were also seen by directly measuring the ionisation parameter $q$ and the ratio
between [\ion{N}{ii}]/H$\beta$ and [\ion{O}{iii}]$\lambda$5007/H$\alpha$ (BPT diagram;
\citealt{Baldwin1981a}), where the overwhelming
majority of H-poor SLSNe were found to be in regions with $\log\,\left[{\rm\ion{O}{iii}}\right]/{\rm H}\beta> 0.5$.
As the equivalent widths of the lines decrease with time after a starburst \citep[e.g.,][]{Leitherer1999},
this evidence strongly points towards very young environments for SLSN-I hosts.\footnote{The relation
between the H$\beta$ equivalent width and the age of the starburst also has a dependence
on metallicity \citep{Inoue2011} and the shape of the star-formation histories \citep[e.g.,][]{Terlevich2004,Lee2009}.}
This led us to propose
that the progenitors of H-poor SLSNe are very young, and are on average more short-lived
than those of GRBs \citep{Leloudas2015a}. Although absolute ages are notoriously difficult to determine, we
identified a very young stellar population with an age of only $\sim3~\rm Myr$ at the explosion
site of PTF12dam, which is the most extreme example in our sample (in terms of emission-line
strength; \citealt{Thoene2015a}).

Recently, \cite{Chen2016a} questioned the importance of young age
for H-poor SLSN progenitors, proposing that metallicity is the only key factor leading to
the production of SLSNe. These authors approximated the effect of age through the sSFR and
by comparing the parameter spaces of their SLSN host samples in the metallicity-sSFR plane
to complete samples of star-forming galaxies in the \textit{local volume} (11HUGS and LVL; \citealt{Kennicutt2008a, Lee2011a}).
However, the two properties are intimately connected through the mass-metallicity-SFR fundamental relation
\citep{Mannucci2011} and can therefore not be easily disentangled. Thus, we expect to see metallicity
and age to drive the SLSN production.
Attributing the dependence of H-poor SLSNe simply on metallicity has led many authors
\cite[e.g.,][]{Lunnan2014a,Chen2016a} to support a magnetar origin for these
explosions, although this explanation is not unique. Acknowledging that young age
plays an important role as well allows models based on more massive progenitors
to remain equally competitive \citep{Leloudas2015a,Thoene2015a}.

In contrast to \citet{Leloudas2015a}, \cite{Perley2016a} argued that the fraction of starbursts
(defined as sSFR $> 10^{-8}~{\rm yr}^{-1}$ in their papers) among SLSN-I hosts is not
exceptionally large and that the starburst fraction among H-poor
SLSN hosts may be explained by the fact that dwarf galaxies tend to have bursty star-formation histories
\citep[e.g.,][]{Guo2016a}.
By using the study of \cite{Lee2009}, we show that the fraction of SLSNe-I
occurring in EELGs in the \citet{Leloudas2015a} sample is significantly increased, even with respect
to dwarf galaxies.
\cite{Lee2009} determined the fraction of starbursts among local
dwarfs in the 11HUGS survey, which is the same survey that \cite{Perley2016a} and
\cite{Chen2016a} used as their main comparison galaxy sample.
Furthermore, \cite{Lee2009} used the same operational definition of starburst that we use
for EELGs (${\rm EW}_{\rm rest}> 100~\rm\AA$), making a direct comparison straightforward.
They determined that only 6\% of dwarf galaxies in the absolute magnitude range of interest
($-19 < M_B < -15$) have ${\rm EW}_{\rm rest}> 100~\rm\AA$ (and only 8\% have ${\rm EW}_{\rm rest}> 80~\rm\AA$).
This means that the probability of attaining the same fraction of EELGs among H-poor SLSN
hosts as in \cite{Leloudas2015a} by chance is $p_{\rm ch}< 10^{-6}$. This might be larger than what is obtained
by comparing with zCOSMOS ($p_{\rm ch} \sim 10^{-12}$), but a chance coincidence
is still extremely unlikely. This can also be understood in the following way: if the duty
cycles in the bursty SFH of dwarf galaxies are 1--2 Gyr, it is very unlikely that we would
happen to catch them by chance so close to an initial starburst, when selecting them through
H-poor SLSNe.

We therefore argue that \textit{both} low metallicity \textit{and} young age play
important roles in the formation of H-poor SLSNe, and that stellar evolution in metal-poor,
starburst environments needs to be better understood to fully appreciate the context.
In particular, mass loss in these
extreme regimes is poorly understood and more effort needs to be put into understanding why
these explosions are H-poor and whether this can be attributed to eruptive mass loss
\citep{Woosley2007a,Quataert2012a}, homogeneous evolution \citep{Yoon2005}, binarity
\citep{Eldridge2008} or another, yet unknown, factor.

\subsection{SLSN host galaxies in the context of other galaxy populations}\label{sec:class_diff}

In the previous sections, we discussed particular aspects of the host populations. In the following,
we compare the host properties to those of other galaxy samples.

\subsubsection{SLSN-I host population}\label{sec:class_diff_hpoor}

\begin{figure}
\includegraphics[width=1\columnwidth]{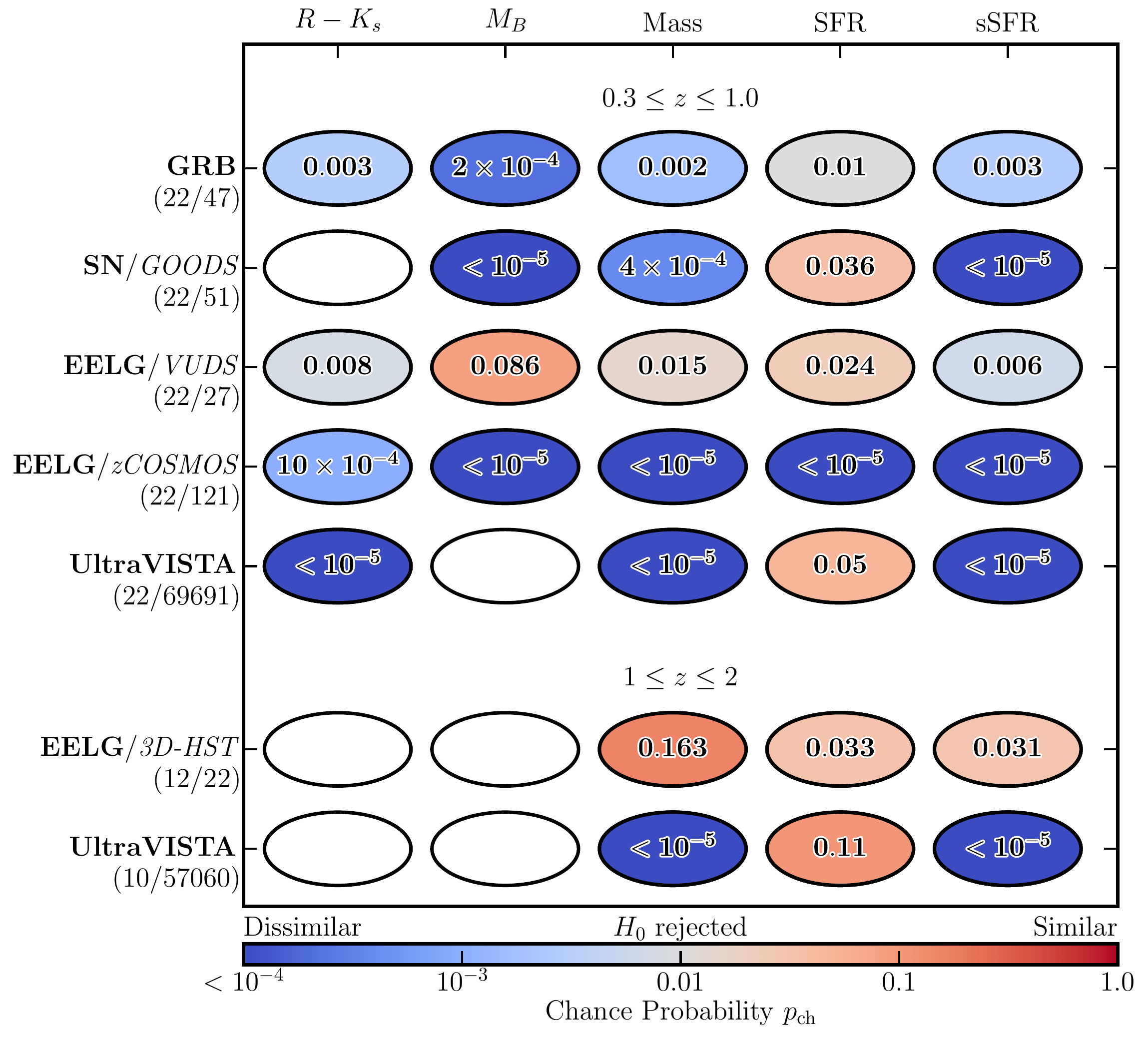}
\caption{
Two-sided Anderson-Darling tests between SLSN-I host galaxies and different galaxy samples
at $0.3\leq z\leq 1.0$ and $1\leq z\leq 2$. The $p$ are reported in the ellipses. The diverging colour
scheme is centred at the $p$-value of 0.01, where we reject the null hypothesis that the two samples
have the same parent distribution. For all tests, we required that the redshift distributions are
similar ($p_{\rm ch}>0.01$) and that each sample consists of at least 9 objects. The size of the
SLSN-I host sample (first) and of the galaxy sample (last) are given below each sample.
}
\label{fig:statcomp_1}
\end{figure}

\begin{figure}
\includegraphics[angle=0, width=0.99\columnwidth]{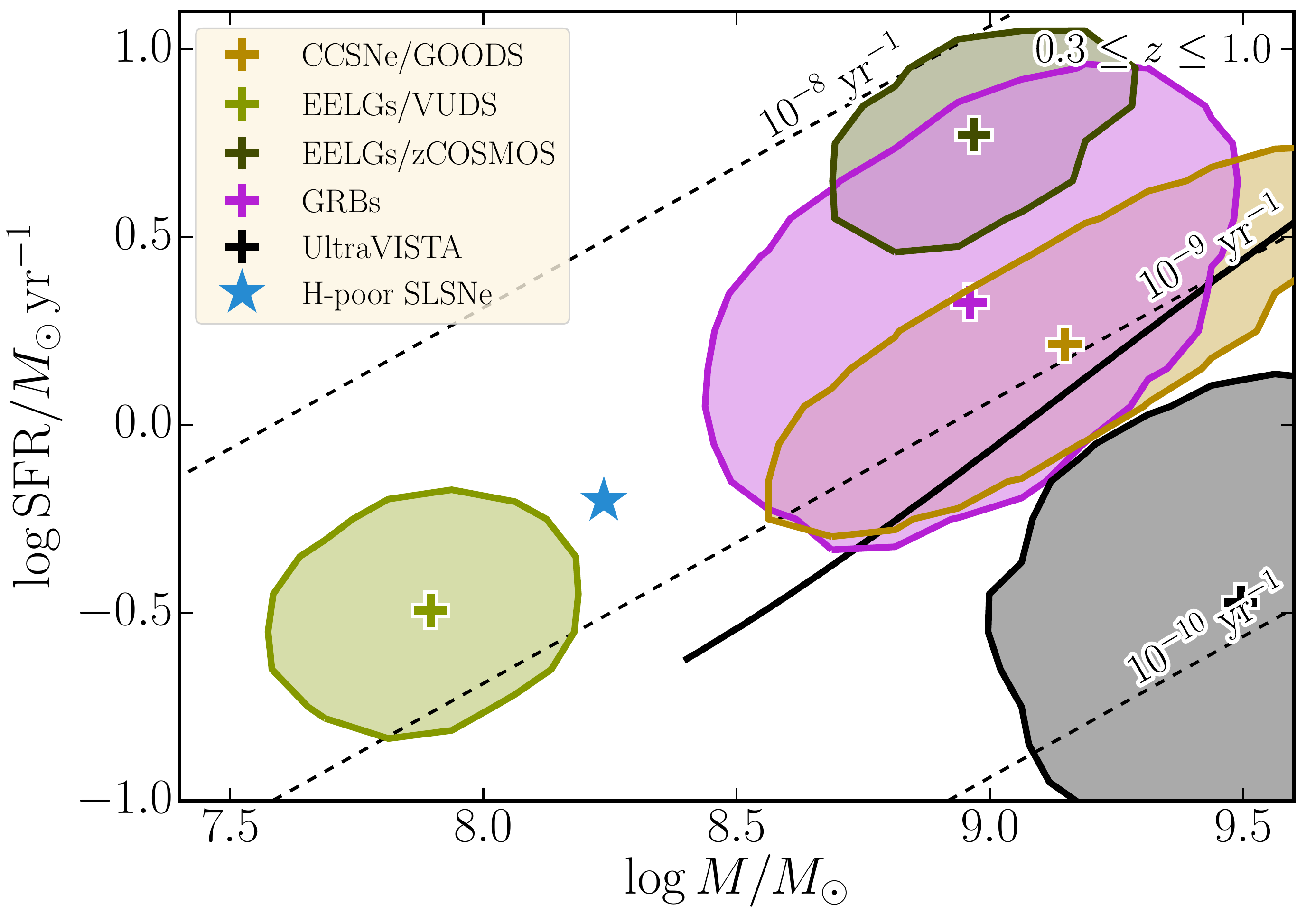}
\includegraphics[angle=0, width=0.99\columnwidth]{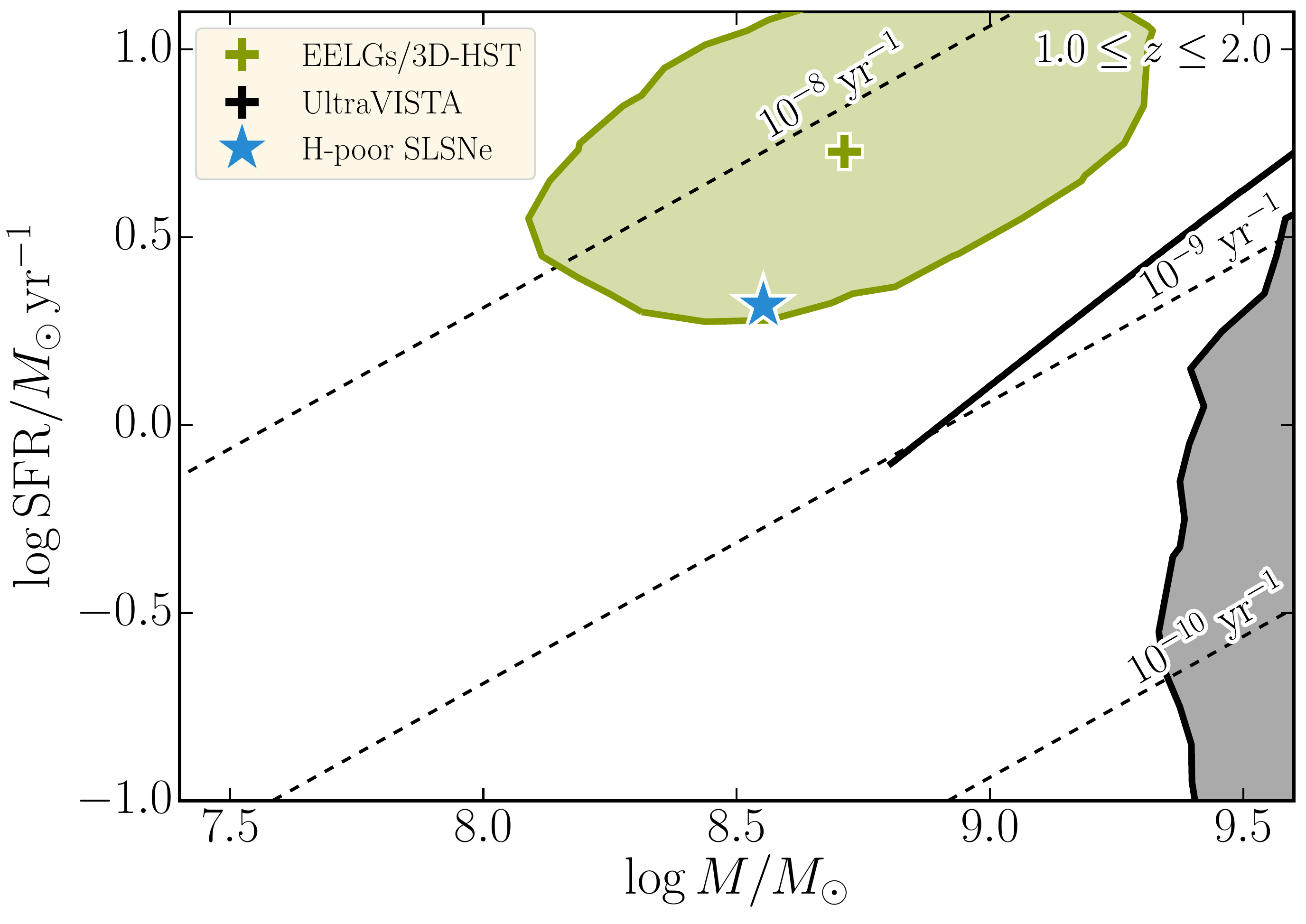}
\caption{
Statistical tests in the mass-SFR plane between SLSN-I host galaxies and various galaxy samples
at $0.3\leq z \leq 1.0$ (top) and $1\leq z \leq 2$ (bottom). The mean mass and SFR of the SLSN-I
hosts is indicated by the ``$\star$".
To assess how SLSN-I hosts differ from other galaxy samples, we bootstrap each galaxy sample
30\,000 times, randomly draw 22 objects (the size of the SLSN-I host sample), and compute the mean SFR
and mass. The barycentre of each distribution is indicated by a ``+''. For the sake of clarity,
these values are not displayed for the UltraVISTA sample. The shaded areas display
the regions that encompass 99\% of all realisations. The mean SFR and the mean mass of SLSN-I host
galaxies cannot be generated from random subsamples of the GRB, EELG and UltraVISTA samples at $z<0.3$.
At $0.3<z<1.0$, the mean SFR and mean mass can be generated from random subsamples of the 3D-HST EELG
sample.The dashed lines show curves of constant specific star-formation rate. The thick line shows the galaxy main sequence.
}
\label{fig:2d-stat}
\end{figure}

Hydrogen-poor SLSNe are preferentially found in blue low-mass dwarf galaxies with high sSFR and
metallicities of $<0.4~~Z_\odot$. These properties are similar to those of extreme emission-line
galaxies and GRB host galaxies. This sparked a long-standing debate on how strong the
similarities actually are \citep[e.g.,][]{Lunnan2014a, Chen2015a, Leloudas2015a, Angus2016a, Japelj2016a}.
The answer to this question was not only of interest to compare the galaxy populations, but also to draw
conclusions on the progenitors of GRBs and SLSNe (see previous
section) and even to propose similarities in the energy source powering these two stellar explosions.

Previous studies were limited to small samples
($\sim10$ objects) or even to the comparison with galaxy samples at different redshifts. In some cases,
selection criteria were introduced that led to non-random sampling of distribution functions, such as excluding
GRB and SLSN host galaxies without $K$-band observations \citep{Japelj2016a}.\footnote{For example,
$\sim30\%$ of our SLSN hosts at $z<1$ are too faint to obtain meaningful $K_{\rm s}$-band constraints,
even with the most efficient instruments.
}
Given the size of our GRB and SLSN samples ($>50$ objects each; Table \ref{tab:comp_samples}),
we attempt to provide a new perspective on this conundrum and to how SLSN hosts
compare to other galaxy samples. We divide our
samples into two redshift intervals: $0.3\leq z \leq 1.0$ and $1.0\leq z \leq 2.0$. Each of these
intervals covers a lookback-time interval of 2.6--4.4~Gyr, which is a compromise between minimising
the impact of the general cosmic evolution of star-forming galaxies and maximising number statistics. For
the GRB sample, we also modelled the SEDs with the same assumptions and the same software
as for the SLSN host galaxies, to minimise systematic errors.

To assess the differences, we apply two distinct tests. We use two-sided
Anderson-Darling tests to ascertain differences in the distribution functions, and we quantify the
frequency of how often the estimator of the mean mass and SFR of SLSN-I host galaxies can be
obtained from the comparison samples by chance (2D test; for details see Sect. \ref{sec:statistics}).
While an AD test compares distribution functions, the 2D test compares multiple parameters
at the same time, namely SFR, mass and indirectly the sSFR. Therefore, its outcome is less sensitive
to the selected properties. The 2D tests are, however, limited to mean values.
We reject the null hypothesis that two distributions are statistically similar, if the chance
probability is $p_{\rm ch}<10^{-2}$ for a given test.

Figure \ref{fig:statcomp_1} summarises the $p$ values of the AD tests for five different properties
($B$-band luminosity, $R-K_{\rm s}$ colour, mass, SFR and sSFR). The AD tests between distributions of
SLSN-I and GRB hosts reveal low $p$ values between $2\times10^{-4}$ ($B$-band luminosity) and 0.01 (SFR).
The statistical tests in the mass-SFR plane, displayed in Fig. \ref{fig:2d-stat}, corroborate these results.
The chance probability to extract an estimator from the GRB sample with a mean mass and SFR similar to SLSN-I hosts
is $\sim10^{-3}$ (equivalent to $3.3\sigma$). Therefore, we reject the null hypothesis that GRB and
SLSN host galaxies are statistically similar. We stress that revealing these differences requires
large and homogenous samples, like the one presented in this paper, which were not available
in previous studies.

\citet{Leloudas2015a} ignited the SLSN-EELG connection by unravelling a high incidence rate of
hosts with intense [\ion{O}{iii}] emission and ionisation conditions comparable to EELGs for our spectroscopy
sample. The comparison between properties of the stellar component is less straightforward. The statistical
tests point to similarities with 3D-HST EELGs at $1<z<2$ ($p_{\rm ch}\sim0.03$--0.16), but to weaker
similarities with VUDS EELGs ($p_{\rm ch}\lesssim0.01$--0.09) and even stark differences to zCOSMOS EELGs
($p_{\rm ch}<10^{-5}$) at lower redshift (Figs. \ref{fig:statcomp_1}, \ref{fig:2d-stat}).

\begin{figure}
\includegraphics[width=1\columnwidth]{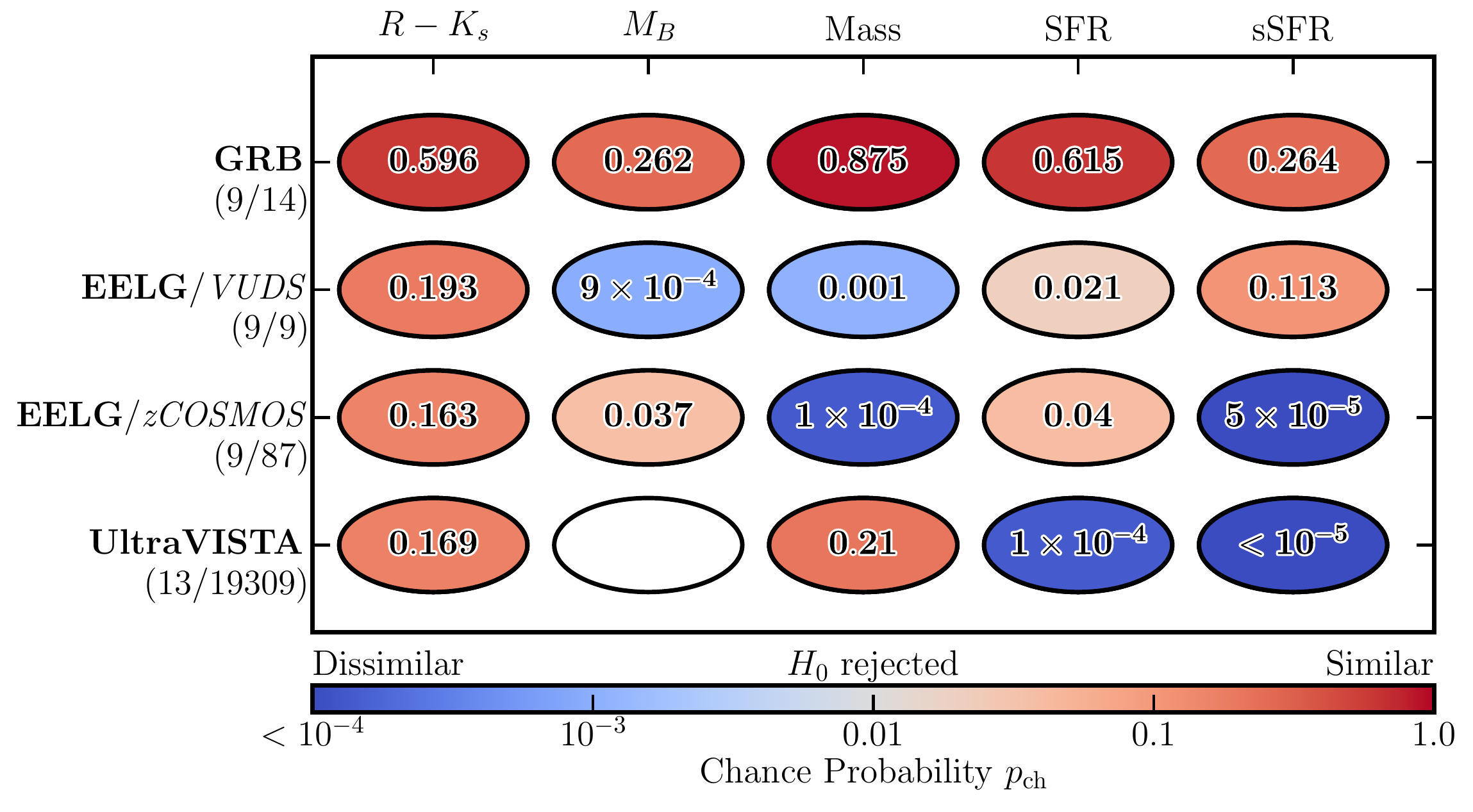}
\caption{
Two-sided Anderson-Darling tests between SLSN-IIn host galaxies and different galaxy samples at $z<0.5$
(similar to Fig. \ref{fig:statcomp_1}). The sizes of the SLSN-IIn host sample (first) and of
the galaxy sample (last) are given below each sample.
}
\label{fig:statcomp_2}
\end{figure}

These findings can be reconciled with the definition of EELGs and how they are identified in galaxy
surveys. EELGs are defined \textit{spectroscopically} by ${\rm EW}_{\rm rest}([{\rm \ion{O}{iii}]} \lambda5007)>100$~\AA\ (a
measure of the recent star-formation activity normalised to the light from all stars). Furthermore,
the VUDS EELGs were originally pooled from a galaxy sample with a brightness of $25<I{\rm(AB)}<23$,
whereas the zCOSMOS sample was limited to bright and therefore more massive EELG candidates
[$I{\rm (AB)}<22$~mag; Tables \ref{tab:comp_samples}, \ref{tab:statresult_non_SLSN}]. In contrast,
the average $R$-band brightness of SLSN-I hosts at $0.3<z<1.0$ is $\sim24.6$~mag, similar to VUDS EELGs
but $>2.5$~mag fainter than the $I$-band magnitude limit of the zCOSMOS sample. This immediately explains why the properties of the
stellar-component of SLSN-I host galaxies and zCOSMOS EELGs are so distinct. The stellar component
in SLSN-I host galaxies is more similar to VUDS EELGs, though the statistical tests are inconclusive
as to whether they are indeed statistically similar or distinct.
Differences between the stellar components of SLSN-I host galaxies and EELG samples are expected
because the properties of the ionised gas for a larger number of
SLSN-I hosts is not as extreme as that of EELGs. Furthermore, the different EELG samples show that this
ephemeral and transformative phase in galaxy evolution is observed in galaxies over a wide range of masses.

The AD and the 2D tests (Figs. \ref{fig:statcomp_1}, \ref{fig:2d-stat}) show that the properties of
the SLSN-I host population are more extreme and in stark contrast to the general population of
star-forming galaxies in the UltraVISTA survey and the hosts galaxies of regular core-collapse SNe
from the GOODS survey.

\subsubsection{SLSN-IIn host population}

\begin{figure}
\includegraphics[width=1\columnwidth]{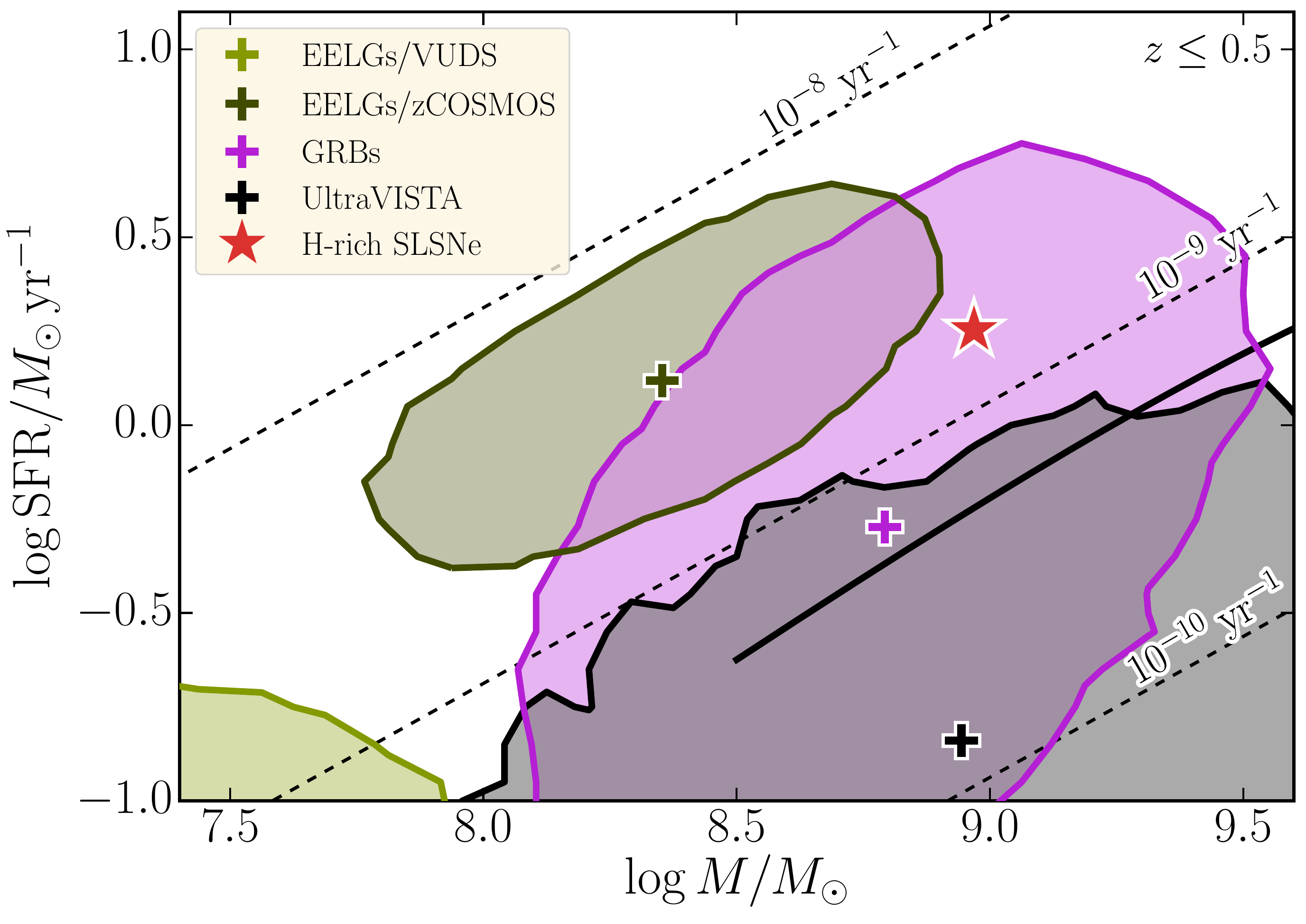}
\caption{
Statistical tests in the mass-SFR plane between SLSN-IIn host galaxies and various galaxy samples
at $z<0.5$ (similar to Fig. \ref{fig:2d-stat}). The mean mass and mean SFR of the SLSN-IIn host sample
(indicated by `$\star$') can be generated by random subsamples of the GRB sample, but is inconsistent
with the EELG and UltraVISTA samples.
}
\label{fig:2d-stat_2}
\end{figure}

The host population of SLSNe-IIn is characterised by a rich diversity:
\textit{i}) the mass
and luminosity distributions have dispersions that are a factor of 1.5--2 larger compared to any other
class of star-forming galaxies discussed in this paper;
\textit{ii}) hosts with stellar masses of more than $10^{10}~M_\odot$ are scarce, despite the
large dispersion in galaxy mass;
\textit{iii}) the $R-K_{\rm s}$ colour has a mean and a dispersion that is similar to star-forming galaxies; and
\textit{iv}) the sSFRs are shifted by 0.6~dex towards higher sSFR with respect to the main sequence
of star-forming galaxies in the mass-sSFR plane (Fig. \ref{fig:ssfr_vs_mass}).

The large dispersion measurements are difficult to map to a single progenitor system of SLSNe-IIn.
Type IIn SNe are primarily powered by the interaction of the SN ejecta with the circumstellar
material expelled prior to explosion. If the interaction is strong, the signature of the original
SN gets washed out. In the most extreme cases of CSM interaction, even different types of
CCSNe as well as thermonuclear Type Ia SNe could give rise to Type IIn SNe \citep[e.g.,][]{Leloudas2015c}.
The fact that all hosts show evidence for recent star-formation and have very high sSFRs suggests that
the contamination by Type Ia SNe is low. This implies that the diversity is primarily
due to different progenitor channels (see also \citealt{Angus2016a}).

Similar to the SLSN-I host population, we perform AD tests (Fig. \ref{fig:statcomp_2}) and the tests in the mass-SFR plane (Fig. \ref{fig:2d-stat_2}) to
put the SLSN-IIn host population in context with other galaxy samples. Despite the limited number
statistics, the SLSN-IIn host population is clearly distinct from the general population of star-forming
galaxies in the UltraVISTA survey. While the distribution functions are broader than those of other galaxy
samples, the lack of massive hosts suggests some dependence on environment properties. The similarities
of the distribution functions to GRBs as well as the locus in the mass-SFR plane suggests that their
hosts are similar. The lack of massive host galaxies would suggests a stifled production efficiency at metallicities
higher than $Z\sim0.8~Z_\odot$, the metallicity above which the GRB production efficiency is significantly
reduced (Sect. \ref{sec:cutoff}). However, the small number of SLSN-IIn in conjunction with their rich diversity precludes
drawing a firm conclusion,
yet.

\subsubsection{SLSN-II host population}

The family of type II SLSNe is the rarest class among SLSNe. In constrast to SLSN-IIn, the emission of
SLSNe-II is not powered by strong interaction of the SN ejecta with the circum-stellar material.
Only 3 events among the 29 H-rich SLSNe known today belong to this class.\footnote{This number was compiled from
the sample presented here and in \citet{Perley2016a}, and also includes two H-rich SLSNe that were
reported in the literature but not discussed in these papers.} Their host properties seem to
be distinct from
the average properties of the SLSN-IIn family. Type II SLSNe occupy the lower to bottom
half of the distribution functions. Two of three hosts are even among the least massive galaxies
in our sample ($10^6$--$10^7~M_\odot$). Those masses are comparable to the least massive
dwarf galaxies in the local Universe. According to the parameterisation
of the mass-metallicity relation in \citet{Andrews2013a}, their masses point to galaxies
with metallicity of $\lesssim0.3~Z_\odot$.

Intriguingly, \citet{Yan2015a, Yan2017a} revealed that an increasing number of SLSNe-I showed episodic hydrogen
emission at late phases. The properties of these hydrogen emission lines are similar to those of
CSS121015, SN2008es and SN2013hx.
\citet{Yan2015a, Yan2017a} attributed this feature to pulsational instabilities, where the outer H-rich envelope
is expelled during a violent mass-loss episode. As the SN ejecta traverses the circumstellar material,
shocks between the ejecta and the circumstellar material produce episodic hydrogen emission. Alternatively,
these authors proposed that the progenitor retained a thin layer of hydrogen where recombination lines
emerge only after the SN ejecta cooled down. Hence, it is possible that SLSNe-II are more closely
connected to SLSNe-I.

\citet{Inserra2016a} noted that the spectroscopic and photometric properties of SN2013hx showed
similarities to brighter regular Type II SNe. However, even these brighter regular
Type II SNe are still significantly less luminous than SLSNe. It is not clear how stars with
an extended hydrogen envelope could produce such high luminosities. Larger samples are
needed to better understand how the SLSN II population compares to different classes of SNe and SLSNe.

\subsection{Selection biases}

Our conclusions could be affected by various selection biases, such as publication bias, target
selection bias and classification bias. Moreover, the SUSHIES sample is compiled from
different SN surveys, which makes it even more difficult to quantify the effective bias.

To examine whether our sample has the same level of bias as the PS1 and PTF samples, we perform two-sided
AD tests between the distributions of the host properties. If the probability
of randomly drawing a distribution from the PS1/PTF samples, which is at least as extreme
as the SUSHIES sample, is larger than 1\%, we reject the hypothesis that the level of bias
in SUSHIES is different from the PS1 and PTF samples. For a fair comparison, we remove common
objects and split our sample into two redshift intervals to take the redshift domains
of the PS1 and the PTF samples into account: $z<0.5$ for the PTF sample and $z>0.5$ for the PS1 sample.

The AD tests between the $B$-band luminosity, mass and SFR distributions of 20 SLSN-I hosts
from our sample and 16 SLSN-I hosts from the PTF sample give a high chance probability of agreement
of $>19\%$. For SLSN-II/IIn
hosts, the chance probability of $>27\%$ is even substantially higher (SUSHIES: 13 objects, PTF: 14 objects).
A similar result can be obtained from the comparison with the PS1 sample ($p_{\rm ch}>8\%$;
SUSHIES: 11 objects, PS1: 15 objects).

In conclusion, the heterogeneous SUSHIES sample has a similar effective bias to the PS1
and the PTF samples. A detailed discussion about possible selection effects biasing the PS1 and PTF
samples is presented in \citet{Lunnan2014a} and \citet{Perley2016a}.

\section{Summary}\label{sec:conclusion}

We present the photometric properties of 53 H-poor and 16 H-rich SLSNe, detected before 2015
and publicly announced before mid 2015. Among those are four new SLSNe (two of each type),
found in the ASIAGO SN catalogue, with a peak luminosity significantly brighter than
$M_{V}=-21$~mag. Each host is a target of deep imaging campaigns that probe the rest-frame
UV to NIR. In addition, we incorporate radio data from wide-field surveys and JVLA observations
to put limits on the total star-formation activity. By modelling the spectral energy
distributions, we derive physical properties, such as mass, SFR and luminosity, and build
distribution functions to ascertain the influence of these properties on the SLSN population.
Our main conclusions are:

\begin{enumerate}
\item H-poor SLSNe are preferentially found in very blue low-mass dwarf galaxies. Their sSFRs
are on average 0.5~dex larger compared to the main sequence of star-forming galaxies and they
populate a part of the sSFR-mass parameter space that is typically occupied by EELGs.

\item The host population of SLSNe-IIn shows very complex properties: 1) the mass
and luminosity distributions have dispersions that are a factor of 1.5--2 larger compared to all
comparison samples;
2) the $R-K_{\rm s}$ colour has a mean and a dispersion which is similar to star-forming galaxies;
and 3) the sSFRs are on average a factor of 10 larger than of regular star-forming
galaxies discussed in this paper. These properties argue for a massive star origin of all SLSNe-IIn
in our sample but to a low dependency on integrated host properties. Because the luminosity of SLSNe-IIn
is determined by the strength of the interaction and not by a particular type of stellar
explosion, this diversity suggests multiple progenitor channels.

\item The hosts of the three Type II SLSNe are at the bottom of any distribution function.
Two out of three Type II SLSNe exploded in the least massive host galaxies in our sample
($10^6$--$10^7~M_\odot$). Their hosts are similar to those of H-poor SLSNe. Their
preference for low-mass and hence low-metallicity galaxies hints to different progenitors
from Type IIn SLSNe. Larger samples are needed to draw a conclusion on this question.

\item The scarcity of hosts above $10^{10}~M_\odot$ for SLSNe-I and SLSNe-IIn can be attributed
to a metallicity bias above which the production efficiency is stifled. Assuming an exponential
cut-off, the best-fit cut-off metallicity of H-poor SLSNe at $z<1$ is $12+\log\,{\rm O/H}=8.31^{+0.16}_{-0.31}$
($Z\sim0.4~Z_\odot$), which is 0.4 dex lower than for GRBs. The similarities between the mass
distributions of SLSN-IIn and GRB host galaxies suggest a metallicity cut-off at $\sim0.8$ solar
metallicity.

\item A growing population of SLSN hosts have masses between $10^6$ and $10^7~M_\odot$.
Those objects are among the least massive star-forming galaxies known to date and could
represent environments similar to those of starburst galaxies in the early Universe.

\item The redshift evolution of the SLSN-I host population is consistent with the general
cosmic evolution of star-forming galaxies. After detrending the data, the galaxy mass shows
evidence for differential evolution at $3.8\sigma$ confidence, while differential evolution
in the $B$-band and FUV luminosity can be excluded at $3\sigma$ confidence. The evolution
of the mass distribution of SLSN-I hosts is similar to the evolution of the mass-metallicity
relation, supporting connecting the dearth of massive hosts to a metallicity bias.

\item
Multiple statistical tests between the host properties of SLSN-I and GRB host galaxies reveal
differences at $>3\sigma$ confidence. H-poor SLSNe are found in less massive (and therefore more
metal-poor) hosts than GRBs. To conclusively show that SLSN-I and GRB host galaxies are different
on average, large samples with well sampled SEDs are needed.

\item SLSN-I hosts and EELGs show similarities, even in broad-band properties. This suggests that environmental
conditions in EELGs play a very important role in the formation of SLSNe-I.
We conclude that metallicity is \textit{not} the sole ingredient regulating the SLSN-I production
and suggest that a young age plays an important role in the formation of H-poor SLSNe as well.

\item The class of H-poor SLSNe comprises of fast- and slow-declining SLSNe.
A sub-sample of 21 SLSNe-I have measured decline time scales: 14 fast and 7 slow
declining SLSNe-I. We find no differences between both host populations. However, larger samples
of SLSNe with measured decay time scales are needed to draw a firm conclusion.

\item No host is detected in wide-field radio surveys. At $z<0.5$, the $4\sigma$ limits
on the total SFR are a factor of 20 larger than the SFRs derived from SED modelling, ruling
out truly obscured star-formation missed by optical diagnostics. This result is consistent
with the lack of high-obscured hosts and SLSNe. The deep radio observation of the solar-metallicity
host of the H-poor SLSN MLS121104 reveals no difference to the SED-derived SFR.

\end{enumerate}

\section*{Acknowledgments}

We acknowledge with sadness the unexpected passing of our esteemed colleague, co-author and
friend Javier Gorosabel. His support of and contributions to this work and astronomy in
general are greatly appreciated.

We thank the referee Sandra Savaglio for a careful reading of the manuscript and for lots of
helpful comments that improved this paper.

We thank R. Quimby for sharing an explosion image of SN2005ap, P. Vresswijk and D.
A. Perley for the host image of PTF13ajg, T.-W. Chen for an SN image of MLS121104, and
A. Uldaski for an SN image of SN2003ma. SSchulze thanks P. Pietrukowicz, D. Whalen and A.
Gal-Yam for fruitful discussions.

SSchulze acknowledges support from the CONICYT-Chile FONDECYT Postdoctorado fellowship
3140534 and the Feinberg Graduate School. SSchulze and FEB acknowledge support from
Basal-CATA PFB-06/2007, and Project IC120009 ``Millennium Institute of Astrophysics (MAS)''
of Iniciativa Cient\'{\i}fica Milenio del Ministerio de Econom\'{\i}a, Fomento y Turismo.
TK acknowledges support through the Sofja Kovalevskaja Award to P. Schady from the Alexander
von Humboldt Foundation of Germany.
AdUP and CT acknowledge support from the Ram\'on y Cajal fellowships and the Spanish Ministry of
Economy and Competitiveness through project AyA2014-58381-P.
RA acknowledges support from the European Research Council (ERC) Advanced Grant 695671 `QUENCH'.

This paper is based partly on observations made with: ESO Telescopes at the La Silla Paranal
Observatory; the 6.5-m Magellan Telescopes located at the Las Campanas Observatory, Chile;
the Gran Telescopio Canarias (GTC), installed in the Spanish Observatorio del Roque de los Muchachos of the Instituto de
Astrof\'isica de Canarias, in the island of La Palma;
the Centro Astron\'omico Hispano Alem\'an (CAHA) at Calar Alto, Spain, operated jointly by
the Max-Planck Institut f\"ur Astronomie and the Instituto de Astrof\'isica de Andaluc\'ia
(CSIC); the Nordic Optical Telescope, operated
by the Nordic Optical Telescope Scientific Association at the Observatorio del Roque
de los Muchachos, La Palma, Spain, of the Instituto de Astrof\'isica de Canarias; and
Karl G. Jansky Very Large Array, New Mexico, United States of America.
This research draws upon data provided by Cypriano as distributed by the NOAO
Science Archive. NOAO is operated by the Association of Universities for Research in
Astronomy (AURA) under a cooperative agreement with the National Science Foundation.
This publication makes use of data products from the Wide-field Infrared Survey
Explorer (\wise), which is a joint project of the University of California, Los Angeles,
and the Jet Propulsion Laboratory/California Institute of Technology, funded by the
National Aeronautics and Space Administration.
\galex\ (Galaxy Evolution Explorer) is a NASA Small Explorer, launched in April 2003.
We gratefully acknowledge NASA's support for construction, operation, and science analysis
for the \galex\ mission, developed in cooperation with the Centre National d'Etudes
Spatiales of France and the Korean Ministry of Science and Technology.
Based in part on data collected at the Subaru Telescope, Hawaii, United States of America, which is operated by the National
Astronomical Observatory of Japan.
The National Radio Astronomy Observatory (NRAO) is a facility of the National Science Foundation
operated under cooperative agreement by Associated Universities, Inc.
Part of the funding for GROND was generously granted from the Leibniz-Prize to Prof. G.
Hasinger (DFG grant HA 1850/28-1).

Funding for SDSS-III has been provided by the Alfred P. Sloan Foundation, the
Participating Institutions, the National Science Foundation, and the U.S. Department
of Energy Office of Science. The SDSS-III web site is \href{http://www.sdss3.org/}{http://www.sdss3.org/}.
SDSS-III is managed by the Astrophysical Research Consortium for the Participating
Institutions of the SDSS-III Collaboration, including the University of Arizona,
the Brazilian Participation Group, Brookhaven National Laboratory, University of
Cambridge, Carnegie Mellon University, University of Florida, the French Participation
Group, the German Participation Group, Harvard University, the Instituto de Astrof\'isica
de Canarias, the Michigan State/Notre Dame/JINA Participation Group, Johns Hopkins
University, Lawrence Berkeley National Laboratory, Max Planck Institute for
Astrophysics, Max Planck Institute for Extraterrestrial Physics, New Mexico State
University, New York University, Ohio State University, Pennsylvania State University,
University of Portsmouth, Princeton University, the Spanish Participation Group,
University of Tokyo, University of Utah, Vanderbilt University, University of
Virginia, University of Washington, and Yale University.
Based on observations obtained with MegaPrime/MegaCam, a joint project of CFHT and CEA/IRFU,
at the Canada-France-Hawaii Telescope (CFHT), which is operated by the National Research
Council (NRC) of Canada, the Institut National des Science de l'Univers of the Centre
National de la Recherche Scientifique (CNRS) of France and the University of Hawaii.
This work is partly based on data products produced at Terapix available at the Canadian
Astronomy Data Centre as part of the Canada-France-Hawaii Telescope Legacy Survey, a
collaborative project of NRC and CNRS.

This project used public archival data obtained with the the Dark Energy Camera (DECam) by the Dark
Energy Survey (DES). Funding for the DES Projects has been provided by
the DOE and NSF (USA),  MISE (Spain),  STFC (UK), HEFCE (UK),  NCSA (UIUC),  KICP (U. Chicago),
CCAPP (Ohio State),  MIFPA (Texas A\&M),  CNPQ, FAPERJ, FINEP (Brazil),  MINECO (Spain),
DFG (Germany) and the collaborating institutions in the Dark Energy Survey, which are
Argonne Lab,  UC Santa Cruz,  University of Cambridge,  CIEMAT-Madrid,  University of Chicago,
University College London,  DES-Brazil Consortium,  University of Edinburgh,  ETH Z{\"u}rich,
Fermilab,  University of Illinois,  ICE (IEEC-CSIC), IFAE Barcelona,  Lawrence Berkeley Lab,
LMU M{\"u}nchen and the associated Excellence Cluster Universe,  University of Michigan,
NOAO,  University of Nottingham,  Ohio State University,  University of Pennsylvania,
University of Portsmouth,  SLAC National Lab,  Stanford University,  University of Sussex,
and Texas A\&M University.

This research has made use of the NASA/ IPAC Infrared Science Archive, which is operated by
the Jet Propulsion Laboratory and the California Institute of Technology, under contract with
the National Aeronautics and Space Administration.

This research made use of \texttt{Astropy} \citep{Astropy2013a}, \texttt{matplotlib}
\citep{Hunter2007a}, \texttt{NumPy} \citep{NumPy} and \texttt{SciPy} \citep{SciPy}.
The results in this paper were obtained using \texttt{R} version 3.3.2 with the packages
\texttt{kSamples} version 1.2.4. \texttt{R} itself and all packages used are available from the
Comprehensive R Archive Network (CRAN)
at \href{http://CRAN.R-project.org/}{http://CRAN.R-project.org/}.

\noindent
$^{1}$ Department of Particle Physics and Astrophysics, Weizmann Institute of Science, Rehovot 7610001, Israel. \\
$^{2}$ Instituto de Astrof\'isica, Facultad de F\'isica, Pontificia Universidad Cat\'olica de Chile, Vicu\~{n}a Mackenna 4860, 7820436 Macul, Santiago, Chile.\\
$^{3}$ Millennium Institute of Astrophysics, Vicu\~{n}a Mackenna 4860, 7820436 Macul, Santiago, Chile.\\
$^{4}$ Max-Planck-Institut f\"ur extraterrestrische Physik, Gie{\ss}enbachstra{\ss}e, 85748 Garching, Germany. \\
$^{5}$ Dark Cosmology Centre, Niels Bohr Institute, University of Copenhagen, Juliane Maries Vej 30, 2100 Copenhagen, Denmark. \\
$^{6}$Unidad Asociada Grupo Ciencia Planetarias UPV/EHU-IAA/CSIC, Departamento de F\'isica Aplicada I, E.T.S. Ingenier\'ia, \\
~Universidad del Pa\'is-Vasco UPV/EHU, Alameda de Urquijo s/n, E-48013 Bilbao, Spain.\\
$^{7}$Ikerbasque, Basque Foundation for Science, Alameda de Urquijo 36-5, E-48008 Bilbao, Spain.\\
$^{8}$Instituto de Astrof\' isica de Andaluc\' ia (IAA-CSIC), Glorieta de la Astronom\' ia s/n, E-18008, Granada, Spain.\\
$^{9}$ESO, Alonso de Cordova 3107, Vitacura, Santiago de Chile, Chile.\\
$^{10}$Instituto de F\'isica y Astronom\'ia, Universidad de Valpara\'iso, Avda. Gran Breta\~na 1111, Valpara\'iso, Chile.\\
$^{11}$Cavendish Laboratory, University of Cambridge, 19 JJ Thomson Avenue, Cambridge, CB3 0HE, UK.\\
$^{12}$Kavli Institute for Cosmology, University of Cambridge, Madingley Road, Cambridge CB3 0HA, UK.\\
$^{13}$Space Science Institute, 4750 Walnut Street, Suite 205, Boulder, Colorado 80301, USA.\\
$^{14}$Mullard Space Science Laboratory - University College London, Holmbury Rd, Dorking RH5 6NT, United Kingdom.\\
$^{15}$INAF $-$ INAF $-$ Osservatorio Astrofisico di Arcetri, Largo Enrico Fermi 5, 50125 Firenze, Italy.\\
$^{16}$Las Campanas Observatory, Carnegie Observatories, Casilla 601, La Serena, Chile.\\
$^{17}$Heidelberger Institut fur Theoretische Studien, Schloss-Wolfsbrunnenweg 35, 69118 Heidelberg, Germany.\\
$^{18}$Institute for Astronomy, University of Hawaii, 2680 Woodlawn Drive, Honolulu, HI 96822, USA.\\
$^{19}$Smithsonian Astrophysical Observatory, 60 Garden Street, Cambridge, MA 02138, USA.\\
$^{20}$Department of Astronomy, University of Texas at Austin, Austin, TX 78712, USA.

\clearpage
\appendix

\section{Data table}
\clearpage
\newpage
\begin{table*}
\caption{List of host observations and their photometries.}
{\centering
\begin{tabular}{lcccccc}
\toprule
\parbox{3cm}{Survey/}	& \parbox{1.5cm}{\multirow{2}{*}{Instrument}}	& \multirow{2}{*}{Filter}	& Brightness		& \multirow{2}{*}{Date}	& Exposure	 & \multirow{2}{*}{Reference}	\\
Telescope		&						& 				& (mag$_{\rm AB}$)	&			& time (s)	 &				\\
\midrule
\multicolumn{7}{l}{\textbf{CSS100217} [SLSN-IIn, $z = 0.147$, $E(B-V)_{\rm MW}= 0.01~{\rm mag} $]}\\
\midrule
2MASS		&	 		&$	H		$&$	17.17	\pm	0.08	$& \nodata			& \nodata	& This work \\
CAHA		& Omega2000	&$	J		$&$	17.34	\pm	0.02	$& 2014-01-10		& $10\times60$	& This work \\
CAHA		& Omega2000	&$	K_{\rm s}		$&$	16.38	\pm	0.02	$& 2014-01-11		& $10\times60$	& This work \\
SDSS		& 			&$	u'		$&$	18.21	\pm	0.01	$& 2003-03-26		& \nodata	& This work \\
SDSS		& 			&$	g'		$&$	17.88	\pm	0.01	$& 2003-03-26		& \nodata	& This work \\
\bottomrule
\end{tabular}
}
\label{tab:data}
\end{table*}

\begin{table*}
\contcaption{List of host observations and their photometries.}
{\centering
\begin{tabular}{lcccccc}
\toprule
\parbox{3cm}{Survey/}	& \parbox{1.5cm}{\multirow{2}{*}{Instrument}}	& \multirow{2}{*}{Filter}	& Brightness		& \multirow{2}{*}{Date}	& Exposure	 & \multirow{2}{*}{Reference}	\\
Telescope		&						& 				& (mag$_{\rm AB}$)	&			& time (s)	 &				\\
\midrule
\multicolumn{7}{l}{\textbf{DES14X3taz} [SLSN-I, $z = 0.608$, $E(B-V)_{\rm MW}= 0.02~{\rm mag} $]}\\
\midrule
DES/Blanco	&	DeCam	&$	g'	$&$	26.16	\pm	0.39	$& \nodata	&  \nodata	& [2]	 	\\
DES/Blanco	&	DeCam	&$	r'	$&$	25.07	\pm	0.13	$& \nodata	&  \nodata	& [2]	 	\\
DES/Blanco	&	DeCam	&$	i'	$&$	24.95	\pm	0.13	$& \nodata	&  \nodata	& [2]	 	\\
DES/Blanco	&	DeCam	&$	z'	$&$	25.00	\pm	0.18	$& \nodata	&  \nodata	& [2]	 	\\
VIRMOS/VLT	&	VIMOS	&$	B	$&$	25.82	\pm	0.19	$& \nodata	&  \nodata	& [3]	 	\\
\bottomrule
\end{tabular}
}
\end{table*}

\begin{table*}
\contcaption{List of host observations and their photometries.}
{\centering
\begin{tabular}{lcccccc}
\toprule
\parbox{3cm}{Survey/}	& \parbox{1.5cm}{\multirow{2}{*}{Instrument}}	& \multirow{2}{*}{Filter}	& Brightness		& \multirow{2}{*}{Date}	& Exposure	 & \multirow{2}{*}{Reference}	\\
Telescope		&						& 				& (mag$_{\rm AB}$)	&			& time (s)	 &				\\
\midrule
\multicolumn{7}{l}{\textbf{MLS121104} [SLSN-I, $z = 0.303$, $E(B-V)_{\rm MW}= 0.15~{\rm mag} $]}\\
\midrule
CAHA		& Omega2000	&$	K_{\rm s}	$&$	20.37	\pm	0.55	$& 2014-08-10	& $9\times60$	& This work	\\
Magellan$^\dagger$& FourStar	&$	J	$&$	20.39	\pm	0.10	$& 2013-12-18	& \nodata	& [6]		\\
Magellan$^\dagger$& FourStar	&$	K_{\rm s}	$&$	19.63	\pm	0.12	$& 2013-12-18	& \nodata	& [6]		\\
SDSS		& 		&$	u'	$&$	26.62	\pm	35.33	$& 2005-06-12	& \nodata	& This work	\\
&		&		&$ (>21.76)			$&		&		&		\\
\bottomrule
\end{tabular}
}
\end{table*}

\begin{table*}
\contcaption{List of host observations and their photometries.}
{\centering
\begin{tabular}{lcccccc}
\toprule
\parbox{3cm}{Survey/}	& \parbox{1.5cm}{\multirow{2}{*}{Instrument}}	& \multirow{2}{*}{Filter}	& Brightness		& \multirow{2}{*}{Date}	& Exposure	 & \multirow{2}{*}{Reference}	\\
Telescope		&						& 				& (mag$_{\rm AB}$)	&			& time (s)	 &				\\
\midrule
\multicolumn{7}{l}{\textbf{PS1-11ap} [SLSN-I, slow declining, $z = 0.524$, $E(B-V)_{\rm MW}= 0.01~{\rm mag} $]}\\
\midrule
\hst$^\dagger$	&		&$	F475W	$&$	24.02	\pm	0.02	$& 2013-10-09		& \nodata	&	[6]	\\
Pan-STARRS$^\dagger$&		&$g_{\rm PS1}	$&$	24.20	\pm	0.15	$& \nodata	& \nodata	&	[6]	\\
Pan-STARRS$^\dagger$&		&$r_{\rm PS1}	$&$	23.32	\pm	0.10	$& \nodata	& \nodata	&	[6]	\\
Pan-STARRS$^\dagger$&		&$i_{\rm PS1}	$&$	22.86	\pm	0.09	$& \nodata	& \nodata	&	[6]	\\
Pan-STARRS$^\dagger$&		&$z _{\rm PS1}	$&$	23.24	\pm	0.13	$& \nodata	& \nodata	&	[6]	\\
\bottomrule
\end{tabular}
}

\end{table*}

\begin{table*}
\contcaption{List of host observations and their photometries.}
{\centering
\begin{tabular}{lcccccc}
\toprule
\parbox{3cm}{Survey/}	& \parbox{1.5cm}{\multirow{2}{*}{Instrument}}	& \multirow{2}{*}{Filter}	& Brightness		& \multirow{2}{*}{Date}	& Exposure	 & \multirow{2}{*}{Reference}	\\
Telescope		&						& 				& (mag$_{\rm AB}$)	&			& time (s)	 &				\\
\midrule
\multicolumn{7}{l}{\textbf{PS1-11bdn} [SLSN-I, $z = 0.738$, $E(B-V)_{\rm MW}= 0.02~{\rm mag} $]}\\
\midrule
CFHTLS/CFHT			&	MegaPrime	&$	u^*	$&$	28.43	\pm	1.92	$& \nodata		& \nodata	& This work	\\
		&				&		&$	(>26.84)			$&				&			& This work	\\
CFHTLS/CFHT			&	MegaPrime	&$	g'	$&$	26.50	\pm	0.24	$& \nodata		& \nodata	& This work	\\
CFHTLS/CFHT			&	MegaPrime	&$	r'	$&$	26.31	\pm	0.31	$& \nodata		& \nodata	& This work	\\
CFHTLS/CFHT			&	MegaPrime	&$	z'	$&$	28.96	\pm	13.39	$& \nodata		& \nodata	& This work	\\
\bottomrule
\end{tabular}
}

\end{table*}

\begin{table*}
\contcaption{List of host observations and their photometries.}
{\centering
\begin{tabular}{lcccccc}
\toprule
\parbox{3cm}{Survey/}	& \parbox{1.5cm}{\multirow{2}{*}{Instrument}}	& \multirow{2}{*}{Filter}	& Brightness		& \multirow{2}{*}{Date}	& Exposure	 & \multirow{2}{*}{Reference}	\\
Telescope		&						& 				& (mag$_{\rm AB}$)	&			& time (s)	 &				\\
\midrule
\multicolumn{7}{l}{\textbf{PS1-13gt} [SLSN-I, $z = 0.884$, $E(B-V)_{\rm MW}= 0.02~{\rm mag} $]}\\
\midrule
NOT		&	ALFOSC	&$	r'	$&$	25.71	\pm	1.88	$& 2015-03-13	& $9\times400$	& This work	\\
&		&		&$	(>24.11)		$&		&		\\
Pan-STARRS$^\dagger$&		&$ g_{\rm PS1}	$&$	>24.50			$& \nodata	& \nodata	&	[6]	\\
Pan-STARRS$^\dagger$&		&$ r_{\rm PS1}	$&$	>24.50			$& \nodata	& \nodata	&	[6]	\\
Pan-STARRS$^\dagger$&		&$ i_{\rm PS1}	$&$	>24.70			$& \nodata	& \nodata	&	[6]	\\
\bottomrule
\end{tabular}
}

\end{table*}

\begin{table*}
\contcaption{List of host observations and their photometries.}
{\centering
\begin{tabular}{lcccccc}
\toprule
\parbox{3cm}{Survey/}	& \parbox{1.5cm}{\multirow{2}{*}{Instrument}}	& \multirow{2}{*}{Filter}	& Brightness		& \multirow{2}{*}{Date}	& Exposure	 & \multirow{2}{*}{Reference}	\\
Telescope		&						& 				& (mag$_{\rm AB}$)	&			& time (s)	 &				\\
\midrule
\multicolumn{7}{l}{\textbf{PTF10vqv} [SLSN-I, $z = 0.452$, $E(B-V)_{\rm MW}= 0.06~{\rm mag} $]}\\
\midrule
CAHA		&	BUSCA	&$	u'	$&$	23.85	\pm	1.43	$&   2012-10-06  	& $15\times500$	& This work		\\
&		&		&$	(>22.45)		$& 	 		&		\\
CAHA		&	BUSCA	&$	g'	$&$	24.01	\pm	0.61	$&   2012-10-06  	& $15\times500$	& This work		\\
&		&		&$	(>23.53)		$&  	  		&		\\
CAHA		&	BUSCA	&$	r'	$&$	23.30	\pm	0.34	$&   2012-10-06  	& $15\times500$	& This work		\\
\bottomrule
\end{tabular}
}

\end{table*}

\begin{table*}
\contcaption{List of host observations and their photometries.}
{\centering
\begin{tabular}{lcccccc}
\toprule
\parbox{3cm}{Survey/}	& \parbox{1.5cm}{\multirow{2}{*}{Instrument}}	& \multirow{2}{*}{Filter}	& Brightness		& \multirow{2}{*}{Date}	& Exposure	 & \multirow{2}{*}{Reference}	\\
Telescope		&						& 				& (mag$_{\rm AB}$)	&			& time (s)	 &				\\
\midrule
\multicolumn{7}{l}{\textbf{SN1999as} [SLSN-I, $z = 0.127$, $E(B-V)_{\rm MW}= 0.03~{\rm mag} $]}\\
\midrule
CAHA		& Omega2000	&$	J	$&$	19.16	\pm	0.09	$& 2014-01-10	& $15\times60$	& This work	\\
CAHA		& Omega2000	&$	K	$&$	19.49	\pm	0.10	$& 2015-05-08	& $30\times60$	& This work	\\
CAHA		& Omega2000	&$	H	$&$	18.71	\pm	0.11	$& 2014-01-10	& $15\times60$	& This work	\\
\galex		&		&$	FUV	$&$	21.31	\pm	0.34	$&		&		& [5]		\\
\galex		&		&$	NUV	$&$	21.05	\pm	0.09	$&		&		& [5]		\\
\bottomrule
\end{tabular}
}

\end{table*}

\begin{table*}
\contcaption{List of host observations and their photometries.}
{\centering
\begin{tabular}{lcccccc}
\toprule
\parbox{3cm}{Survey/}	& \parbox{1.5cm}{\multirow{2}{*}{Instrument}}	& \multirow{2}{*}{Filter}	& Brightness		& \multirow{2}{*}{Date}	& Exposure	 & \multirow{2}{*}{Reference}	\\
Telescope		&						& 				& (mag$_{\rm AB}$)	&			& time (s)	 &				\\
\midrule
\multicolumn{7}{l}{\textbf{SN2006gy} [SLSN-IIn, $z = 0.019$, $E(B-V)_{\rm MW}= 0.14~{\rm mag} $]}\\
\midrule
2MASS		&	 	&$	J	$&$	11.76	\pm	0.02	$&  \nodata	&  \nodata	& This work	\\
2MASS		&	 	&$	H	$&$	11.61	\pm	0.02	$&  \nodata	&  \nodata	& This work	\\
2MASS		&	 	&$	K_{\rm s}	$&$	11.65	\pm	0.01	$&  \nodata	&  \nodata	& This work	\\
CAHA		&	BUSCA	&$	u'	$&$	16.36	\pm	0.01	$& 2012-10-08	& $18\times300$	& This work	\\
CAHA		&	BUSCA	&$	g'	$&$	14.27	\pm	0.01	$& 2012-10-08	& $18\times300$	& This work	\\
\bottomrule
\end{tabular}
}
\end{table*}

\begin{table*}
\contcaption{List of host observations and their photometries.}
{\centering
\begin{tabular}{lcccccc}
\toprule
\parbox{3cm}{Survey/}	& \parbox{1.5cm}{\multirow{2}{*}{Instrument}}	& \multirow{2}{*}{Filter}	& Brightness		& \multirow{2}{*}{Date}	& Exposure	 & \multirow{2}{*}{Reference}	\\
Telescope		&						& 				& (mag$_{\rm AB}$)	&			& time (s)	 &				\\
\midrule
\multicolumn{7}{l}{\textbf{SN2007bw} [SLSN-IIn, $z = 0.140$, $E(B-V)_{\rm MW}= 0.04~{\rm mag} $]}\\
\midrule
CAHA		& Omega2000	&$	J	$&$	18.49	\pm	0.03	$& 2013-07-24	& $15\times60$	& This work	\\
CAHA		& Omega2000	&$	H	$&$	18.20	\pm	0.04	$& 2013-07-24	& $15\times60$	& This work	\\
CAHA		& Omega2000	&$	K_{\rm s}	$&$	18.27	\pm	0.07	$& 2013-07-24	& $15\times60$	& This work	\\
SDSS		&		&$	u'	$&$	20.51	\pm	0.24	$& 2002-05-08	& \nodata	& This work	\\
SDSS		&		&$	g'	$&$	19.20	\pm	0.03	$& 2002-05-08	& \nodata	& This work	\\
\bottomrule
\end{tabular}
}
\end{table*}

\begin{table*}
\contcaption{List of host observations and their photometries.}
{\centering
\begin{tabular}{lcccccc}
\toprule
\parbox{3cm}{Survey/}	& \parbox{1.5cm}{\multirow{2}{*}{Instrument}}	& \multirow{2}{*}{Filter}	& Brightness		& \multirow{2}{*}{Date}	& Exposure	 & \multirow{2}{*}{Reference}	\\
Telescope		&						& 				& (mag$_{\rm AB}$)	&			& time (s)	 &				\\
\midrule
\multicolumn{7}{l}{\textbf{SN2009jh} [SLSN-I, slow declining, $z = 0.349$, $E(B-V)_{\rm MW}= 0.01~{\rm mag} $]}\\
\midrule
\hst		&	WFC3	&$	F390W	$&$	>25.92			$& \nodata	& \nodata	&	[12]	\\
\hst		&	WFC3	&$	F160W	$&$	25.30	\pm	0.15	$& \nodata	& \nodata	&	[12]	\\
VLT		&	FORS2	&$R_{\rm Special}$&$	25.86	\pm	0.13	$& 2013-02-20	& $6\times300$	& This work	\\
VLT		&	FORS2	&$	I	$&$	>24.49			$& 2013-03-17	& $3\times200$	& This work	\\
VLT		&	FORS2	&$z_{\rm Gunn}	$&$	24.18	\pm	0.60	$& 2013-02-20	& $10\times120$	& This work	\\
\bottomrule
\end{tabular}
}
\end{table*}

\begin{table*}
\contcaption{List of host observations and their photometries.}
{\centering
\begin{tabular}{lcccccc}
\toprule
\parbox{3cm}{Survey/}	& \parbox{1.5cm}{\multirow{2}{*}{Instrument}}	& \multirow{2}{*}{Filter}	& Brightness		& \multirow{2}{*}{Date}	& Exposure	 & \multirow{2}{*}{Reference}	\\
Telescope		&						& 				& (mag$_{\rm AB}$)	&			& time (s)	 &				\\
\midrule
\multicolumn{7}{l}{\textbf{SN2011ke$^{d}$} [SLSN-I, fast declining, $z = 0.143$, $E(B-V)_{\rm MW}= 0.01~{\rm mag} $]}\\
\midrule
VLT		&	FORS2	&$ u_{\rm High}	$&$	22.94	\pm	0.08	$& 2013-05-30	& $1\times400$		& This work	\\
VLT		&	FORS2	&$ g_{\rm High}	$&$	22.49	\pm	0.03	$& 2013-05-30	& $1\times300$		& This work	\\
VLT		&	FORS2	&$ R_{\rm Special}$&$	22.43	\pm	0.04	$& 2013-05-30	& $1\times90$		& This work	\\
VLT		&	FORS2	&$	I	$&$	22.24	\pm	0.06	$& 2013-05-30	& $1\times90$		& This work	\\
VLT		&	FORS2	&$ z_{\rm Gunn}	$&$	22.66	\pm	0.19	$& 2013-05-30	& $1\times90$		& This work	\\
\bottomrule
\end{tabular}
}
\end{table*}

\begin{table*}
\contcaption{List of host observations and their photometries.}
{\centering
\begin{tabular}{lcccccc}
\toprule
\parbox{3cm}{Survey/}	& \parbox{1.5cm}{\multirow{2}{*}{Instrument}}	& \multirow{2}{*}{Filter}	& Brightness		& \multirow{2}{*}{Date}	& Exposure	 & \multirow{2}{*}{Reference}	\\
Telescope		&						& 				& (mag$_{\rm AB}$)	&			& time (s)	 &				\\
\midrule
\multicolumn{7}{l}{\textbf{SN1000+0216} [SLSN-I, $z = 3.899$, $E(B-V)_{\rm MW}= 0.02~{\rm mag} $]}\\
\midrule
COSMOS/Subaru	& Suprime-Cam	&$	Bj	$&$	26.82	\pm	0.24	$& \nodata	& \nodata	&	[8]	\\
COSMOS/Subaru	& Suprime-Cam	&$	Vj	$&$	25.65	\pm	0.12	$& \nodata	& \nodata	&	[8]	\\
COSMOS/Subaru	& Suprime-Cam	&$	g+	$&$	26.70	\pm	0.26	$& \nodata	& \nodata	&	[8]	\\
COSMOS/Subaru	& Suprime-Cam	&$	r+	$&$	25.09	\pm	0.08	$& \nodata	& \nodata	&	[8]	\\
COSMOS/Subaru	& Suprime-Cam	&$	i+	$&$	24.60	\pm	0.06	$& \nodata	& \nodata	&	[8]	\\
\bottomrule
\end{tabular}
}
\end{table*}

\begin{table*}
\contcaption{List of host observations and their photometries.}
{\centering
\begin{tabular}{lcccccc}
\toprule
\parbox{3cm}{Survey/}	& \parbox{1.5cm}{\multirow{2}{*}{Instrument}}	& \multirow{2}{*}{Filter}	& Brightness		& \multirow{2}{*}{Date}	& Exposure	 & \multirow{2}{*}{Reference}	\\
Telescope		&						& 				& (mag$_{\rm AB}$)	&			& time (s)	 &				\\
\midrule
\multicolumn{7}{l}{\textbf{SSS120810} [SLSN-I, fast declining, $z = 0.156$, $E(B-V)_{\rm MW}= 0.02~{\rm mag} $]}\\
\midrule
\swift		&	UVOT	&$	uvm2	$&$	23.02	\pm	0.17	$& 2014-11-09--2016-01-28& 8922		& This work	\\
VLT		&	FORS2	&$ g_{\rm High}	$&$	22.71	\pm	0.10	$& 2013-05-30	& $1\times300$	& This work	\\
VLT		&	FORS2	&$R_{\rm Special}$&$	22.61	\pm	0.06	$& 2013-05-30	& $1\times300$	& This work	\\
VLT		&	FORS2	&$	I	$&$	22.47	\pm	0.07	$& 2013-05-30	& $3\times120$	& This work	\\
VLT		&	FORS2	&$ z_{\rm Gunn}	$&$	22.43	\pm	0.14	$& 2013-05-30	& $3\times120$	& This work	\\
VLT		&	HAWK-I	&$	J	$&$	23.11	\pm	0.14	$& 2013-06-02	& $6\times120$	& This work	\\
VLT		&	HAWK-I	&$	K_{\rm s}	$&$	23.38	\pm	0.24	$& 2013-06-02	& $6\times120$	& This work	\\
\bottomrule
\end{tabular}
}
\tablecomments{Data were not corrected for Galactic extinction apart from the data designated by $^\dagger$.
The CFHTLS $y'$ band filter is similar to CFHTLS $i'$. If a measurement has a confidence of $<2\sigma$,
we also report the $3\sigma$ limiting magnitude.\newline
$^a$ An error of 0.15~mag was added in quadrature to the CTIO/$R$-band measurement due to the contamination by a bright star.\newline
$^b$ The brightness was measured with a circular aperture with a diameter of $1.5\times{\rm FWHM}$ of the stellar PSF.\newline
$^c$ The object is on Chip 1 and 2. The measurement is only for Chip 2.\newline
$^d$ The brightness was measured with a circular aperture with a diameter of 7~px of the stellar PSF.
}
\tablerefs{
[1]: \citet{Lawrence2007a};
[2]: \citet{Smith2016a};
[3]: \citet{LeFevre2004a};
[4]: \citet{Nicholl2014a};
[5]: \citet{Bianchi2011a};
[6]: \citet{Lunnan2014a};
[7]: \citet{Lunnan2013a};
[8]: \citet{Ilbert2009a};
[9]: \citet{Hudelot2012a};
[10]: \citet{Jarvis2013a};
[11]: \citet{Vreeswijk2014a};
[12]: \citet{Angus2016a};
[13]: Inserra (priv. comm.);
[14]: \citet{Barbary2009a};
[15]: AllWISE Source Catalog;
[16]: \citet{Kato2007a};
[17]: \citet{Rest2011a};
[18]: \citet{Adami2006a};
[19]: \citet{Papadopoulos2015a};
[20]: \citet{McCracken2012a}
}
\end{table*}

\clearpage
\newpage
\begin{table*}
\caption{Radio observations of SLSN host galaxies.}
{\centering
\begin{tabular}{lcccccccc}
\toprule
\multirow{2}{*}{Object}	& \multirow{2}{*}{Redshift}& Survey/ 	& Observed	& r.m.s			& \multirow{2}{*}{Date}			& \multirow{2}{*}{Reference}	\\
			&							& Telescope	& frequency	& (mJy/beam)	& 										&\\
\midrule
\multicolumn{6}{l}{\textbf{SLSN-I host galaxies}}\\
\midrule
CSS140925			&	0.460	&	NVSS/VLA		&	1.4~GHz		&	0.45	& \nodata	& [1]\\
DES14S2qri			&	1.500	&	FIRST/VLA		&	1.4~GHz		&	0.155	& \nodata	&[2]\\
DES14X2byo			&	0.869	&	FIRST/VLA		&	1.4~GHz		&	0.108	& \nodata	&[2]\\
DES14X3taz			&	0.608	&	FIRST/VLA		&	1.4~GHz		&	0.106	& \nodata	&[2]\\
iPTF13ajg$^\dagger$	&	0.740	&	FIRST/VLA		&	1.4~GHz		&	0.102	& \nodata	&[2]\\
LSQ12dlf$^\ddagger$	&	0.255	&	NVSS/VLA		&	1.4~GHz		&	0.45	& \nodata	&[1]\\
LSQ14an				&	0.163	&	NVSS/VLA		&	1.4~GHz		&	0.45	& \nodata	&[1]\\
LSQ14mo$^\ddagger$	&	0.256	&	NVSS/VLA		&	1.4~GHz		&	0.45	& \nodata	&[1]\\
LSQ14bdq$^\dagger$	&	0.345	&	NVSS/VLA		&	1.4~GHz		&	0.45	& \nodata	&[1]\\
LSQ14fxj			&	0.360	&	FIRST/VLA		&	1.4~GHz		&	0.114	& \nodata	&[2]\\
MLS121104			&	0.303	&	JVLA			&	1.4~GHz		&	0.015	& 2015-07-28 \& & This work\\
		&			&					&				&			& 2015-08-05& This work\\
PS1-10ky			&	0.956	&	FIRST/VLA		&	1.4~GHz		&	0.162	& \nodata	&[2]\\
PS1-10pm			&	1.206	&	FIRST/VLA		&	1.4~GHz		&	0.141	& \nodata	&[2]\\
PS1-10ahf			&	1.158	&	FIRST/VLA		&	1.4~GHz		&	0.11	& \nodata	&[2]\\
PS1-10awh			&	0.909	&	FIRST/VLA		&	1.4~GHz		&	0.105	& \nodata	&[2]\\
PS1-10bzj$^\ddagger$&	0.649	&	NVSS/VLA		&	1.4~GHz		&	0.45	& \nodata	&[1]\\
PS1-11ap$^\dagger$	&	0.524	&	FIRST/VLA		&	1.4~GHz		&	0.144	& \nodata	&[2]\\
PS1-11tt			&	1.283	&	FIRST/VLA		&	1.4~GHz		&	0.151	& \nodata	&[2]\\
PS1-11afv			&	1.407	&	FIRST/VLA		&	1.4~GHz		&	0.162	& \nodata	&[2]\\
PS1-11aib			&	0.997	&	FIRST/VLA		&	1.4~GHz		&	0.139	& \nodata	&[2]\\
PS1-11bam			&	1.565	&	FIRST/VLA		&	1.4~GHz		&	0.139	& \nodata	&[2]\\
PS1-11bdn			&	0.738	&	FIRST/VLA		&	1.4~GHz		&	0.117	& \nodata	&[2]\\
PS1-12zn			&	0.674	&	FIRST/VLA		&	1.4~GHz		&	0.153	& \nodata	&[2]\\
PS1-12bmy			&	1.566	&	NVSS/VLA		&	1.4~GHz		&	0.45	& \nodata	&[1]\\
PS1-12bqf			&	0.522	&	FIRST/VLA		&	1.4~GHz		&	0.123	& \nodata	&[2]\\
PS1-13gt			&	0.884	&	FIRST/VLA		&	1.4~GHz		&	0.16	& \nodata	&[2]\\
PTF09atu			&	0.501	&	FIRST/VLA		&	1.4~GHz		&	0.172	& \nodata	&[2]\\
PTF09cnd$^\dagger$	&	0.258	&	FIRST/VLA		&	1.4~GHz		&	0.141	& \nodata	&[2]\\
PTF10hgi$^\ddagger$	&	0.099	&	NVSS/VLA		&	1.4~GHz		&	0.45	& \nodata	&[1]\\
PTF10vqv			&	0.452	&	FIRST/VLA		&	1.4~GHz		&	0.17	& \nodata	&[2]\\
PTF11rks$^\ddagger$	&	0.190	&	NVSS/VLA		&	1.4~GHz		&	0.45	& \nodata	&[1]\\
PTF12dam$^\dagger$	&	0.107	&	FIRST/VLA		&	1.4~GHz		&	0.14	& \nodata	&[2]\\
SCP06F6$^\ddagger$	&	1.189	&	FIRST/VLA		&	1.4~GHz		&	0.143	& \nodata	&[2]\\
SN1999as			&	0.127	&	FIRST/VLA		&	1.4~GHz		&	0.142	& \nodata	&[2]\\
\multirow{2}{*}{SN2005ap$^\ddagger$}&	0.283&	FIRST/VLA	&1.4~GHz&	0.13	& \nodata	&[2]\\
		&			&	JVLA			&	1.4~GHz		&	0.025	& 2015-09-20&This work\\
SN2006oz			&	0.396	&	FIRST/VLA		&	1.4~GHz		&	0.099	& \nodata	&[2]\\
SN2007bi$^\dagger$	&	0.128	&	FIRST/VLA		&	1.4~GHz		&	0.136	& \nodata	&[2]\\
SN2009de			&	0.311	&	FIRST/VLA		&	1.4~GHz		&	0.149	& \nodata	&[2]\\
SN2009jh$^\dagger$	&	0.349	&	FIRST/VLA		&	1.4~GHz		&	0.145	& \nodata	&[2]\\
SN2010gx$^\ddagger$	&	0.230	&	NVSS/VLA		&	1.4~GHz		&	0.45	& \nodata	&[1]\\
SN2010kd			&	0.101	&	FIRST/VLA		&	1.4~GHz		&	0.159	& \nodata	&[2]\\
SN2011ep			&	0.280	&	FIRST/VLA		&	1.4~GHz		&	0.158	& \nodata	&[2]\\
SN2011ke$^\ddagger$	&	0.143	&	FIRST/VLA		&	1.4~GHz		&	0.158	& \nodata	&[2]\\
SN2011kf$^\ddagger$	&	0.245	&	FIRST/VLA		&	1.4~GHz		&	0.154	& \nodata	&[2]\\
SN2012il$^\ddagger$	&	0.175	&	FIRST/VLA		&	1.4~GHz		&	0.145	& \nodata	&[2]\\
SN2013dg$^\ddagger$	&	0.265	&	FIRST/VLA		&	1.4~GHz		&	0.199	& \nodata	&[2]\\
SN2013hy			&	0.663	&	NVSS/VLA		&	1.4~GHz		&	0.45	& \nodata	&[1]\\
SN2015bn			&	0.110	&	FIRST/VLA		&	1.4~GHz		&	0.147	& \nodata	&[2]\\
SN1000+0216			&	3.899	&	FIRST/VLA		&	1.4~GHz		&	0.135	& \nodata	&[2]\\
SN2213-1745			&	2.046	&	NVSS/VLA		&	1.4~GHz		&	0.45	& \nodata	&[1]\\
SNLS06D4eu			&	1.588	&	NVSS/VLA		&	1.4~GHz		&	0.45	& \nodata	&[1]\\
SNLS07D2bv			&	1.500	&	FIRST/VLA		&	1.4~GHz		&	0.143	& \nodata	&[2]\\
SSS120810$^\ddagger$&	0.156	&	SUMSS			&	843~MHz		&	1.3	 	& \nodata	&[3]\\
\bottomrule
\end{tabular}
}
\label{tab:data_radio}
\end{table*}

\begin{table*}
\contcaption{Radio observations of SLSN host galaxies.}
{\centering
\begin{tabular}{lcccccccc}
\toprule
\multirow{2}{*}{Object}	& \multirow{2}{*}{Redshift}& Survey/ 	& Observed	& r.m.s			& \multirow{2}{*}{Date}			& \multirow{2}{*}{Reference}	\\
			&							& Telescope	& frequency	& (mJy/beam)	& 										&\\
\midrule
\multicolumn{6}{l}{\textbf{SLSN-IIn host galaxies}}\\
\midrule
CSS100217			&	0.147	&	FIRST/VLA		&	1.4~GHz		&	0.15	& \nodata	& [2]\\
PTF10heh			&	0.338	&	FIRST/VLA		&	1.4~GHz		&	0.12	& \nodata	& [2]\\
PTF10qaf			&	0.284	&	FIRST/VLA		&	1.4~GHz		&	0.135	& \nodata	& [2]\\
PTF11dsf			&	0.385	&	FIRST/VLA		&	1.4~GHz		&	0.15	& \nodata	& [2]\\
SN1999bd			&	0.151	&	FIRST/VLA		&	1.4~GHz		&	0.164	& \nodata	& [2]\\
SN2003ma			&	0.289	&	\nodata			&	\nodata		&	\nodata & \\
SN2006gy			&	0.019	&	NVSS/VLA		&	1.4~GHz		&	0.45	& \nodata	& [1]\\
SN2006tf			&	0.074	&	FIRST/VLA		&	1.4~GHz		&	0.132	& \nodata	& [2]\\
SN2007bw			&	0.14	&	FIRST/VLA		&	1.4~GHz		&	0.162	& \nodata	& [2]\\
SN2008am			&	0.233	&	FIRST/VLA		&	1.4~GHz		&	0.144	& \nodata	& [2]\\
\multirow{2}{*}{SN2008fz}&	0.133	&	FIRST/VLA	&	1.4~GHz		&	0.14	& \nodata	& [2]\\
		&			&	JVLA			&	1.4~GHz		&	0.015	& 2015-07-21&  This work\\
SN2009nm			&	0.21	&	FIRST/VLA		&	1.4~GHz		&	0.141	& \nodata	& [2]\\
SN2011cp			&	0.38	&	FIRST/VLA		&	1.4~GHz		&	0.137	& \nodata	& [2]\\
\midrule
\multicolumn{6}{l}{\textbf{SLSN-II host galaxies}}\\
\midrule
CSS121015$^\ddagger$&	0.287	&	FIRST/VLA		&	1.4~GHz		&	0.172	& \nodata	& [2]\\
SN2008es$^\ddagger$&	0.205	&	FIRST/VLA		&	1.4~GHz		&	0.147	& \nodata	& [2]\\
SN2013hx$^\ddagger$&	0.13	&	SUMSS			&	843~MHz		&	1.3	 	& \nodata	& [3]\\
\bottomrule
\end{tabular}
\tablecomments{Objects with decline time-scales smaller/larger than 50 days
are marked by a $^\dagger$/$^\ddagger$.
}
\tablerefs{
[1]: \citet{Condon1998a};
[2]: \citet{Becker1995a};
[3]: \citet{Mauch2003a}
}
}
\end{table*}

\clearpage

\clearpage
\section{Spectral energy distribution fits}

\begin{figure*}
\includegraphics[width=0.27\textwidth]{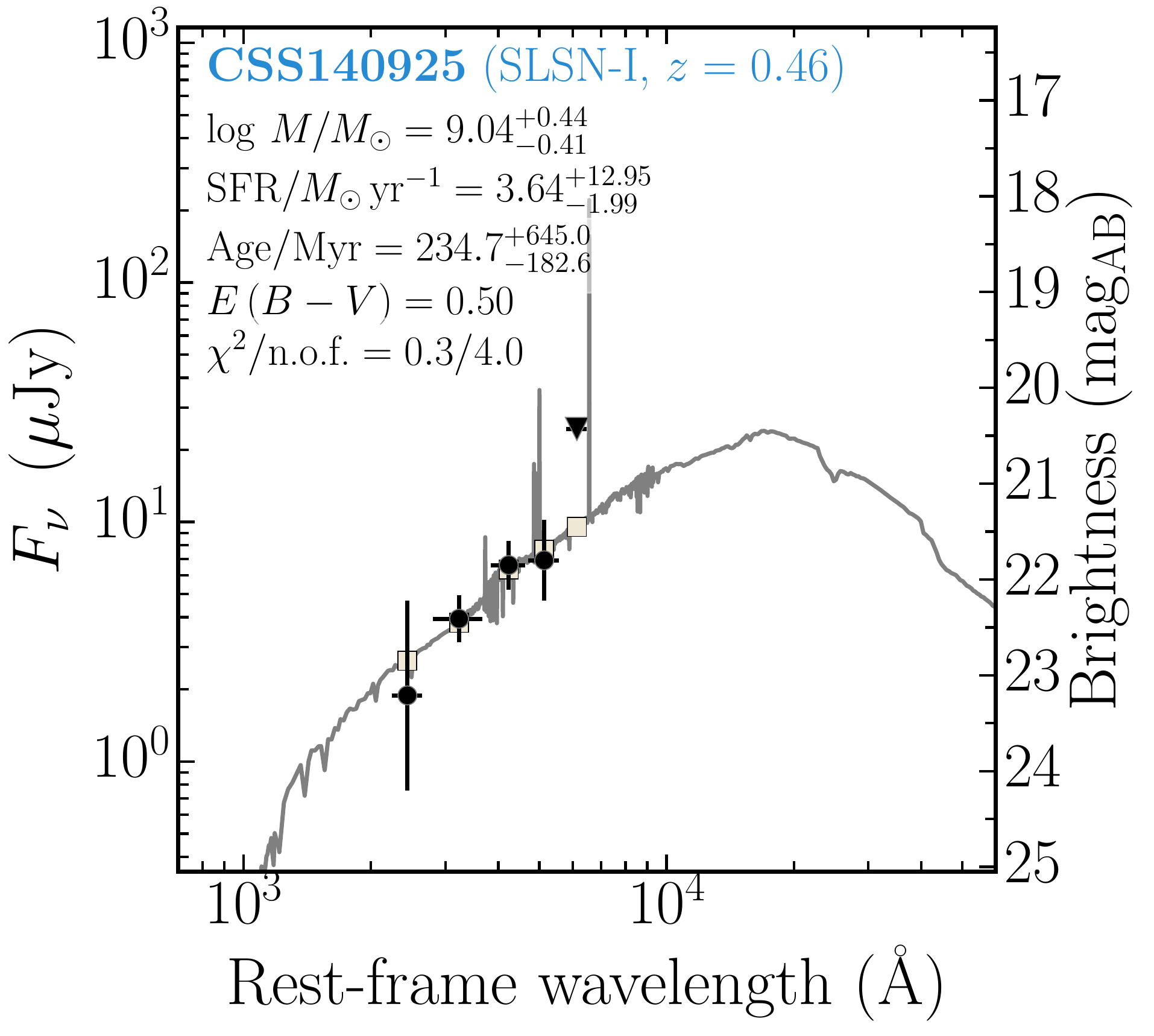}
\includegraphics[width=0.27\textwidth]{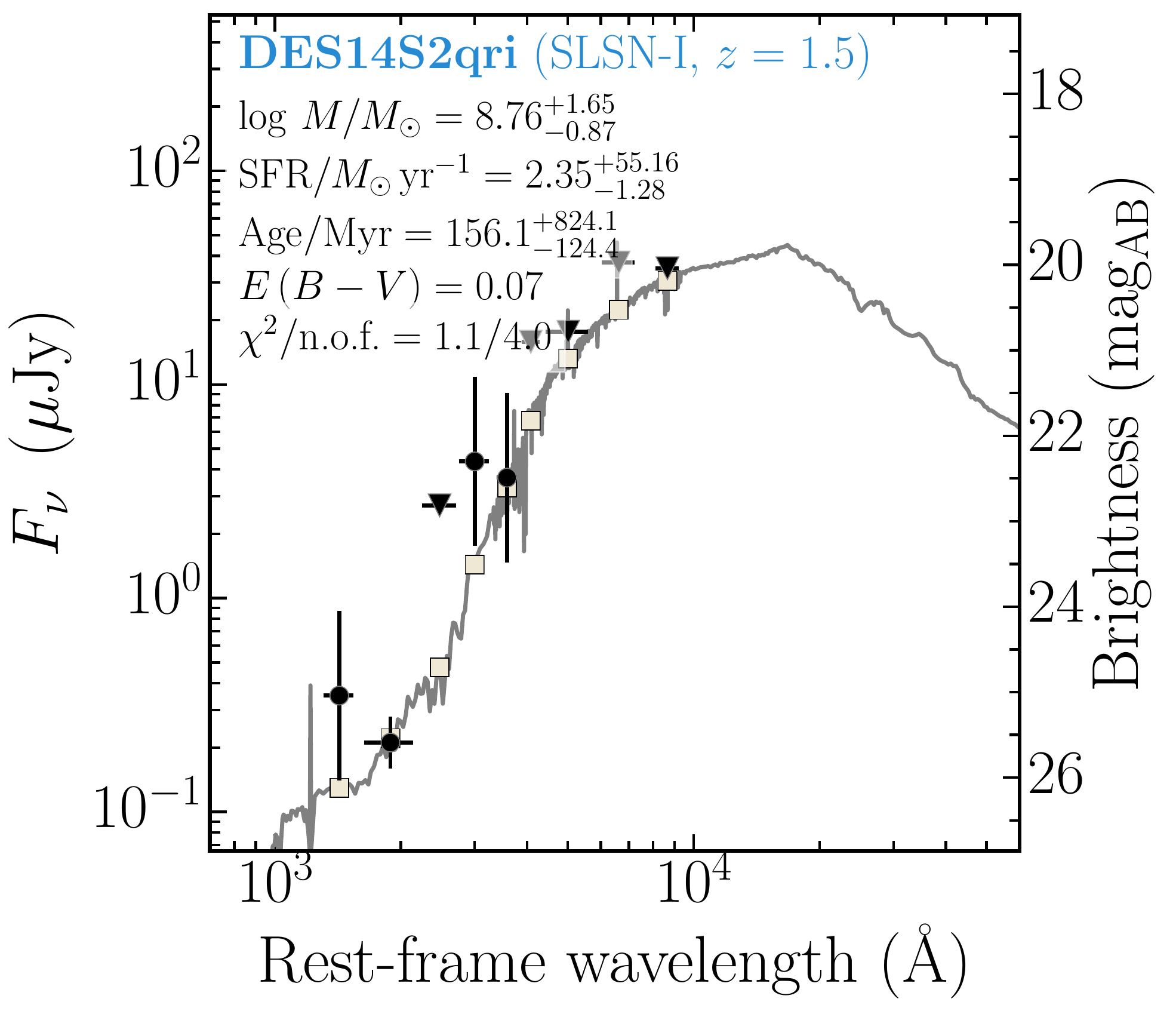}
\includegraphics[width=0.27\textwidth]{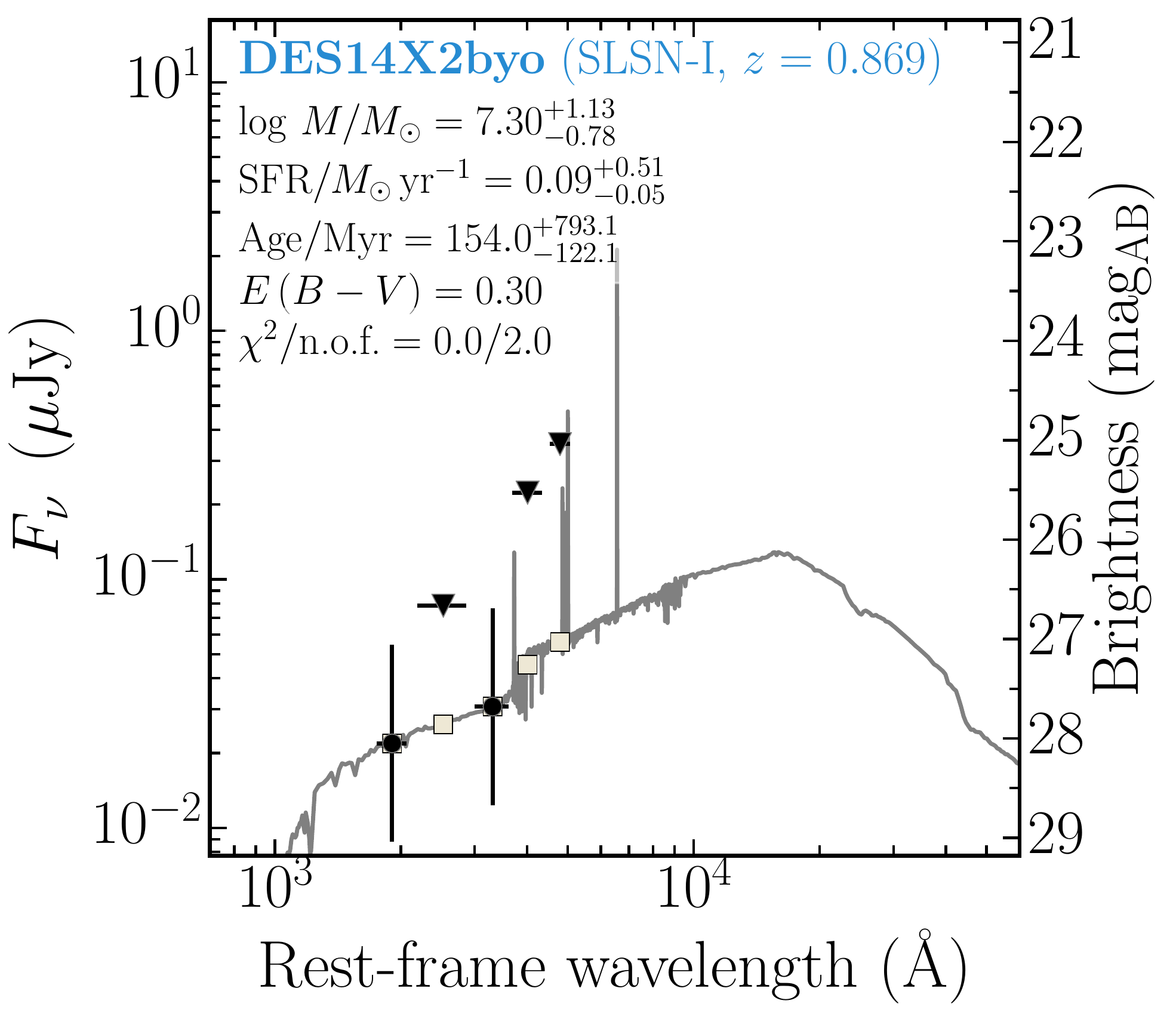}
\includegraphics[width=0.27\textwidth]{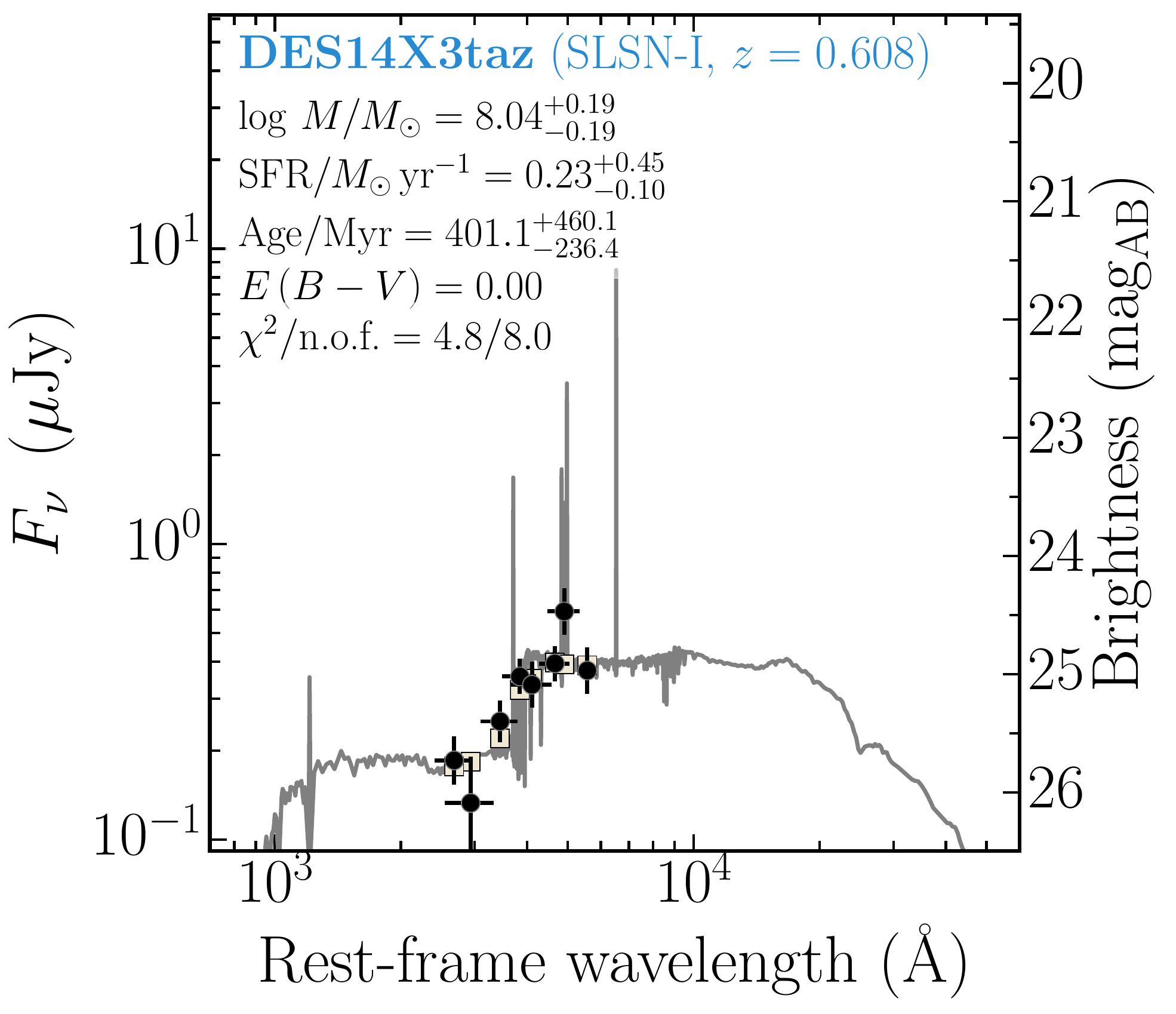}
\includegraphics[width=0.27\textwidth]{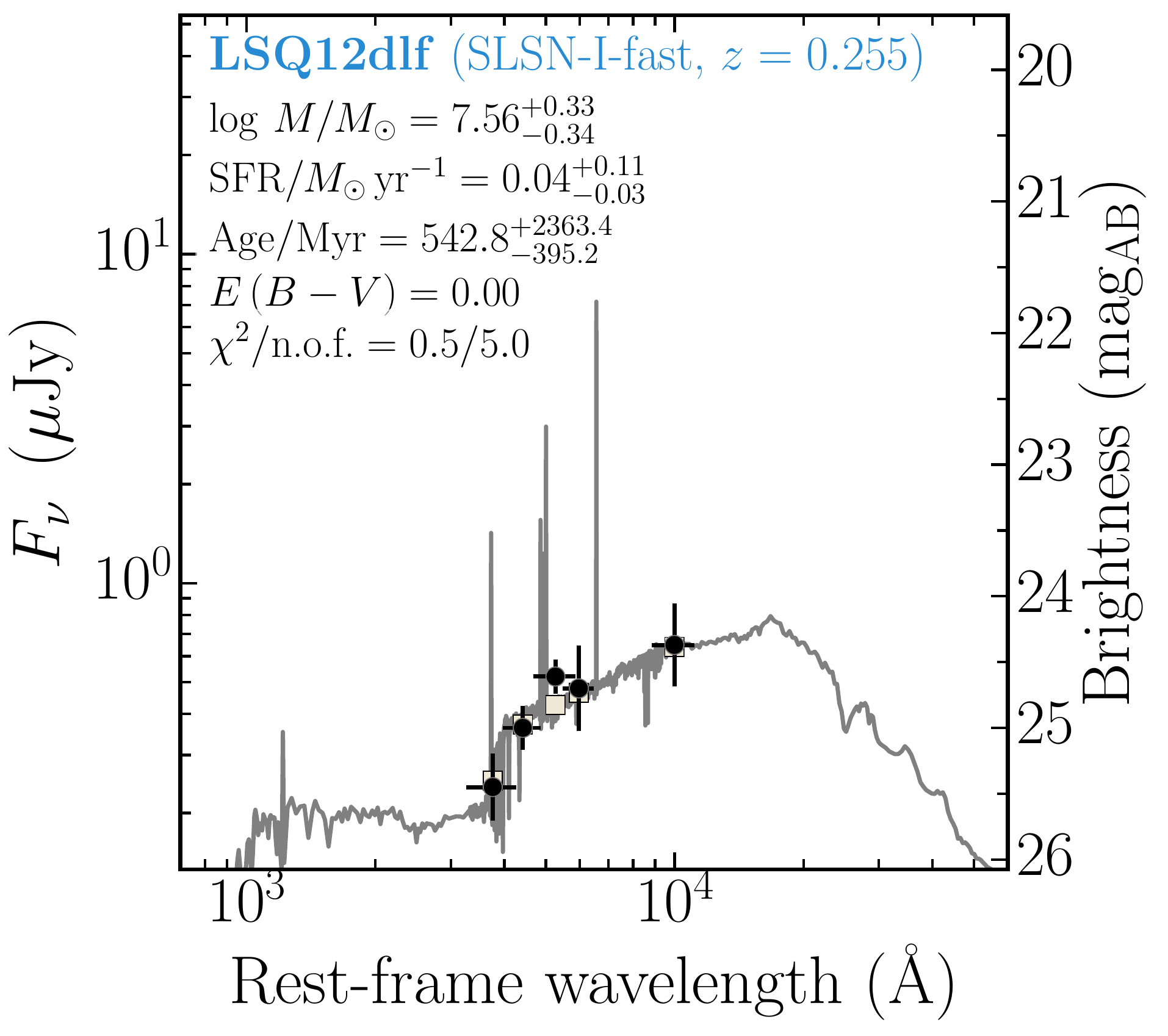}
\includegraphics[width=0.27\textwidth]{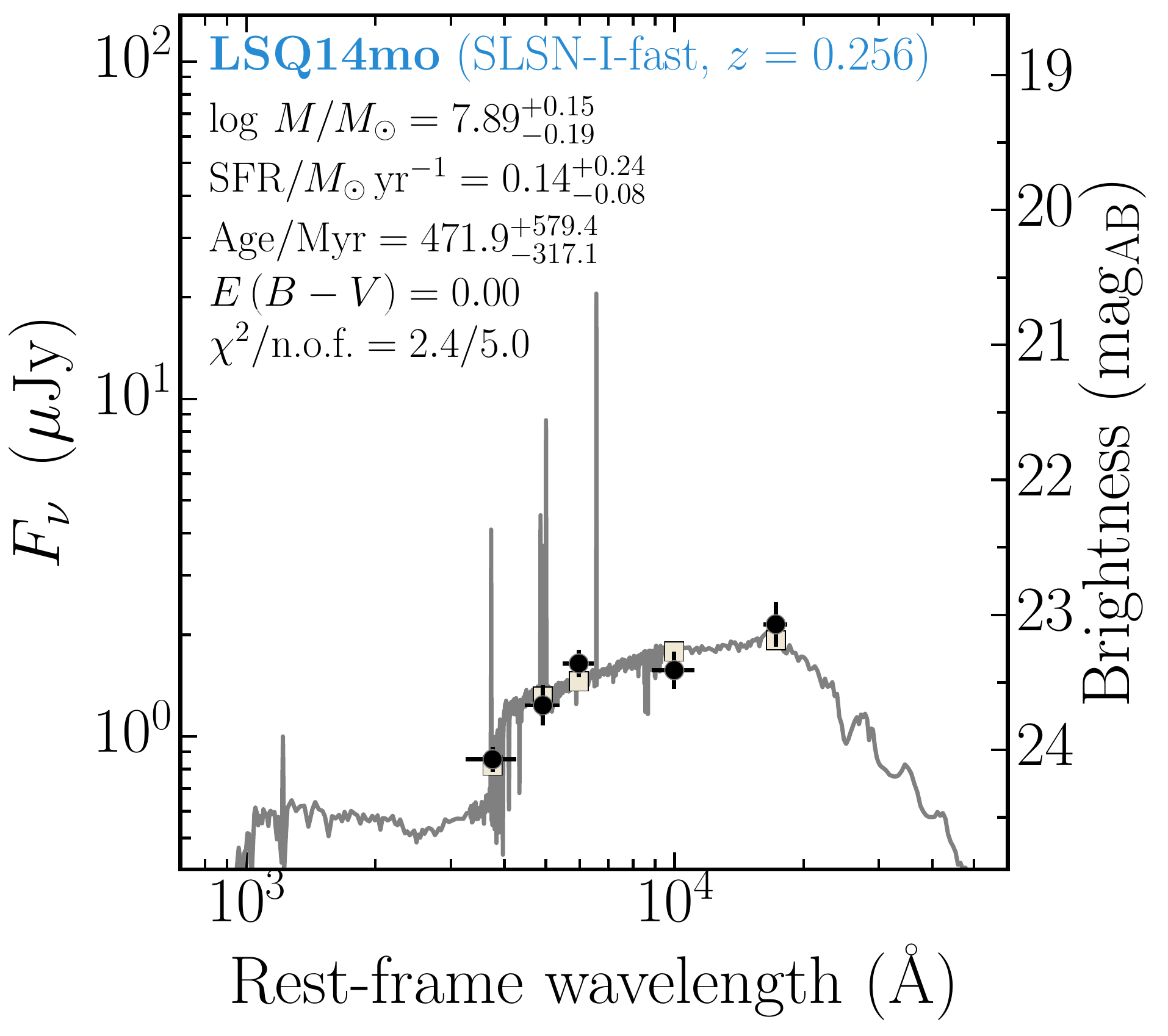}
\includegraphics[width=0.27\textwidth]{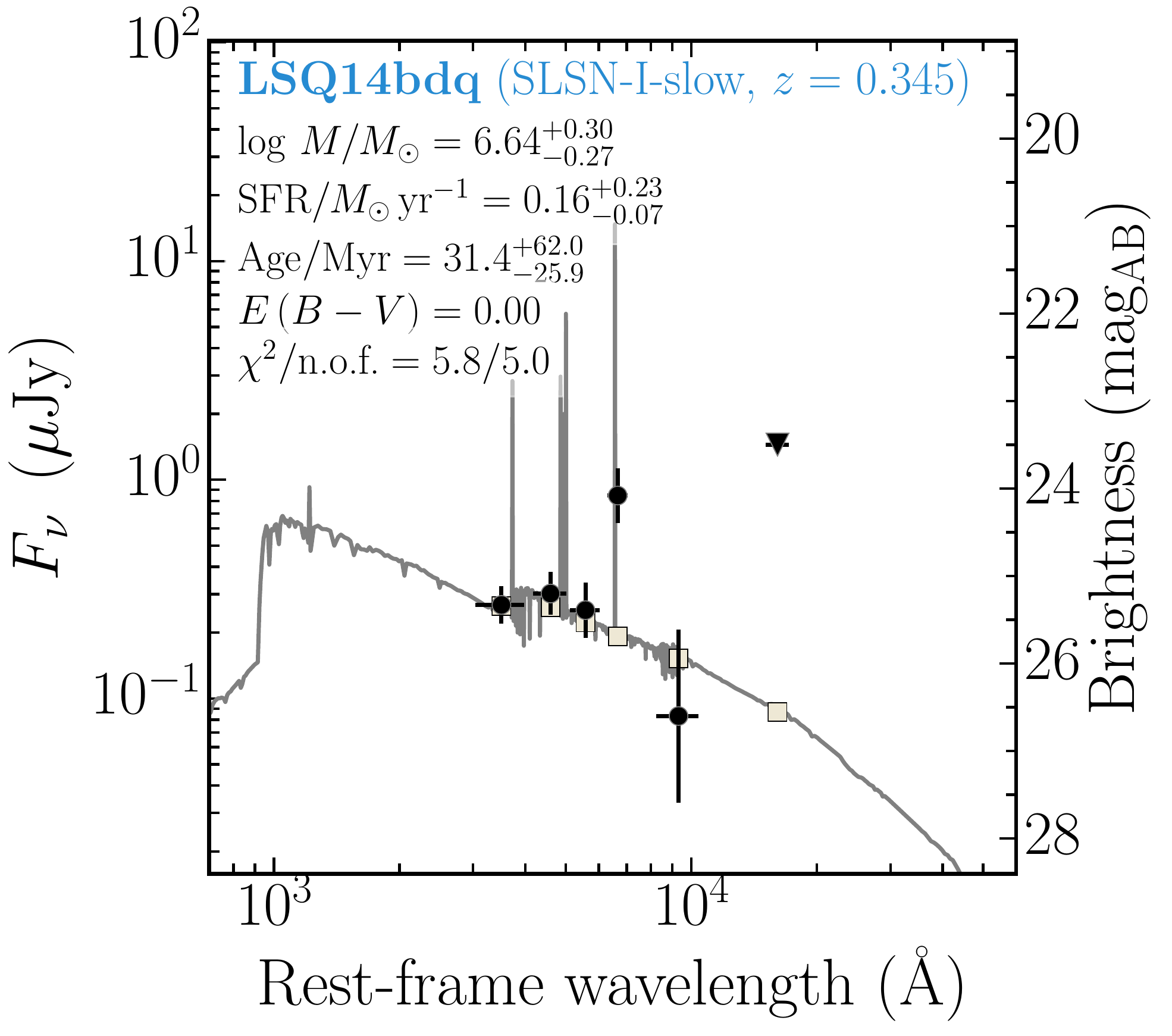}
\includegraphics[width=0.27\textwidth]{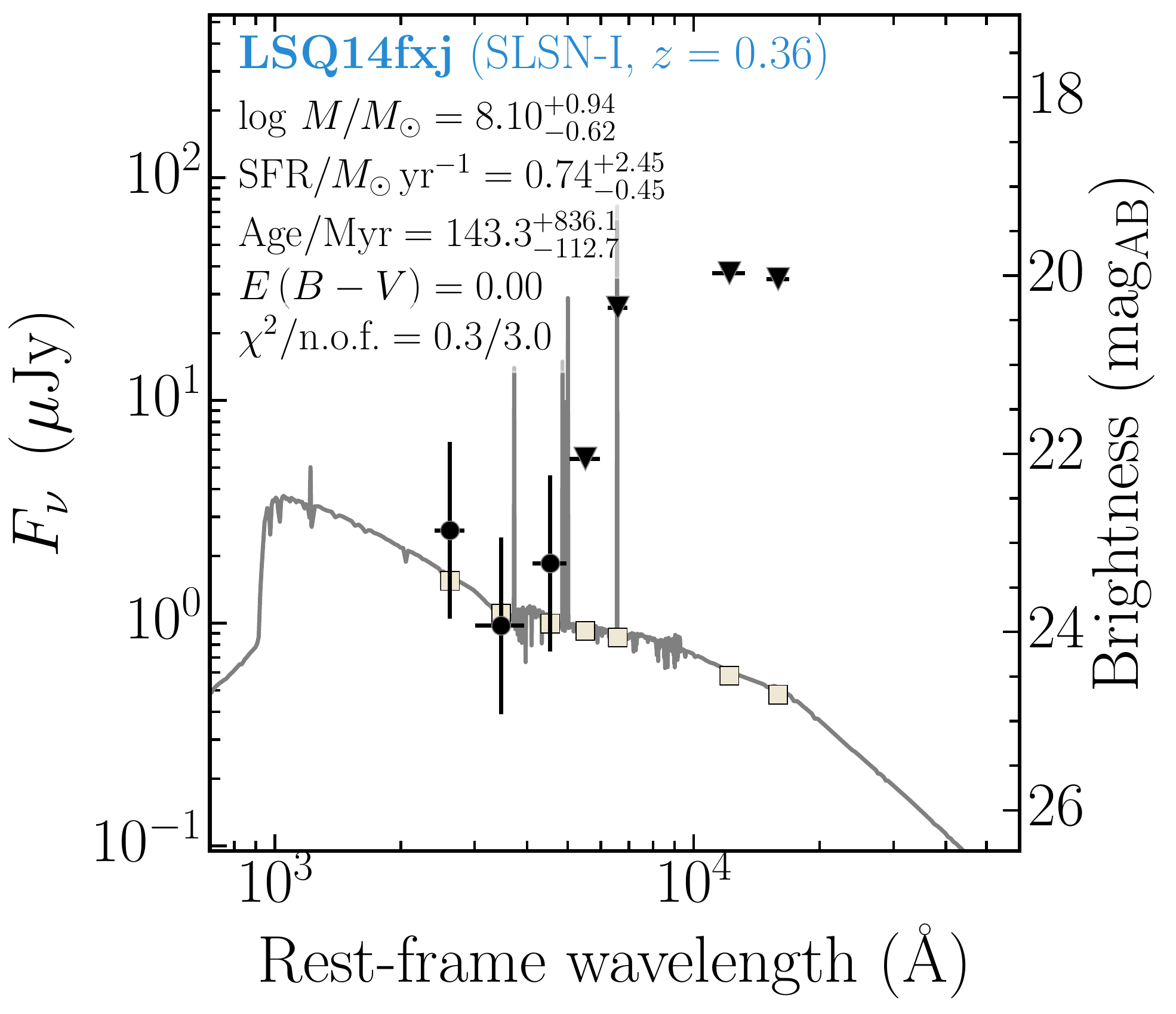}
\includegraphics[width=0.27\textwidth]{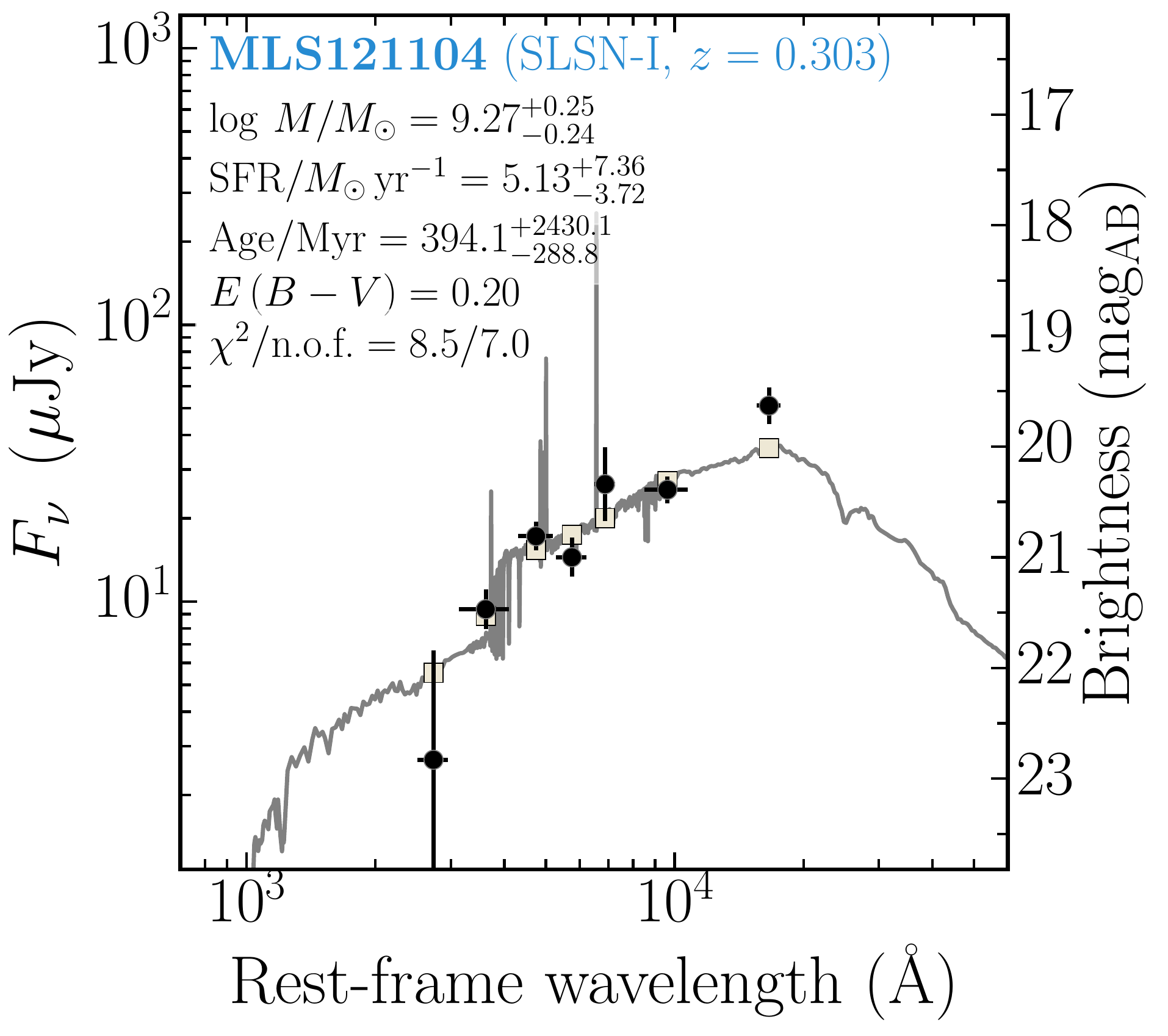}
\includegraphics[width=0.27\textwidth]{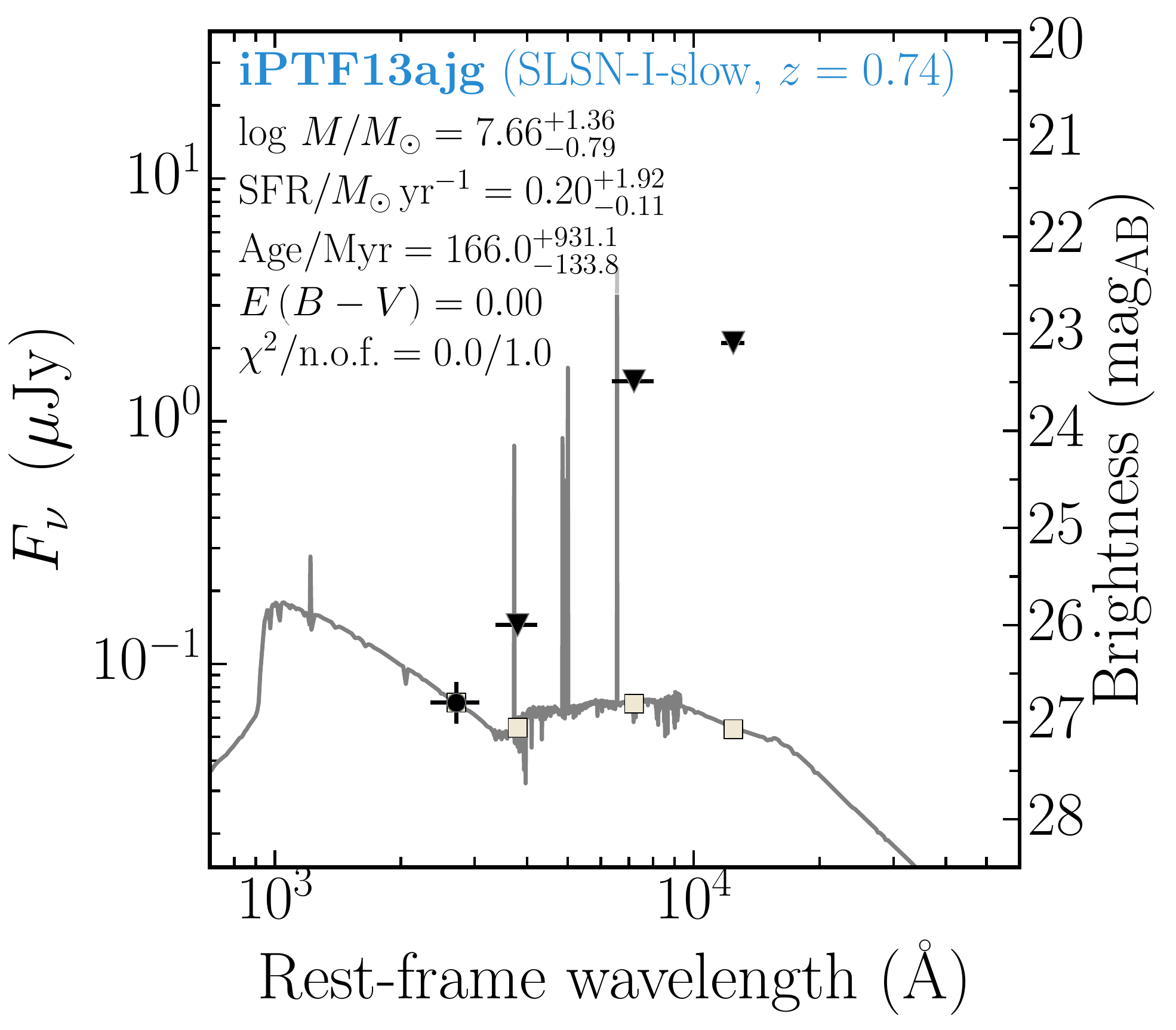}
\includegraphics[width=0.27\textwidth]{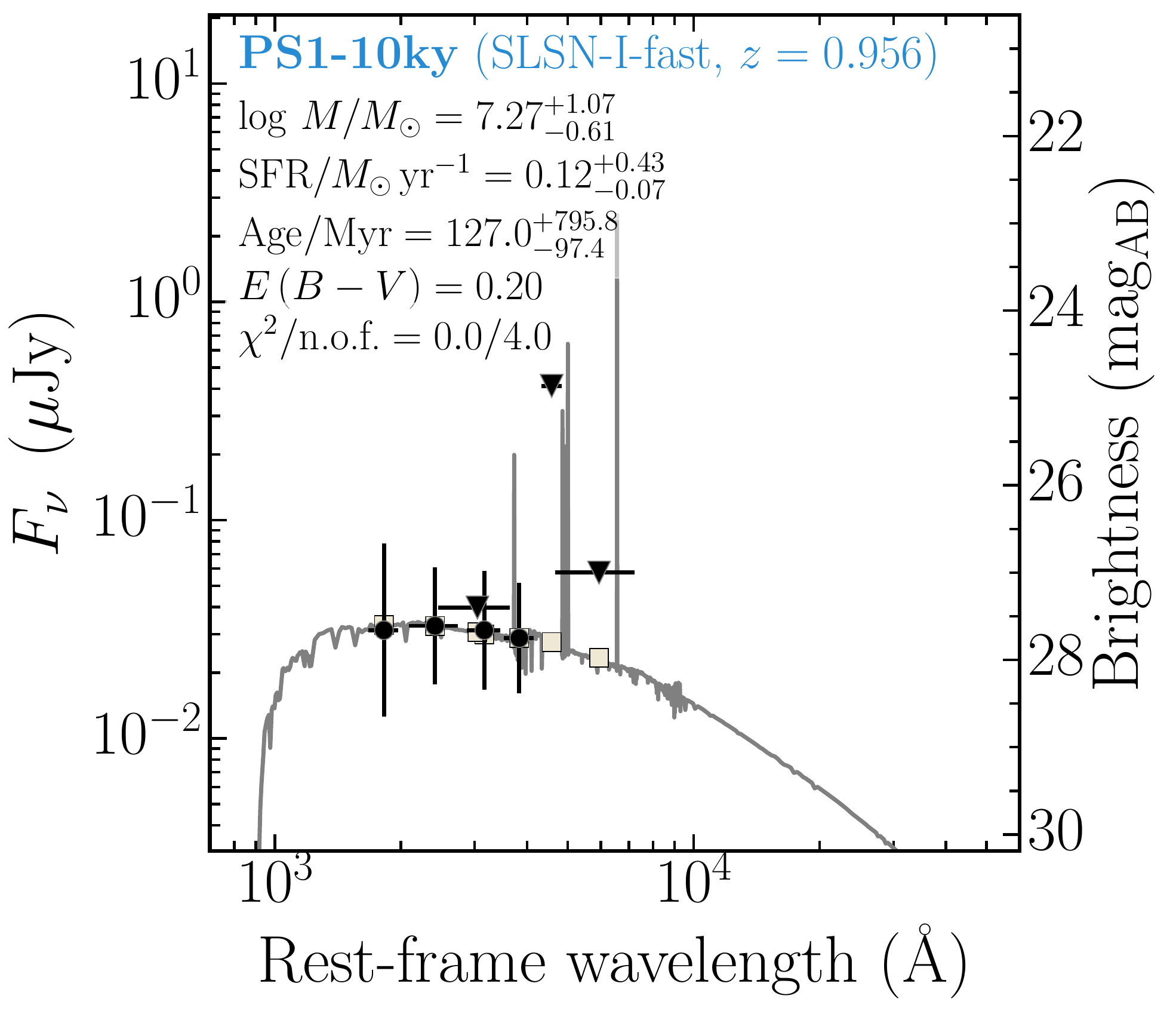}
\includegraphics[width=0.27\textwidth]{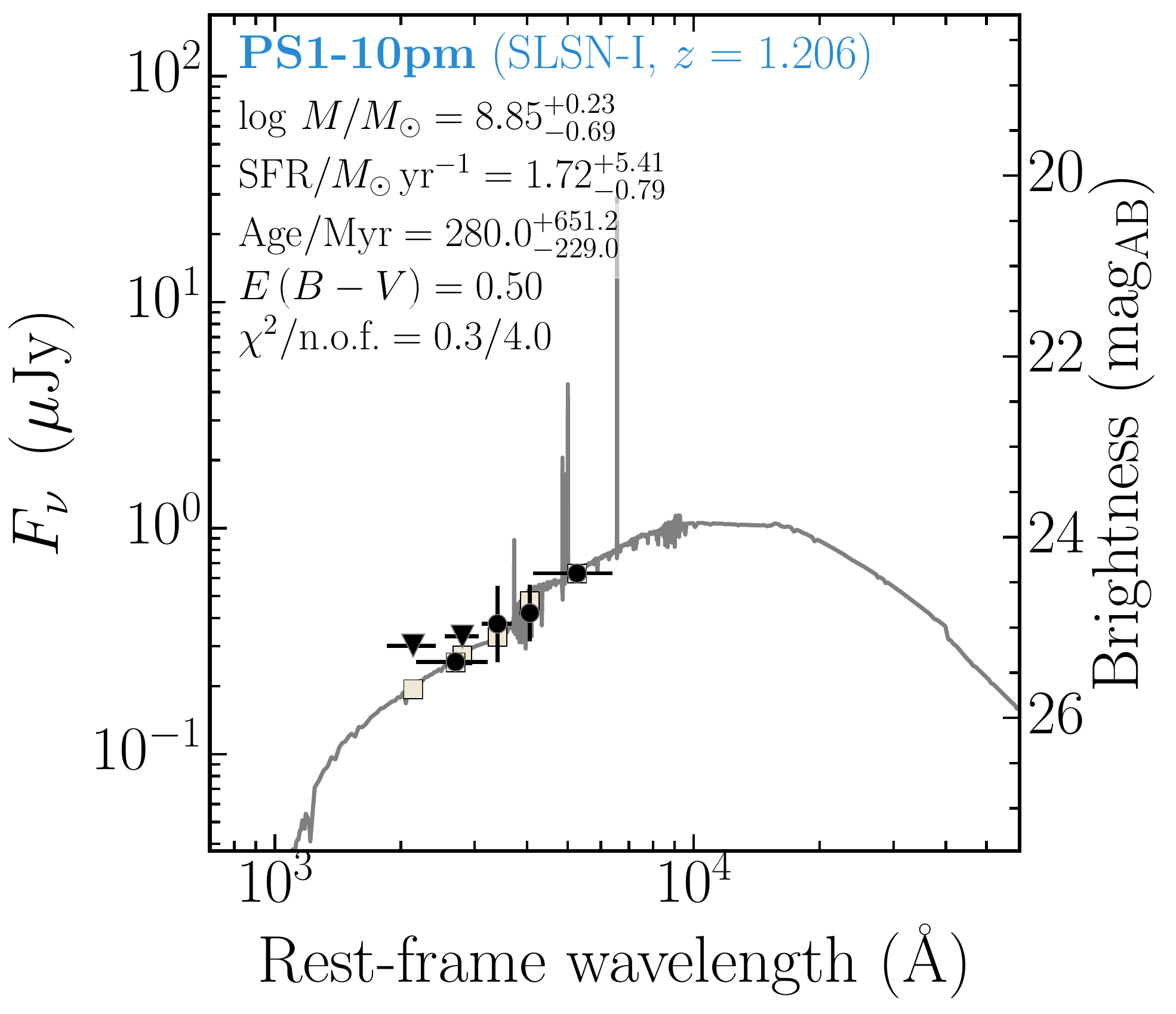}
\includegraphics[width=0.27\textwidth]{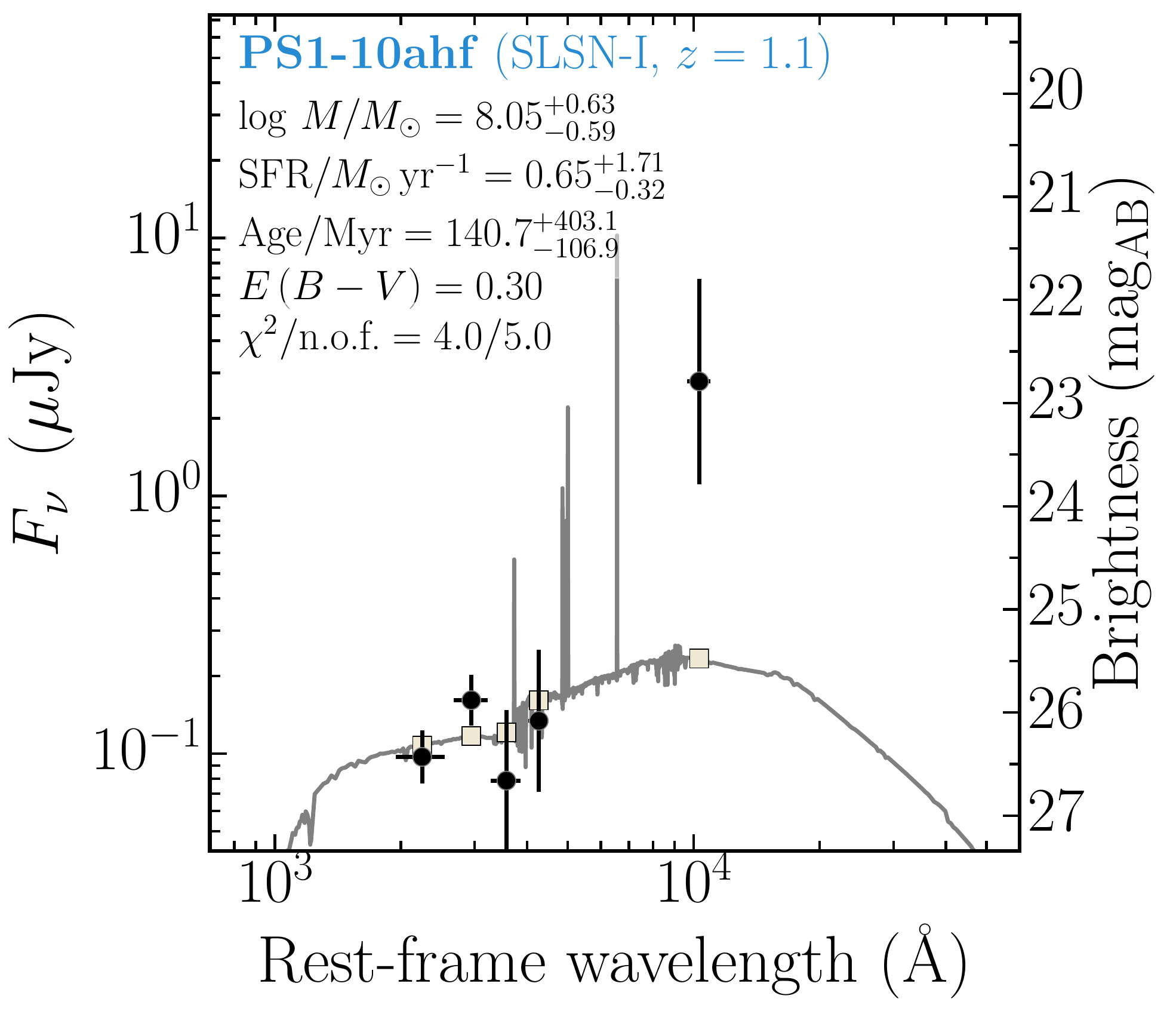}
\includegraphics[width=0.27\textwidth]{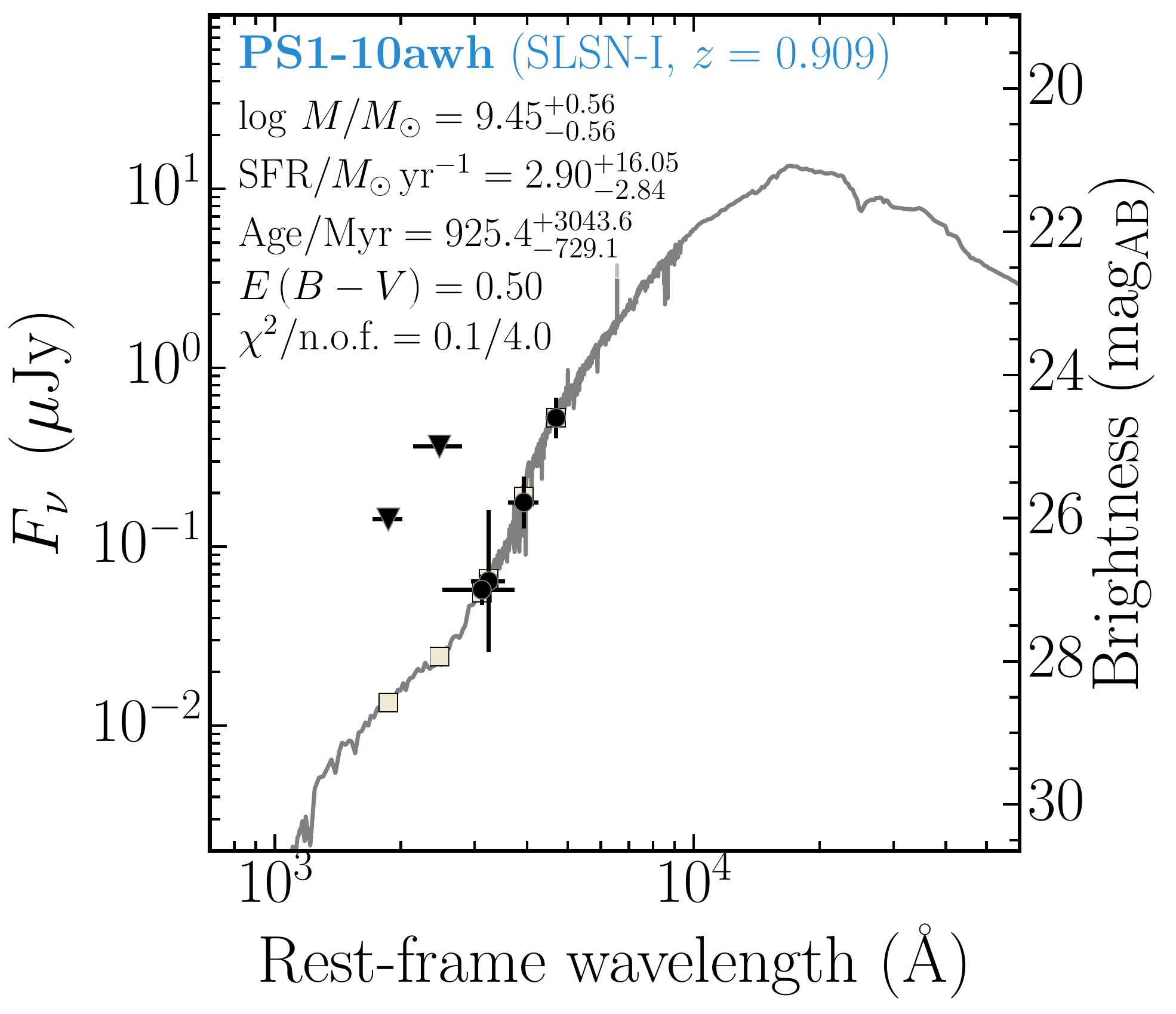}
\includegraphics[width=0.27\textwidth]{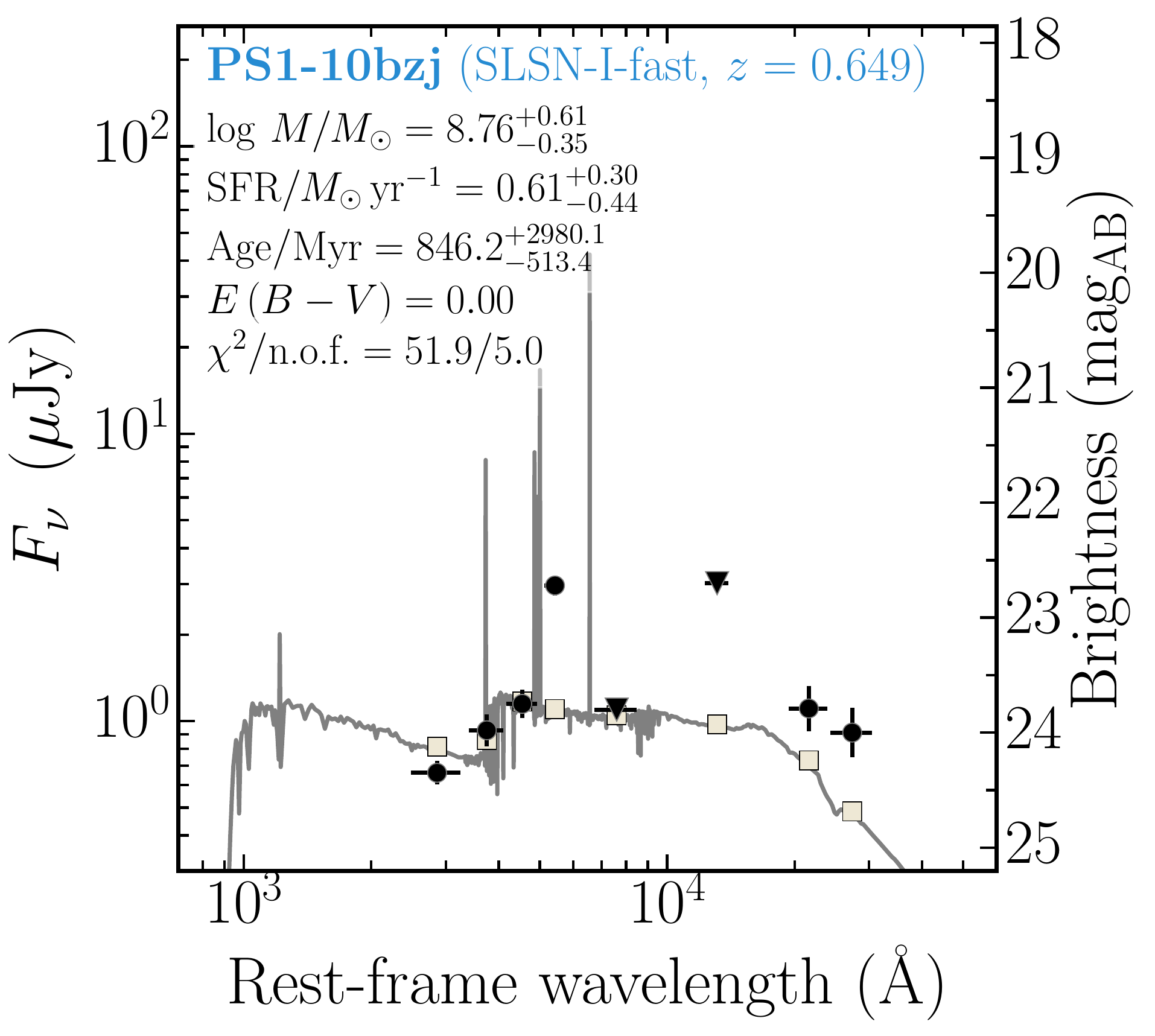}
\caption{Similar to Fig. \ref{fig:sed}. Spectral energy distributions of hosts of
H-poor SLSNe from 1000 to 40000 {\AA} (detections: $\bullet$; upper limits: $\blacktriangledown$).
The solid line displays the best-fit model of the SED with \texttt{Le Phare}. The squares
in a lighter shade are the model predicted magnitudes. Key fitting parameters are displayed
for each SED. See Table \ref{tab:sed_results} and Sect. \ref{sec:sed_fitting} for details.
}
\label{fig:SED_app_1}
\end{figure*}
\clearpage

\begin{figure*}
\ContinuedFloat
\includegraphics[width=0.27\textwidth]{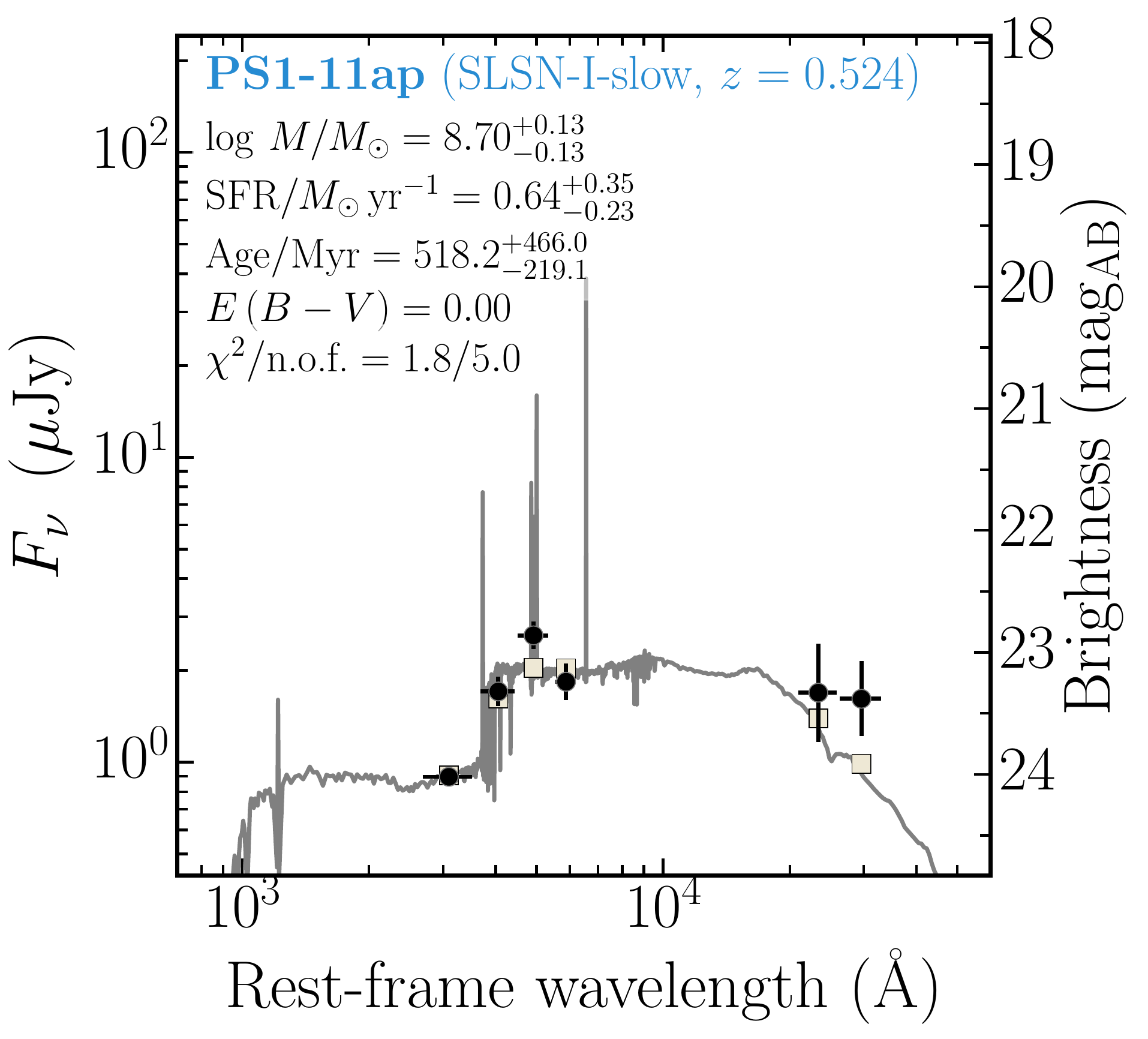}
\includegraphics[width=0.27\textwidth]{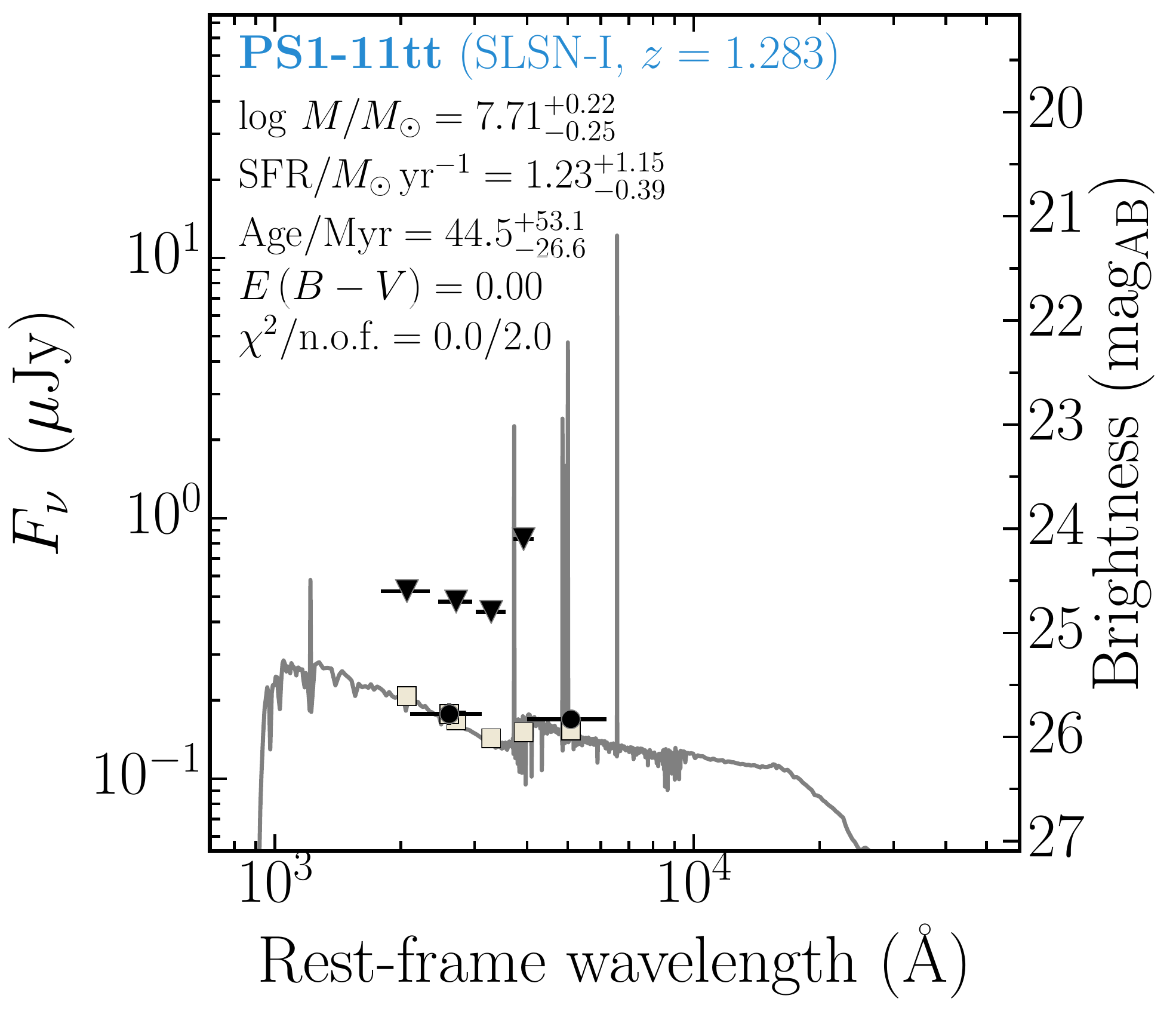}
\includegraphics[width=0.27\textwidth]{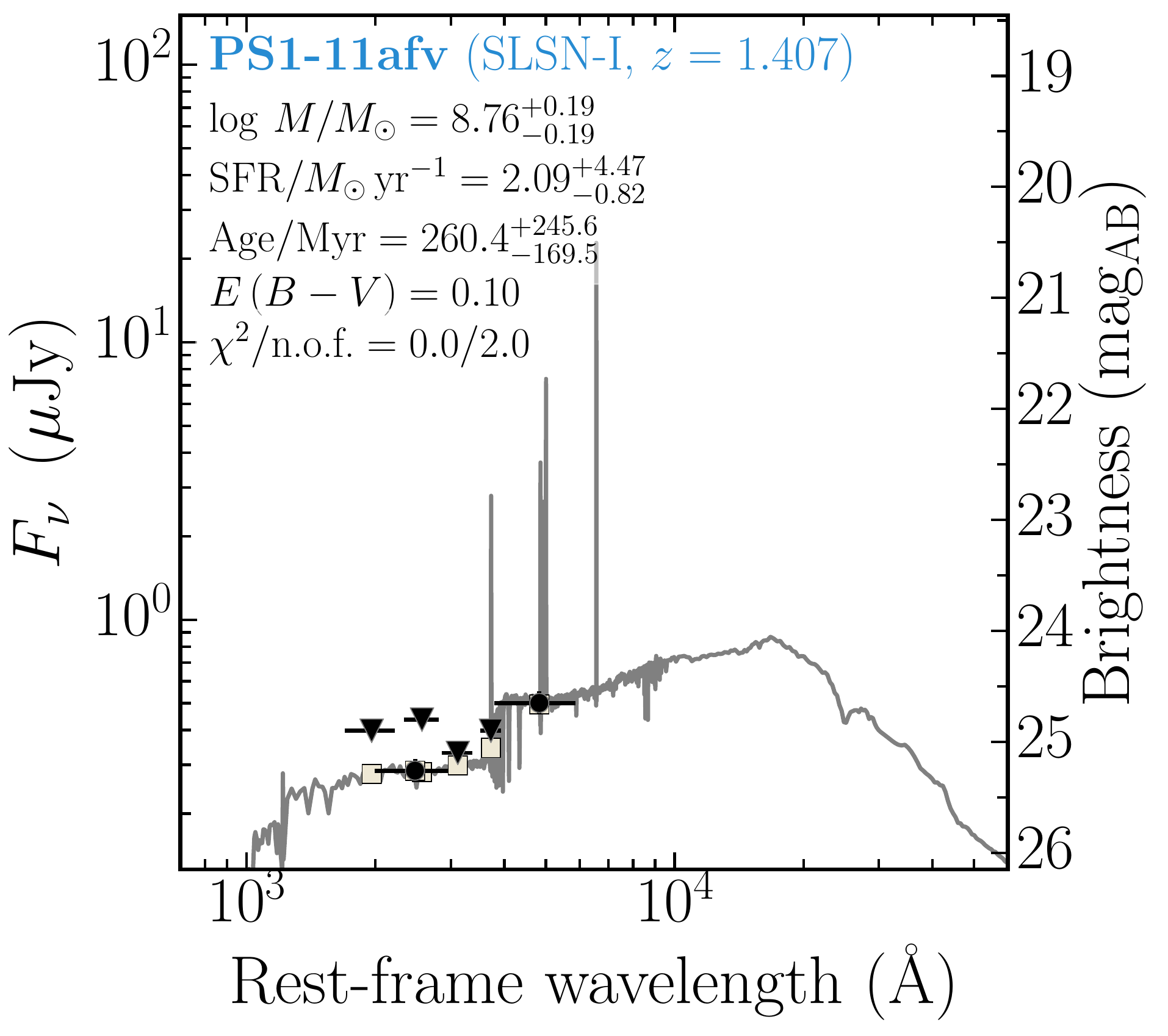}
\includegraphics[width=0.27\textwidth]{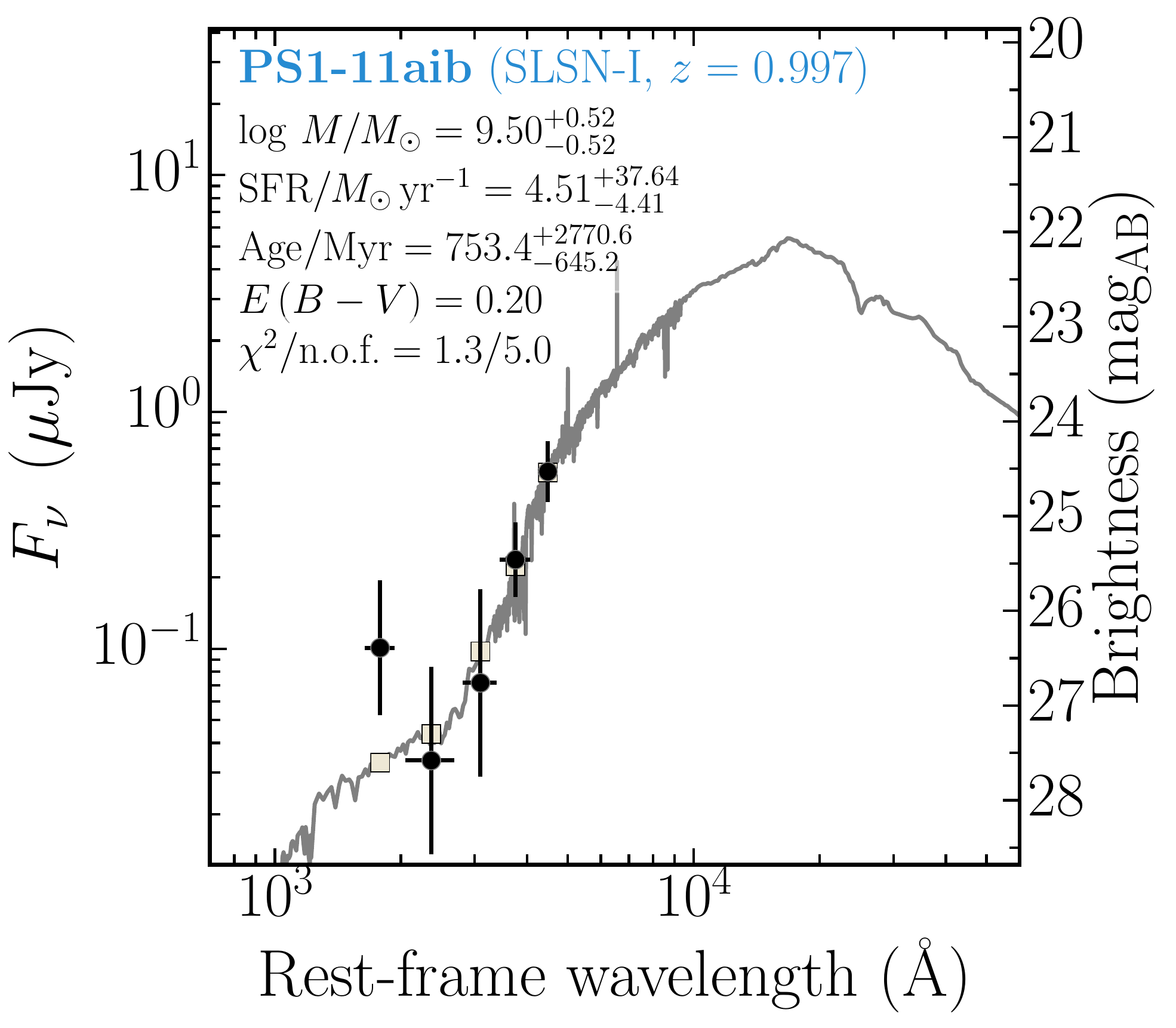}
\includegraphics[width=0.27\textwidth]{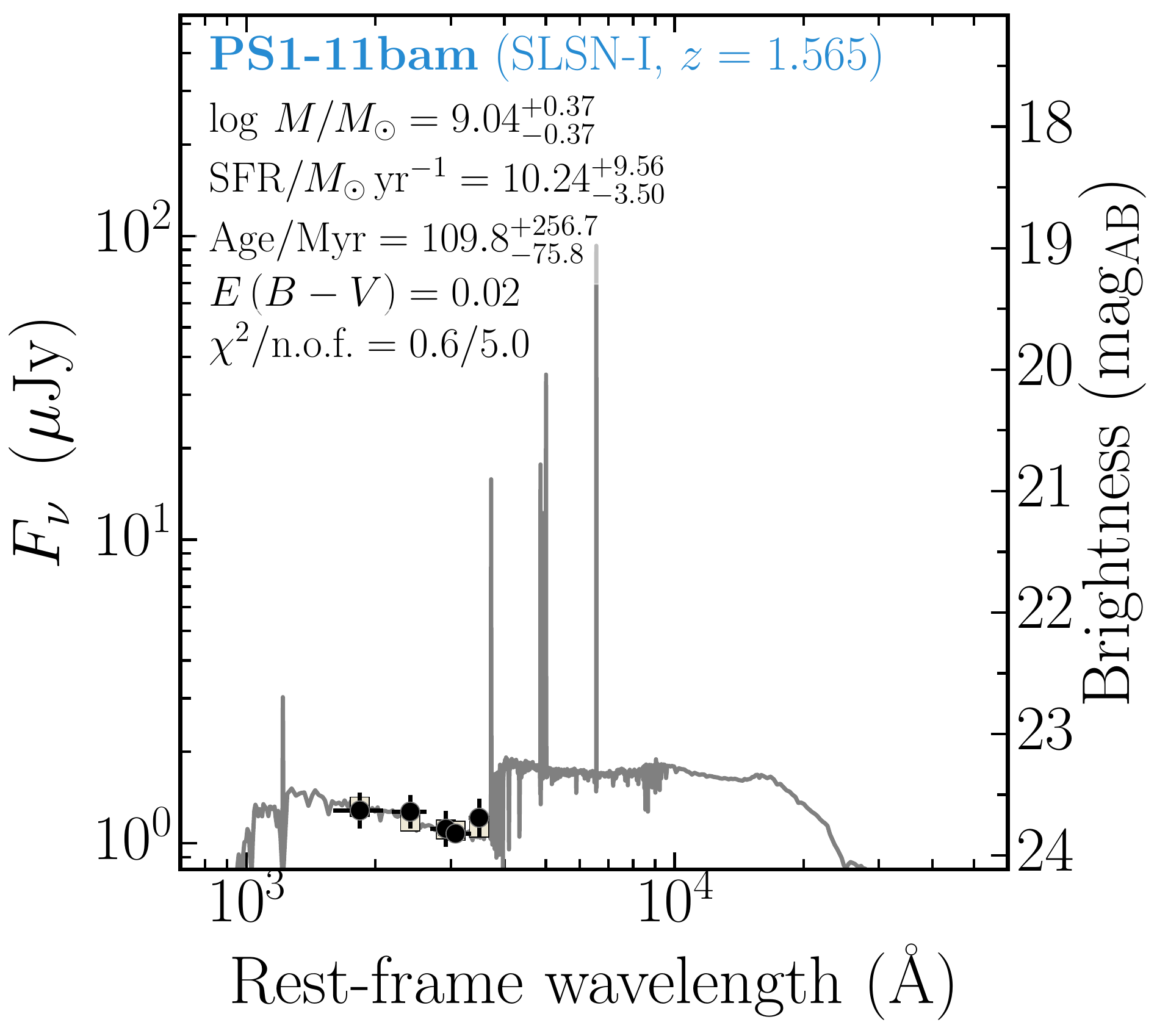}
\includegraphics[width=0.27\textwidth]{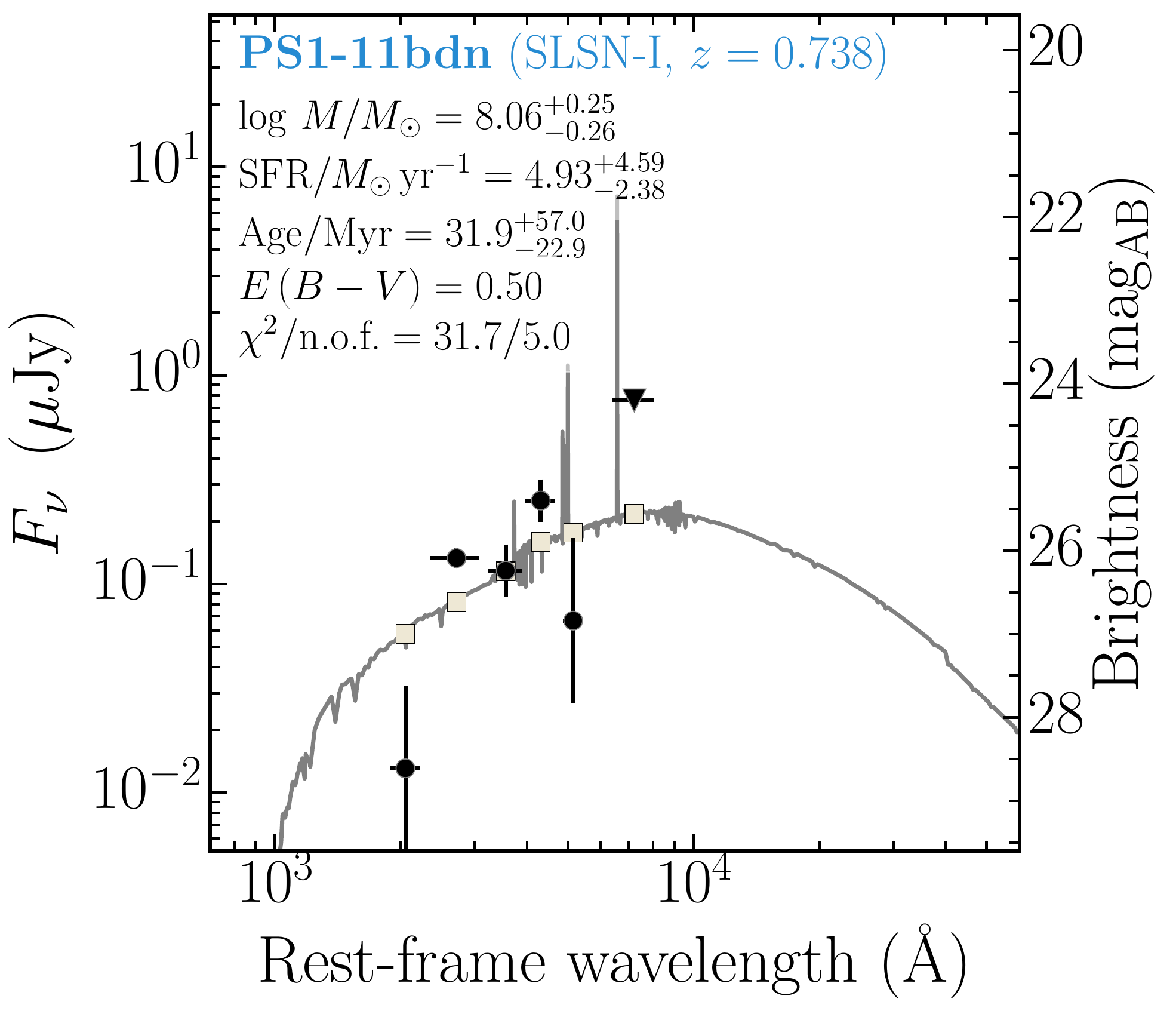}
\includegraphics[width=0.27\textwidth]{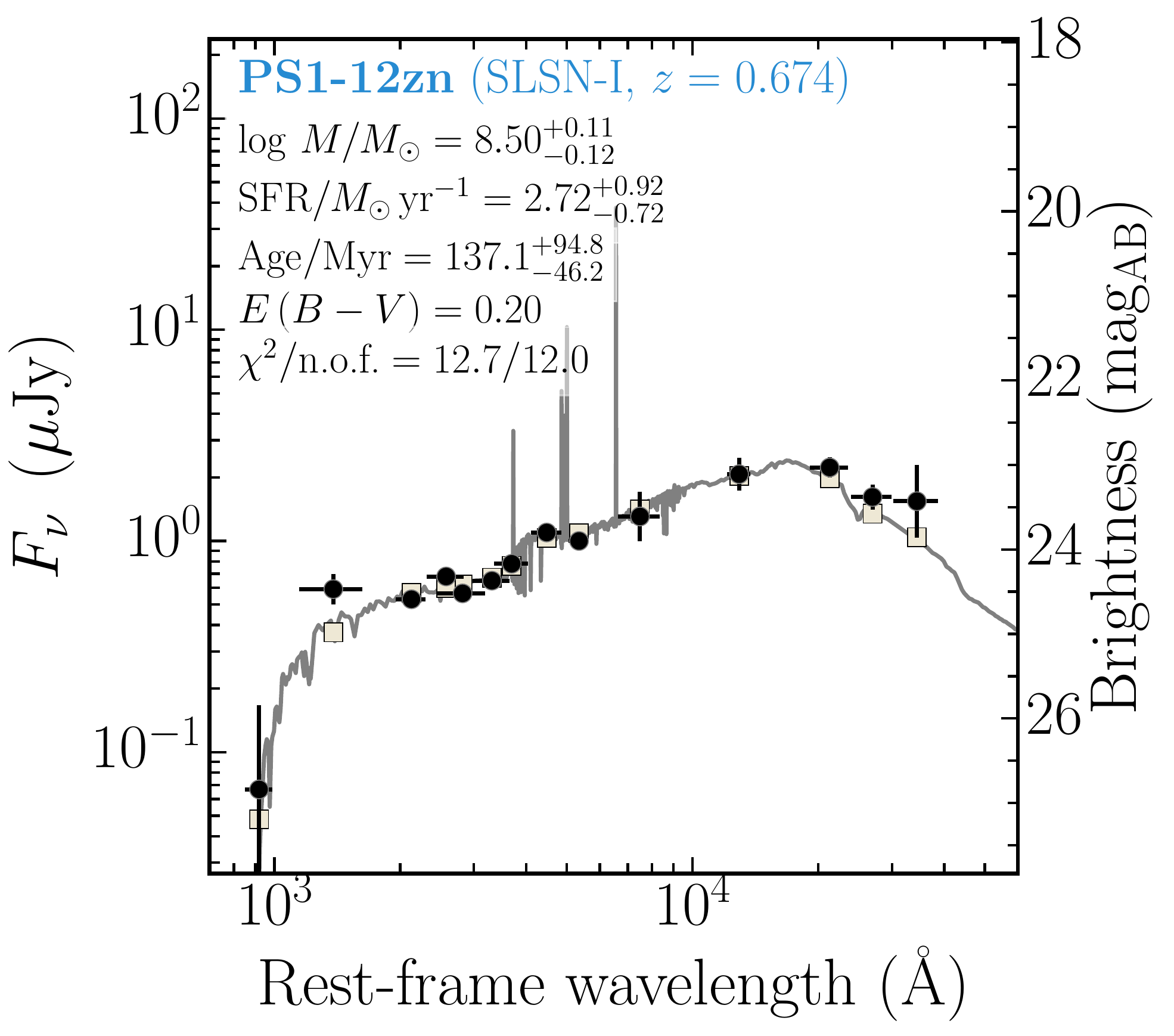}
\includegraphics[width=0.27\textwidth]{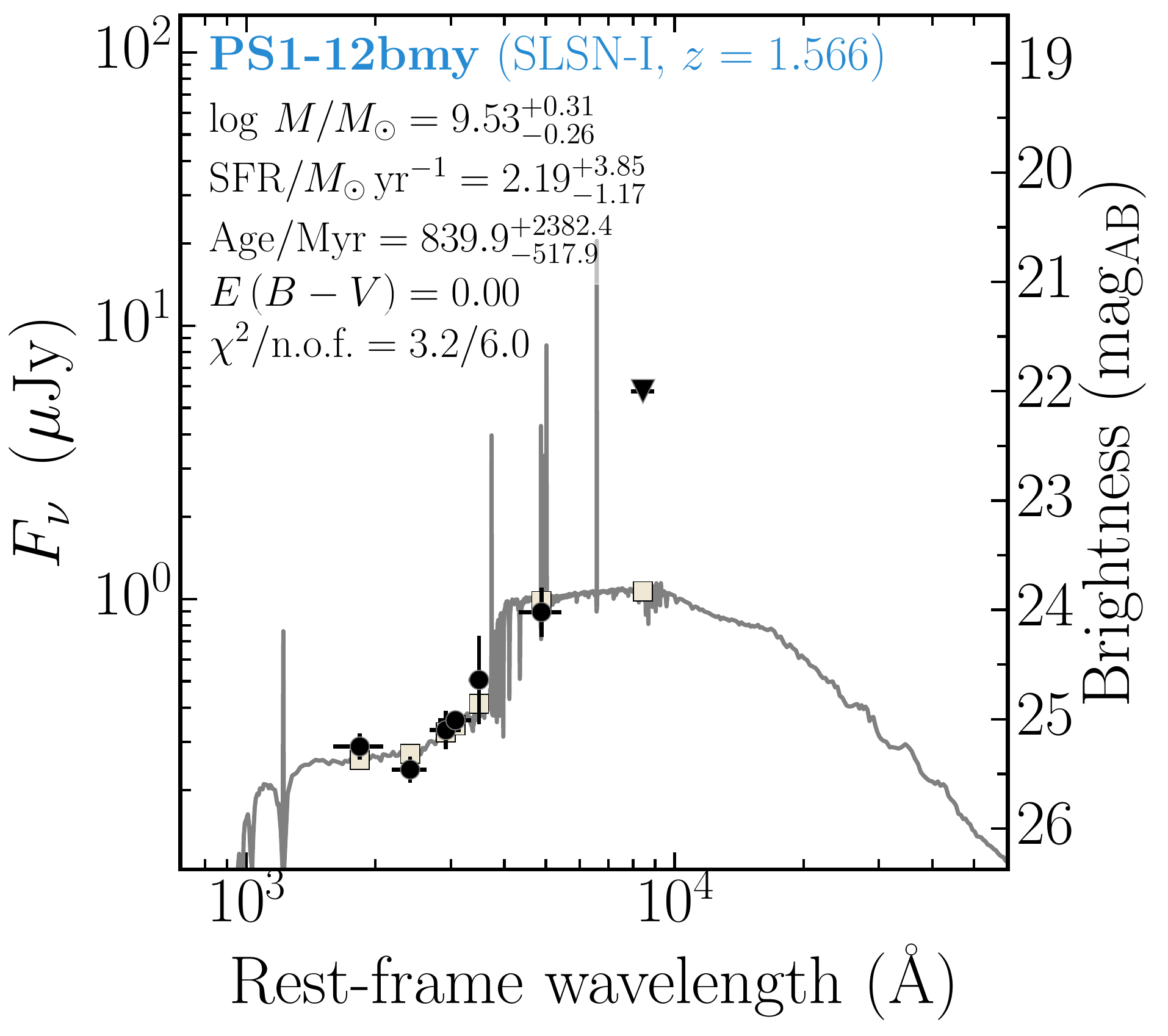}
\includegraphics[width=0.27\textwidth]{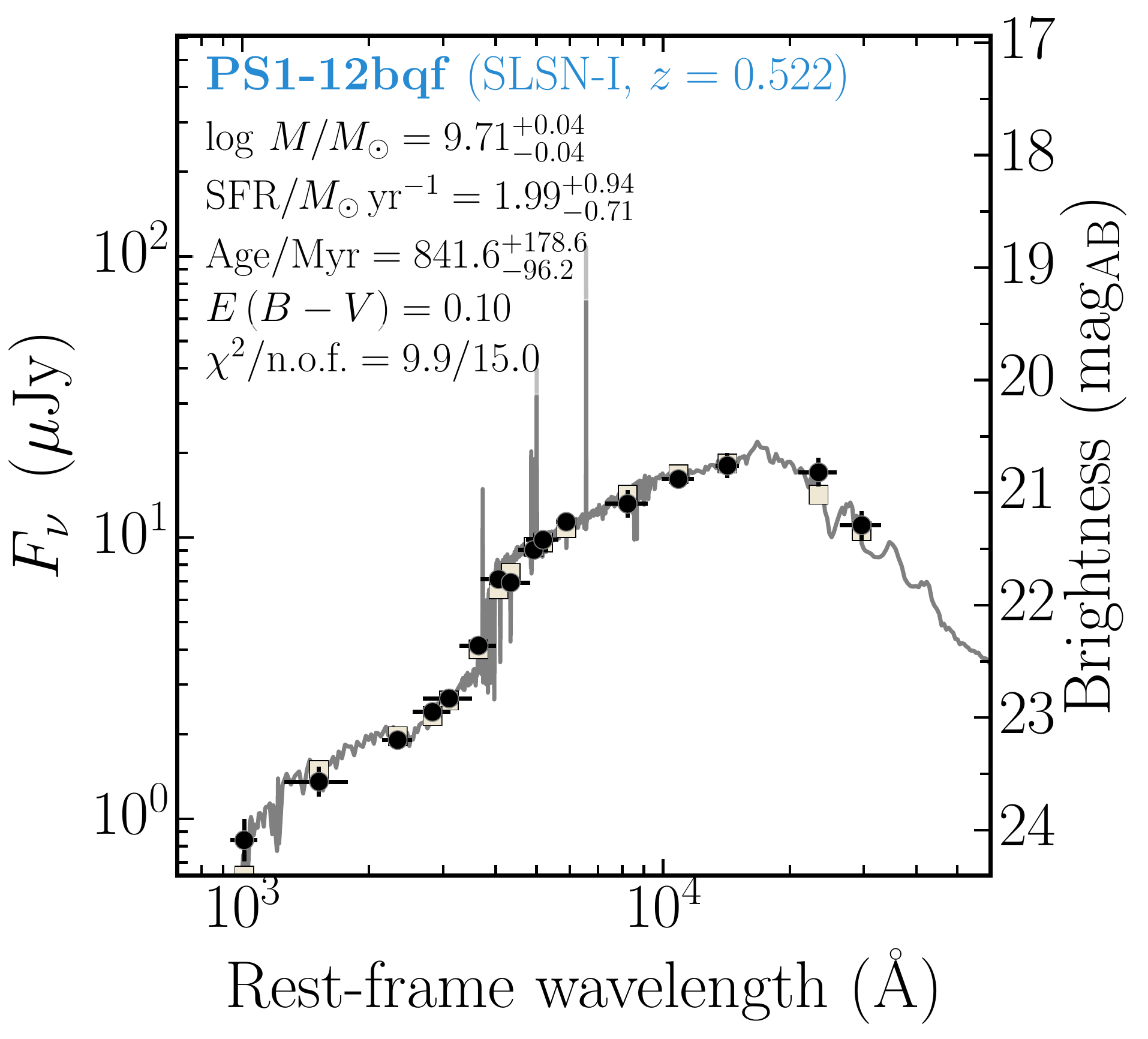}
\includegraphics[width=0.27\textwidth]{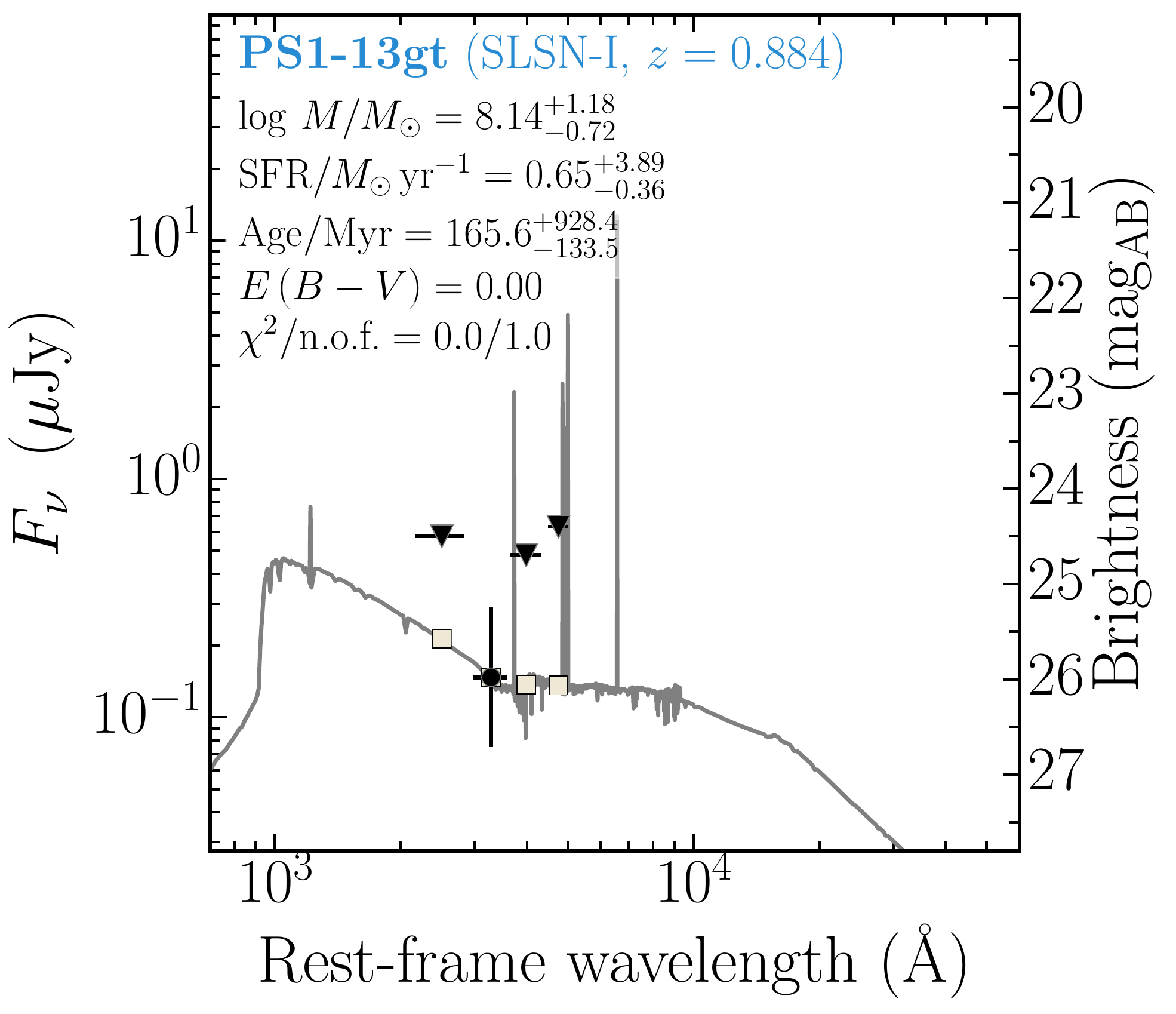}
\includegraphics[width=0.27\textwidth]{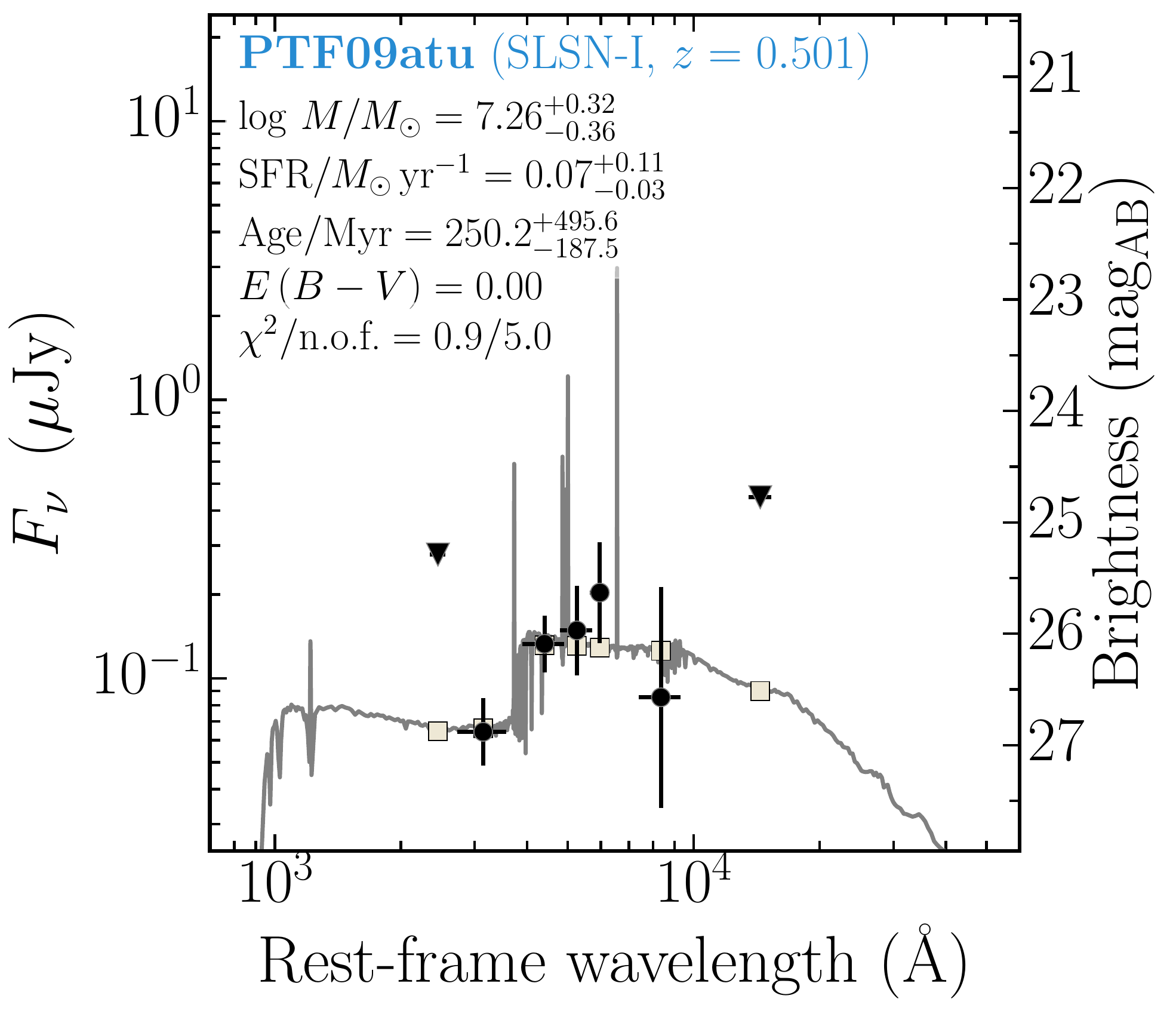}
\includegraphics[width=0.27\textwidth]{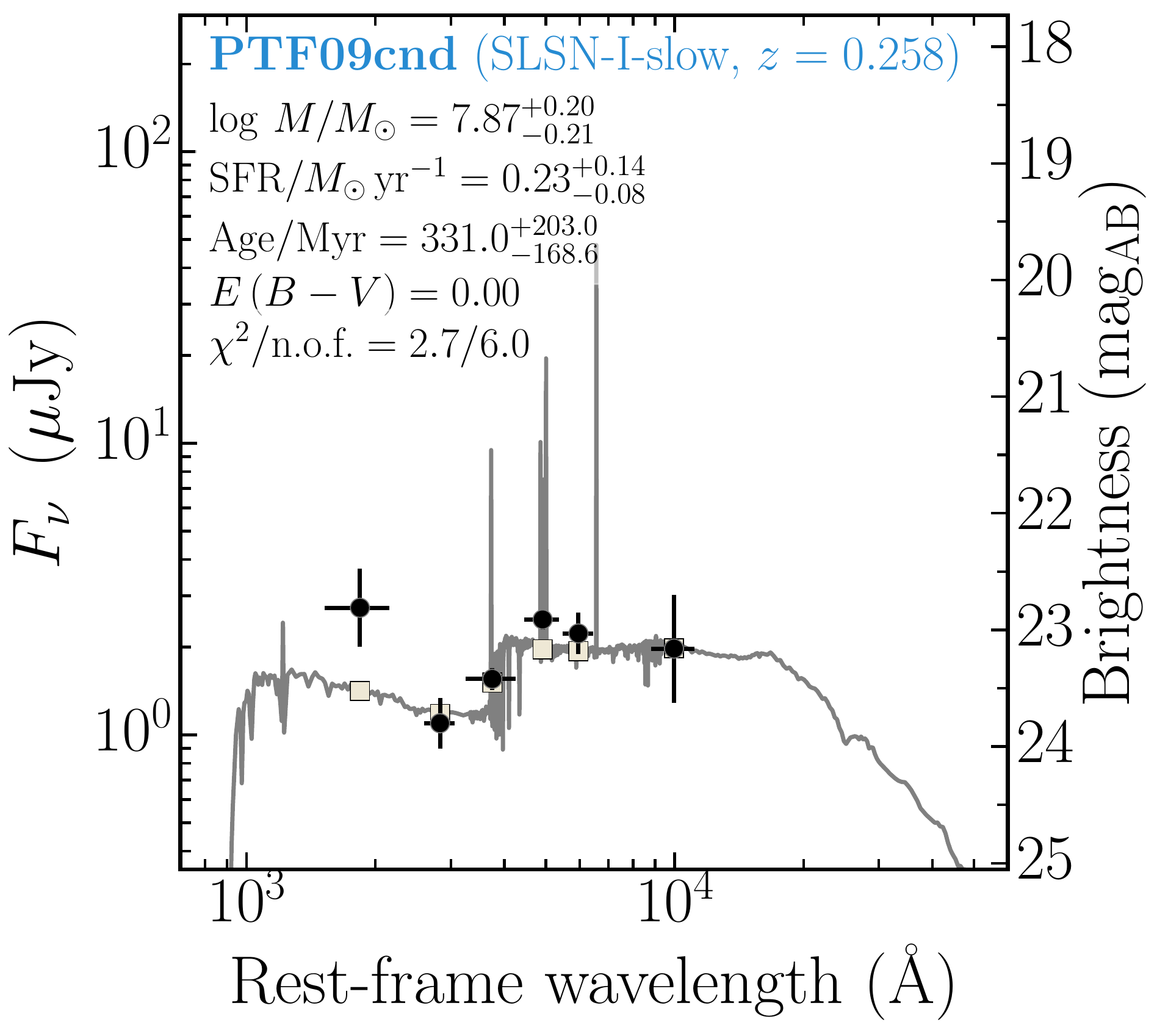}
\includegraphics[width=0.27\textwidth]{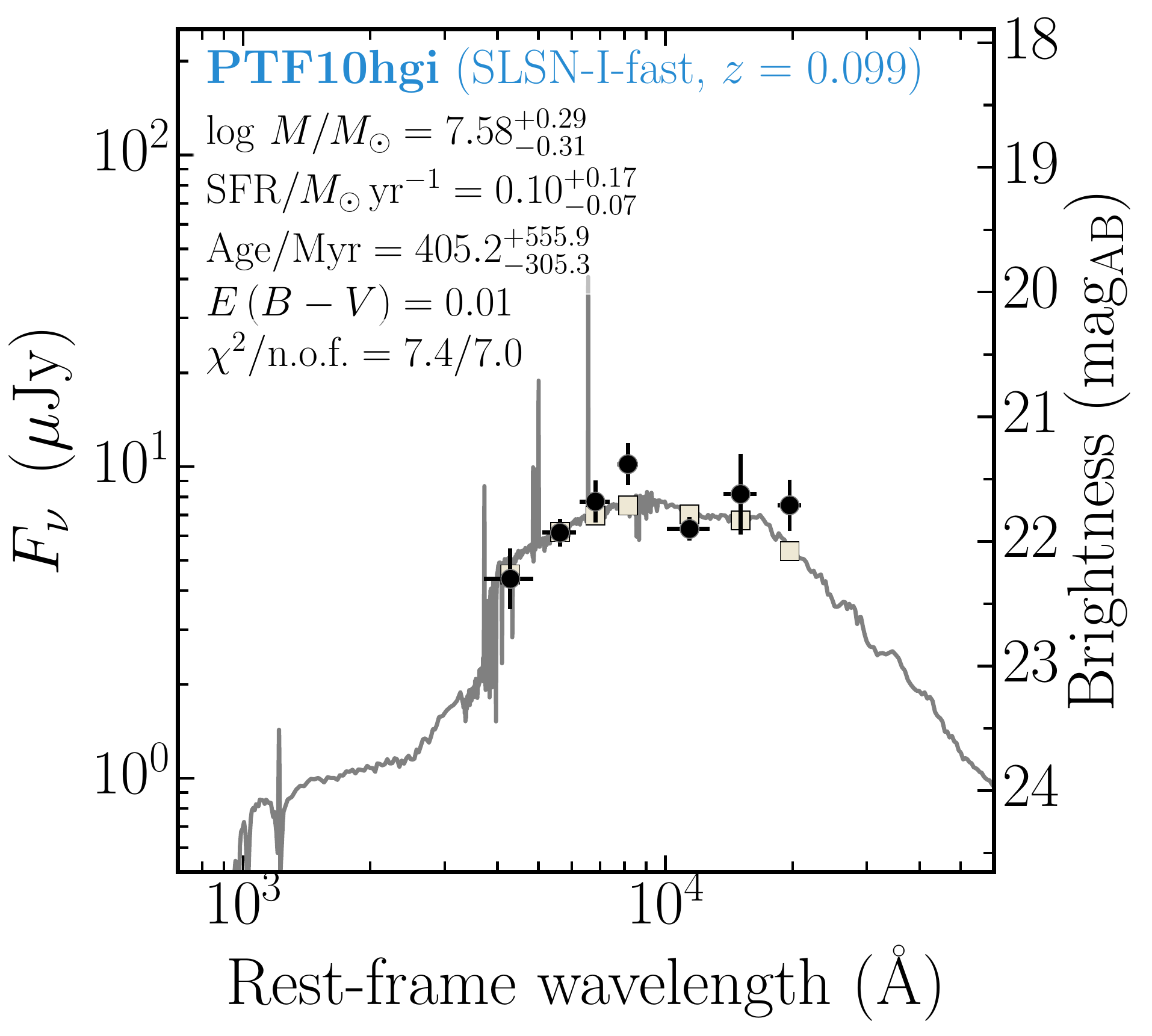}
\includegraphics[width=0.27\textwidth]{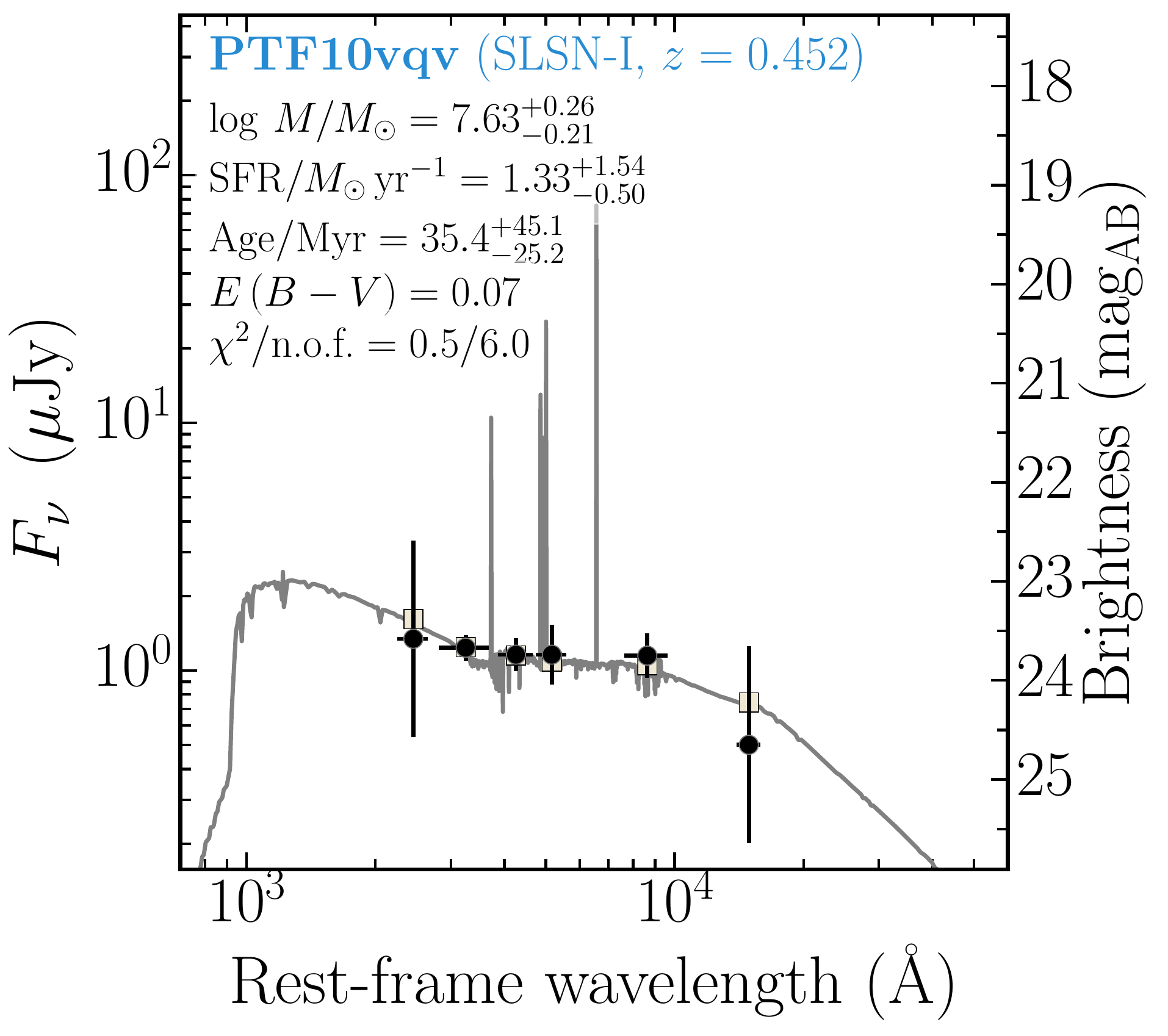}
\includegraphics[width=0.27\textwidth]{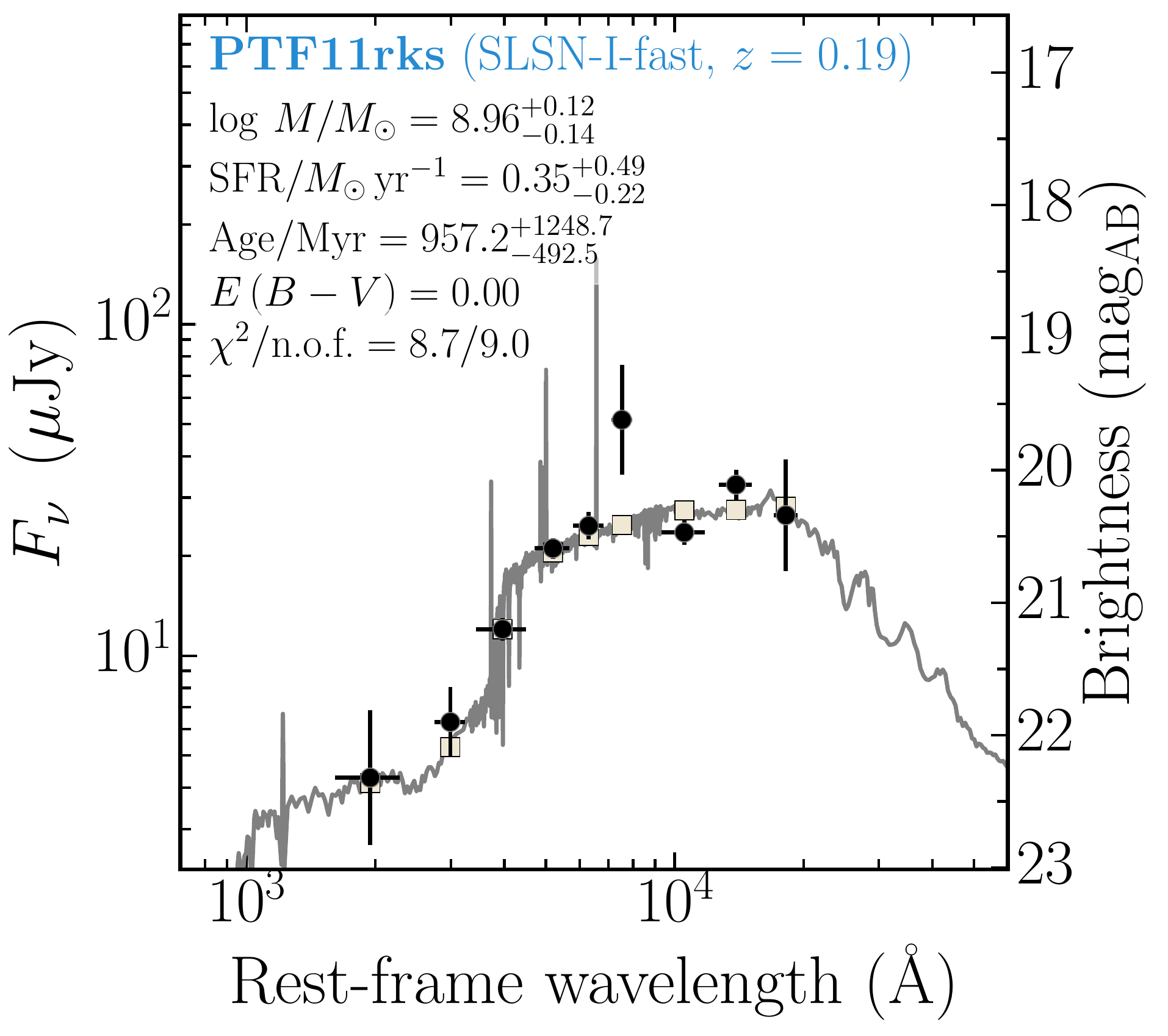}
\caption{
(Continued)
}
\end{figure*}
\clearpage

\begin{figure*}
\ContinuedFloat
\includegraphics[width=0.27\textwidth]{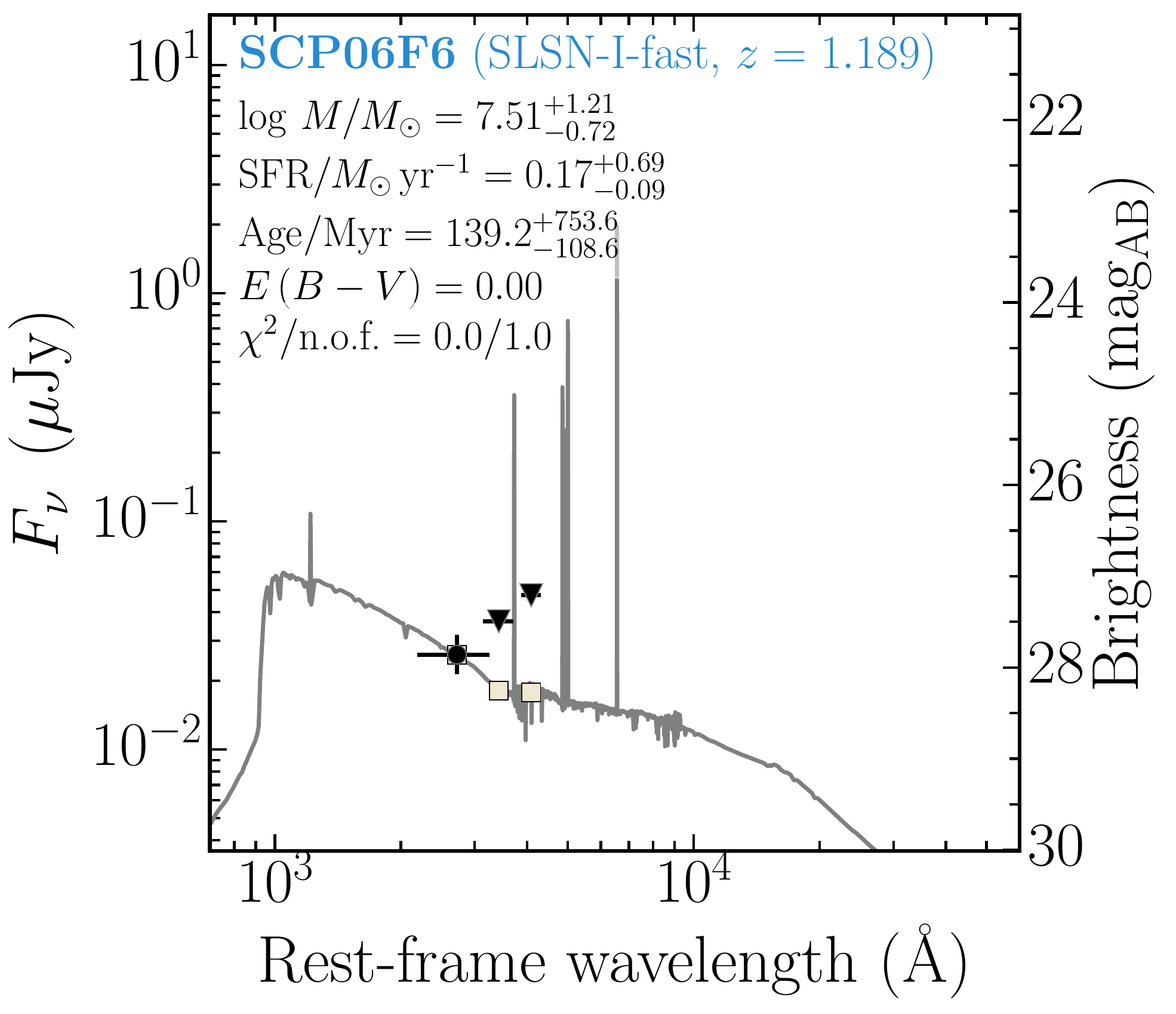}
\includegraphics[width=0.27\textwidth]{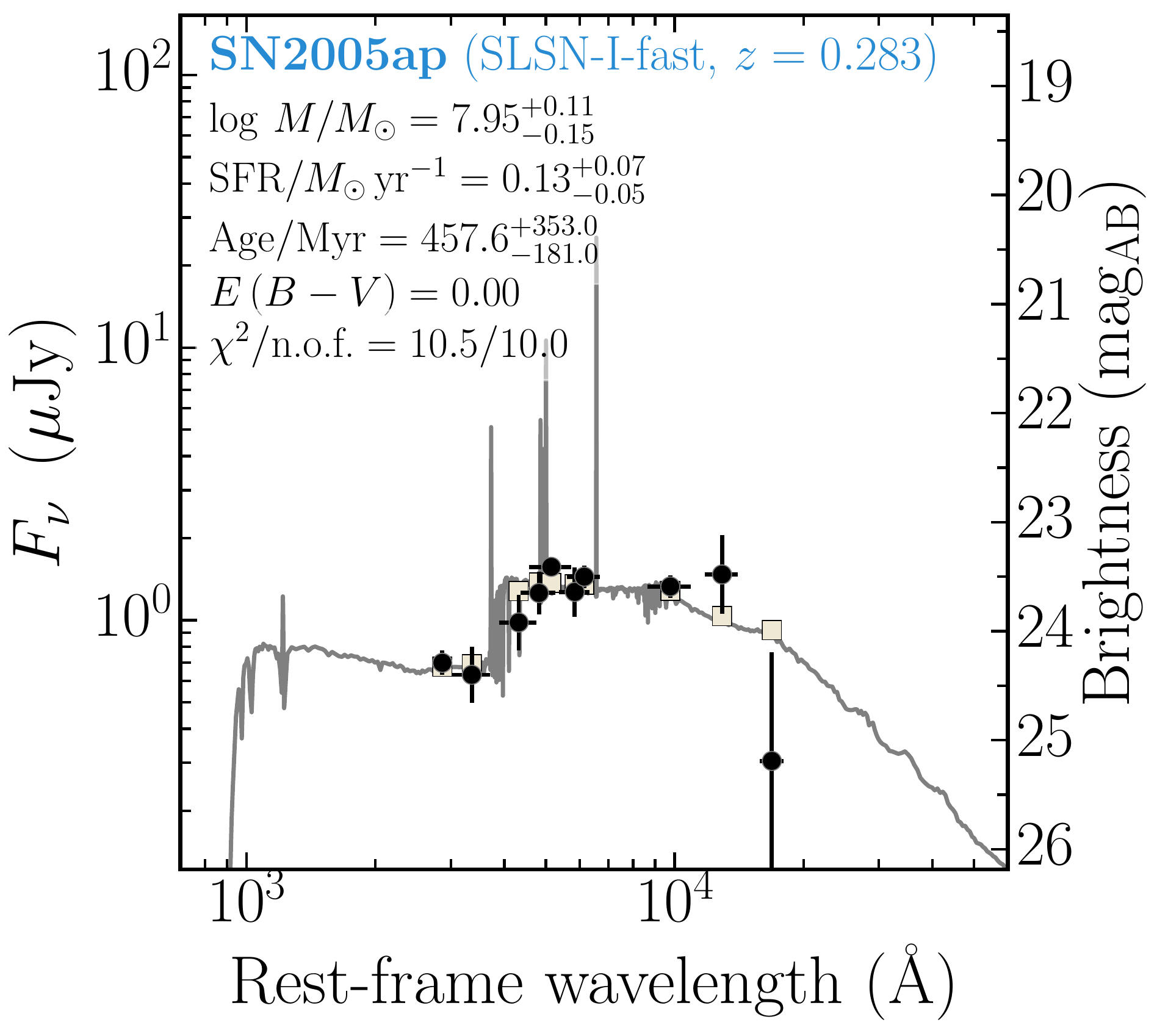}
\includegraphics[width=0.27\textwidth]{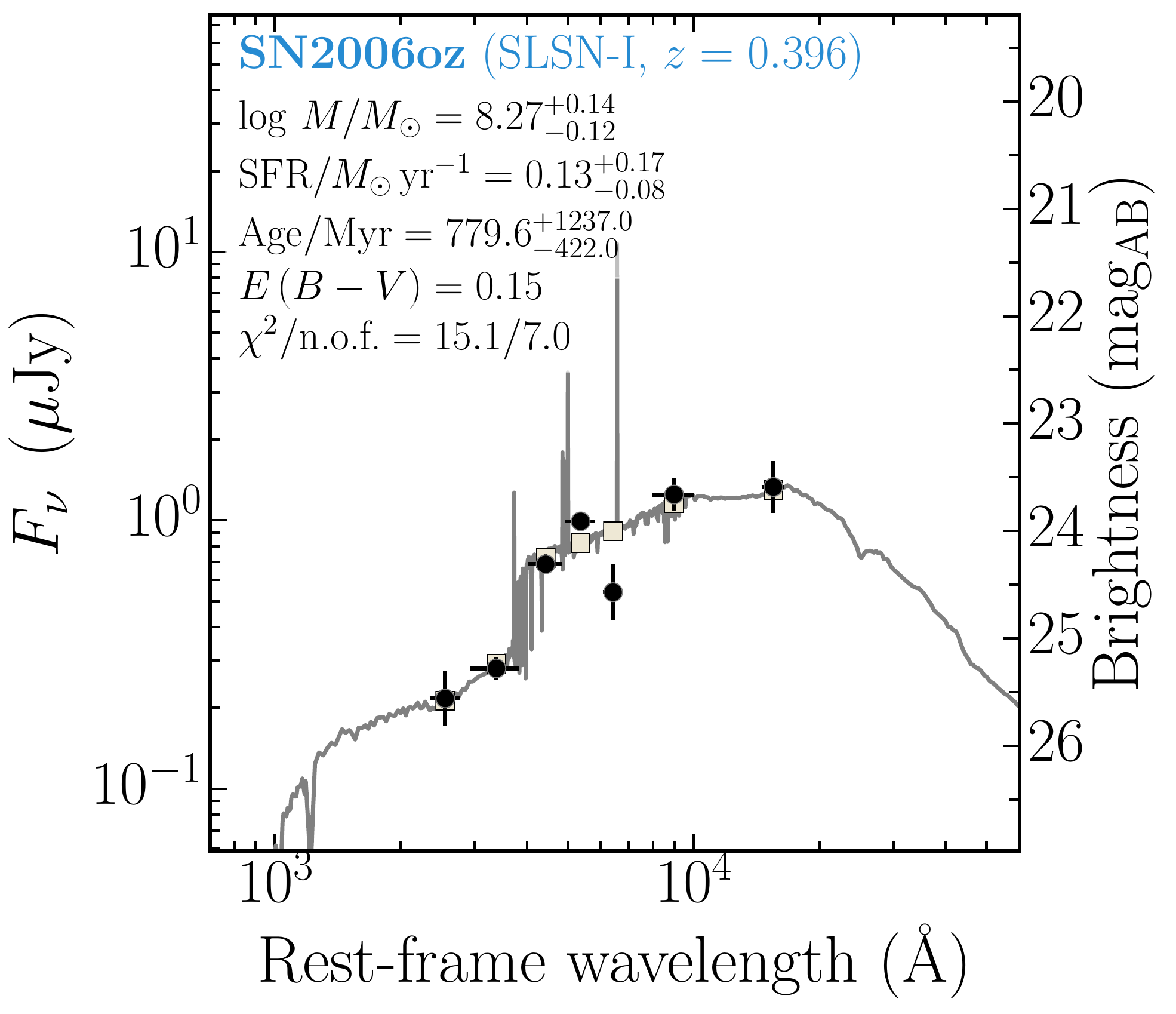}
\includegraphics[width=0.27\textwidth]{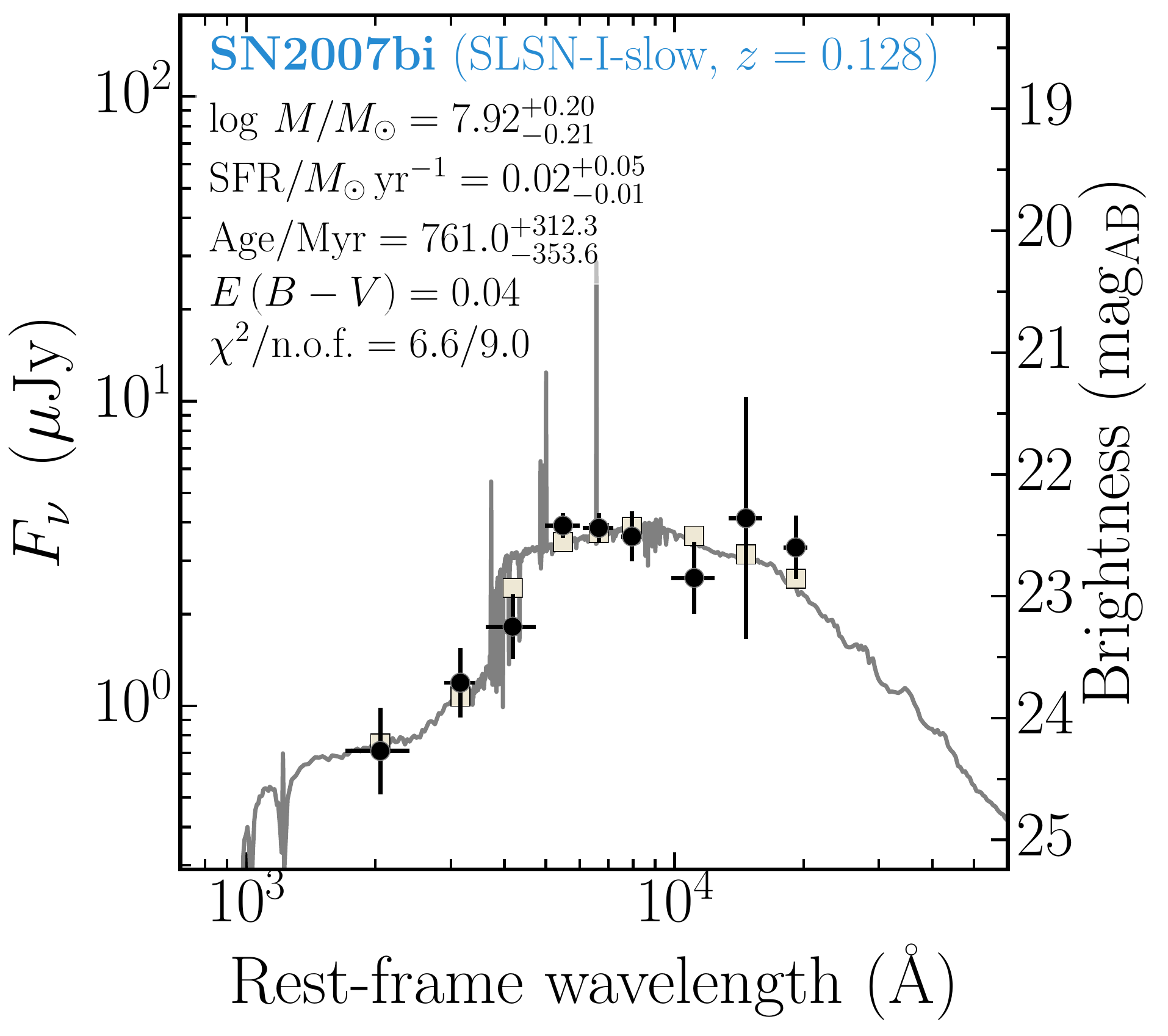}
\includegraphics[width=0.27\textwidth]{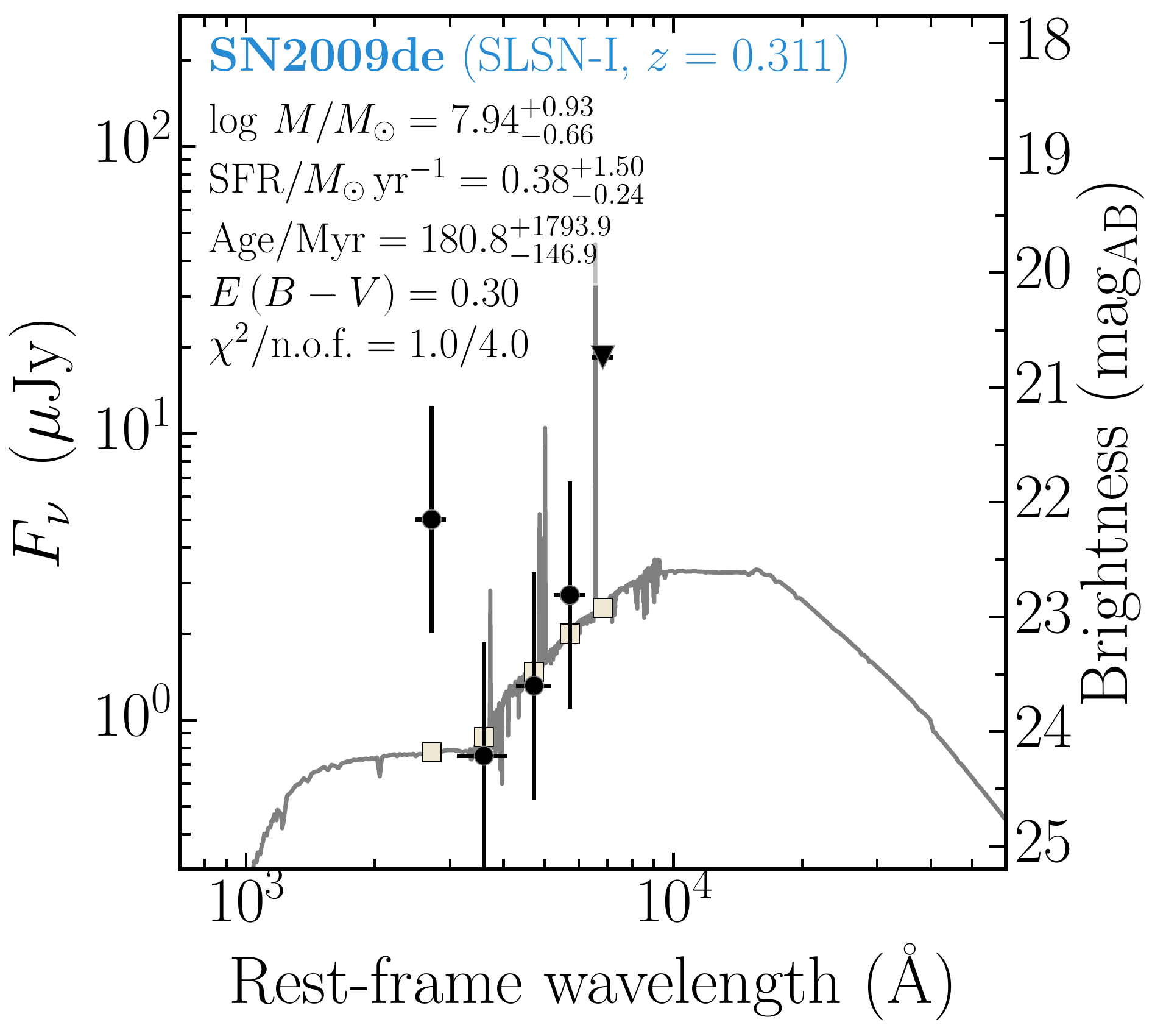}
\includegraphics[width=0.27\textwidth]{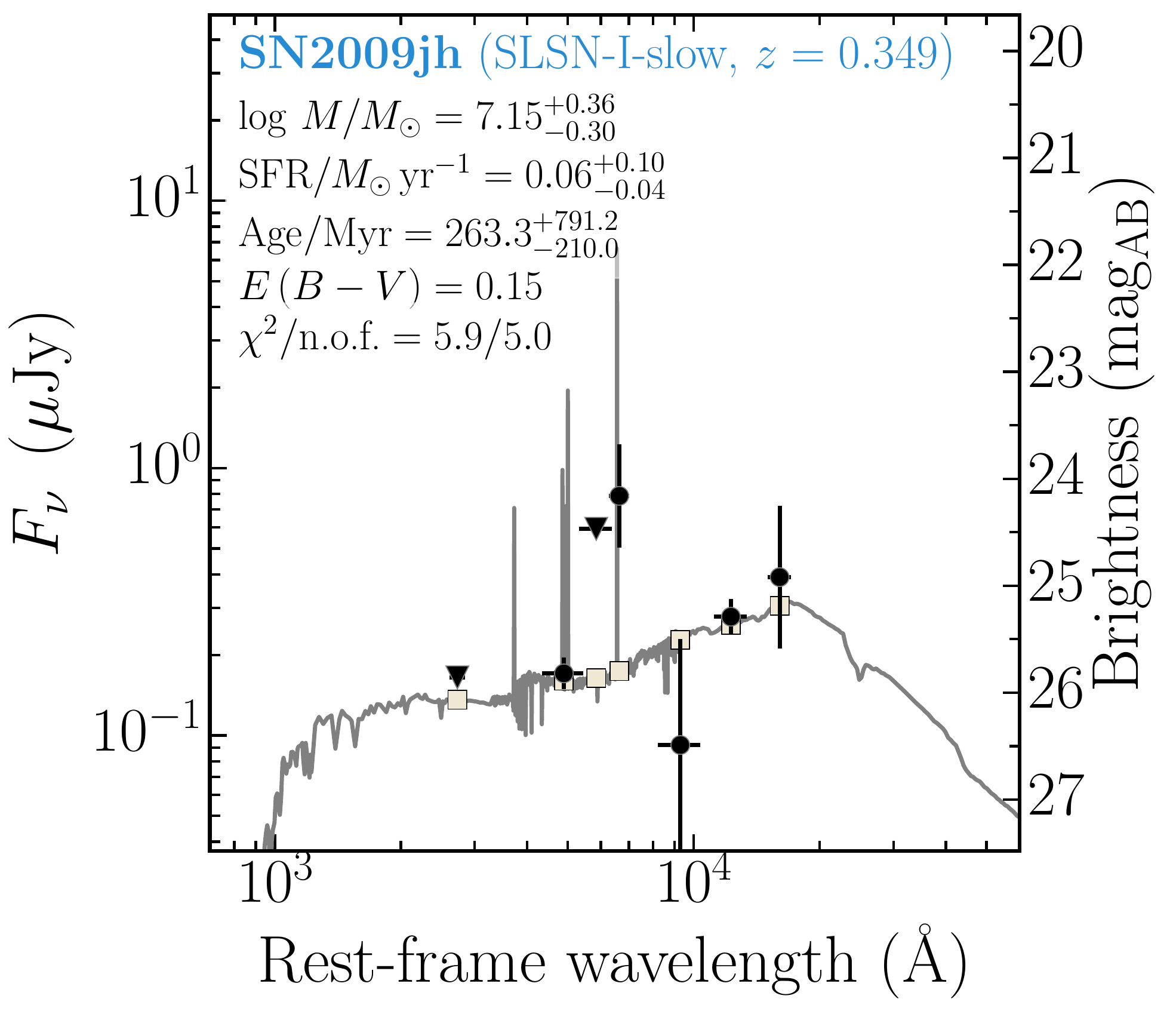}
\includegraphics[width=0.27\textwidth]{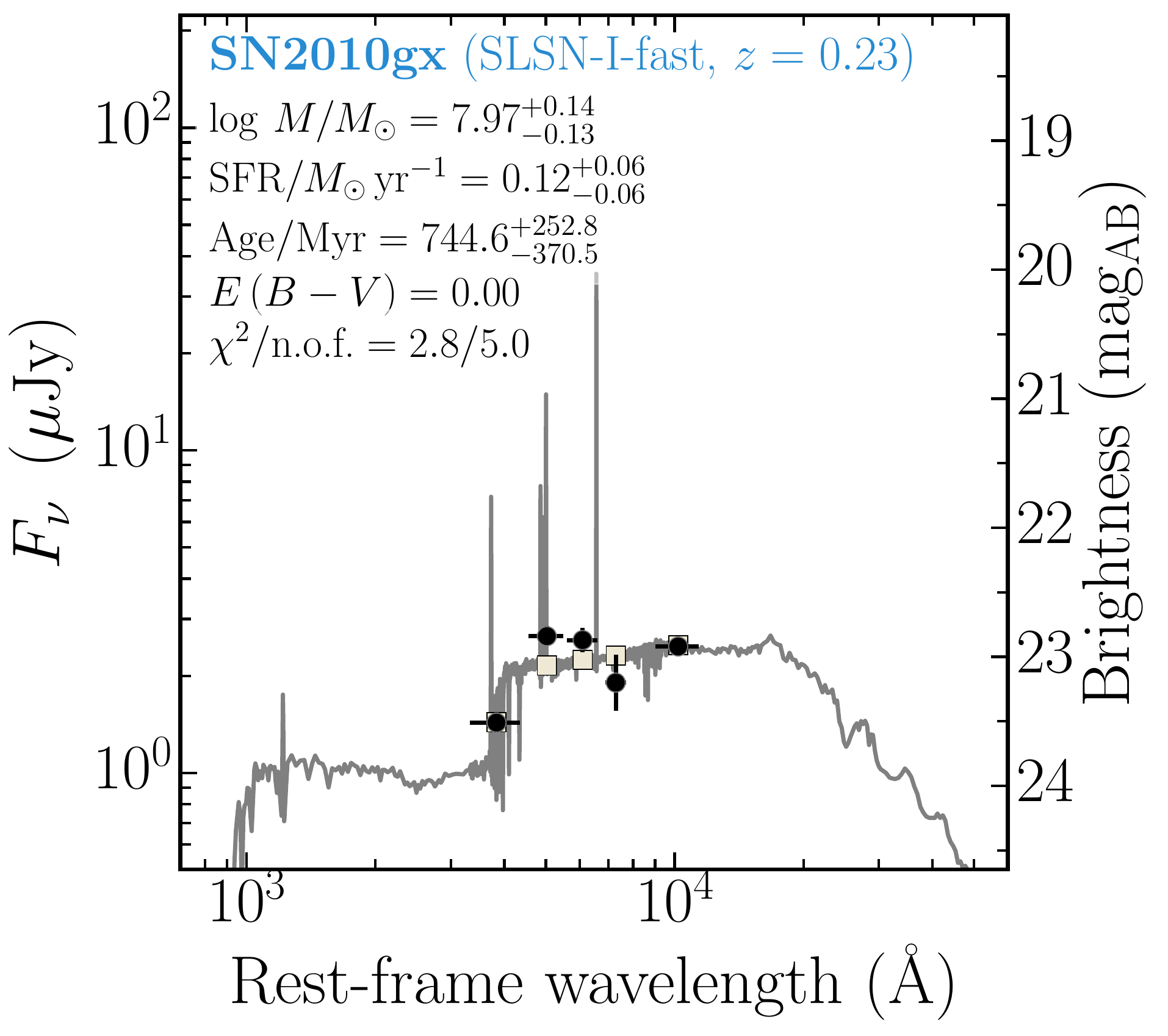}
\includegraphics[width=0.27\textwidth]{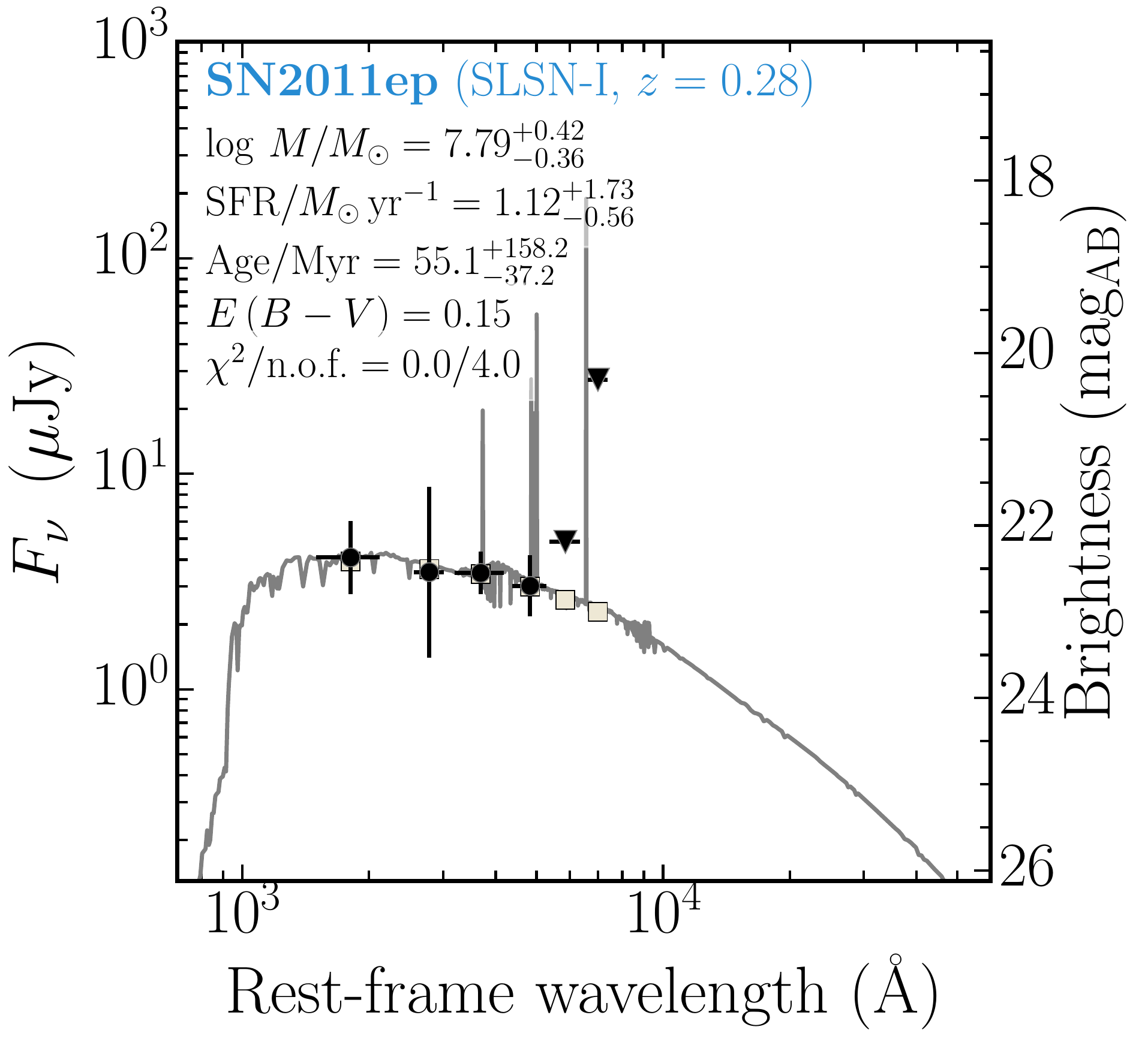}
\includegraphics[width=0.27\textwidth]{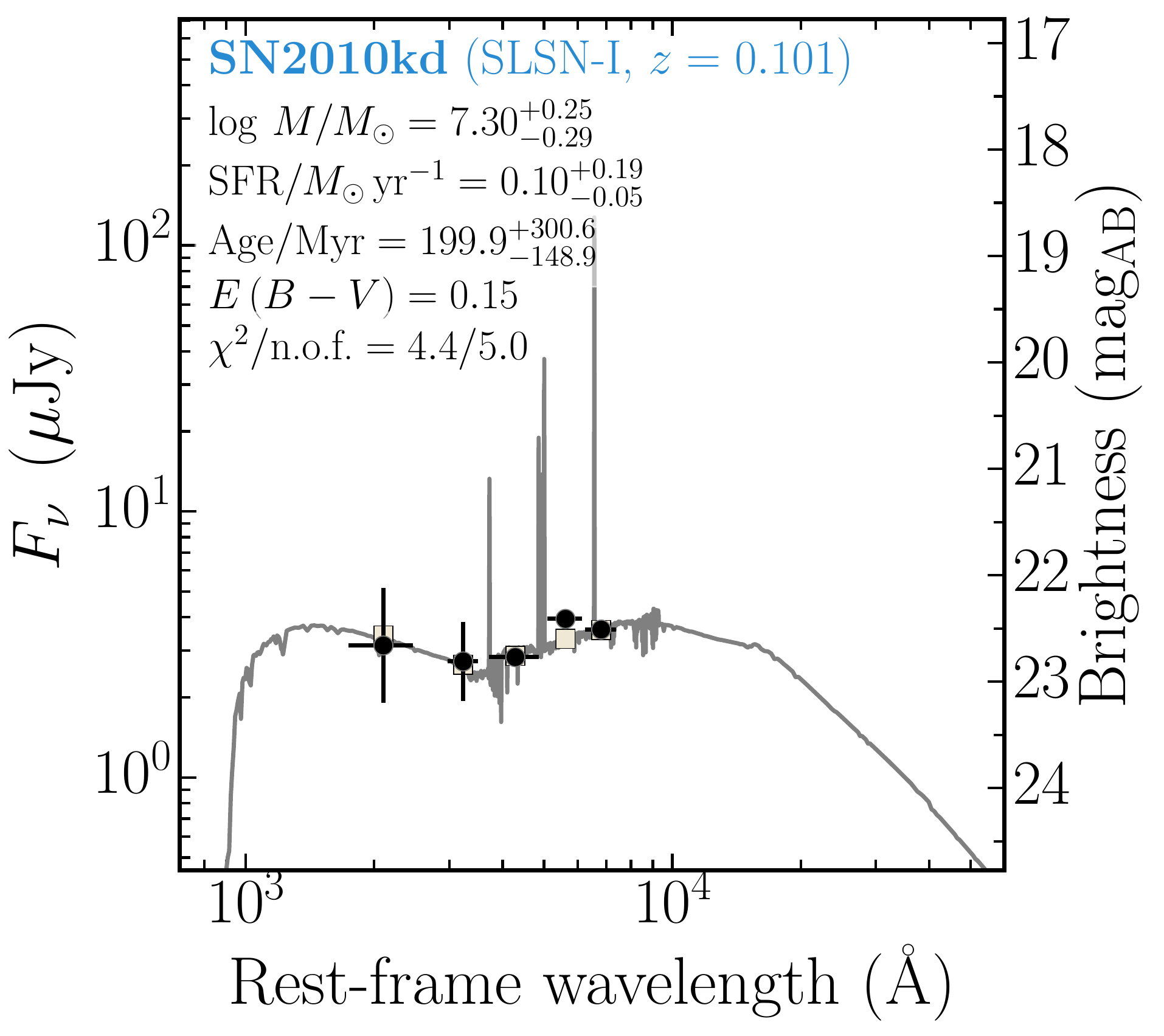}
\includegraphics[width=0.27\textwidth]{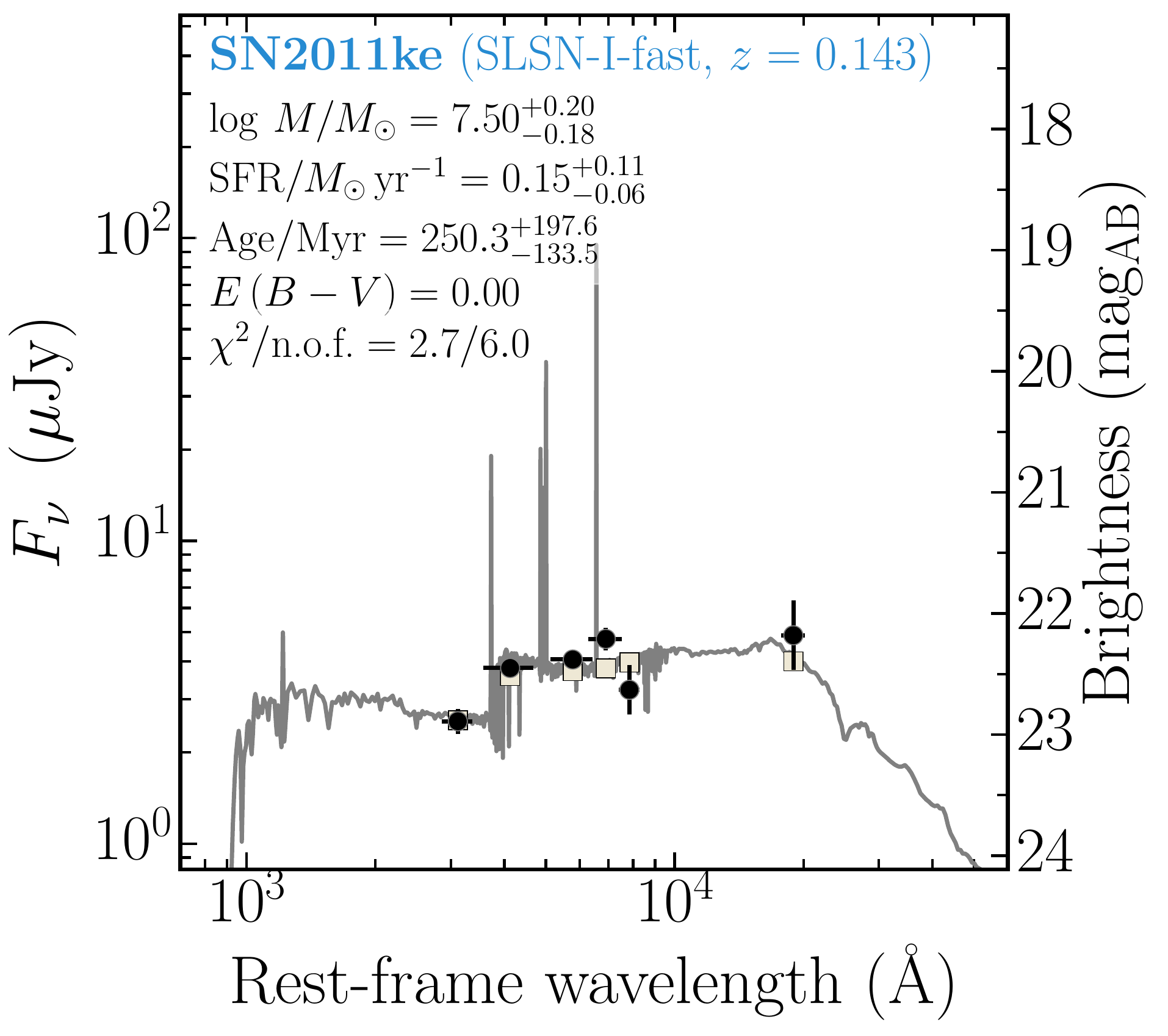}
\includegraphics[width=0.27\textwidth]{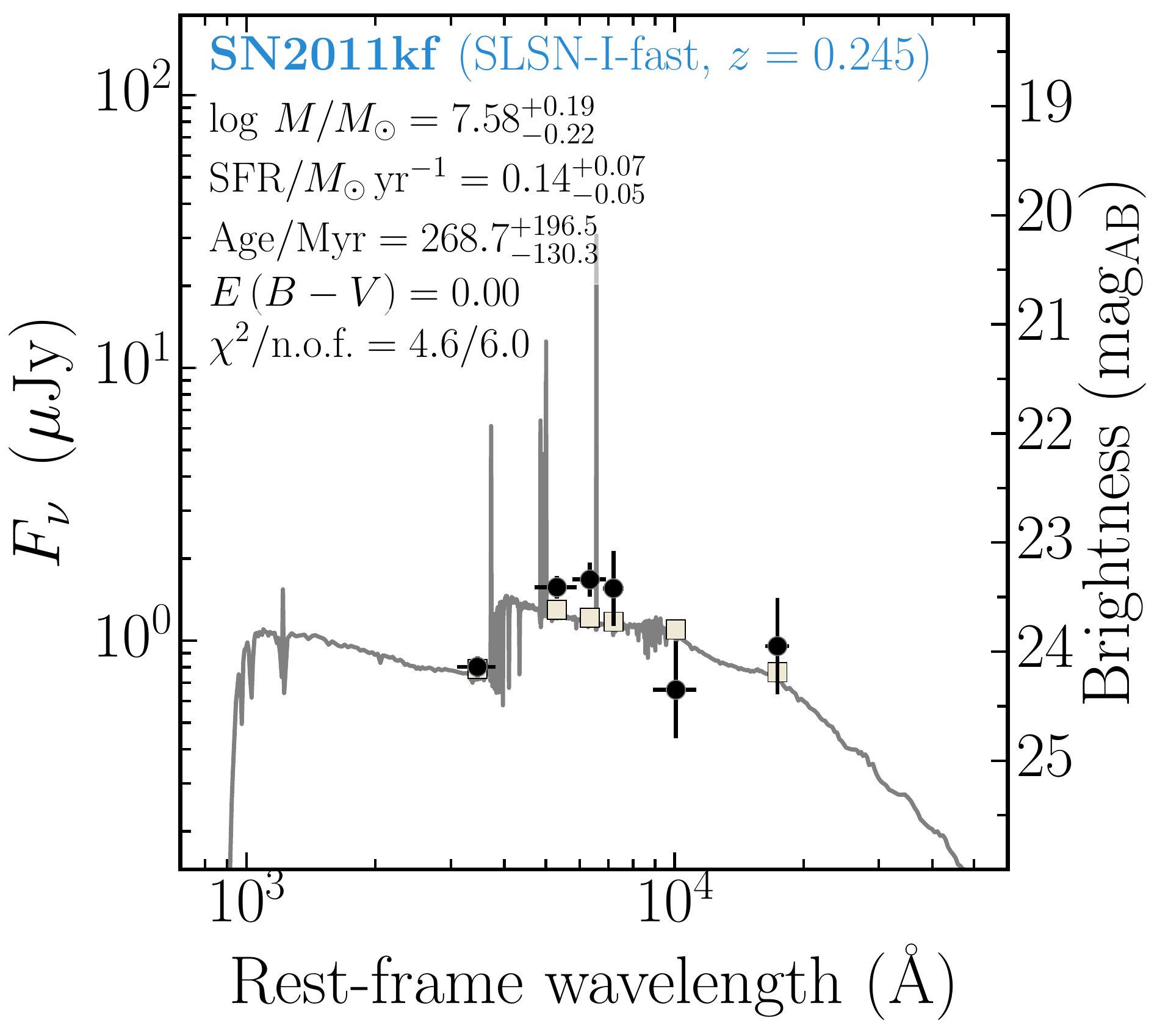}
\includegraphics[width=0.27\textwidth]{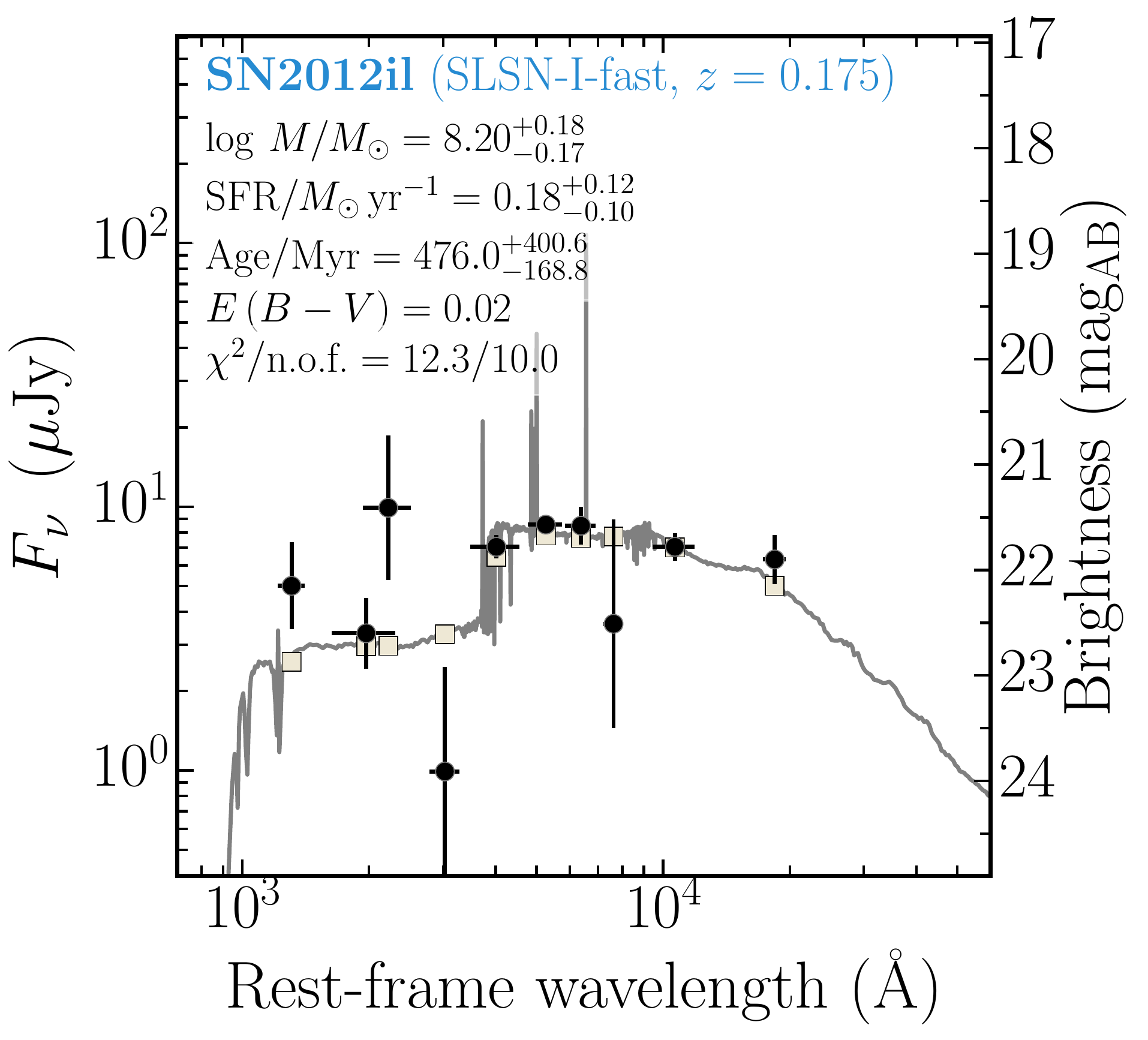}
\includegraphics[width=0.27\textwidth]{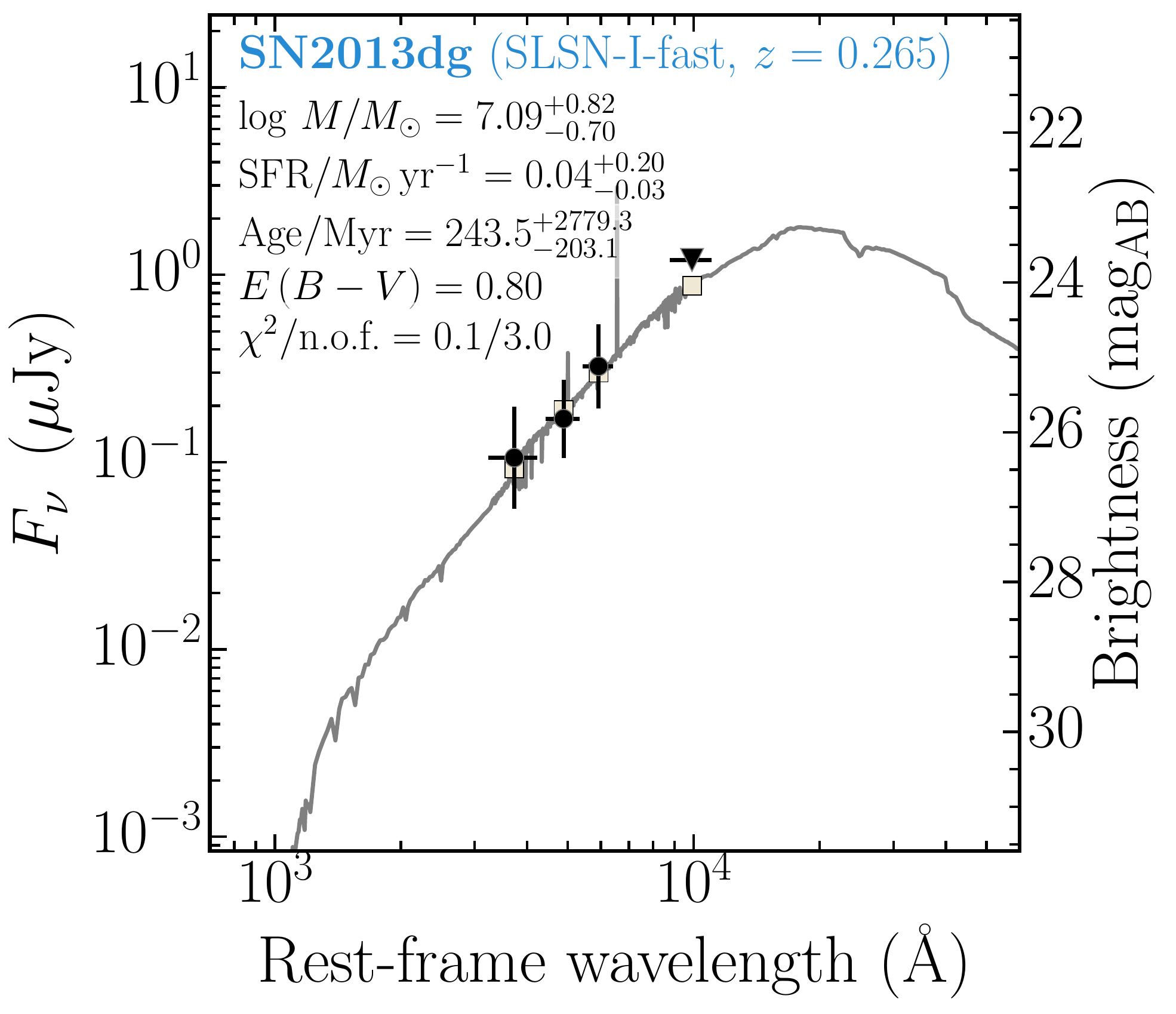}
\includegraphics[width=0.27\textwidth]{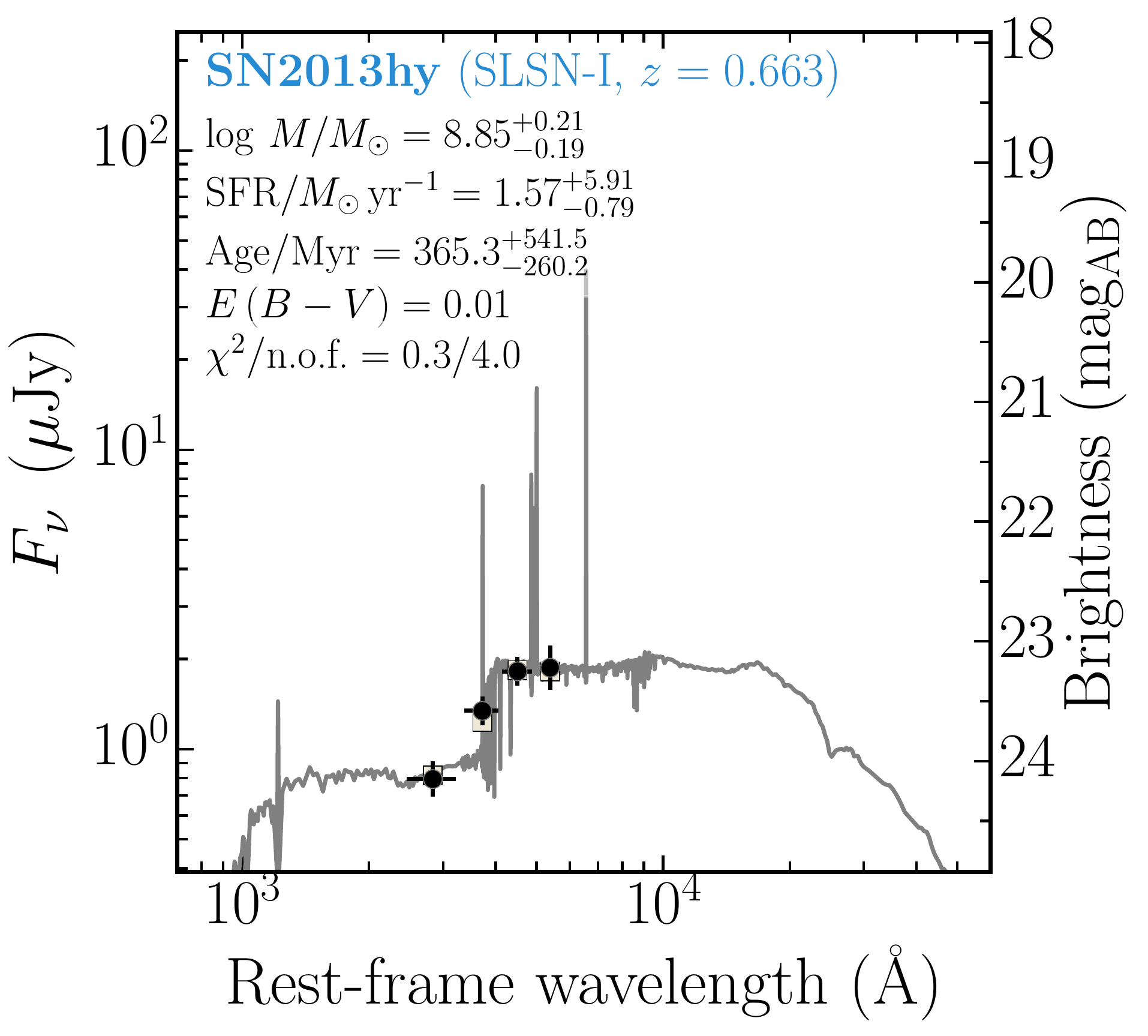}
\includegraphics[width=0.27\textwidth]{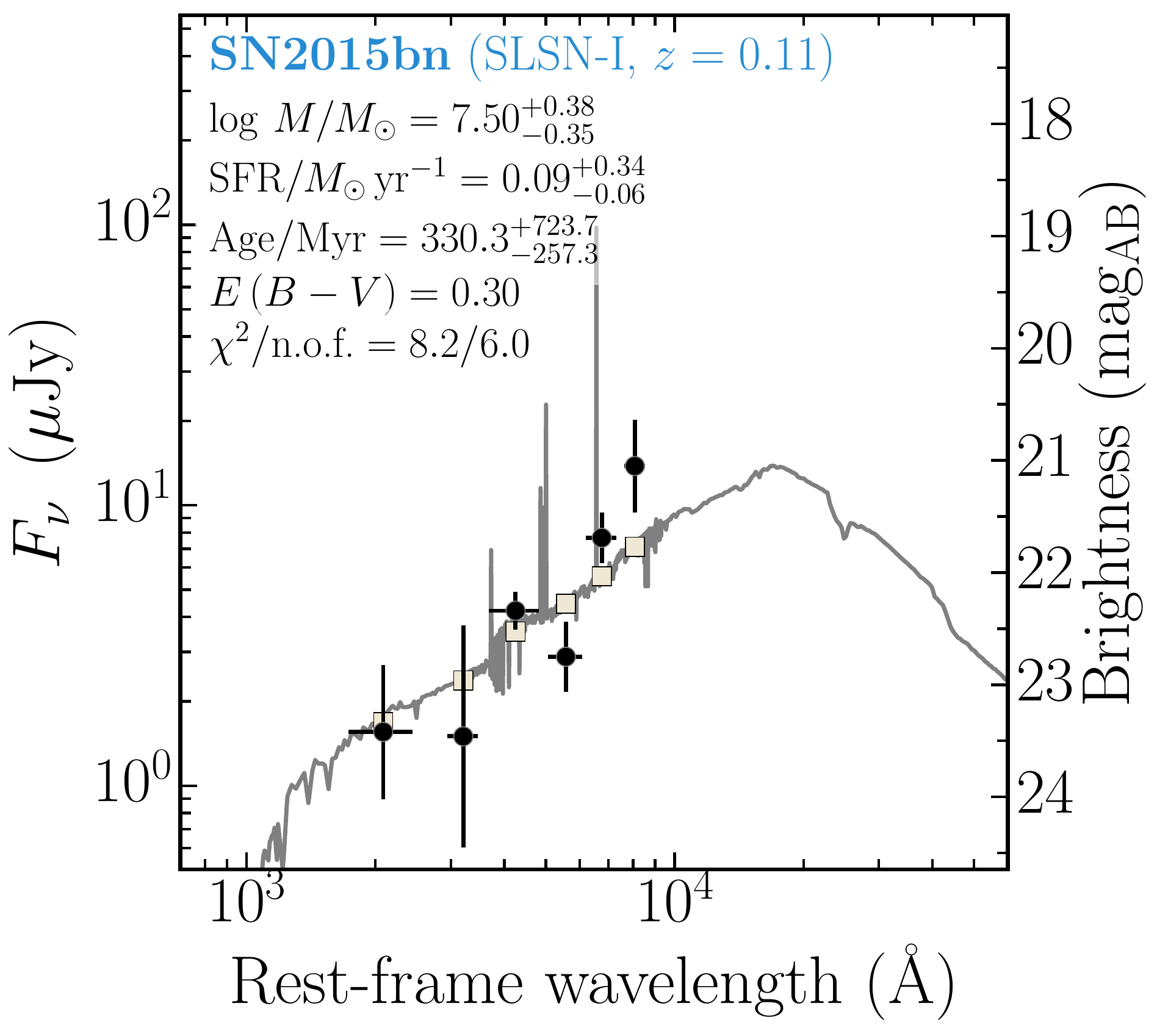}
\caption{
(Continued)
}
\end{figure*}
\clearpage

\begin{figure*}
\ContinuedFloat
\includegraphics[width=0.27\textwidth]{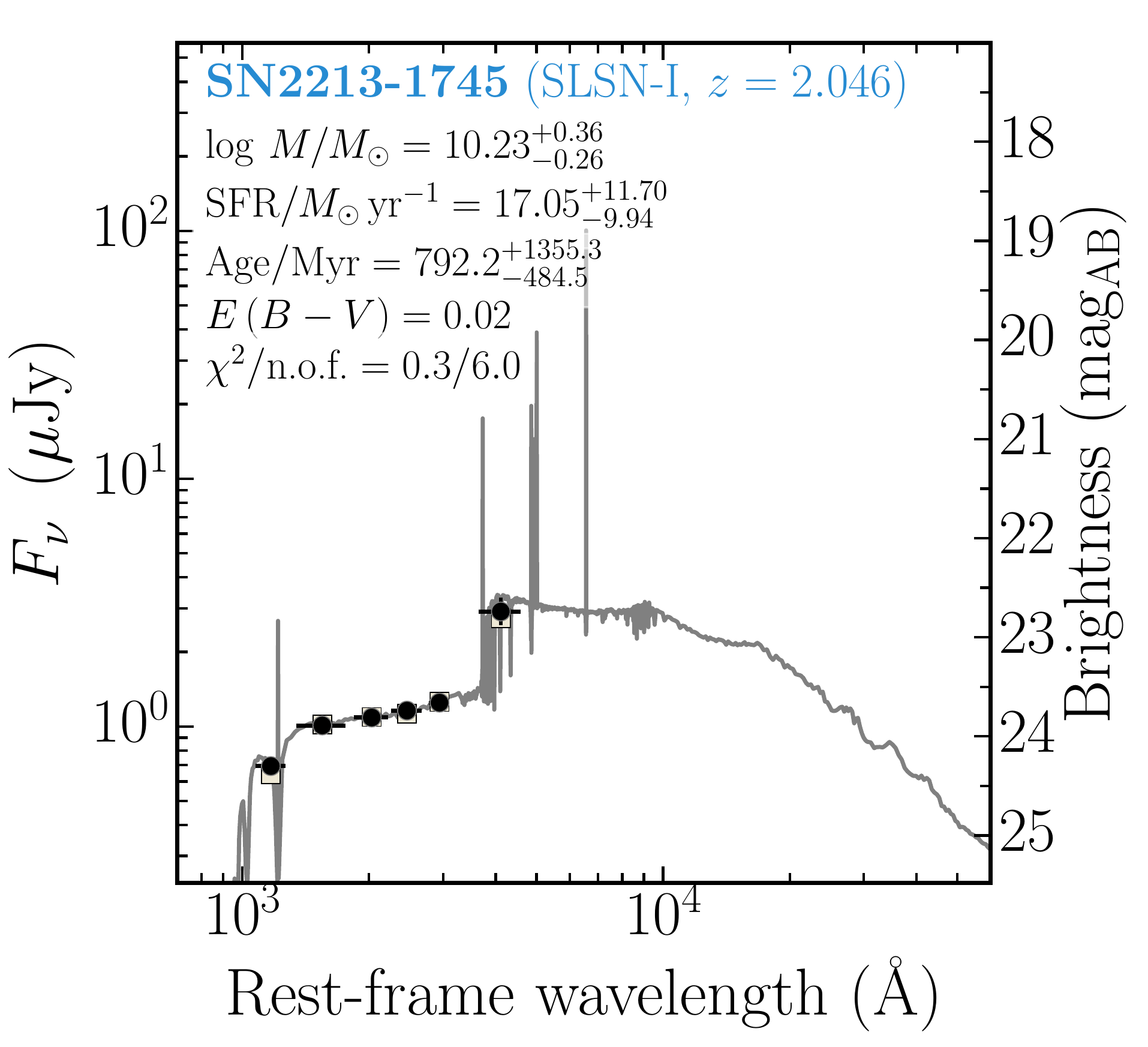}
\includegraphics[width=0.27\textwidth]{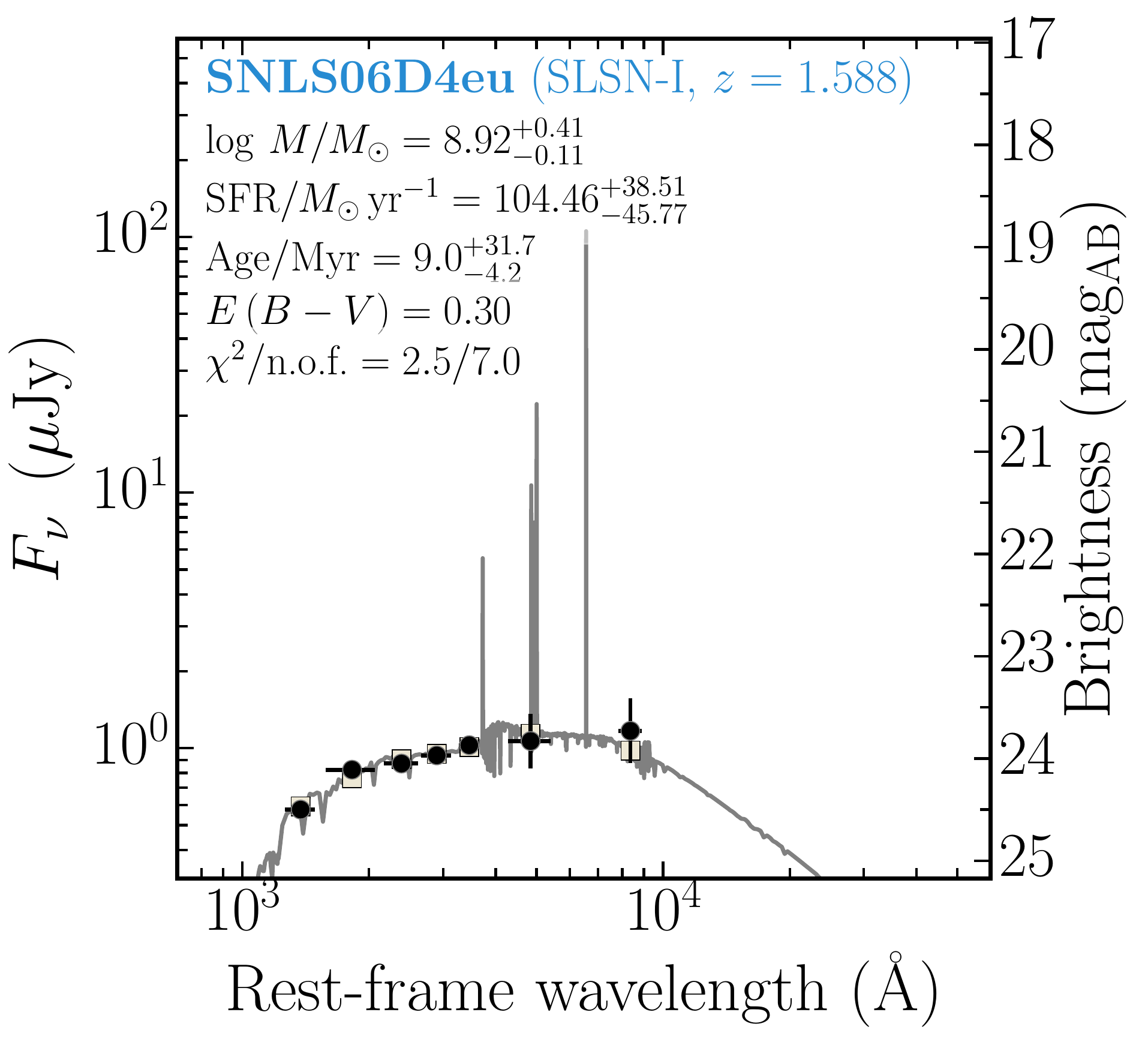}
\includegraphics[width=0.27\textwidth]{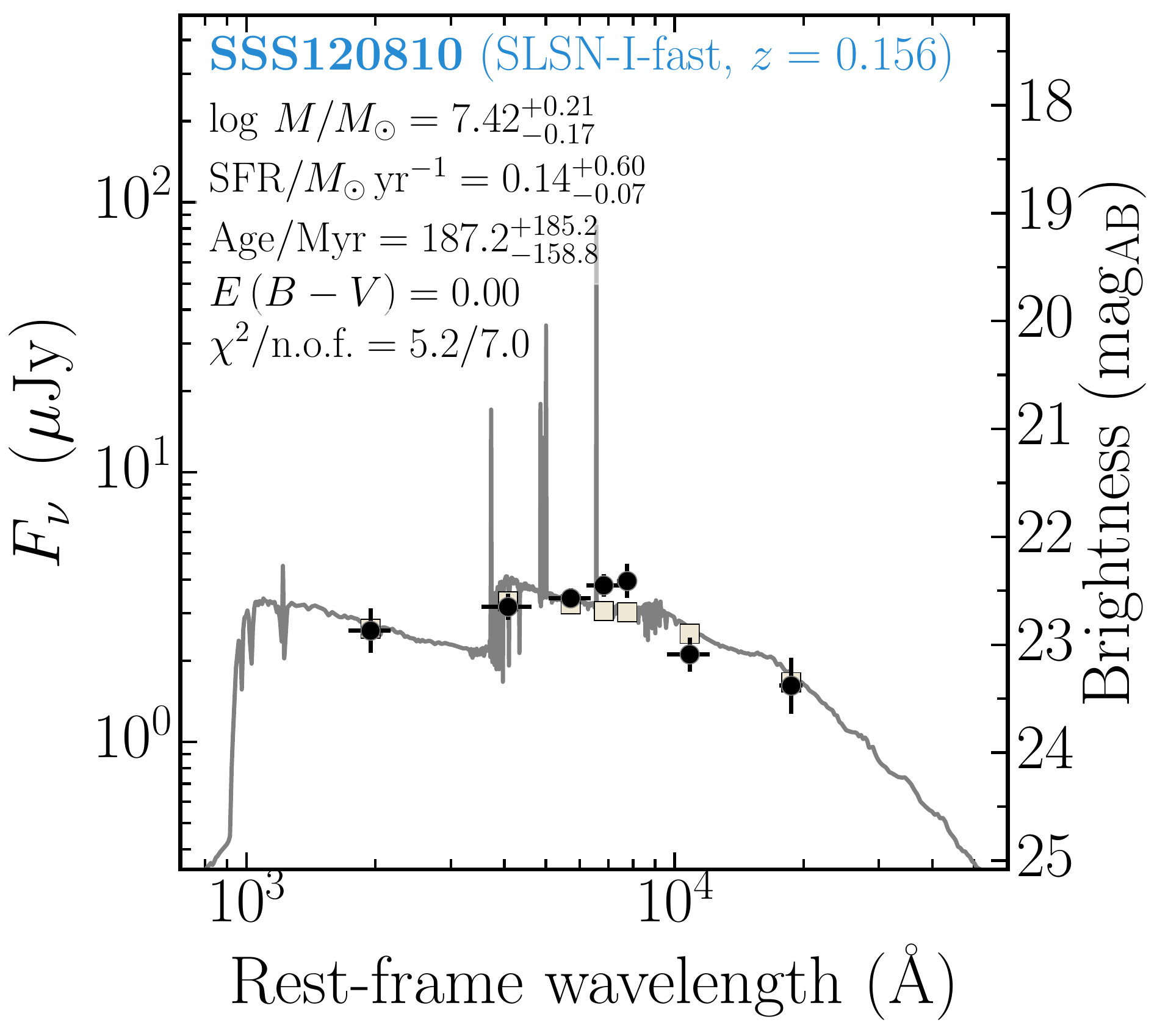}
\caption{
(Continued)
}
\end{figure*}

\begin{figure*}
\includegraphics[width=0.27\textwidth]{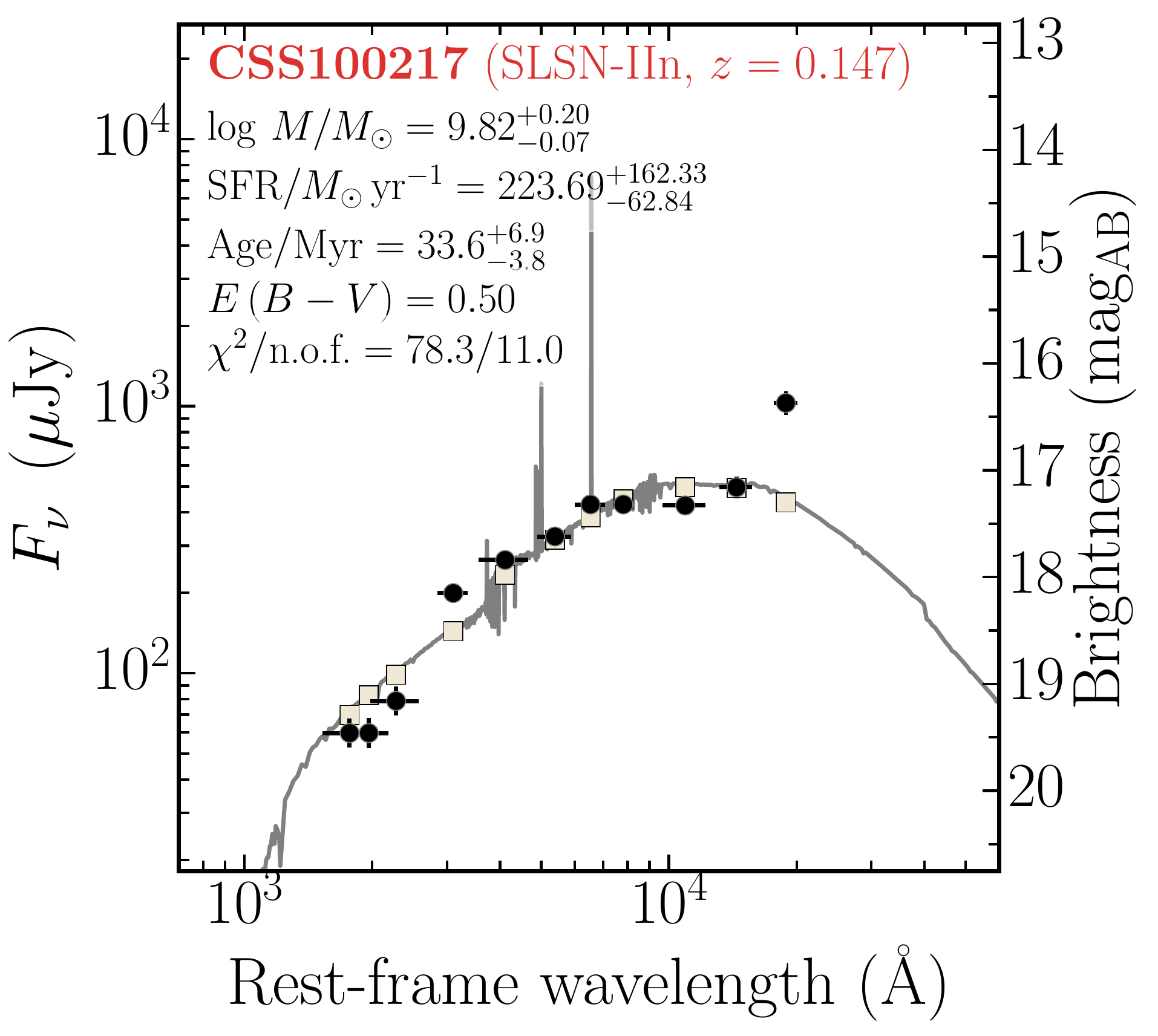}
\includegraphics[width=0.27\textwidth]{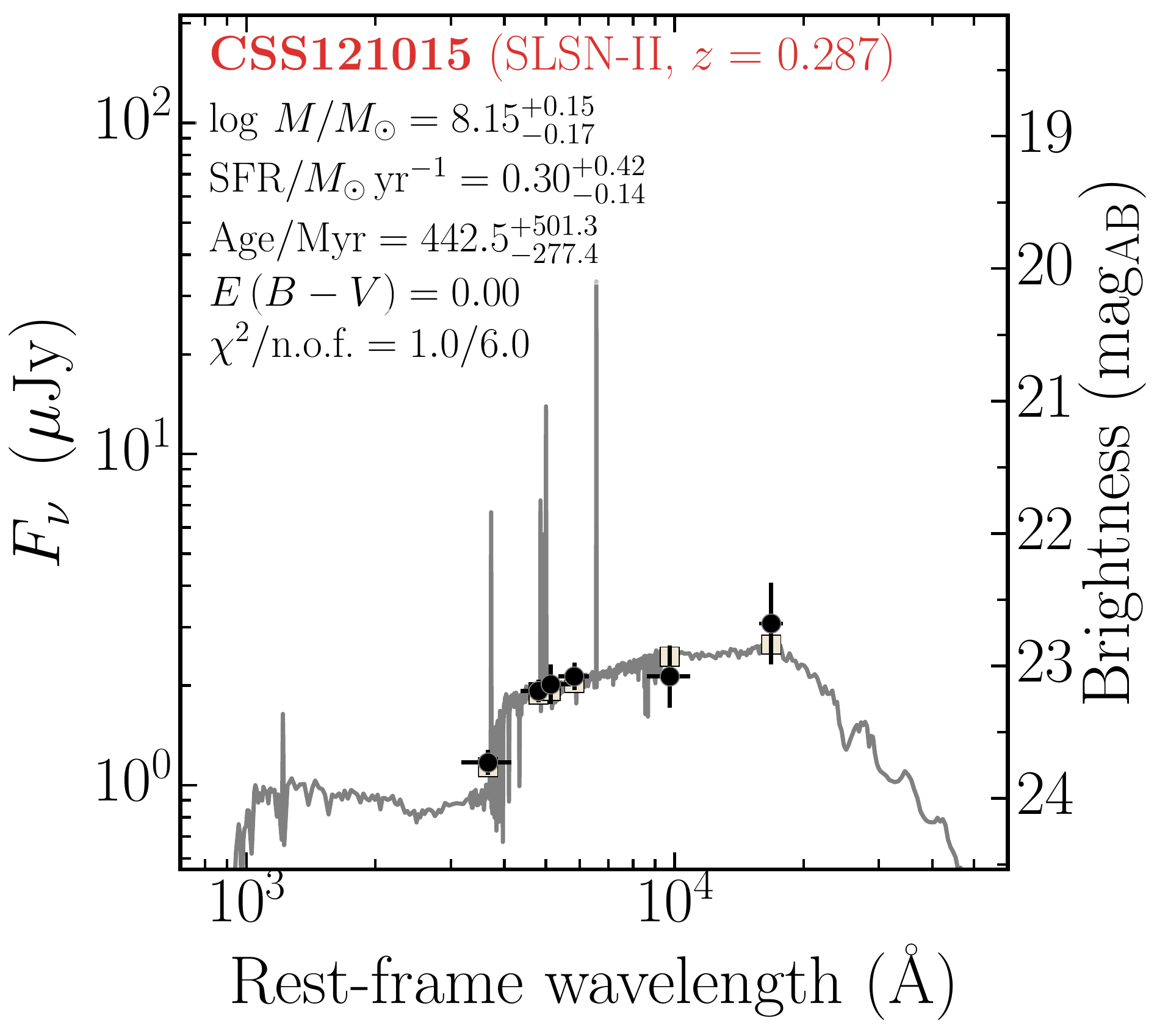}
\includegraphics[width=0.27\textwidth]{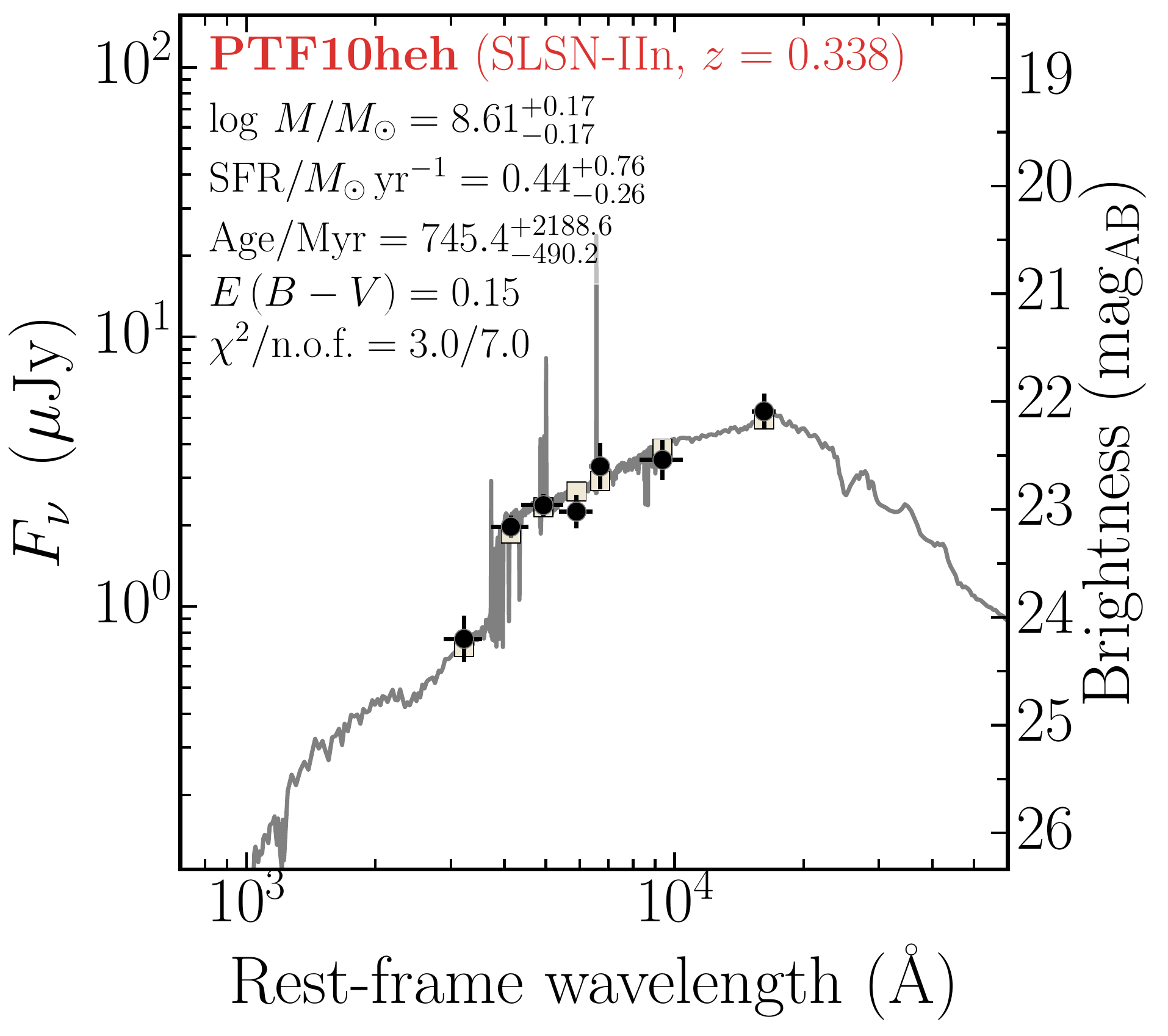}
\includegraphics[width=0.27\textwidth]{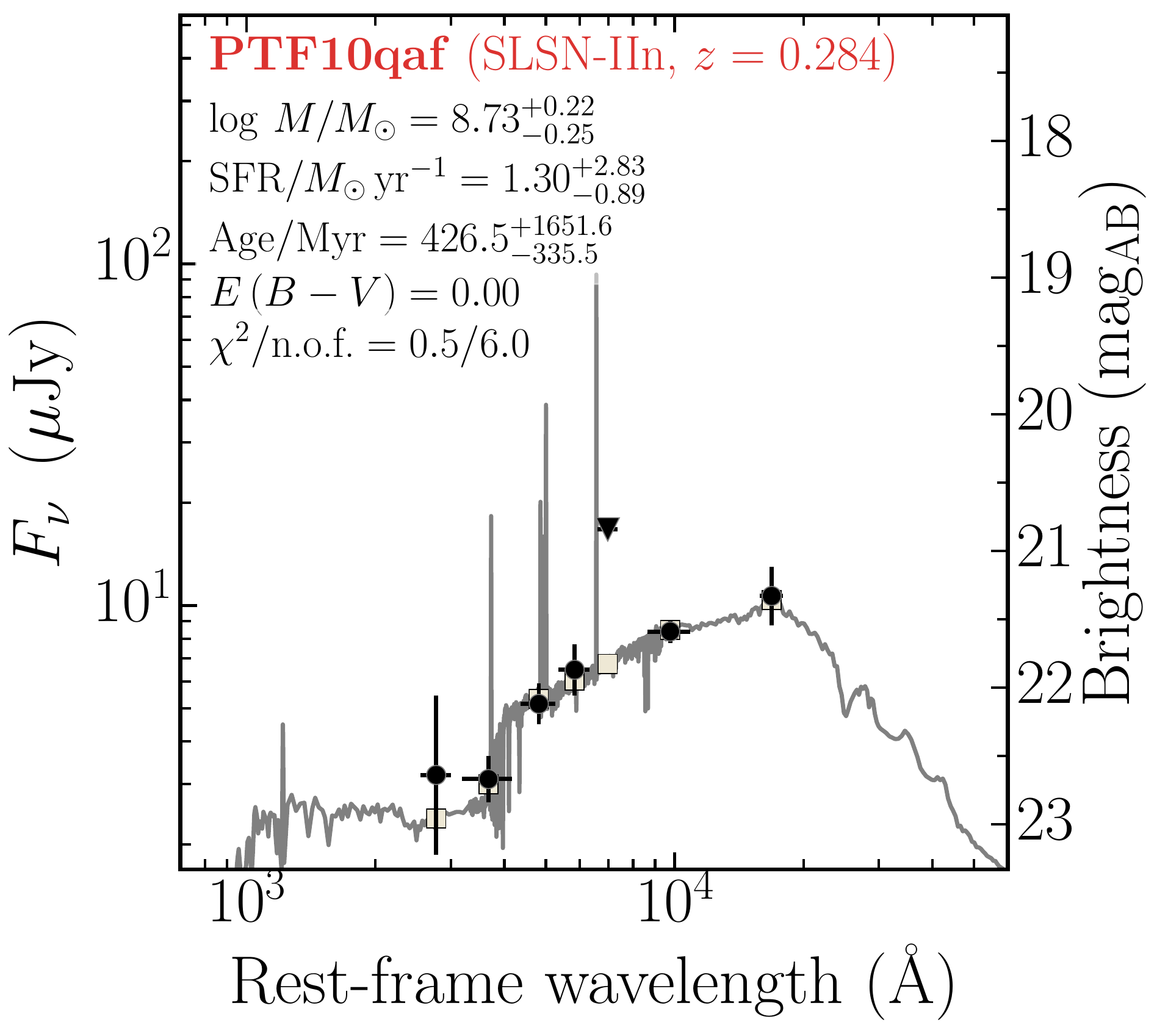}
\includegraphics[width=0.27\textwidth]{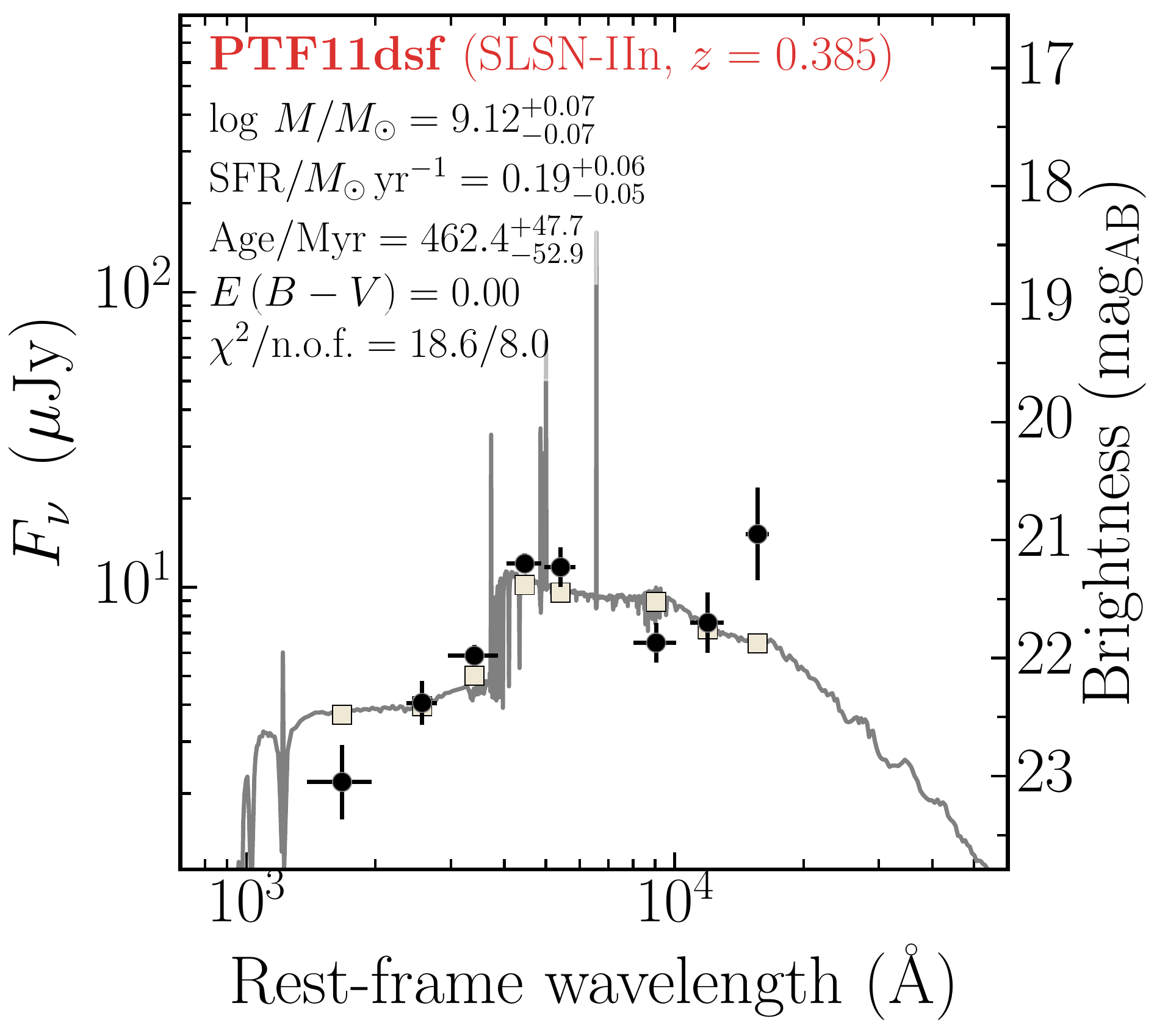}
\includegraphics[width=0.27\textwidth]{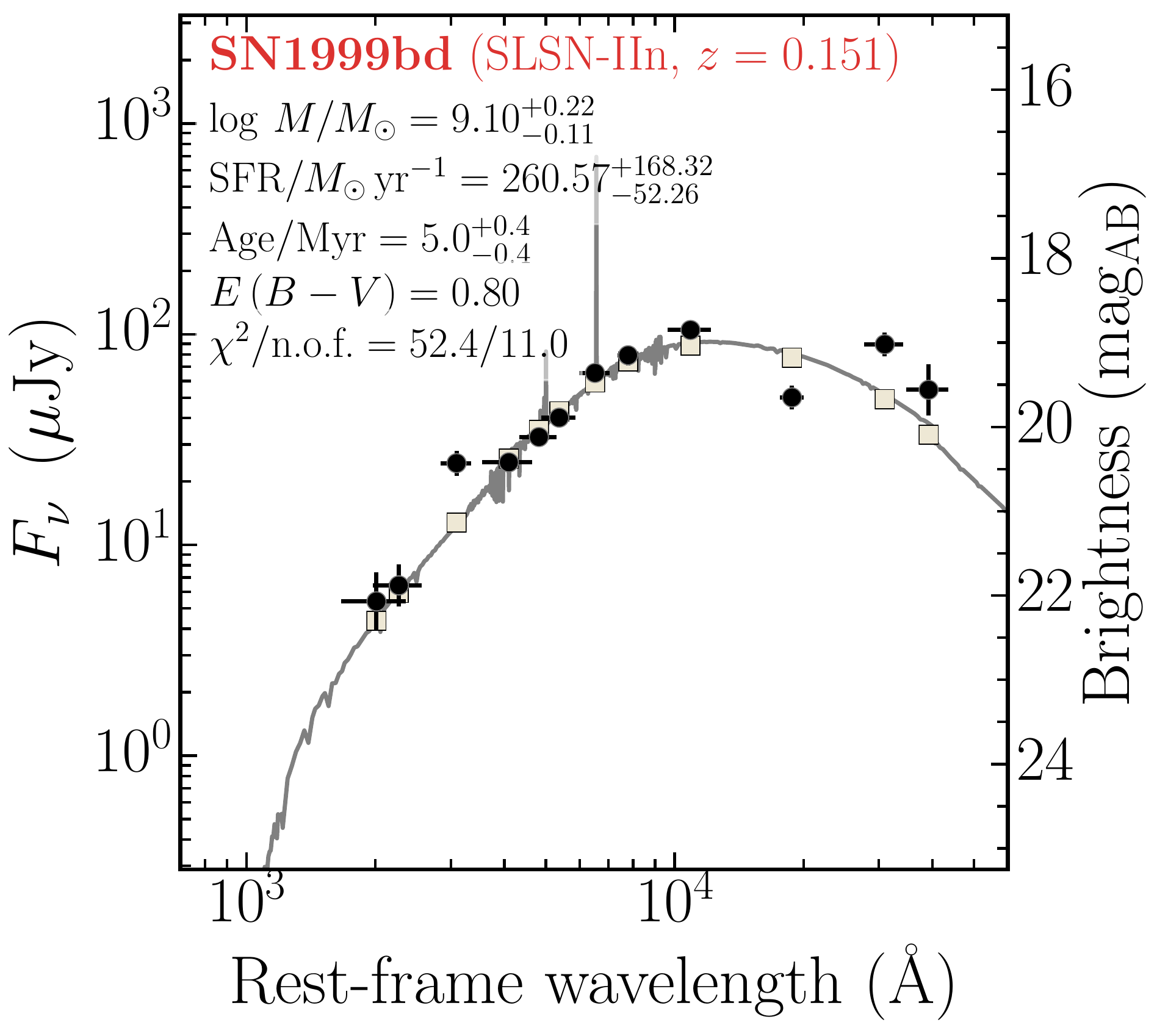}
\includegraphics[width=0.27\textwidth]{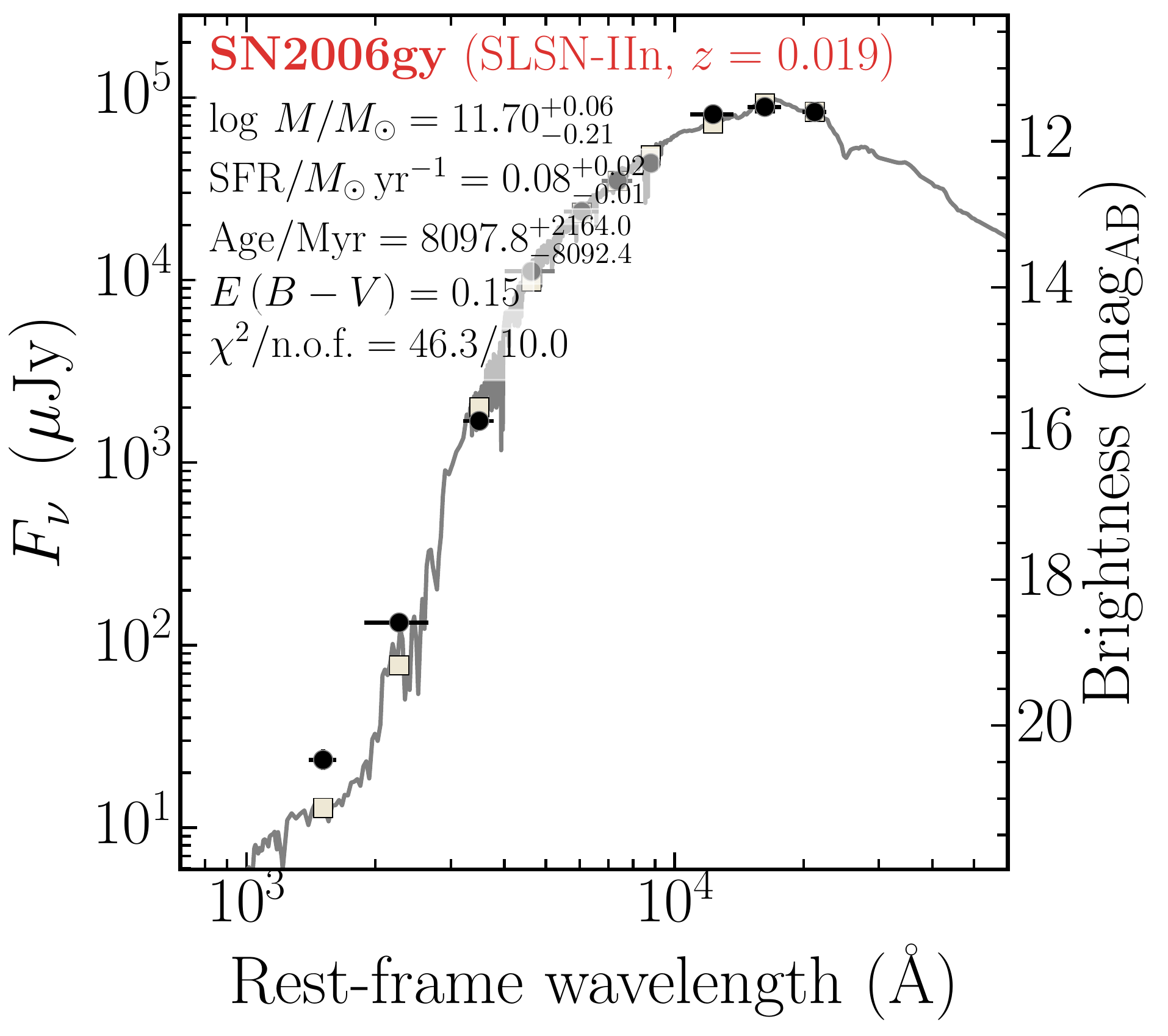}
\includegraphics[width=0.27\textwidth]{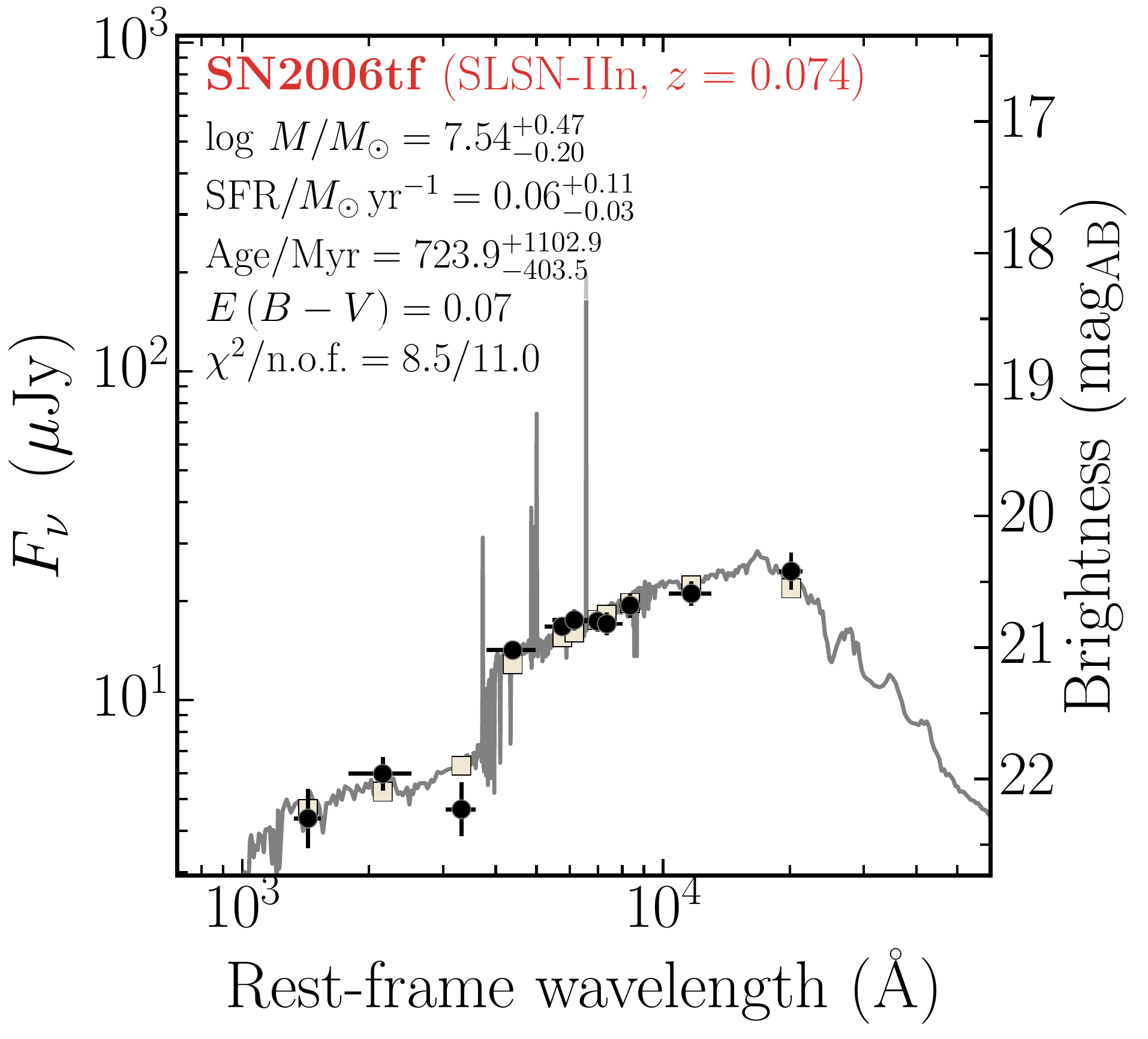}
\includegraphics[width=0.27\textwidth]{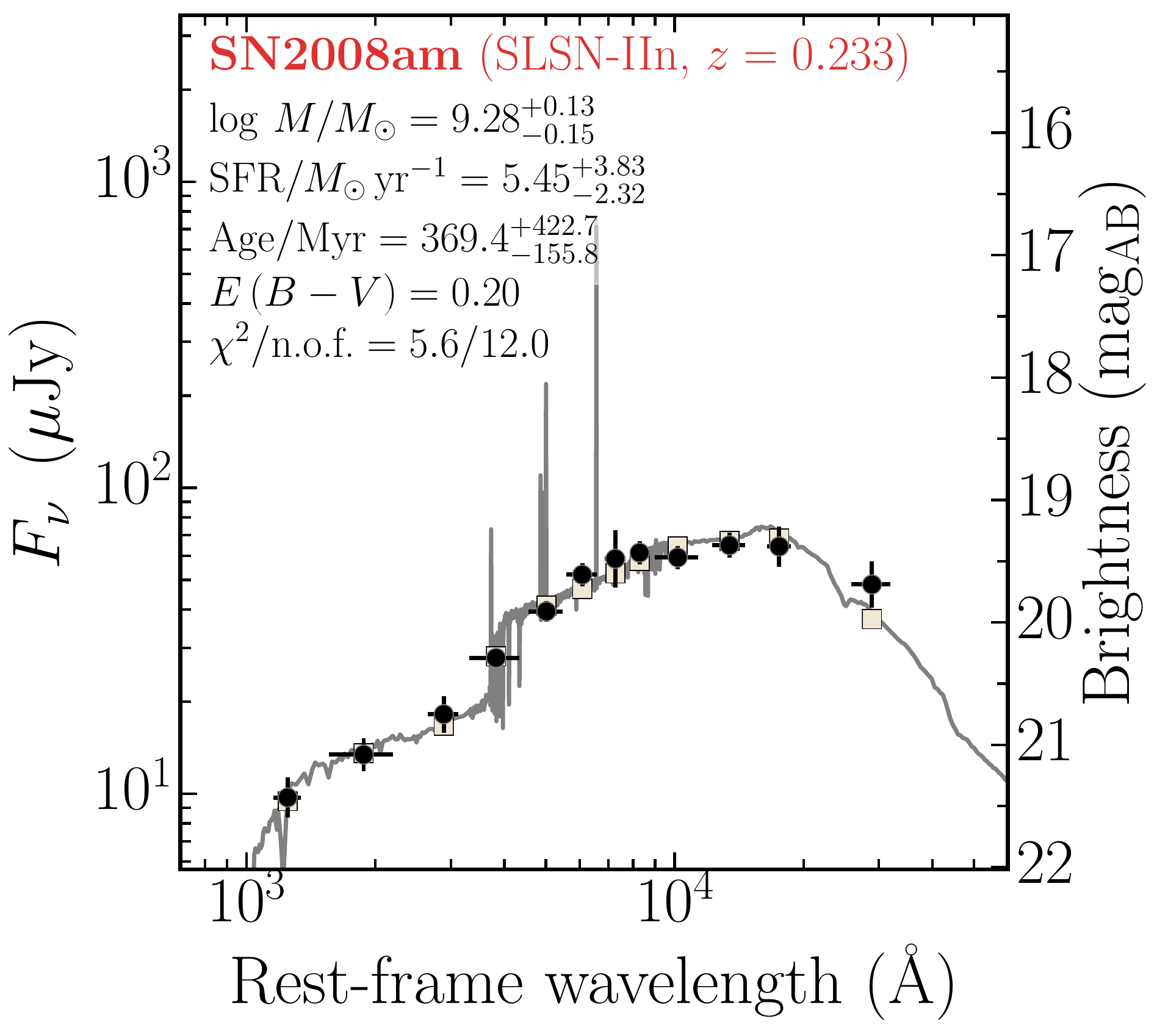}
\includegraphics[width=0.27\textwidth]{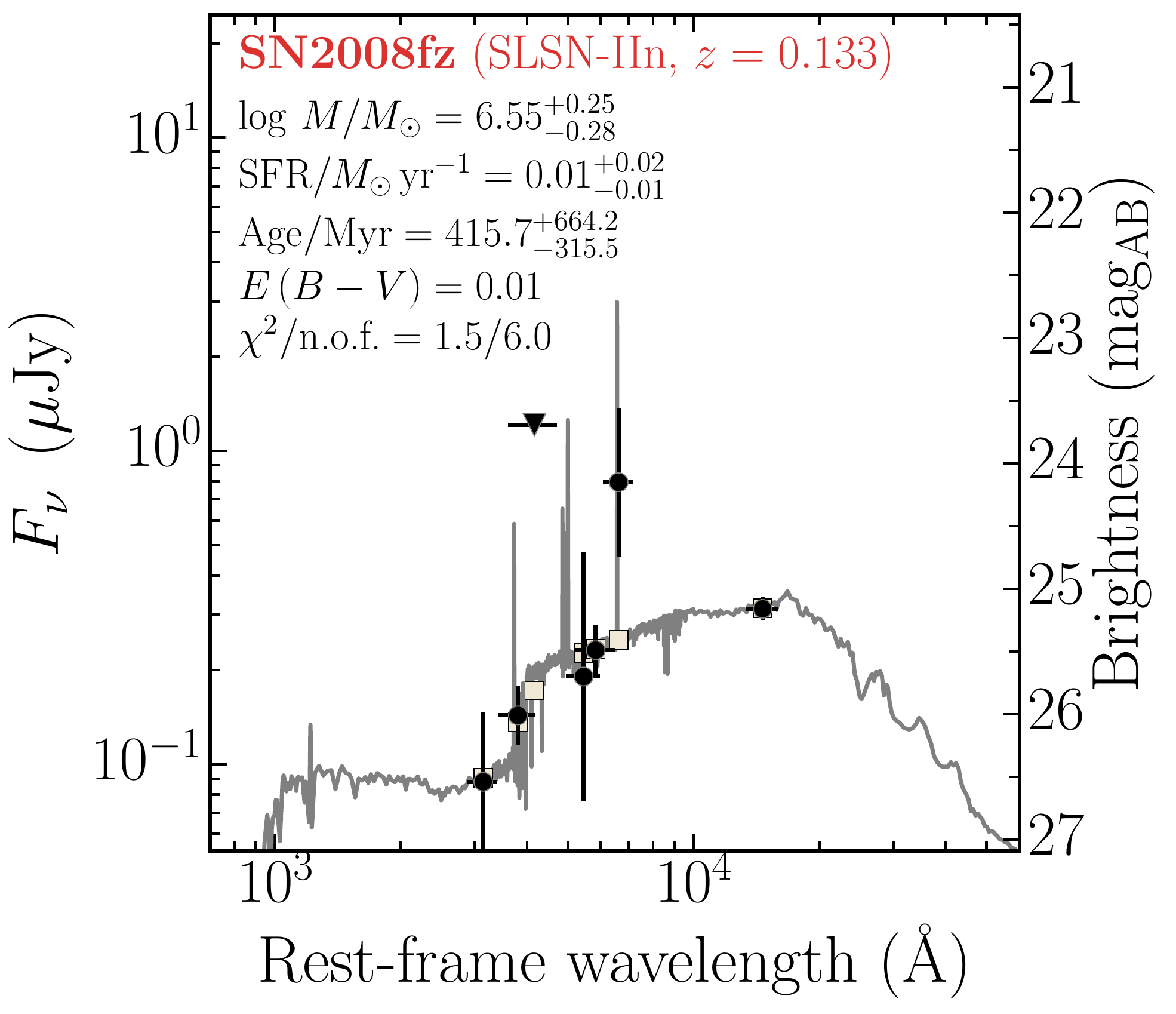}
\includegraphics[width=0.27\textwidth]{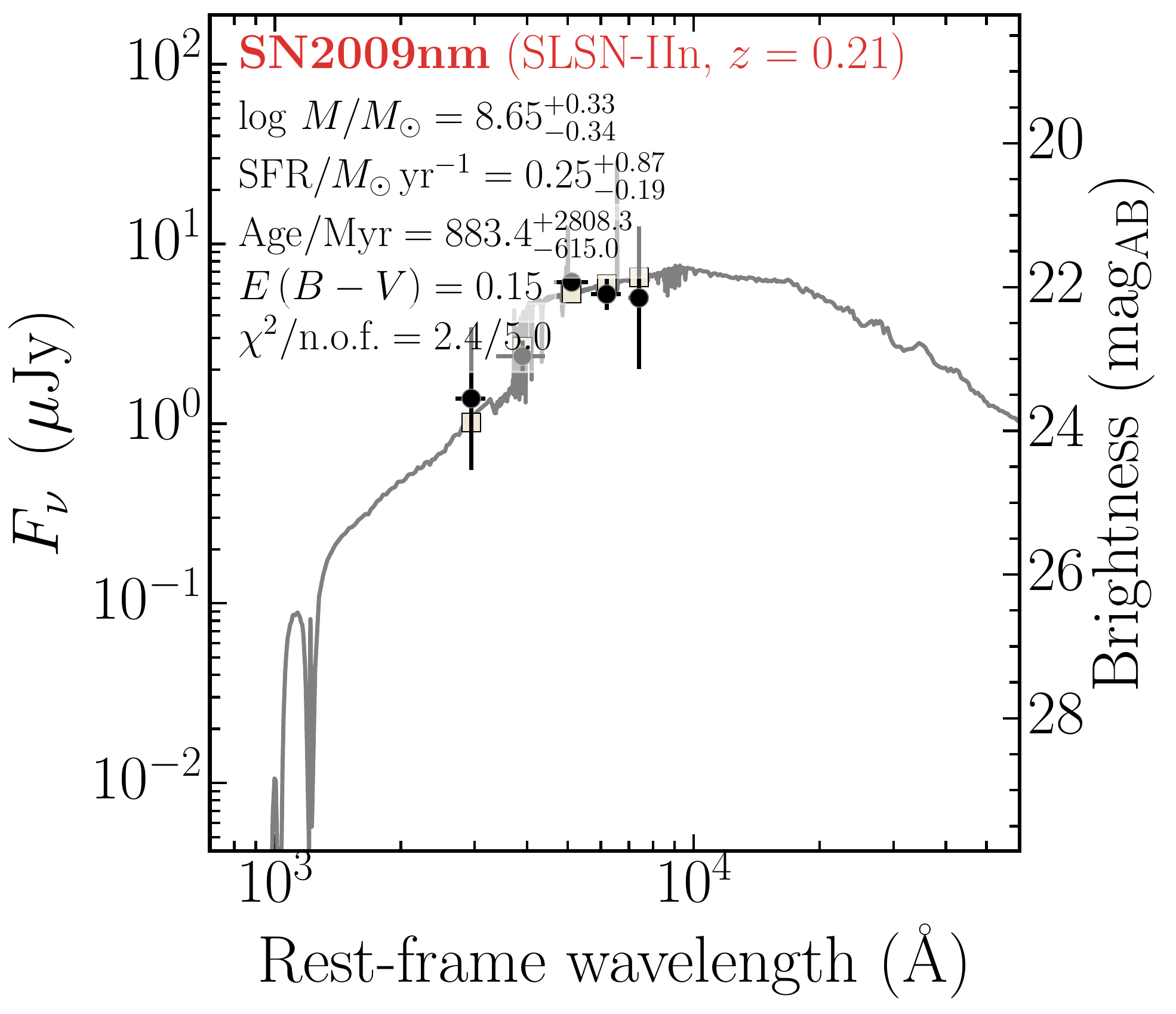}
\includegraphics[width=0.27\textwidth]{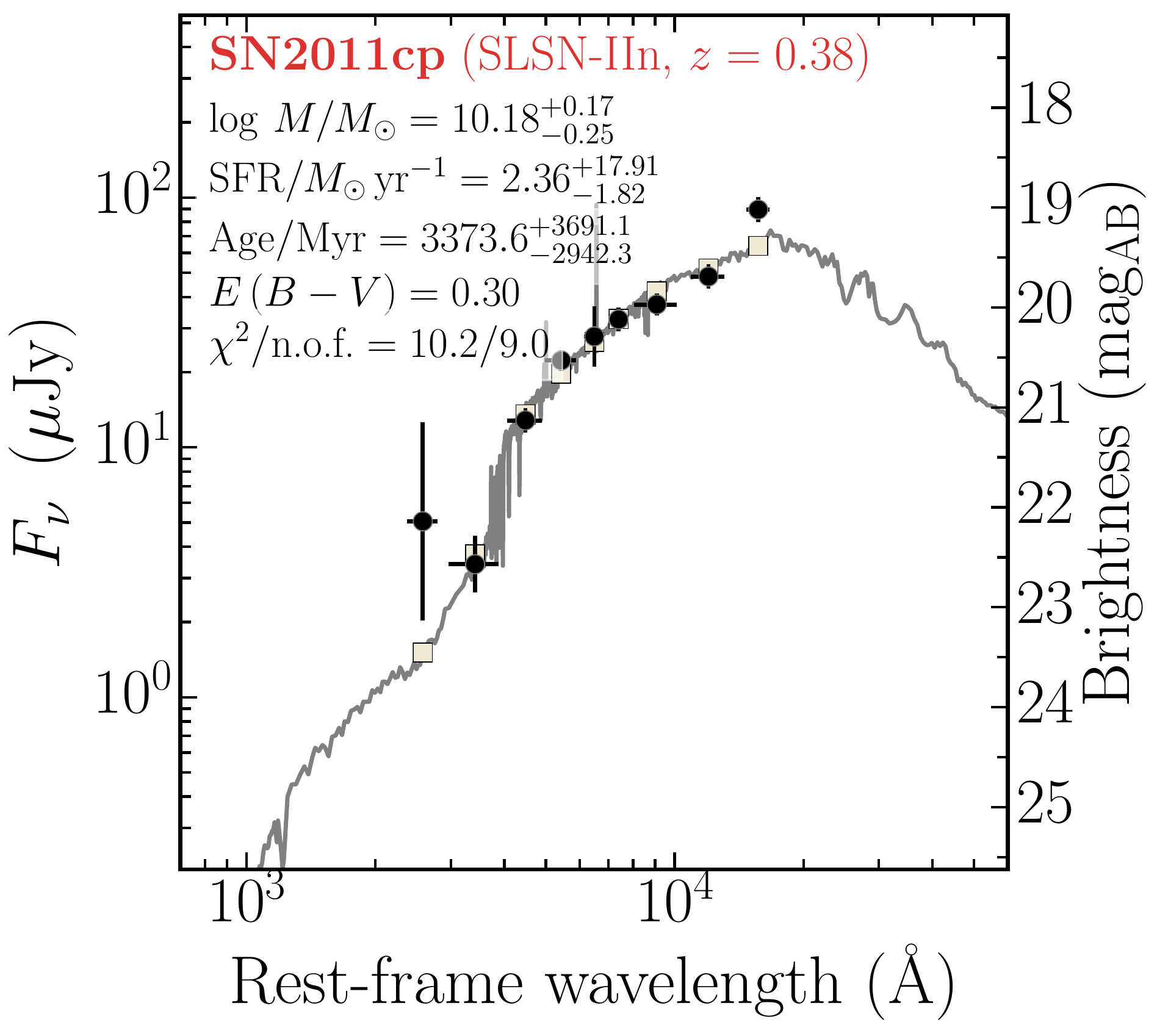}
\includegraphics[width=0.27\textwidth]{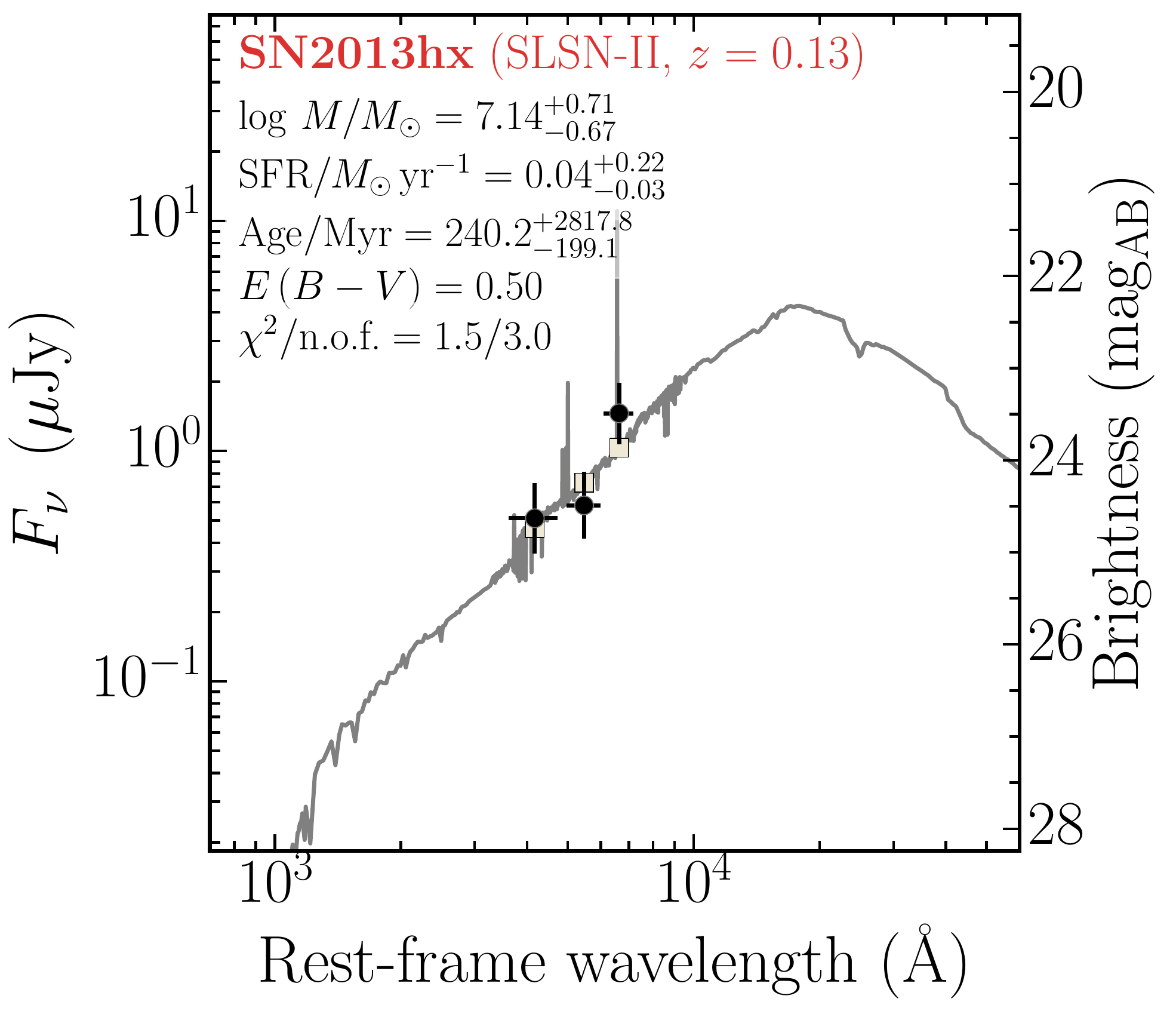}
\caption{
Similar to Figs. \ref{fig:SED} and \ref{fig:SED_app_1} but for H-rich host galaxies.
}
\label{fig:SED_app_2}
\end{figure*}

\clearpage

\section{Postage stamps}

\begin{figure*}
\includegraphics[width=0.19\textwidth]{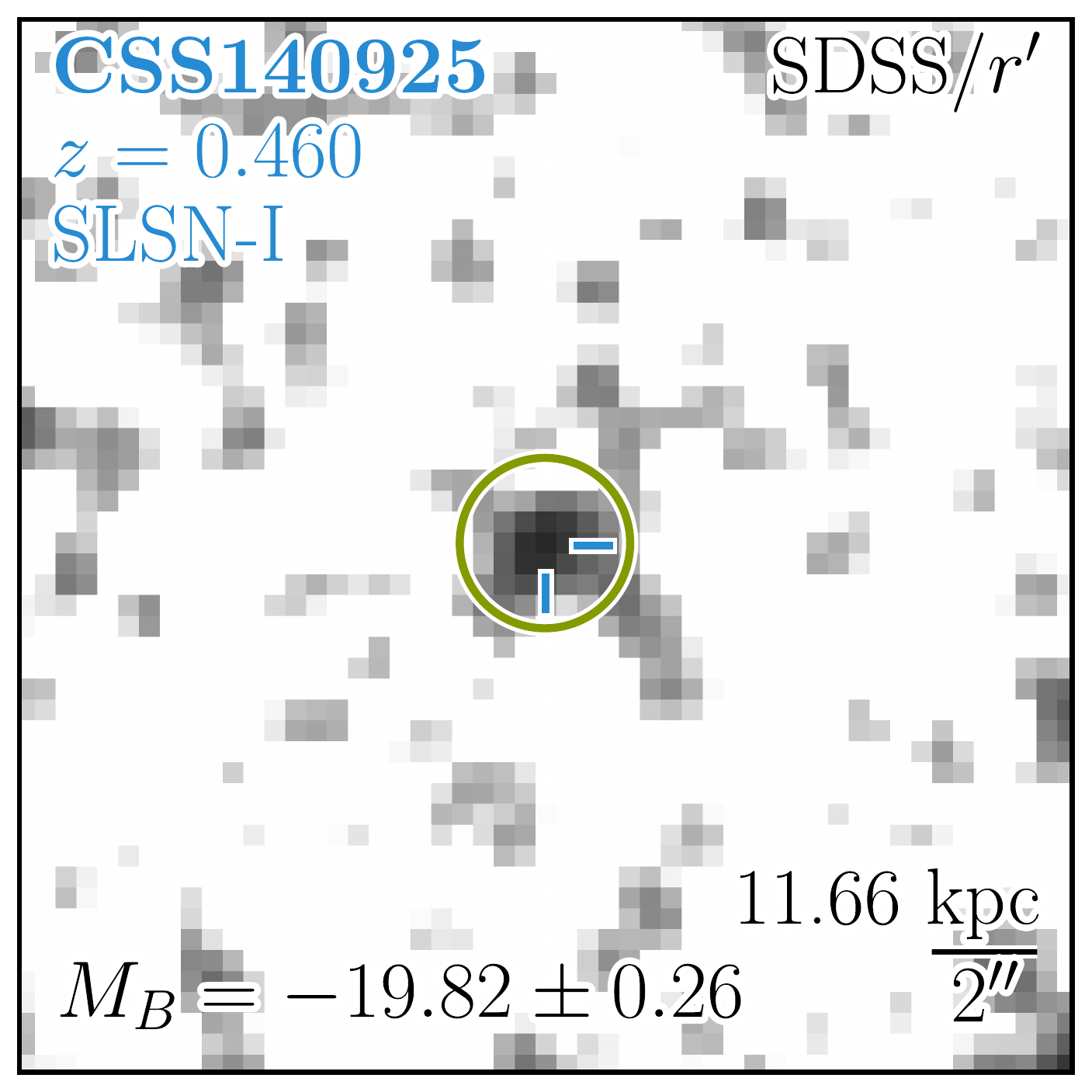}
\includegraphics[width=0.19\textwidth]{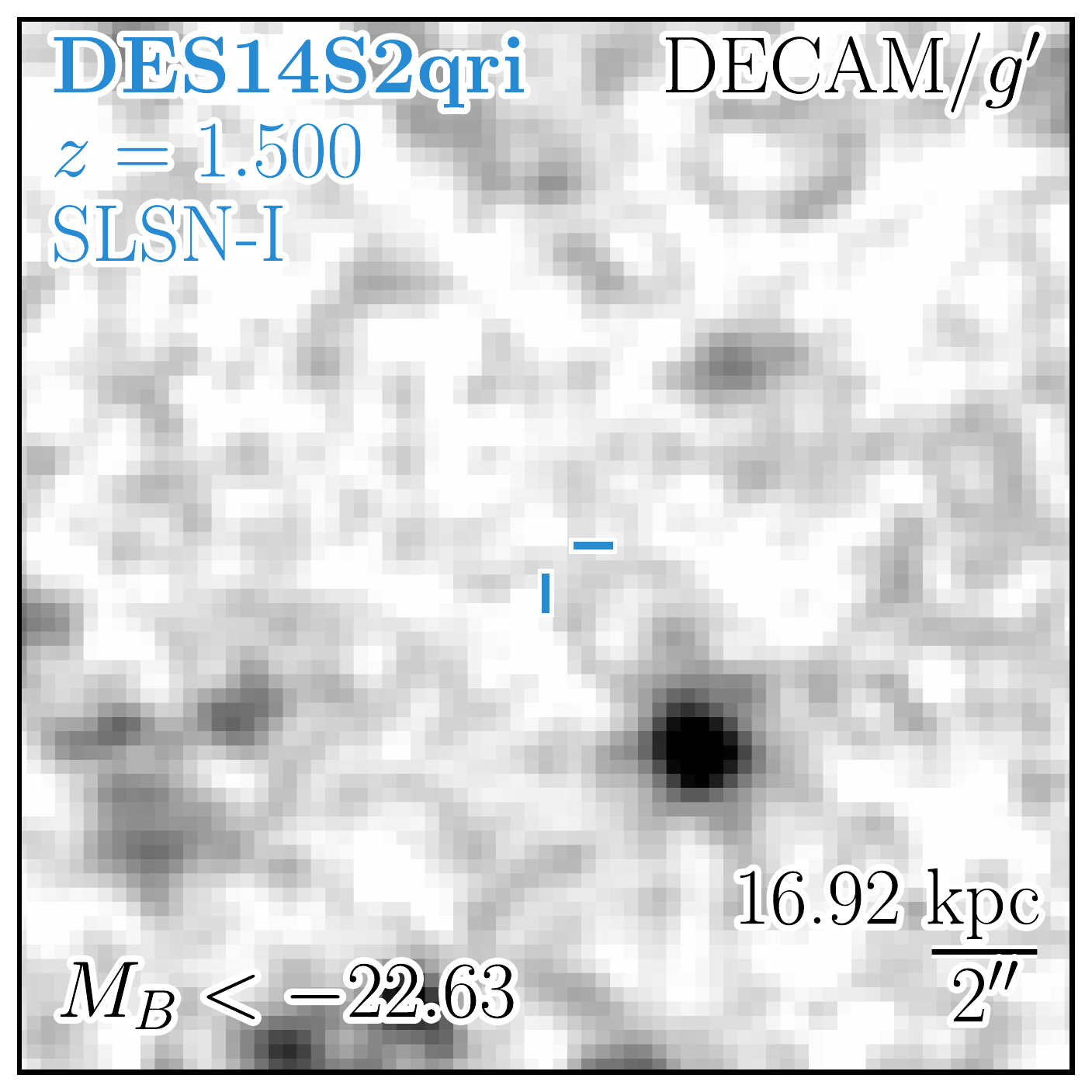}
\includegraphics[width=0.19\textwidth]{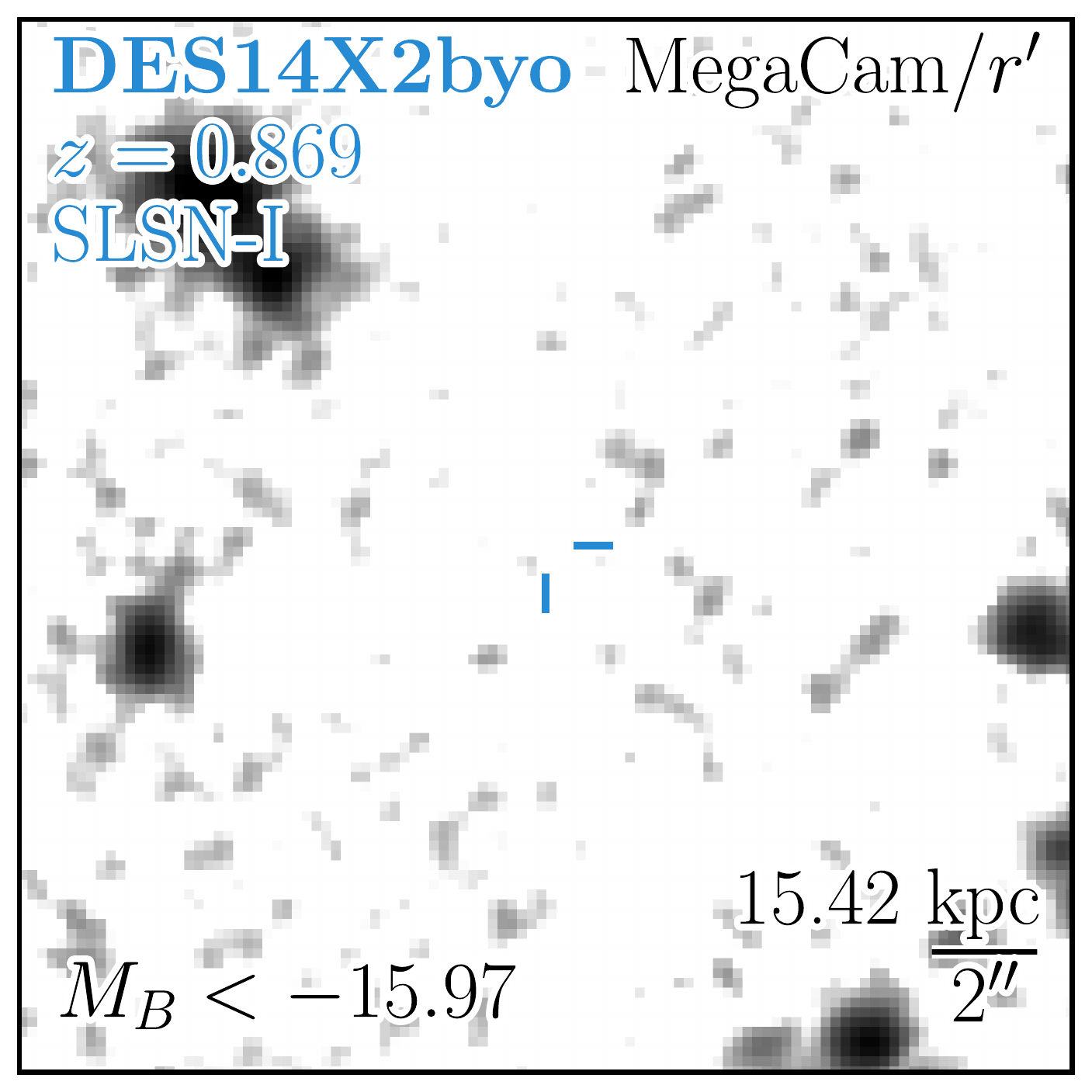}
\includegraphics[width=0.19\textwidth]{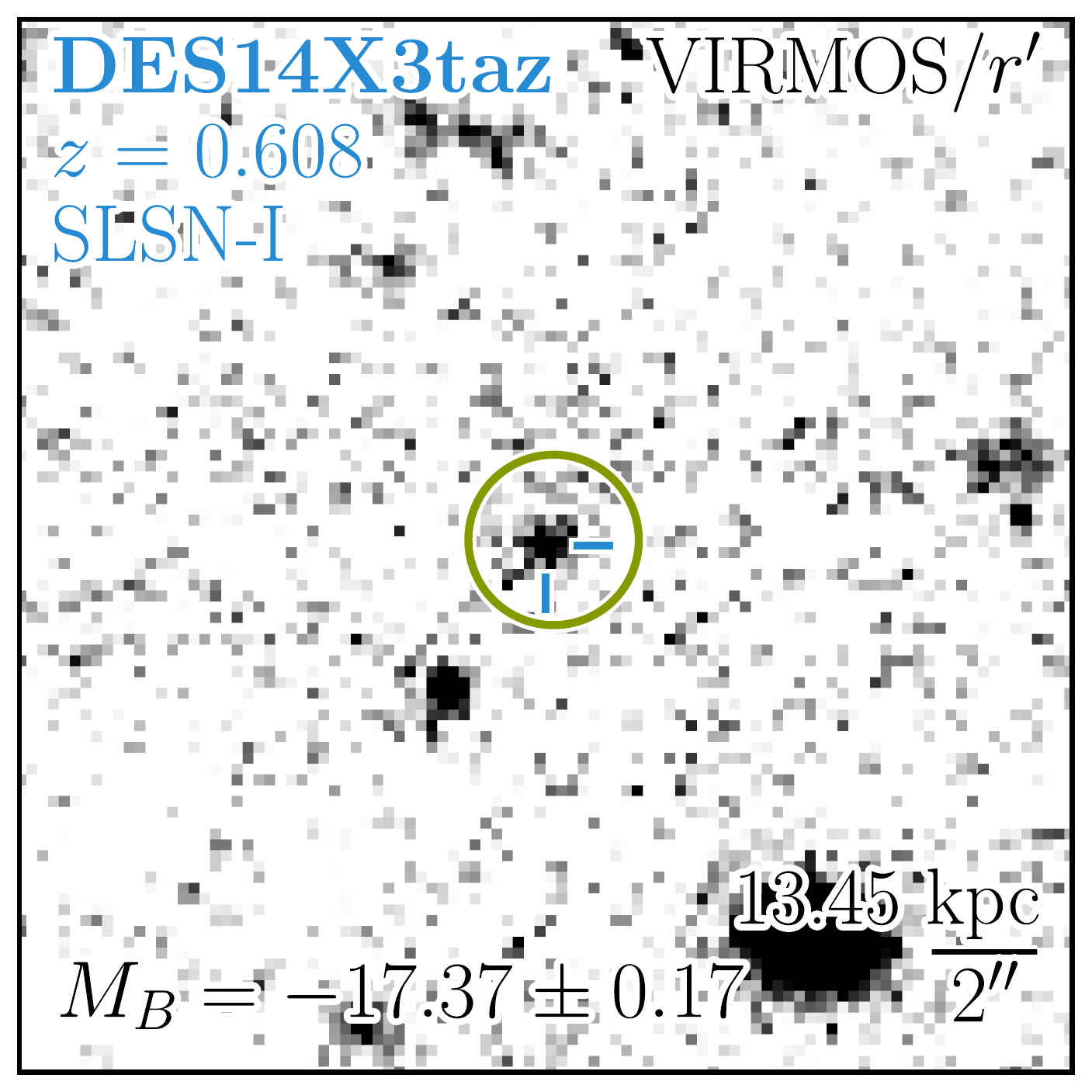}
\includegraphics[width=0.19\textwidth]{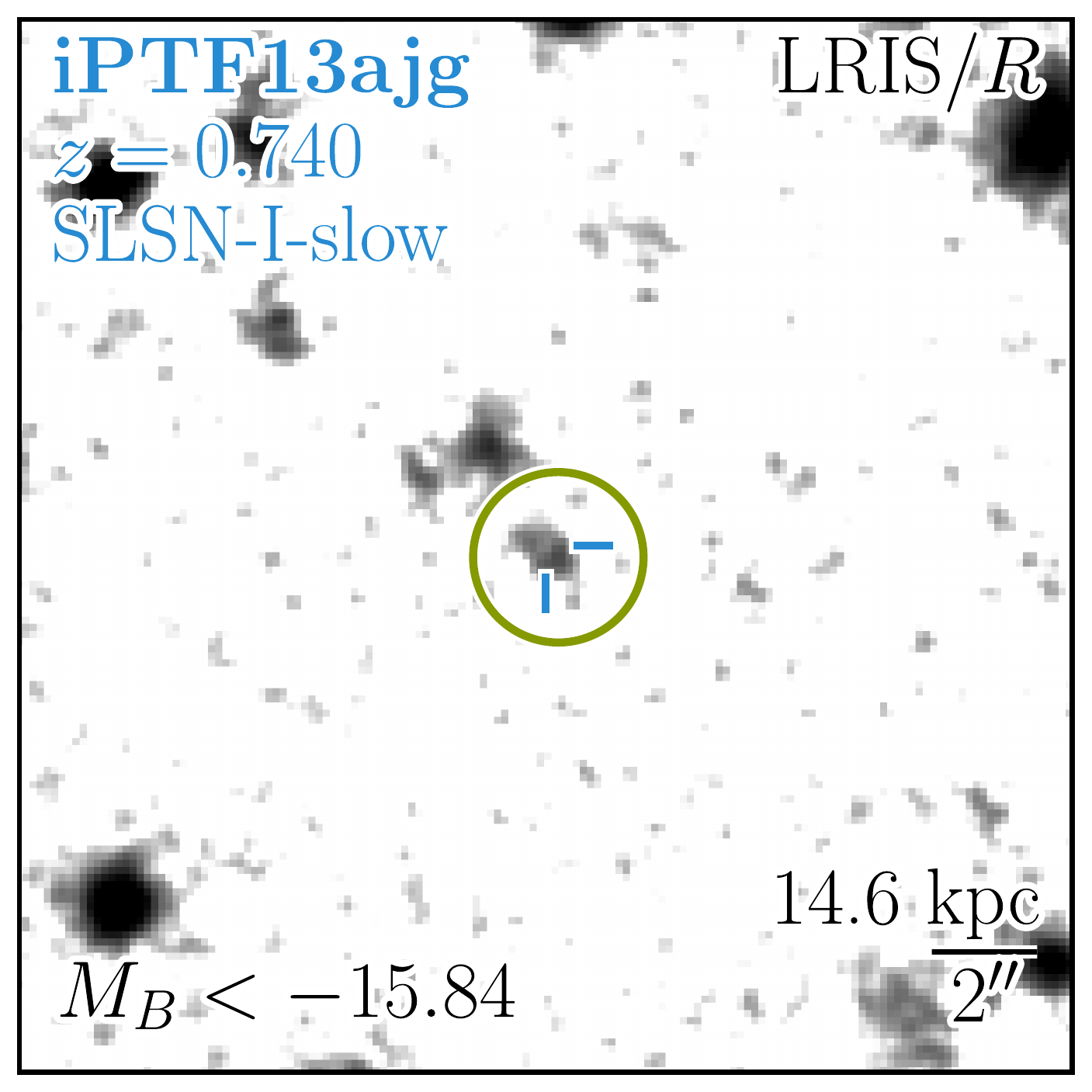}
\includegraphics[width=0.19\textwidth]{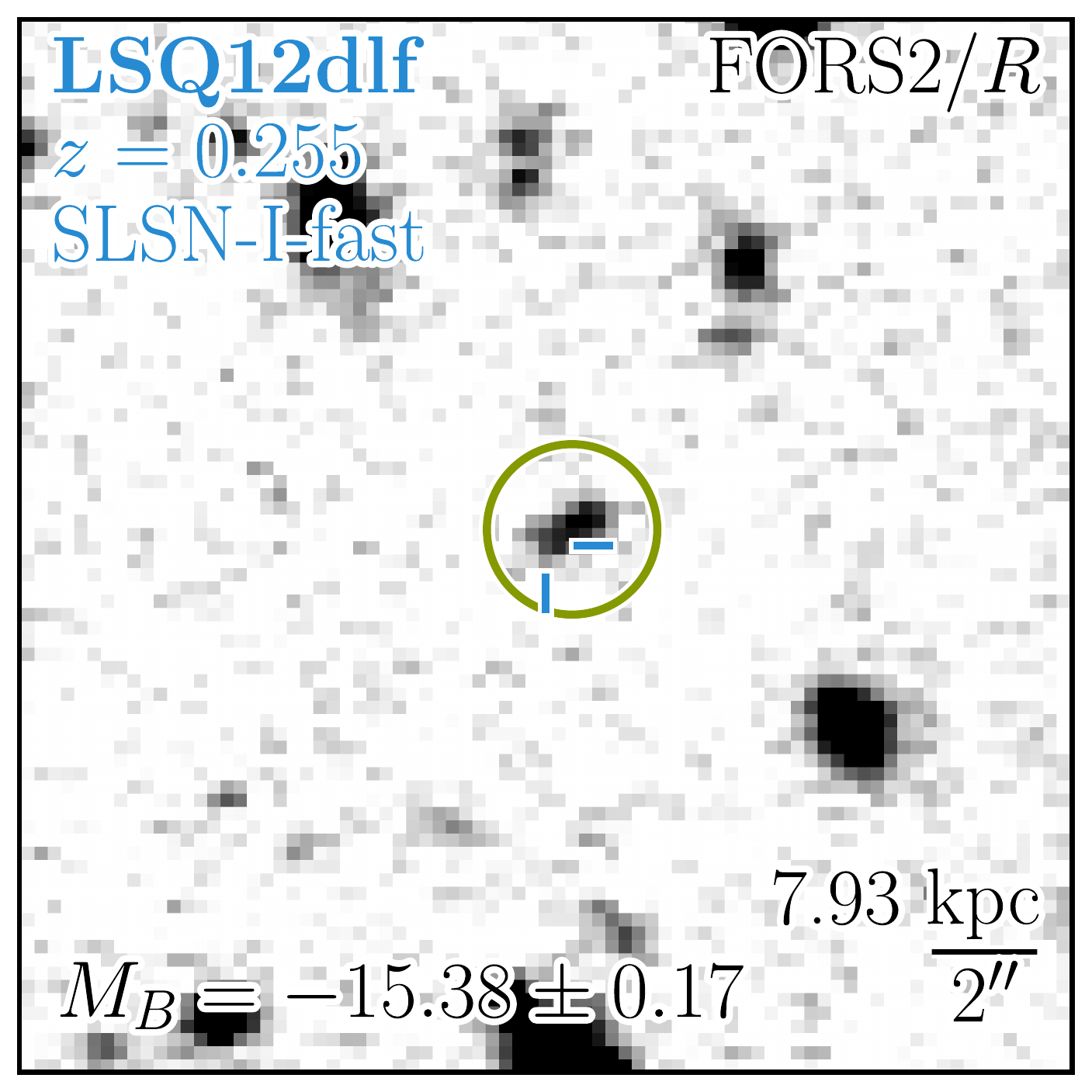}
\includegraphics[width=0.19\textwidth]{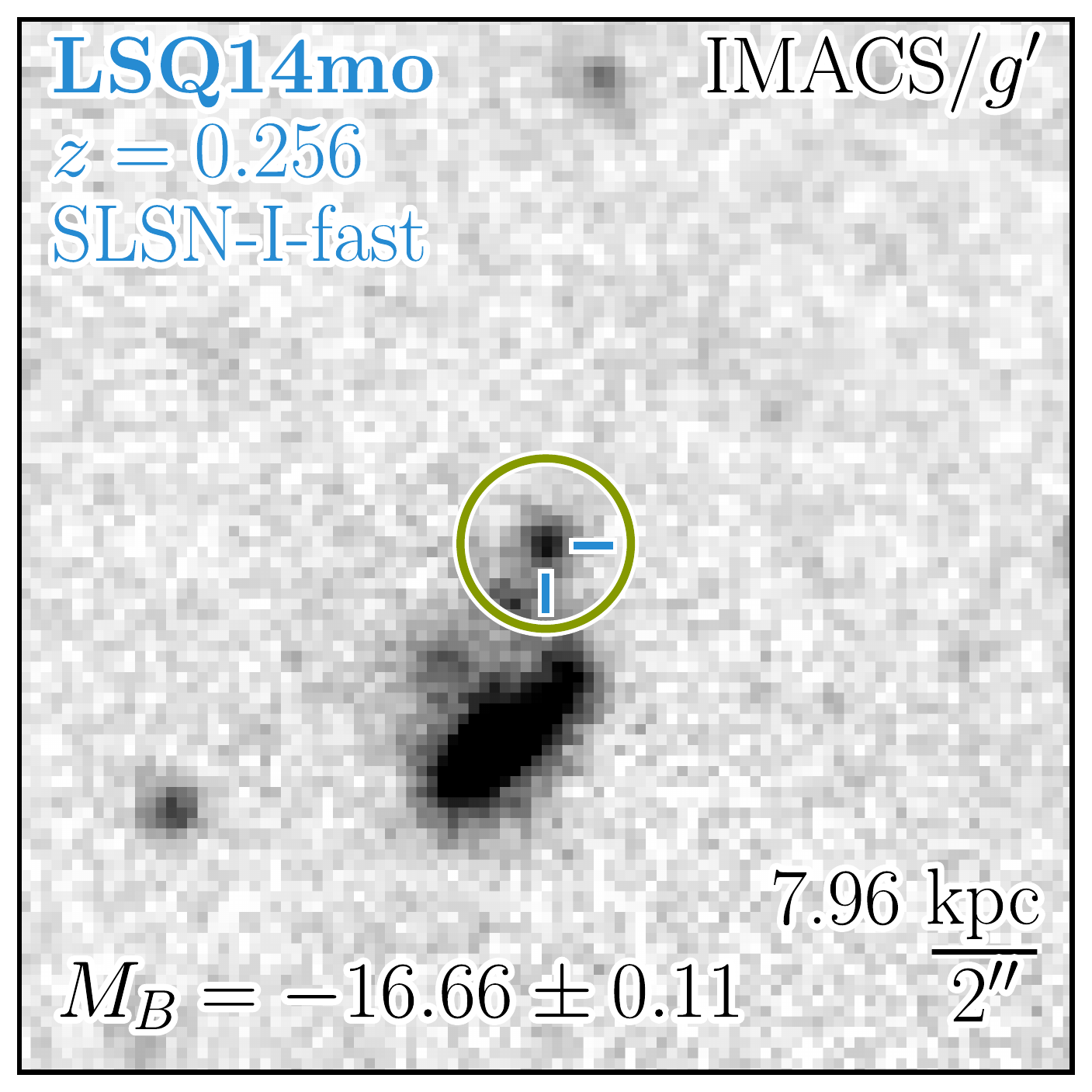}
\includegraphics[width=0.19\textwidth]{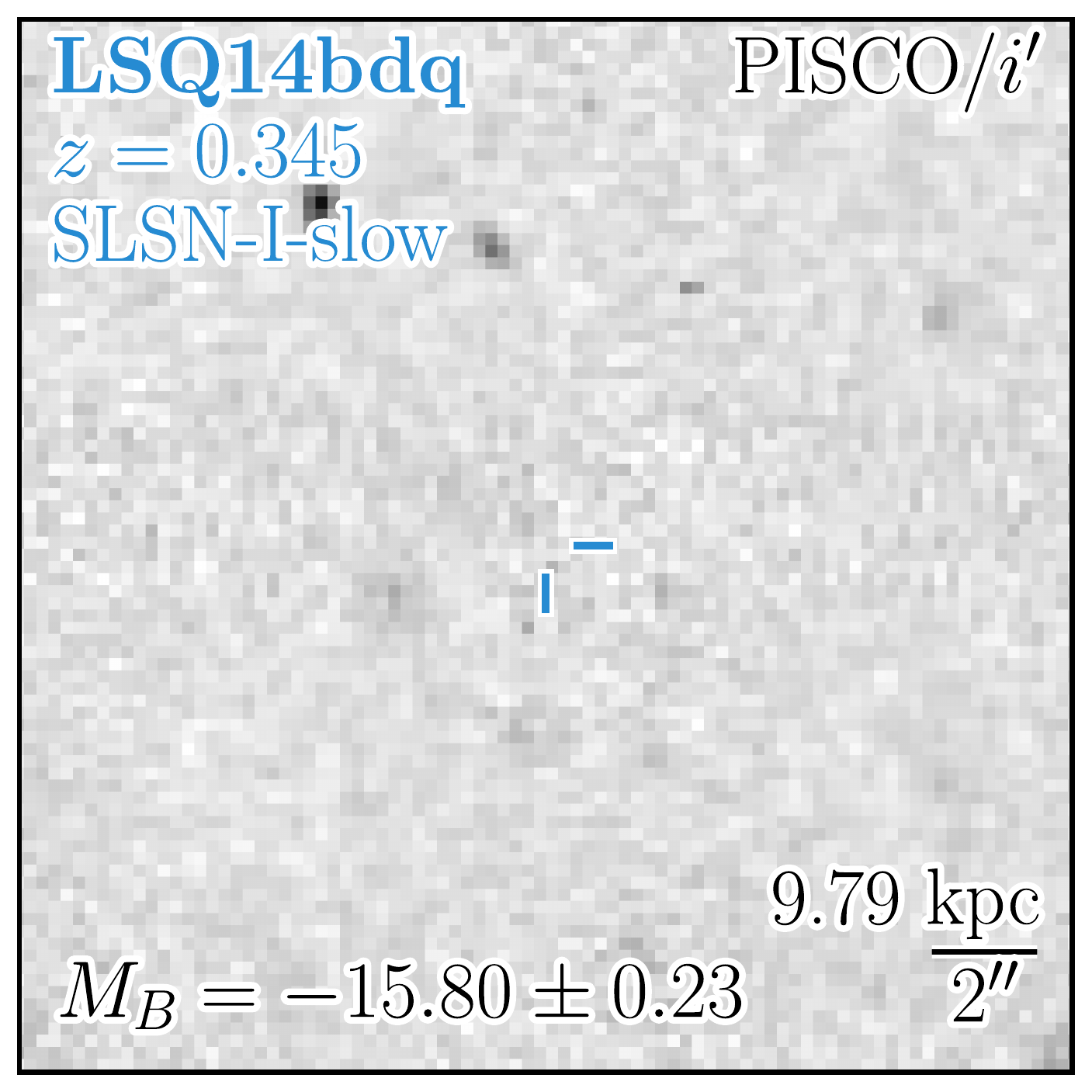}
\includegraphics[width=0.19\textwidth]{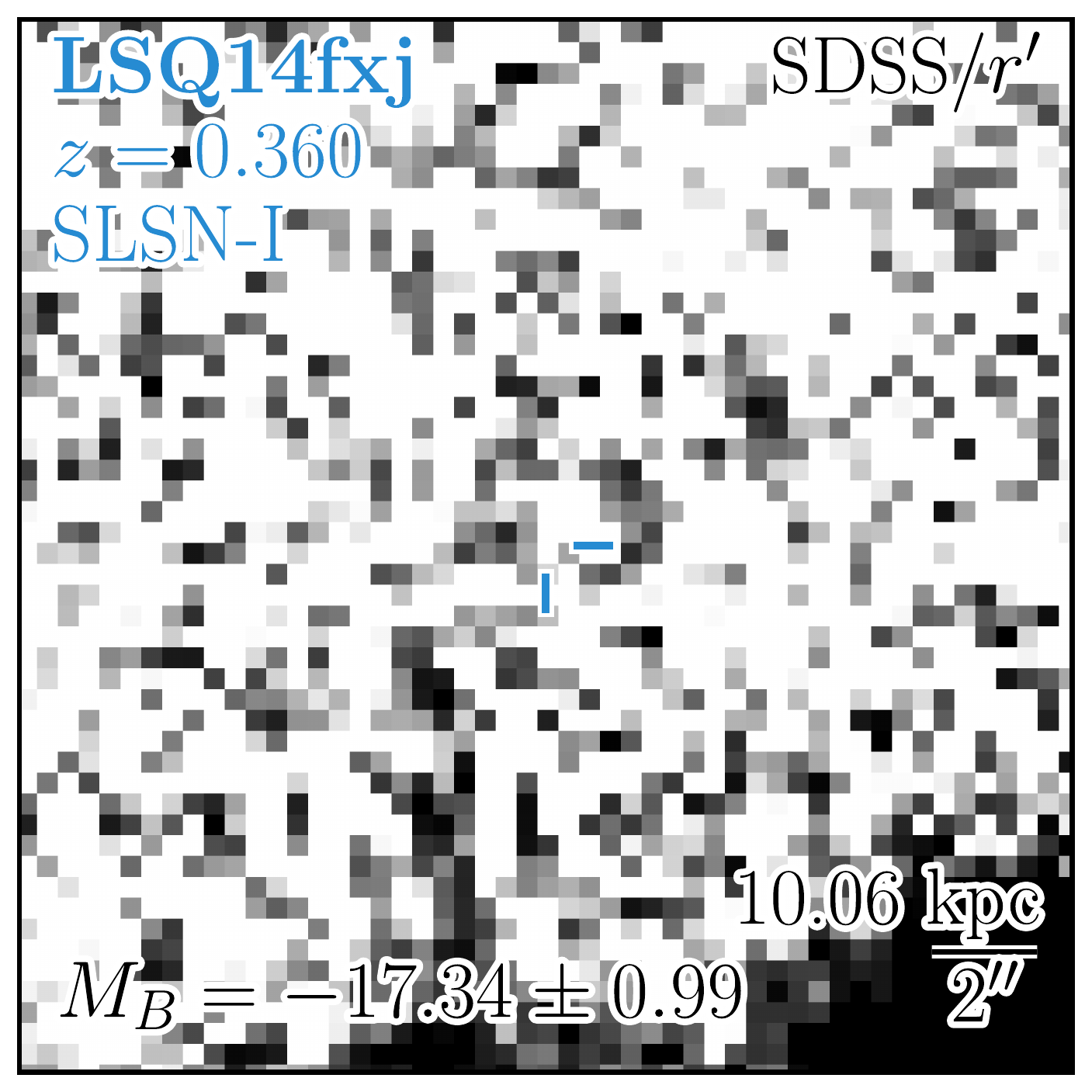}
\includegraphics[width=0.19\textwidth]{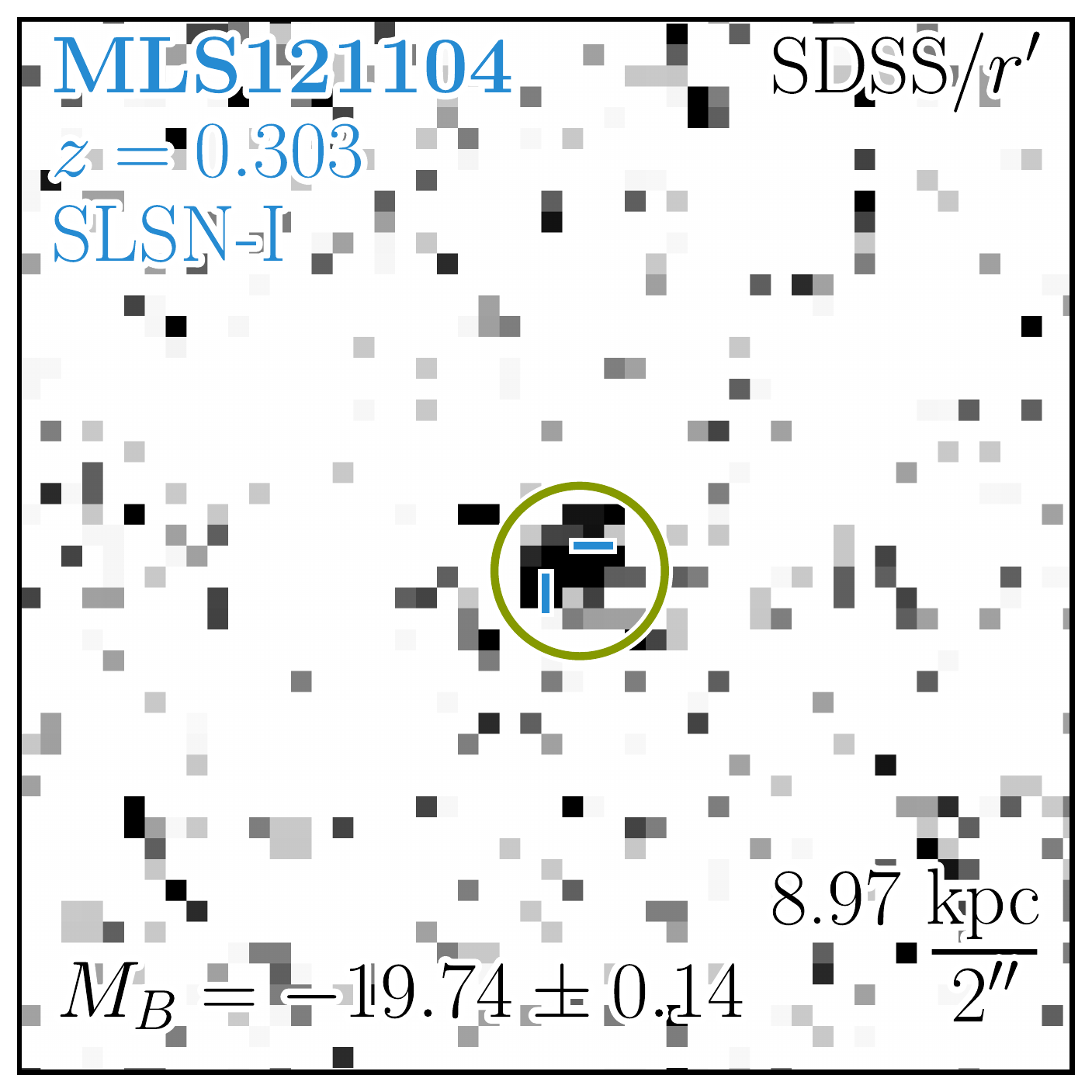}
\includegraphics[width=0.19\textwidth]{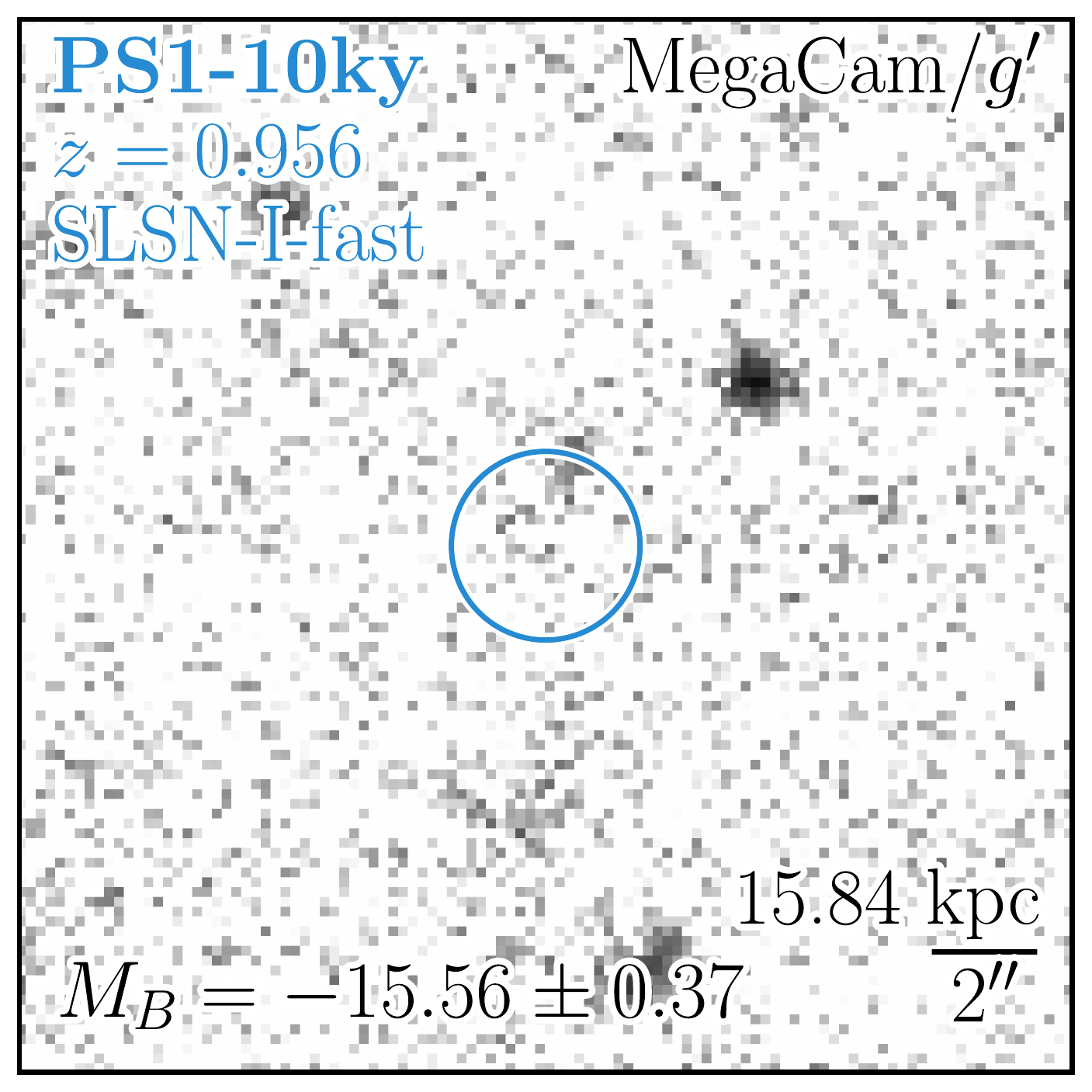}
\includegraphics[width=0.19\textwidth]{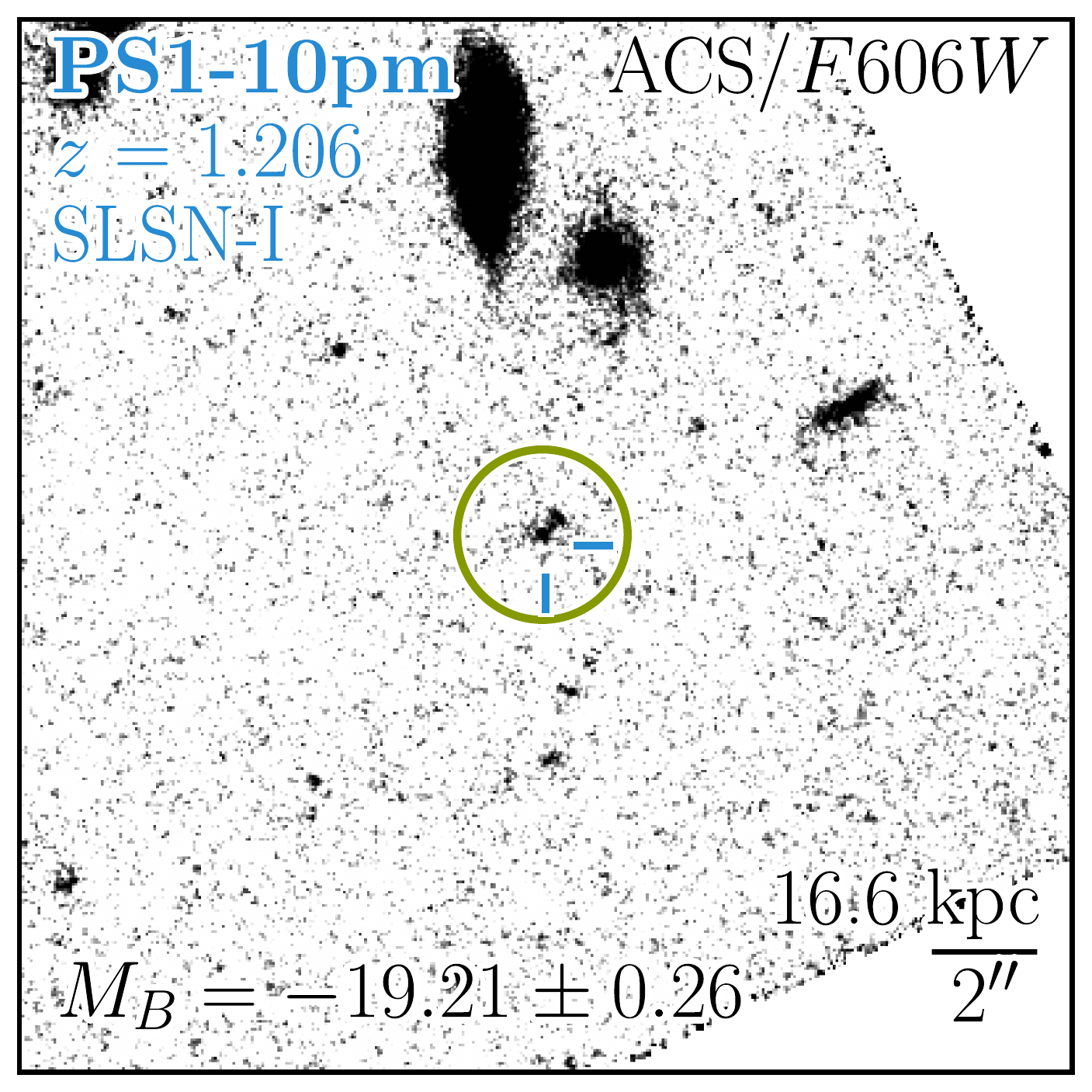}
\includegraphics[width=0.19\textwidth]{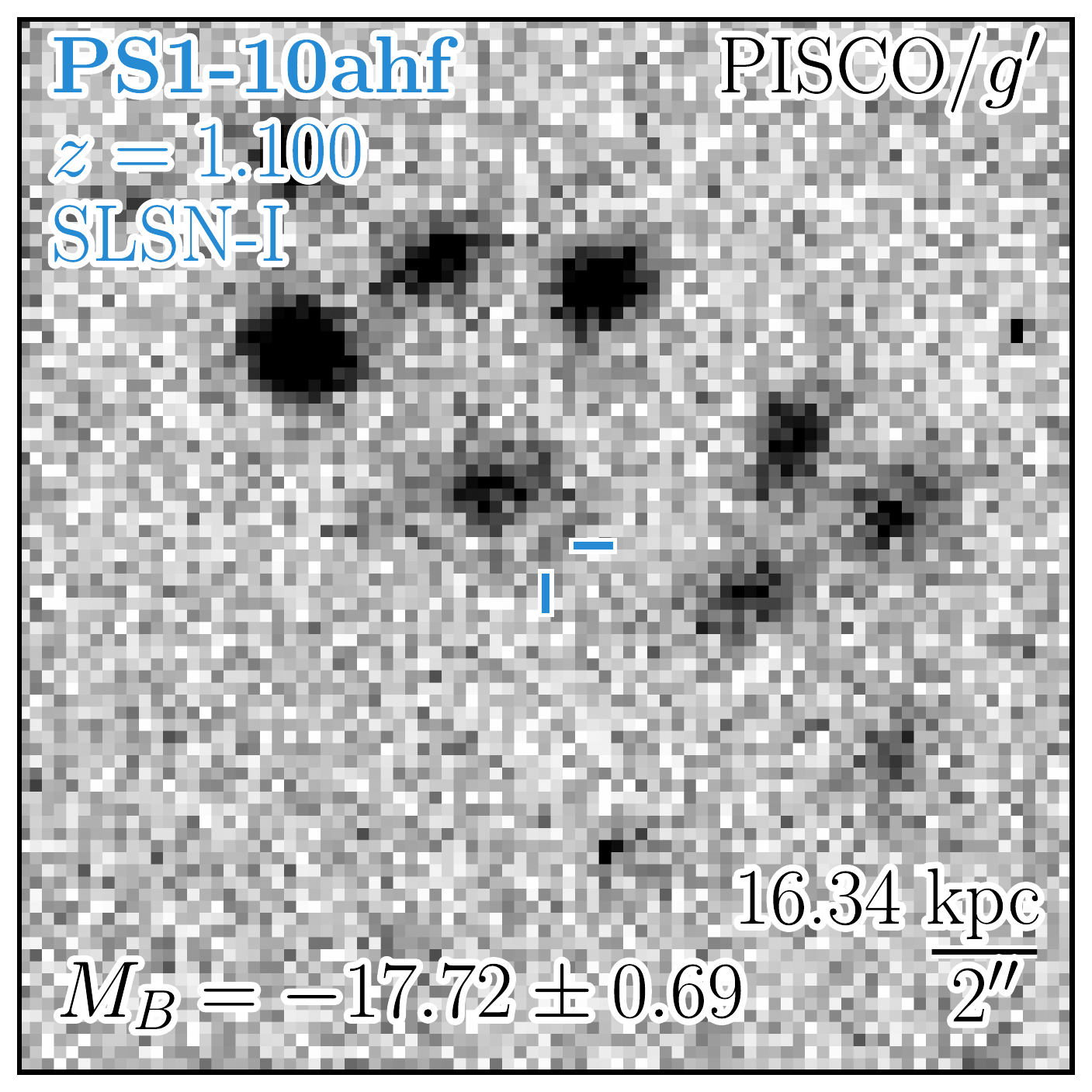}
\includegraphics[width=0.19\textwidth]{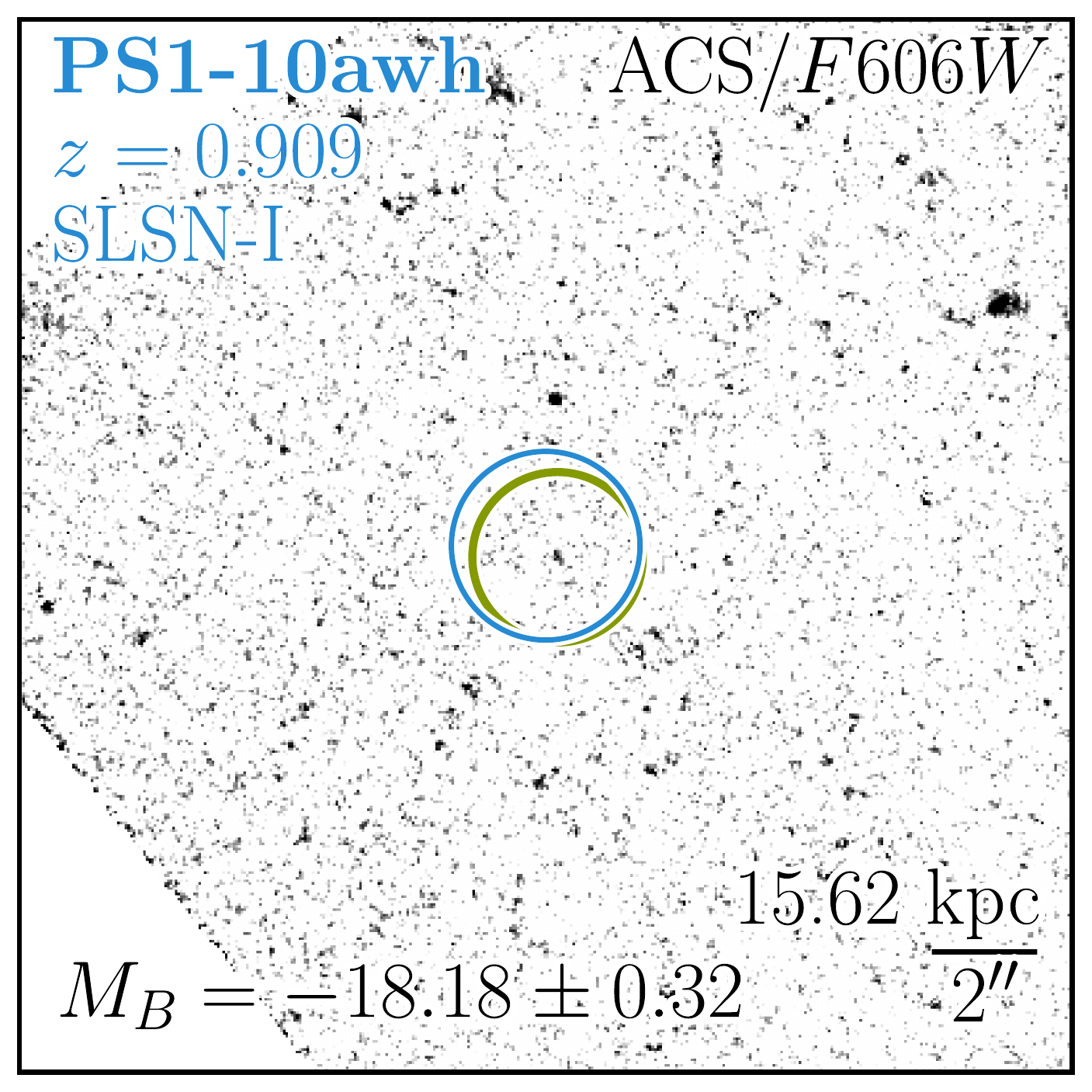}
\includegraphics[width=0.19\textwidth]{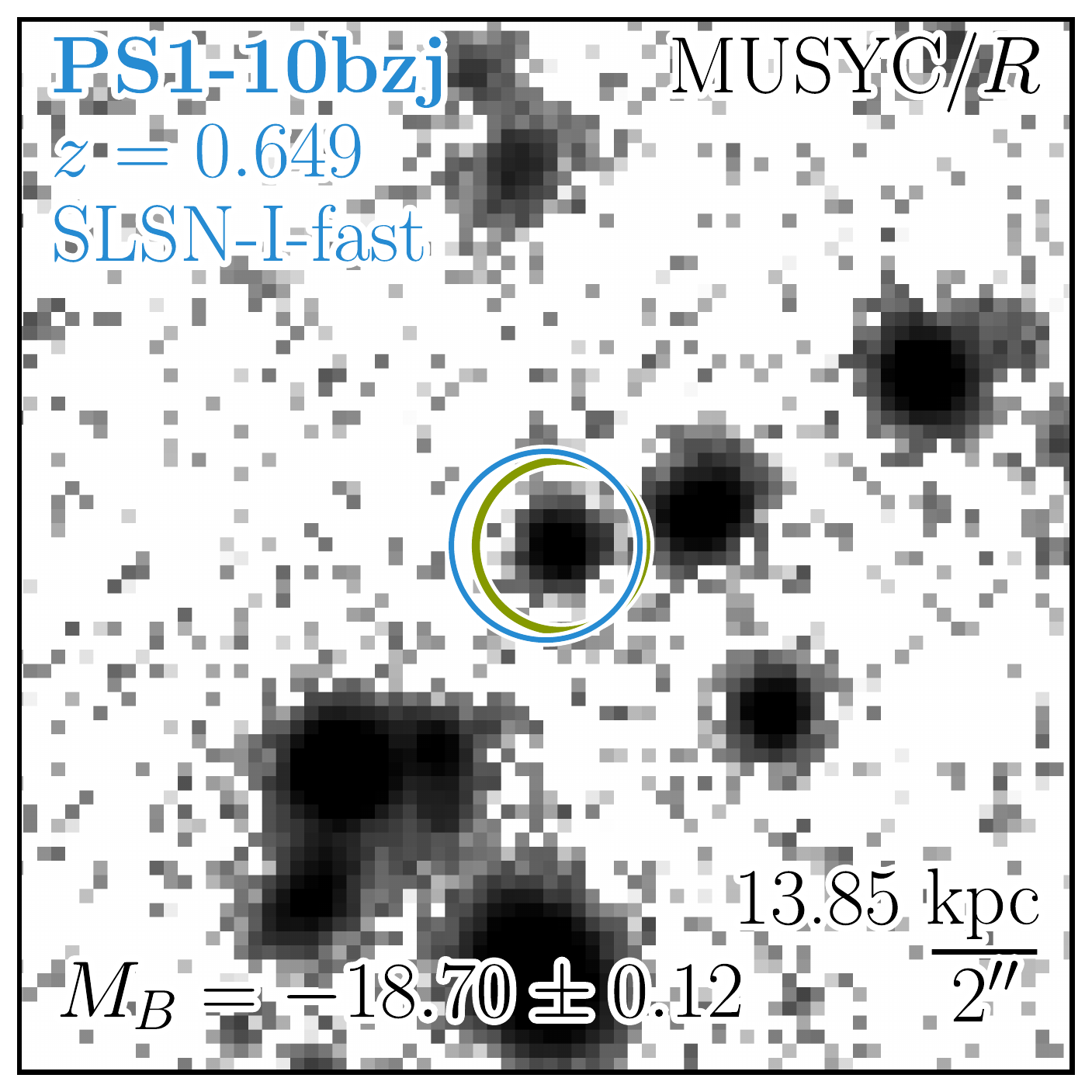}
\includegraphics[width=0.19\textwidth]{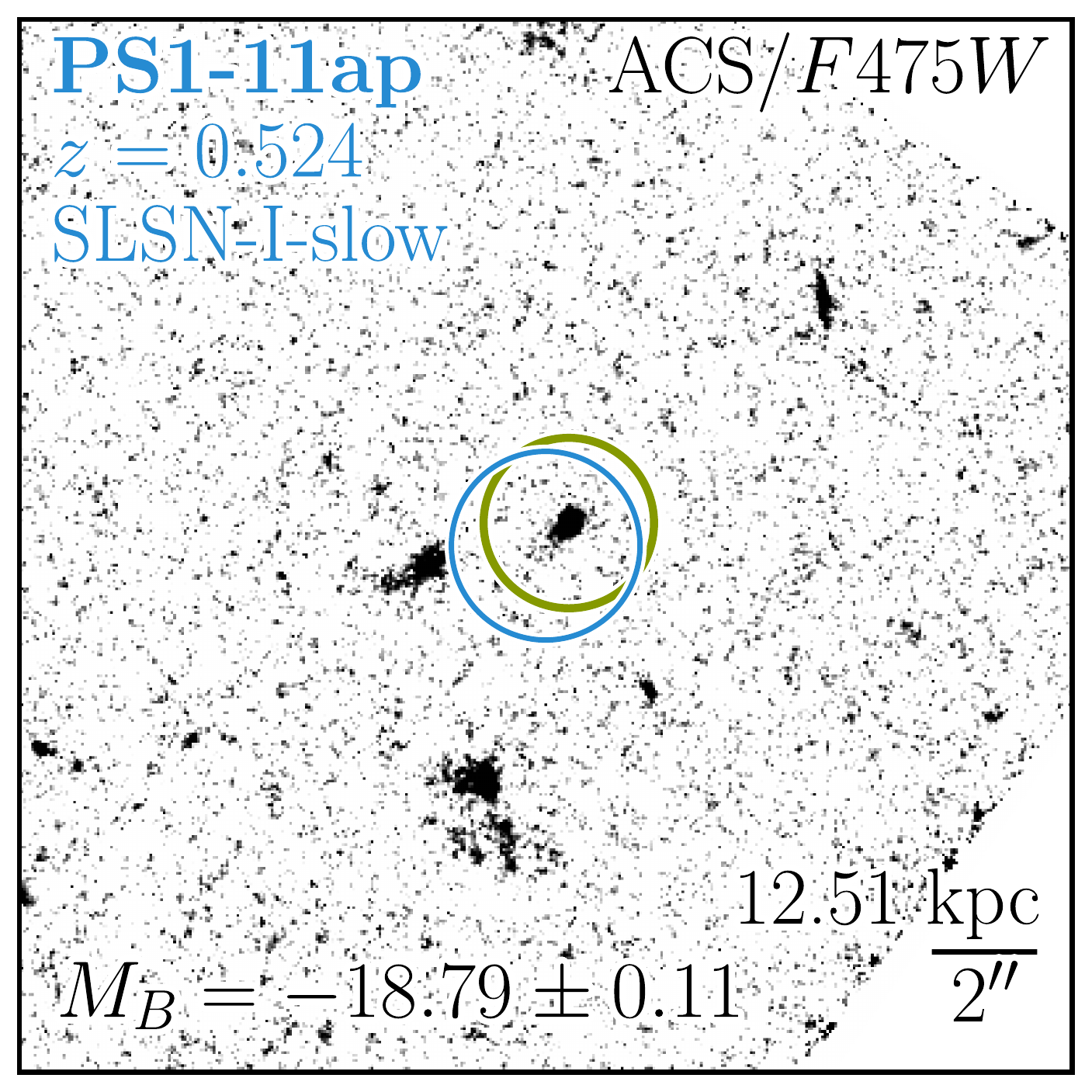}
\includegraphics[width=0.19\textwidth]{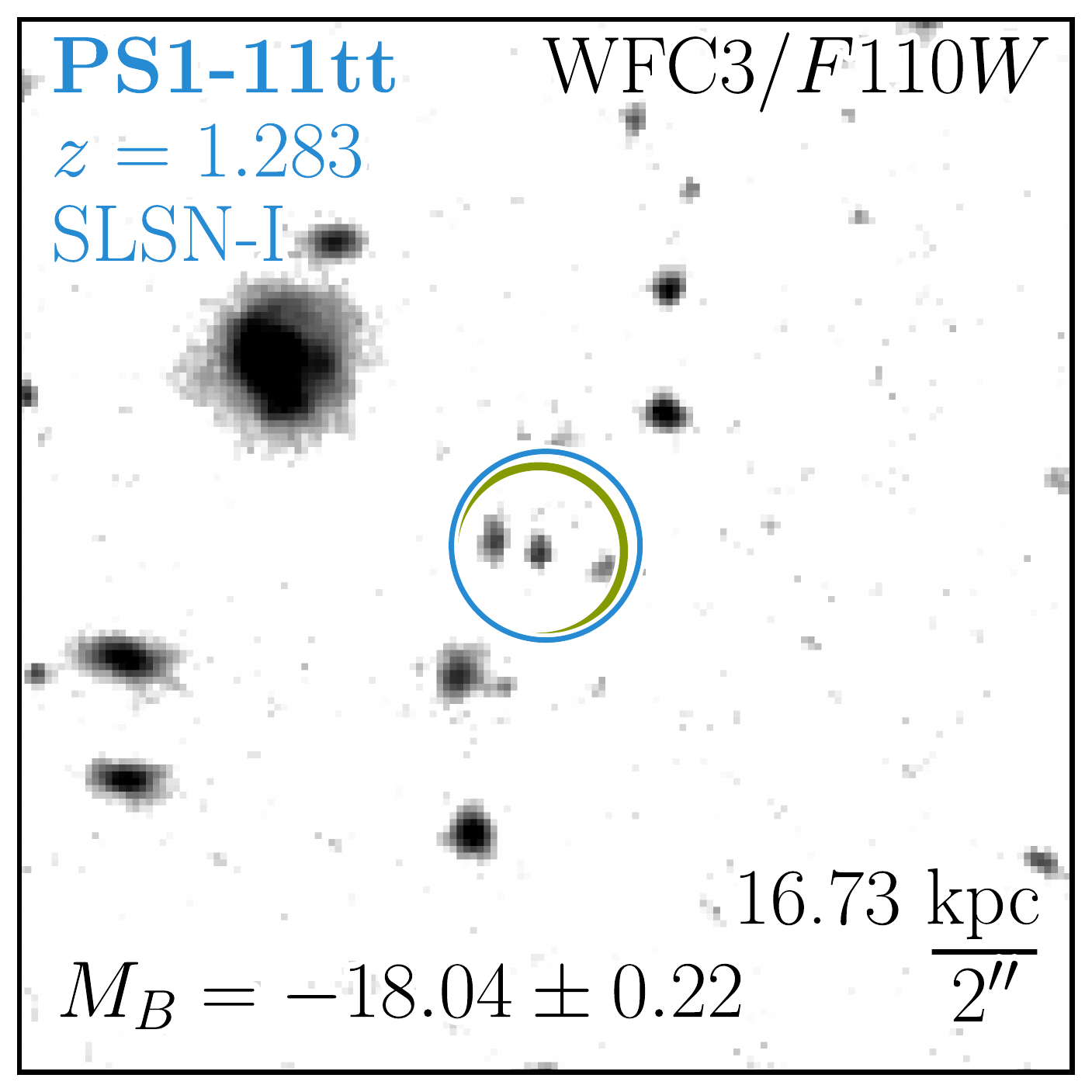}
\includegraphics[width=0.19\textwidth]{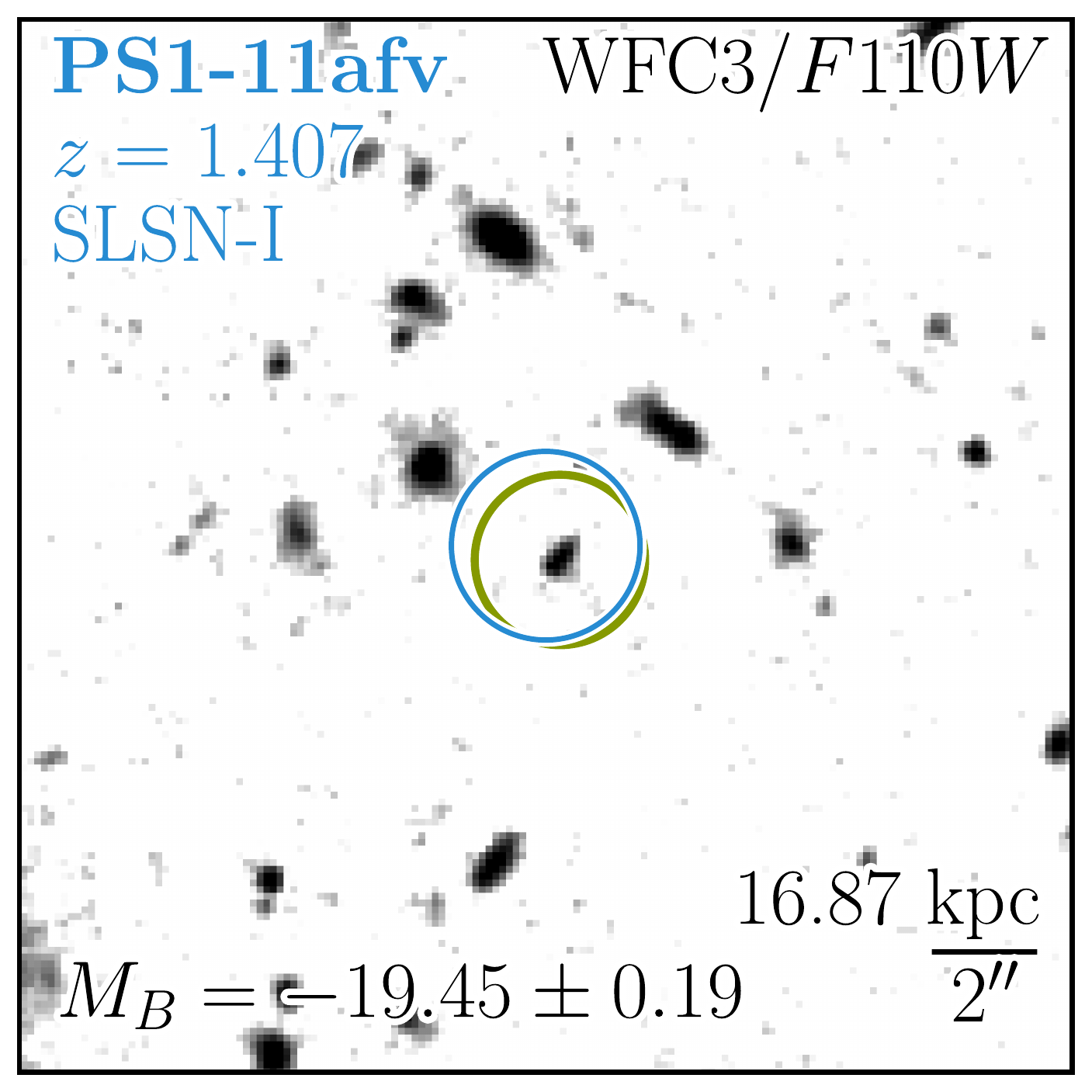}
\includegraphics[width=0.19\textwidth]{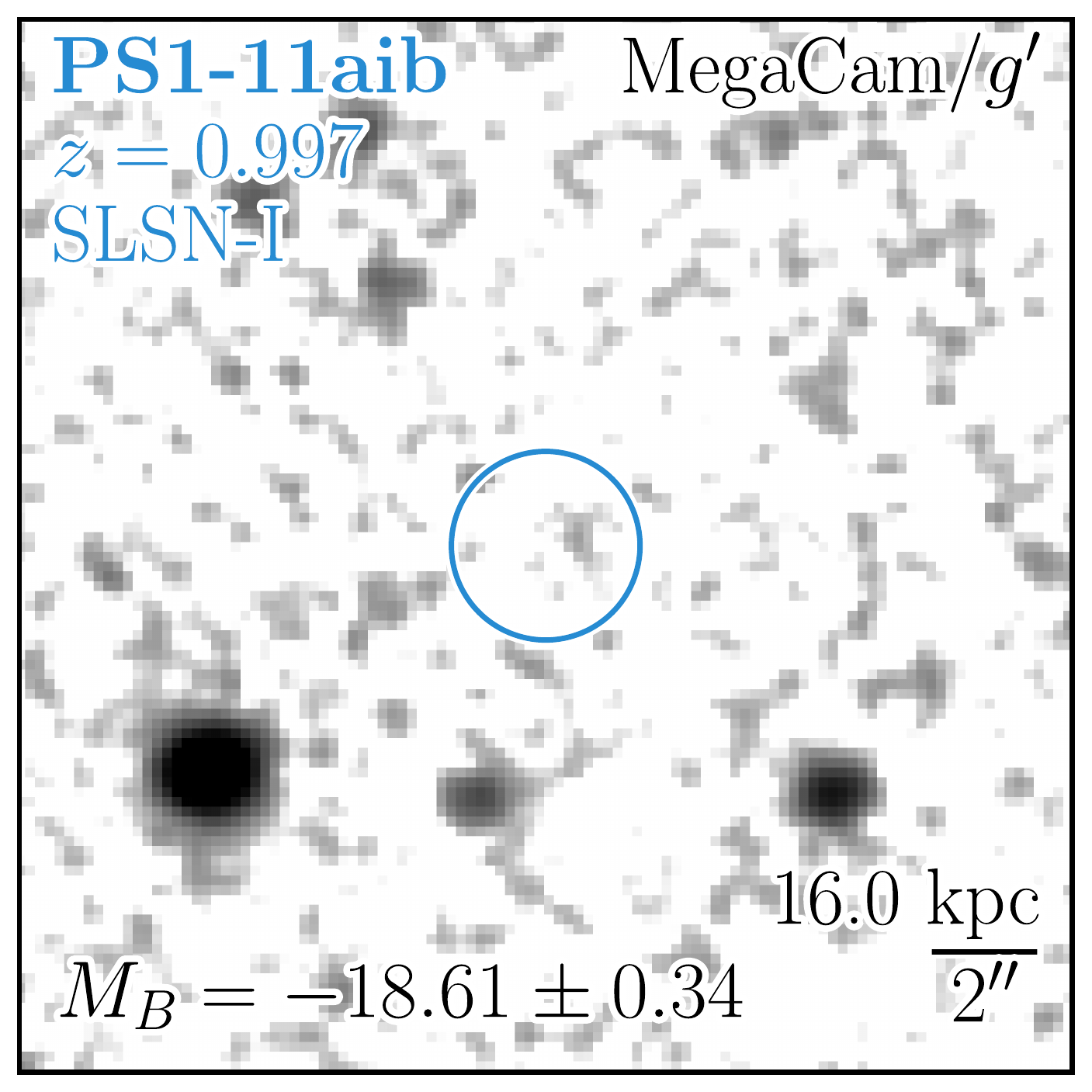}
\includegraphics[width=0.19\textwidth]{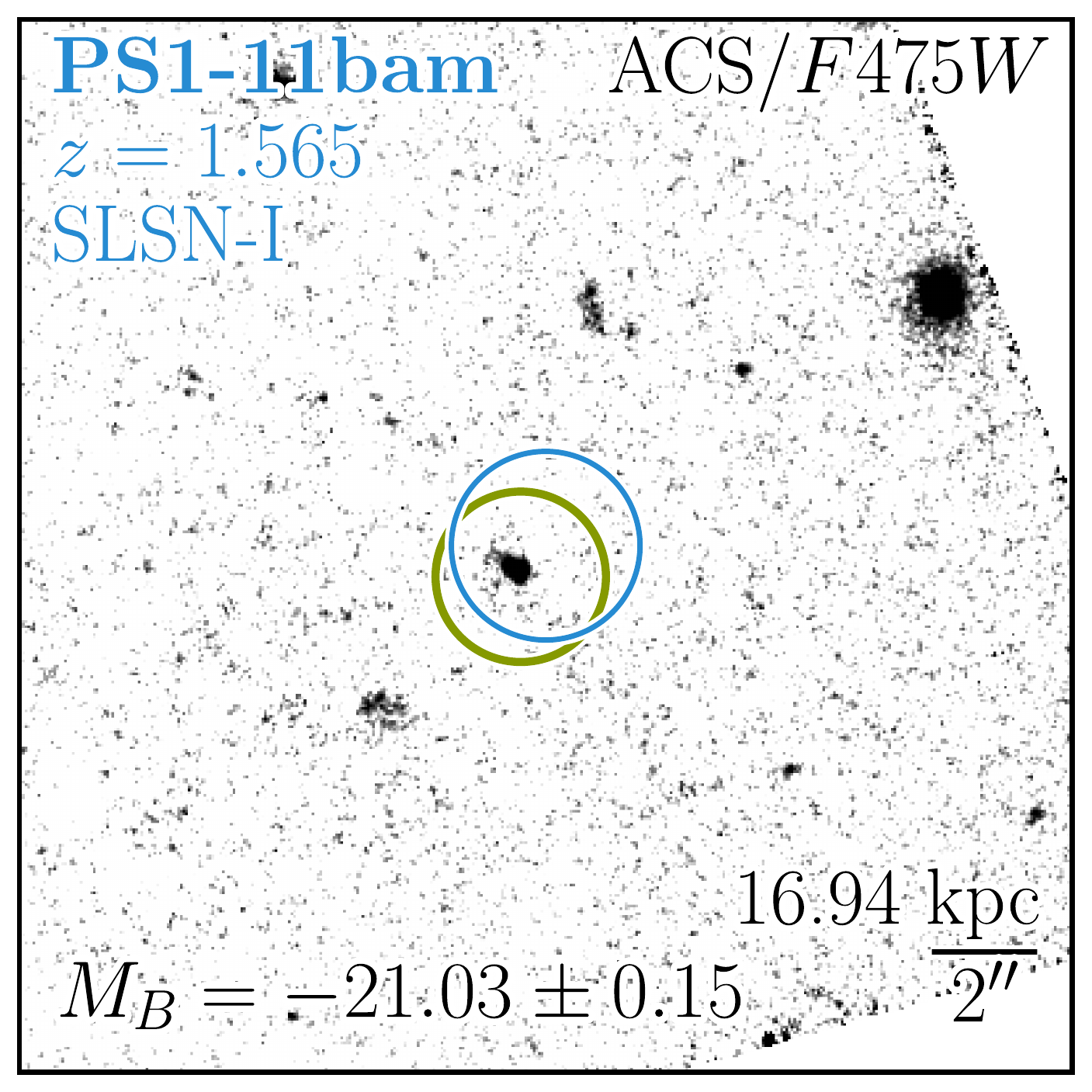}
\includegraphics[width=0.19\textwidth]{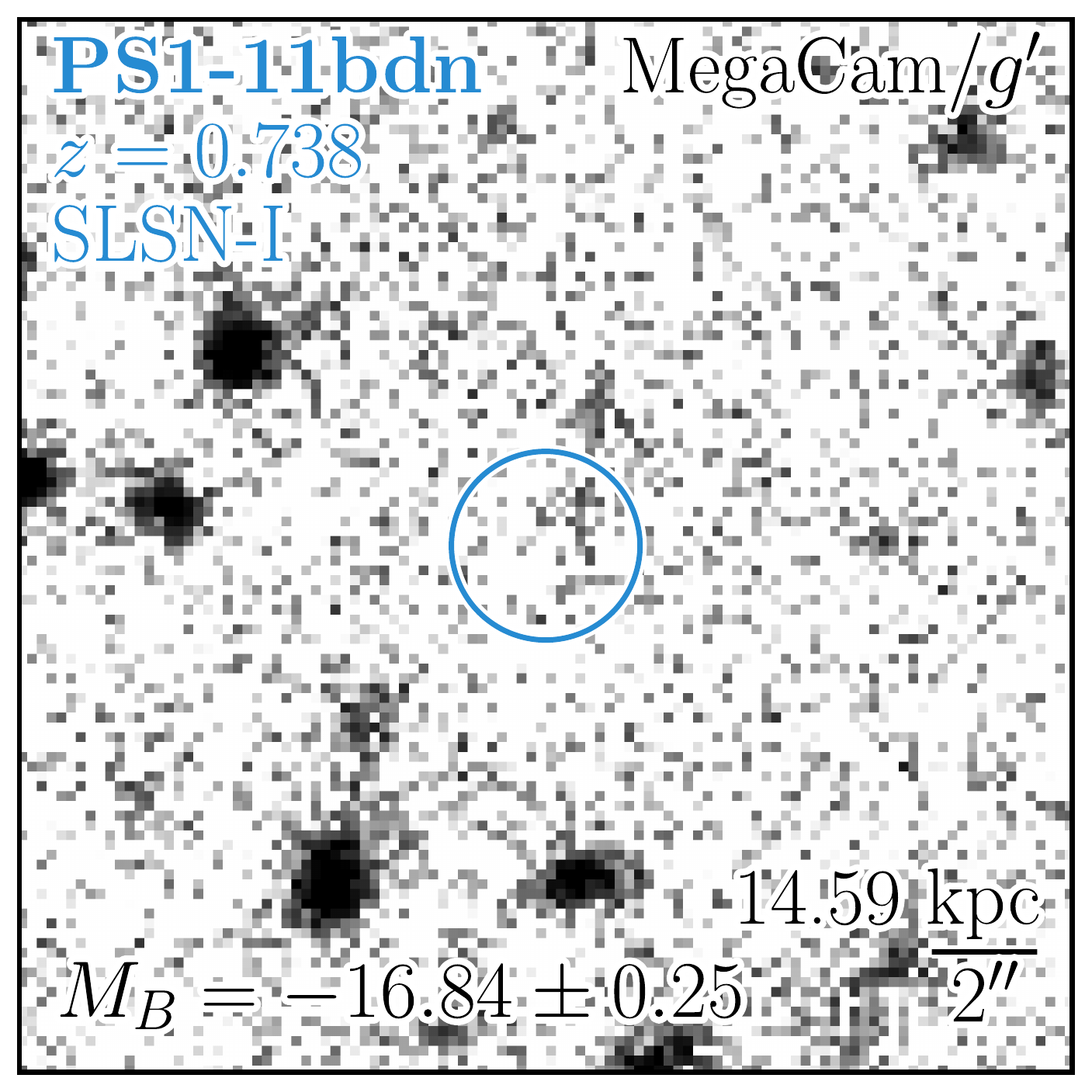}
\includegraphics[width=0.19\textwidth]{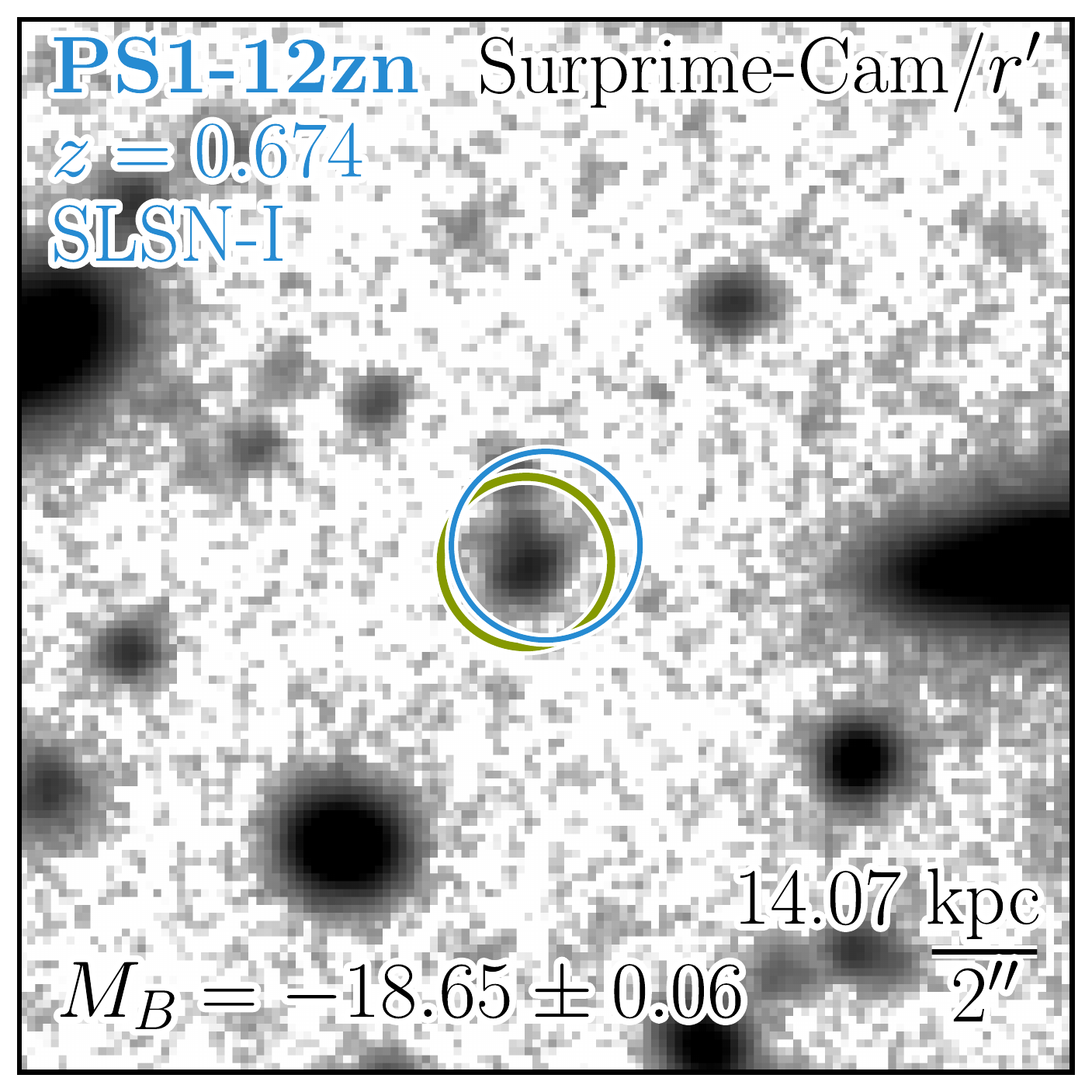}
\includegraphics[width=0.19\textwidth]{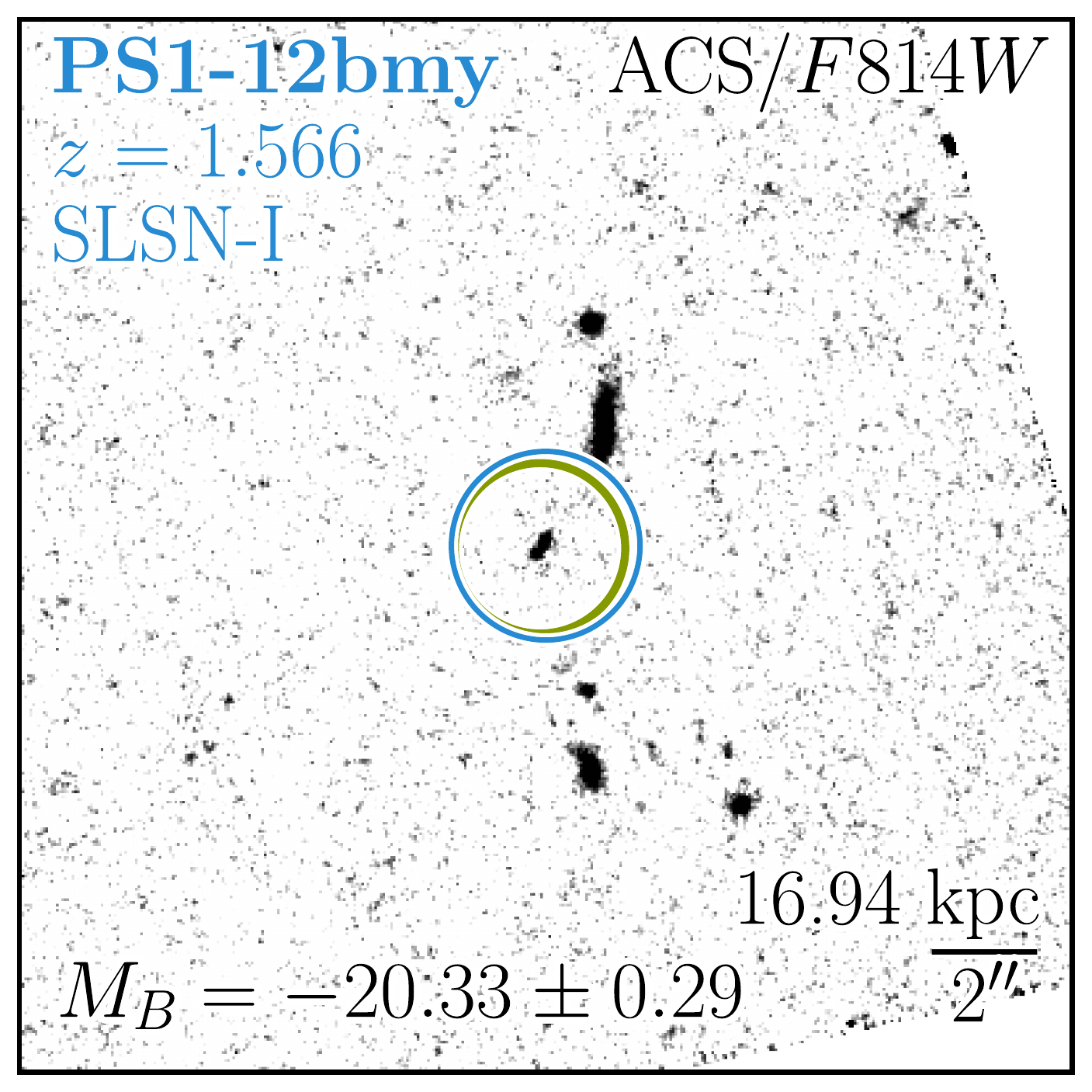}
\includegraphics[width=0.19\textwidth]{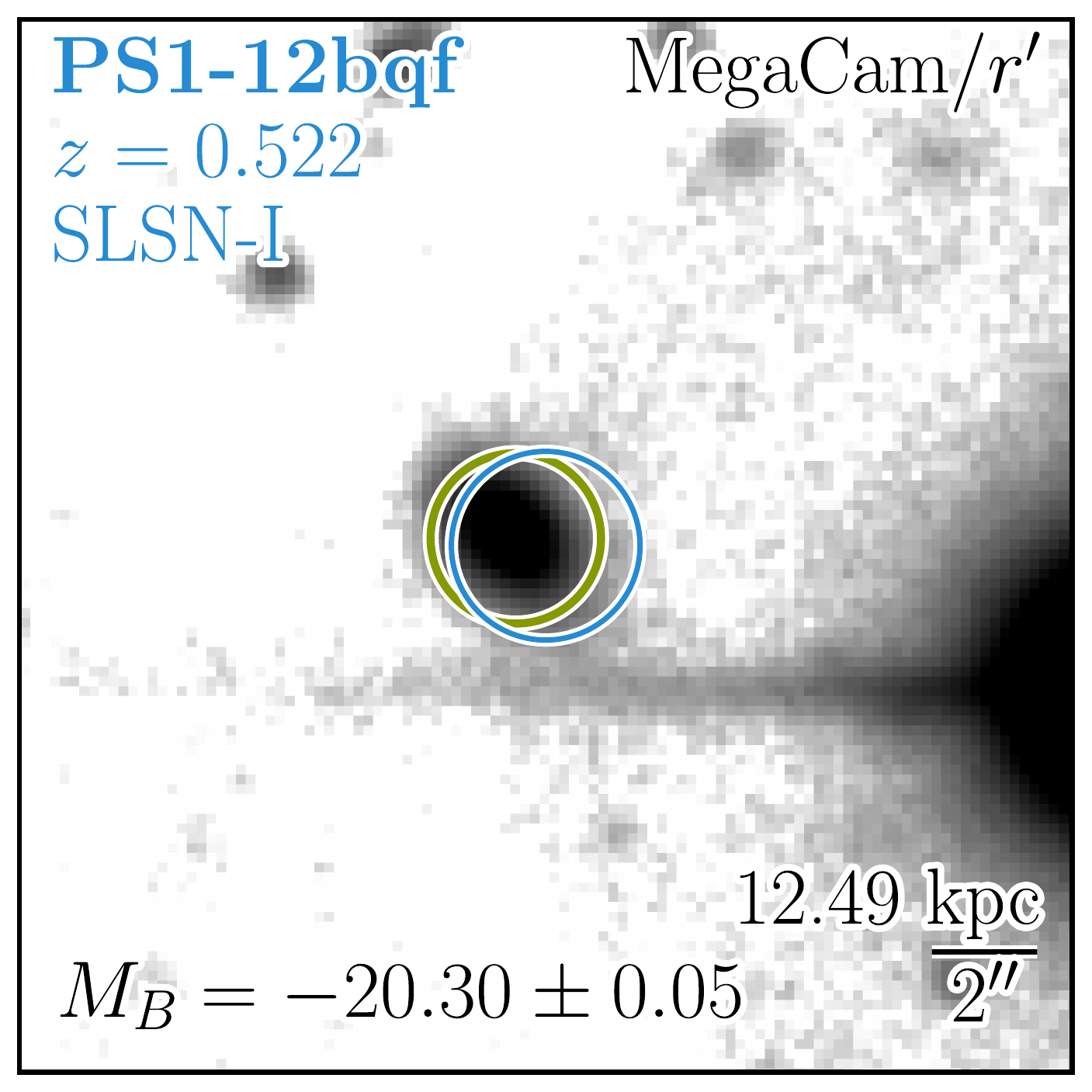}
\includegraphics[width=0.19\textwidth]{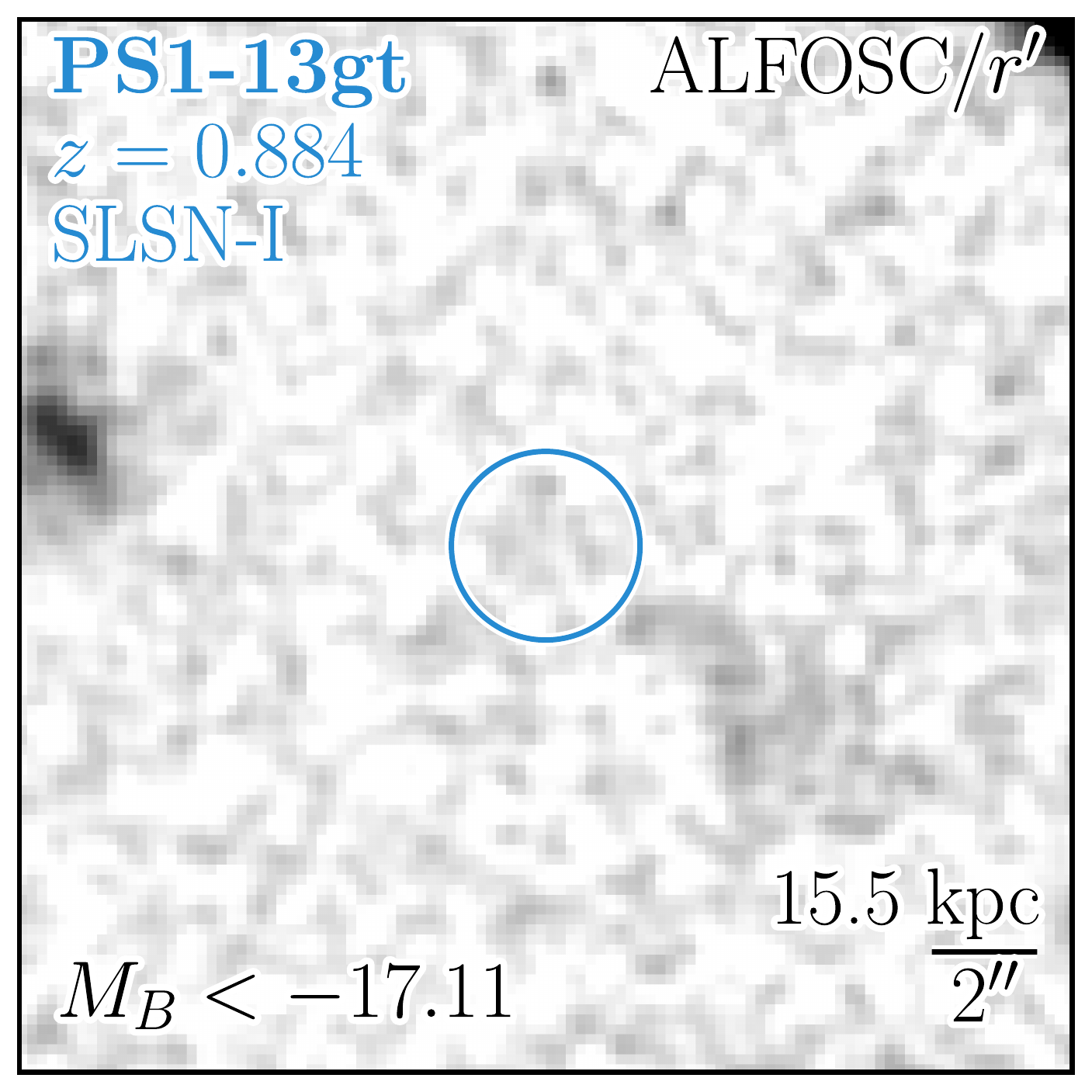}
\includegraphics[width=0.19\textwidth]{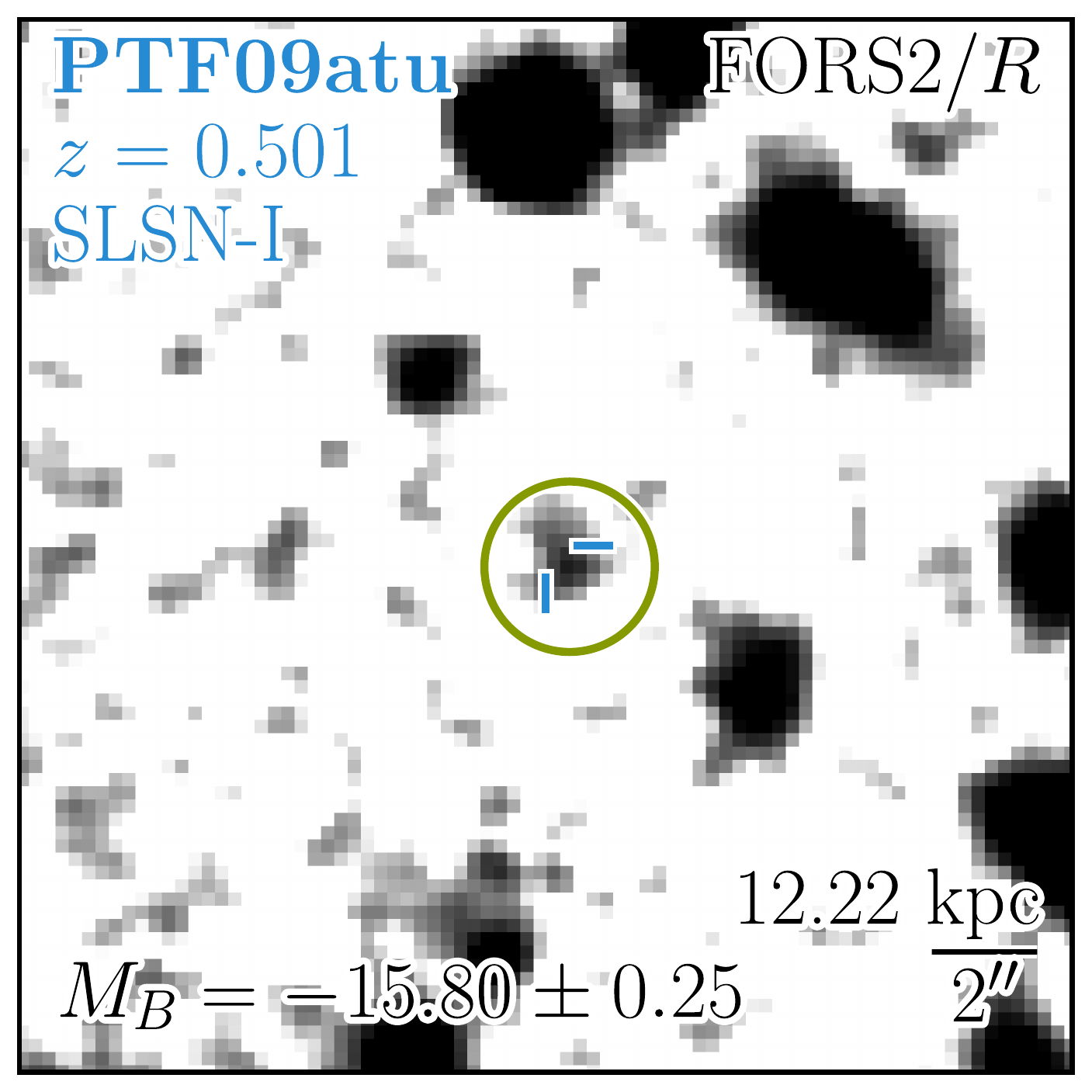}
\includegraphics[width=0.19\textwidth]{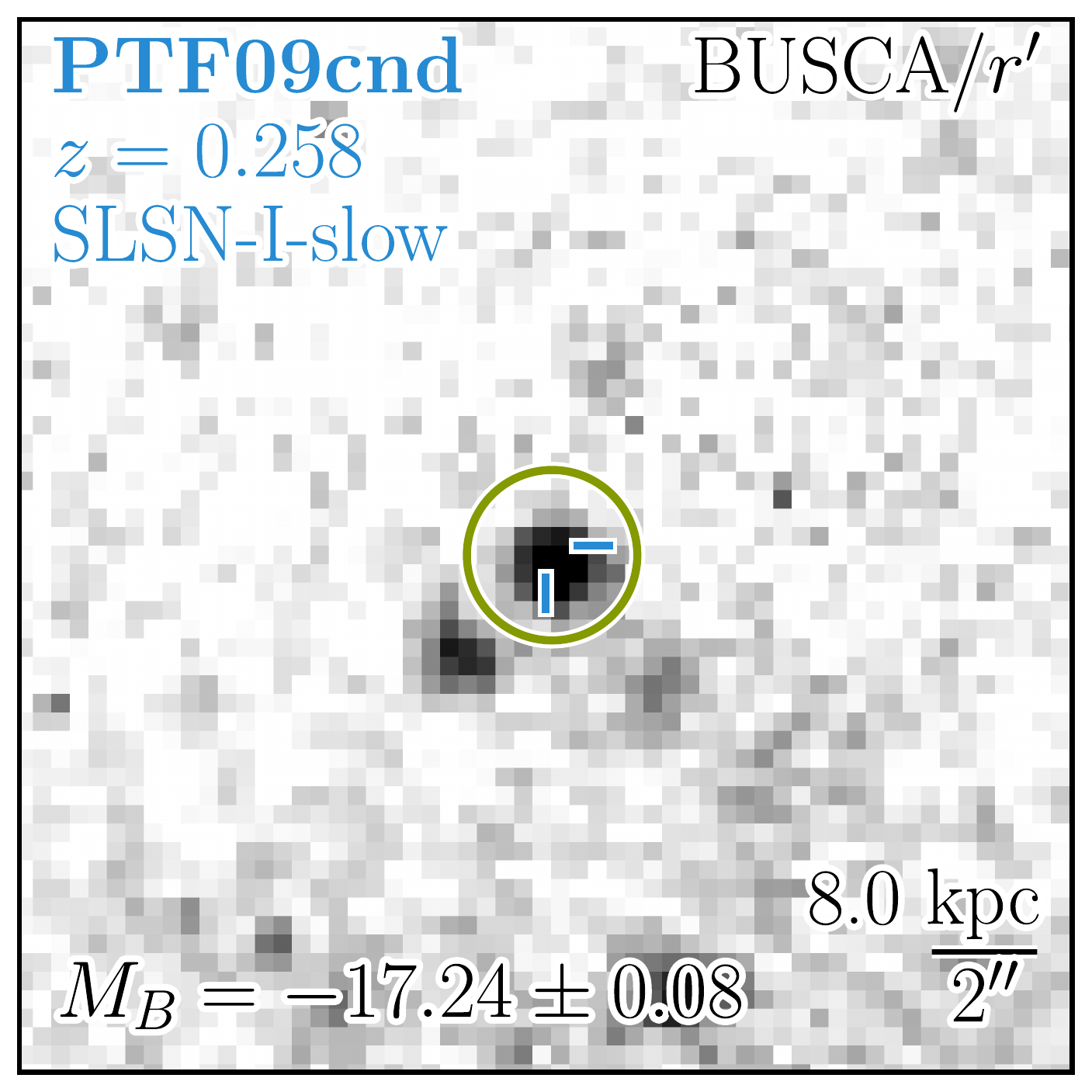}
\includegraphics[width=0.19\textwidth]{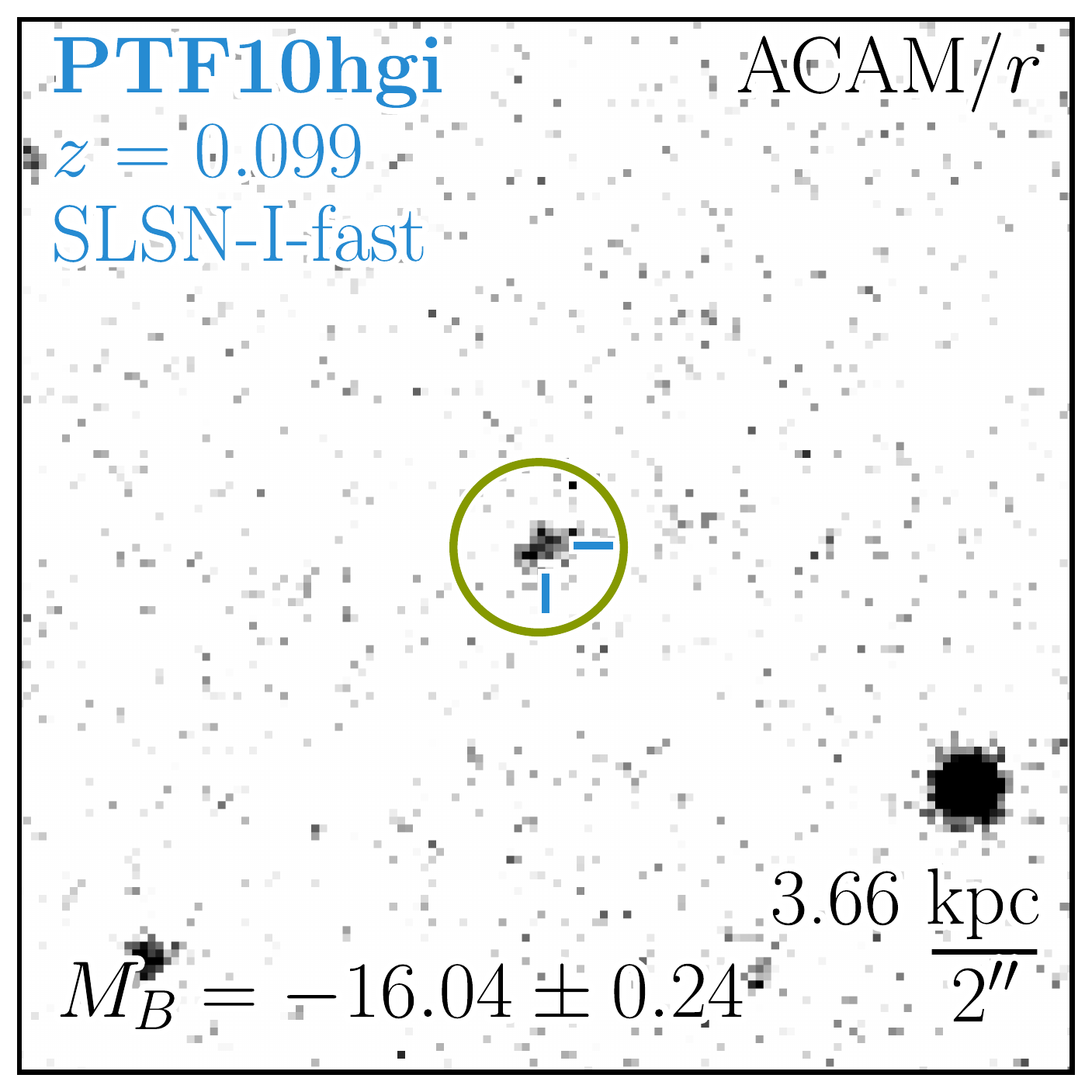}
\includegraphics[width=0.19\textwidth]{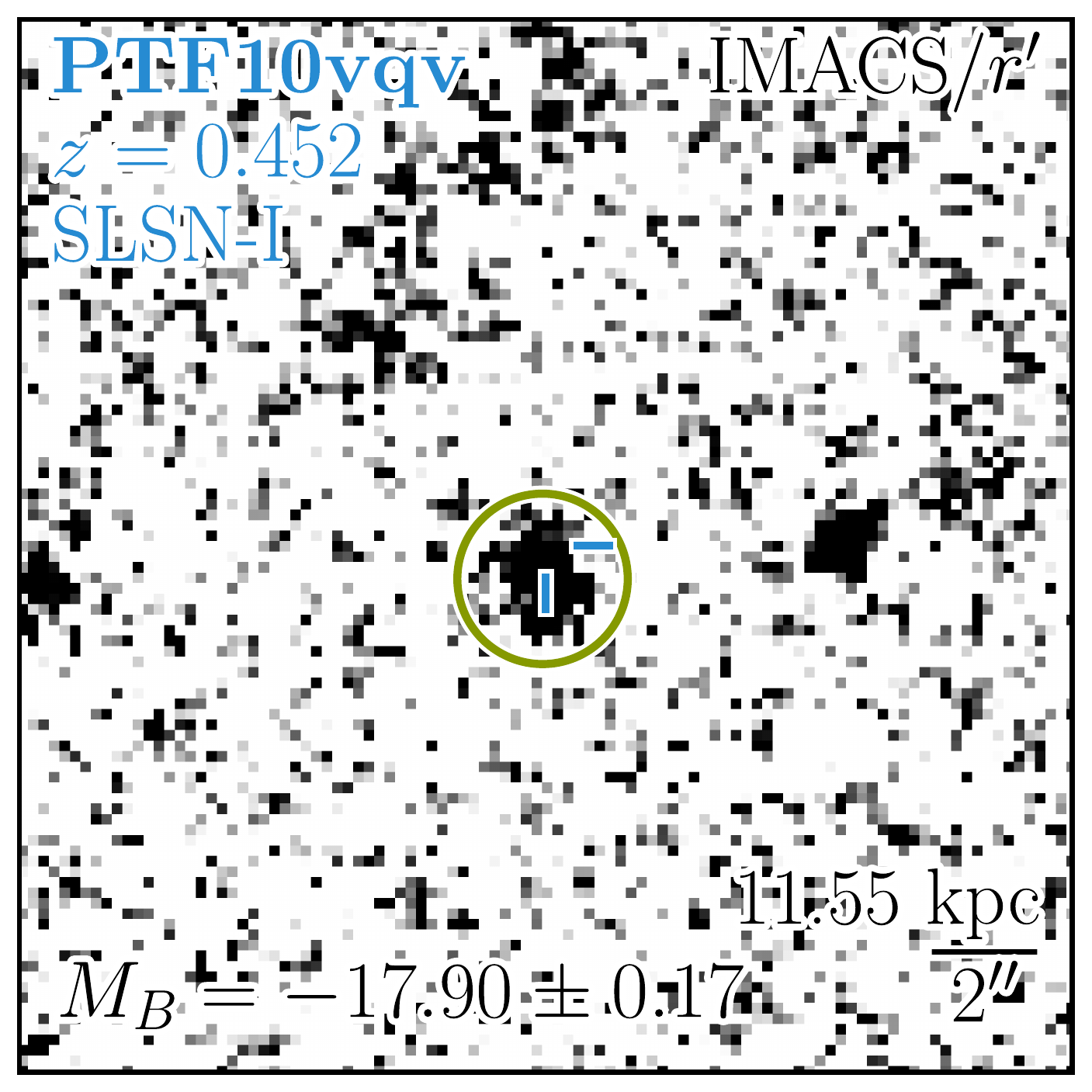}
\includegraphics[width=0.19\textwidth]{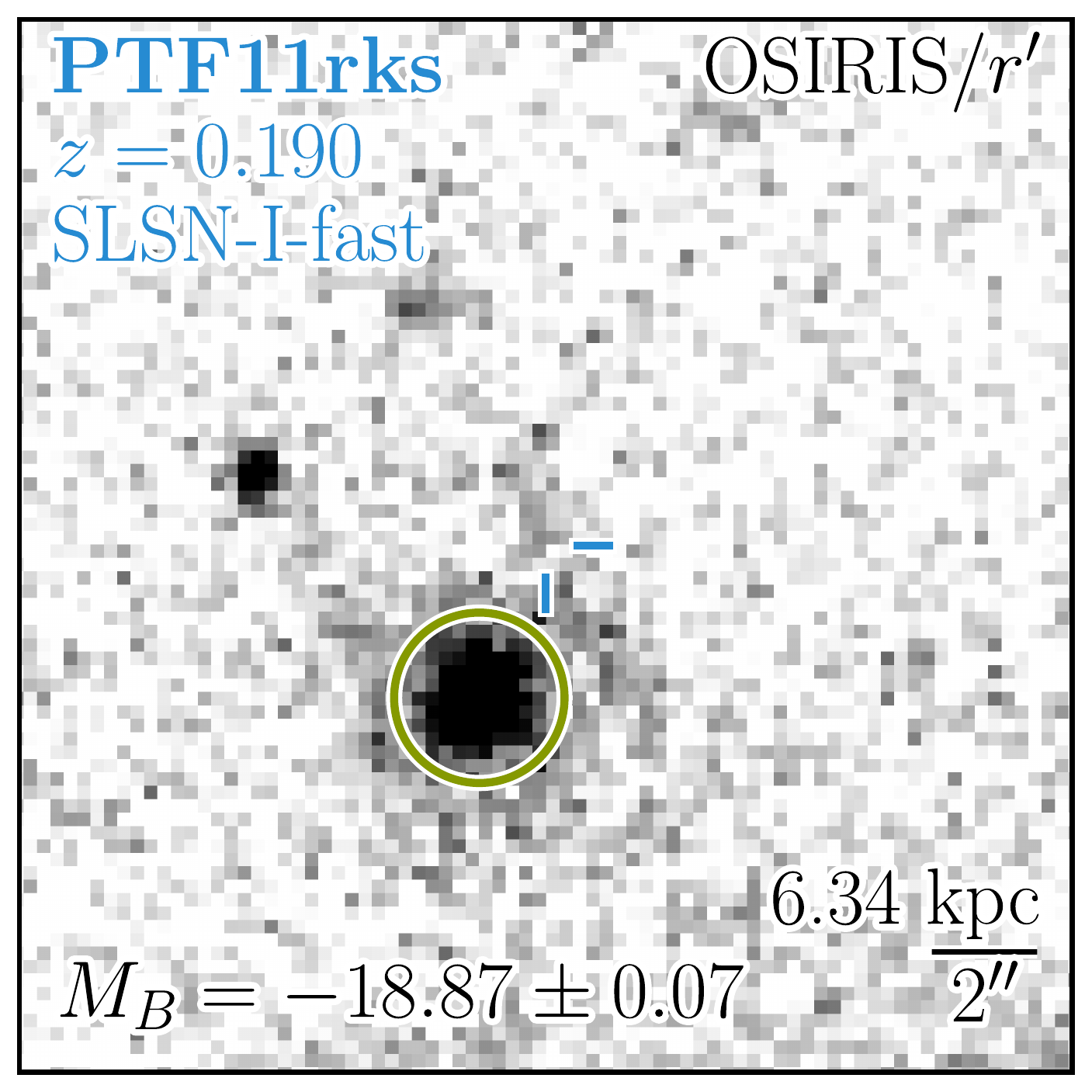}
\caption{Similar to Fig. \ref{fig:poststamp}. Each panel has a size of $20\arcsec\times20\arcsec$
where North is up and East is left. The blue crosshair marks the position of the SNe after
aligning a SN and a host image. If no SN image was available, the blue circle (arbitrary radius)
indicates the SN position reported in the literature. The average alignment error was
$0\farcs17$ but it exceeded $1\farcs0$ in a few cases. 
See Sect. \ref{res:morphologies} for details. The green circle (arbitrary radius) marks the host galaxy.
The observed absolute $B$-band brightness is displayed in the lower left.
The images of CSS140925,
DES14S2qri, DES14X2byo, PS1-11aib, PS1-13gt, PTF09atu, SN2013dg and SN2015bn were smoothed
with a Gaussian kernel (width of 1 px) to improve the visibility of the host.
}
\label{fig:poststamp_app_1}
\end{figure*}
\clearpage

\begin{figure*}
\ContinuedFloat
\includegraphics[width=0.19\textwidth]{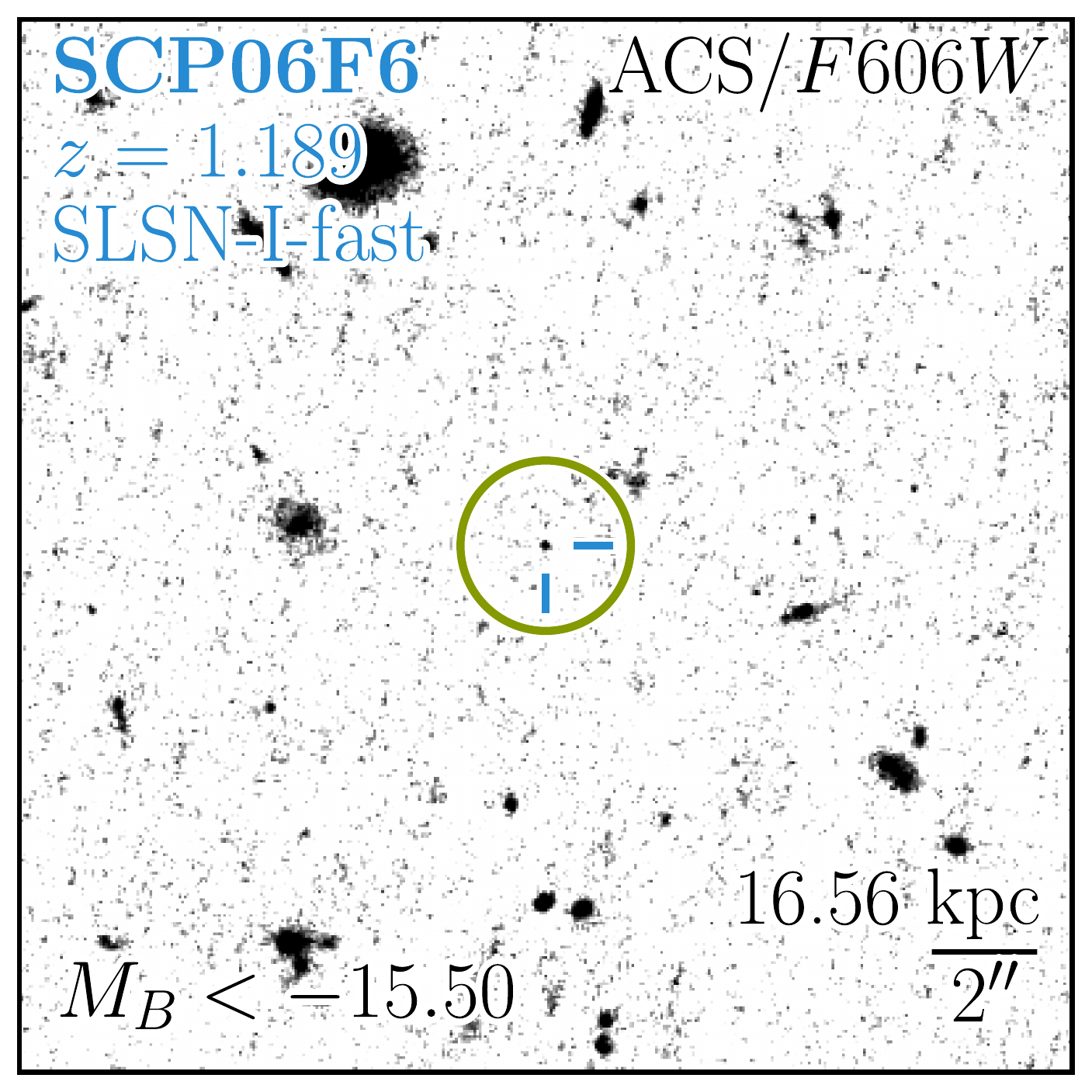}
\includegraphics[width=0.19\textwidth]{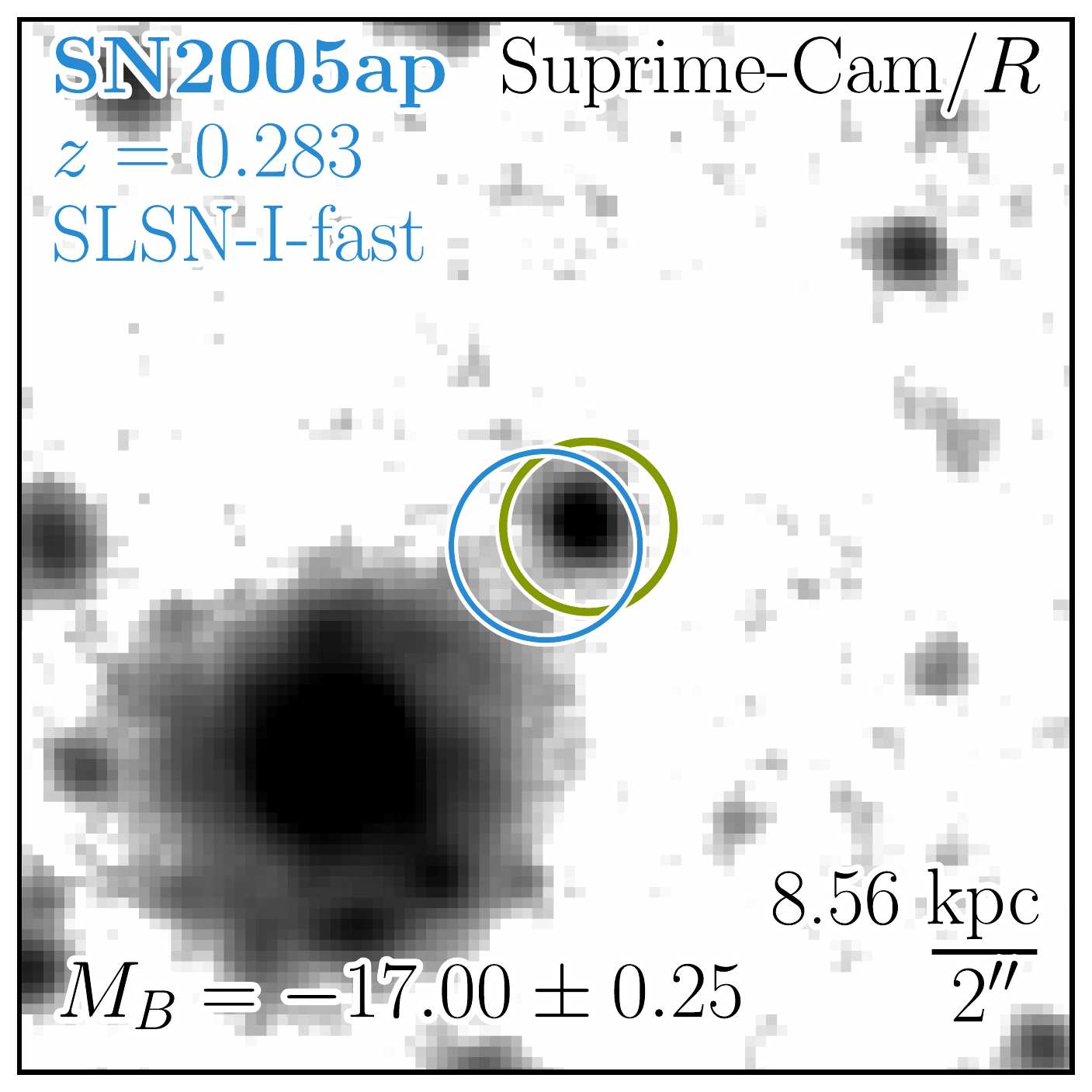}
\includegraphics[width=0.19\textwidth]{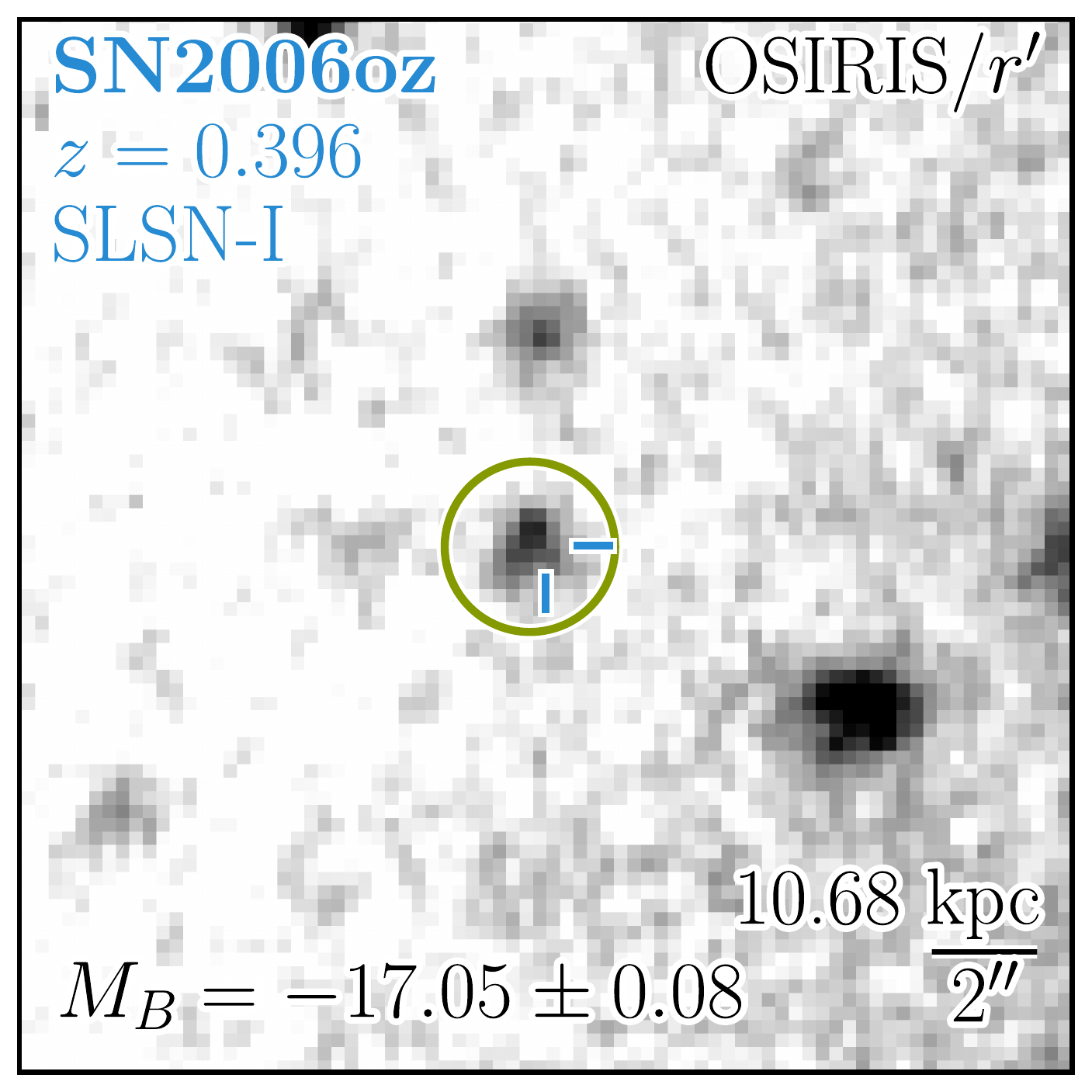}
\includegraphics[width=0.19\textwidth]{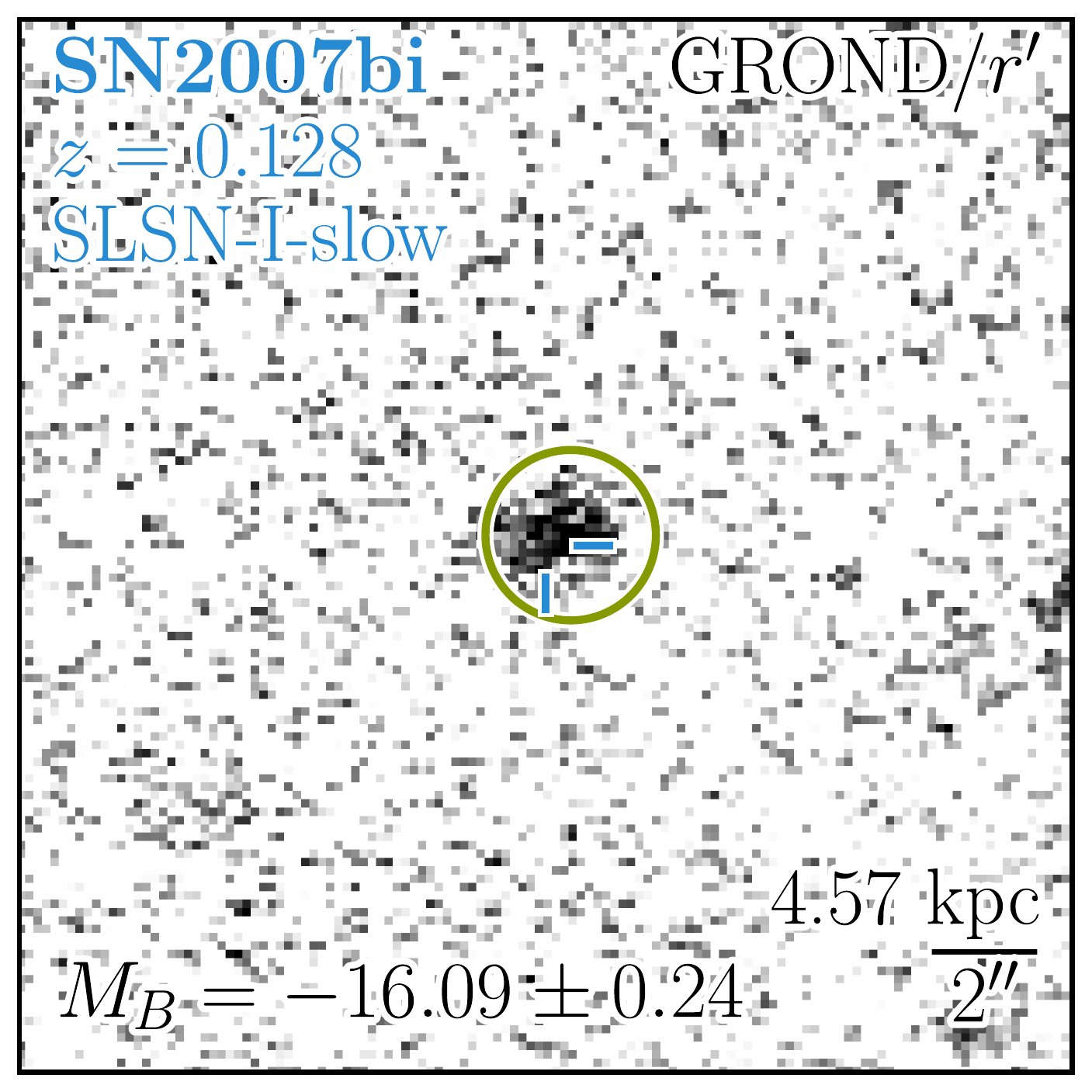}
\includegraphics[width=0.19\textwidth]{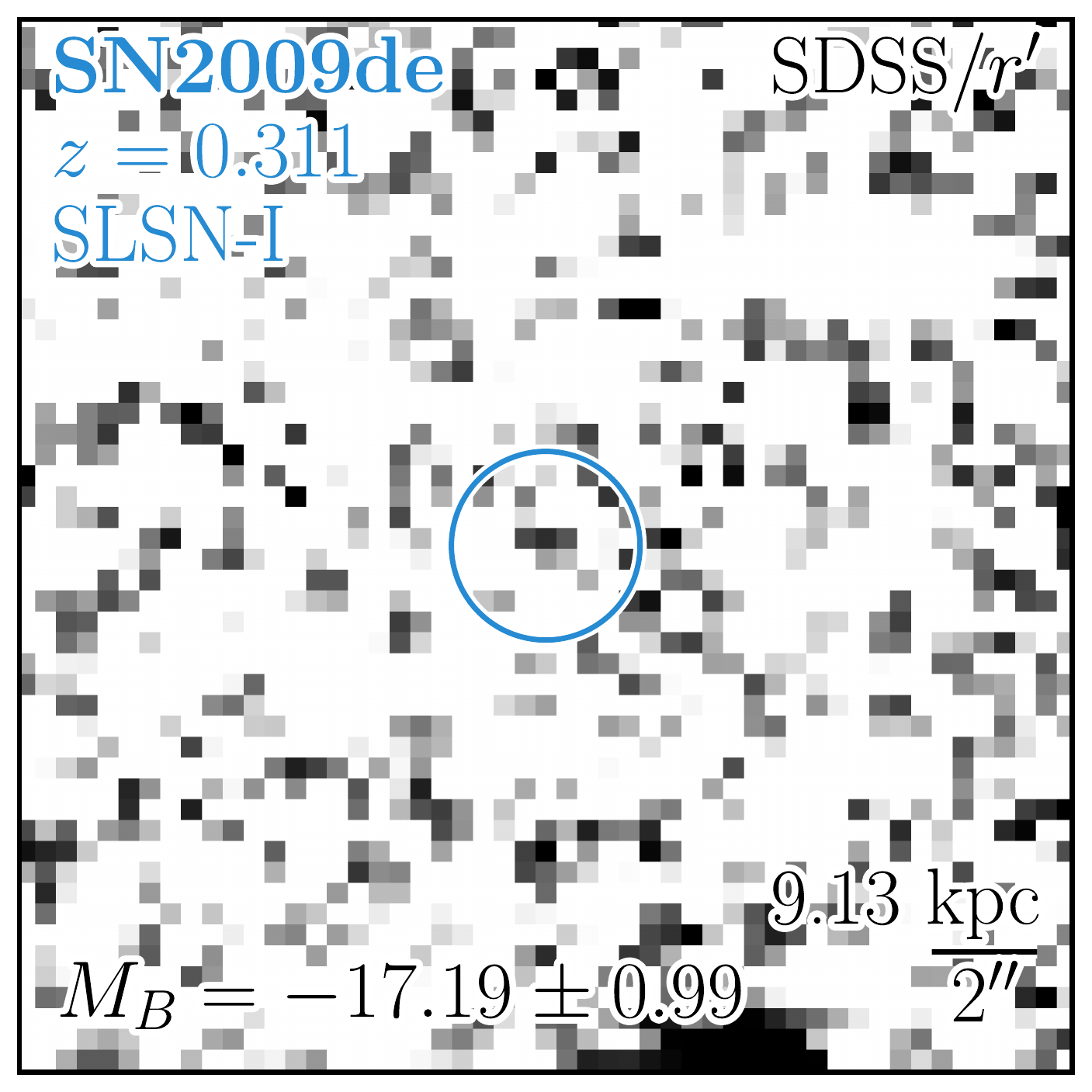}
\includegraphics[width=0.19\textwidth]{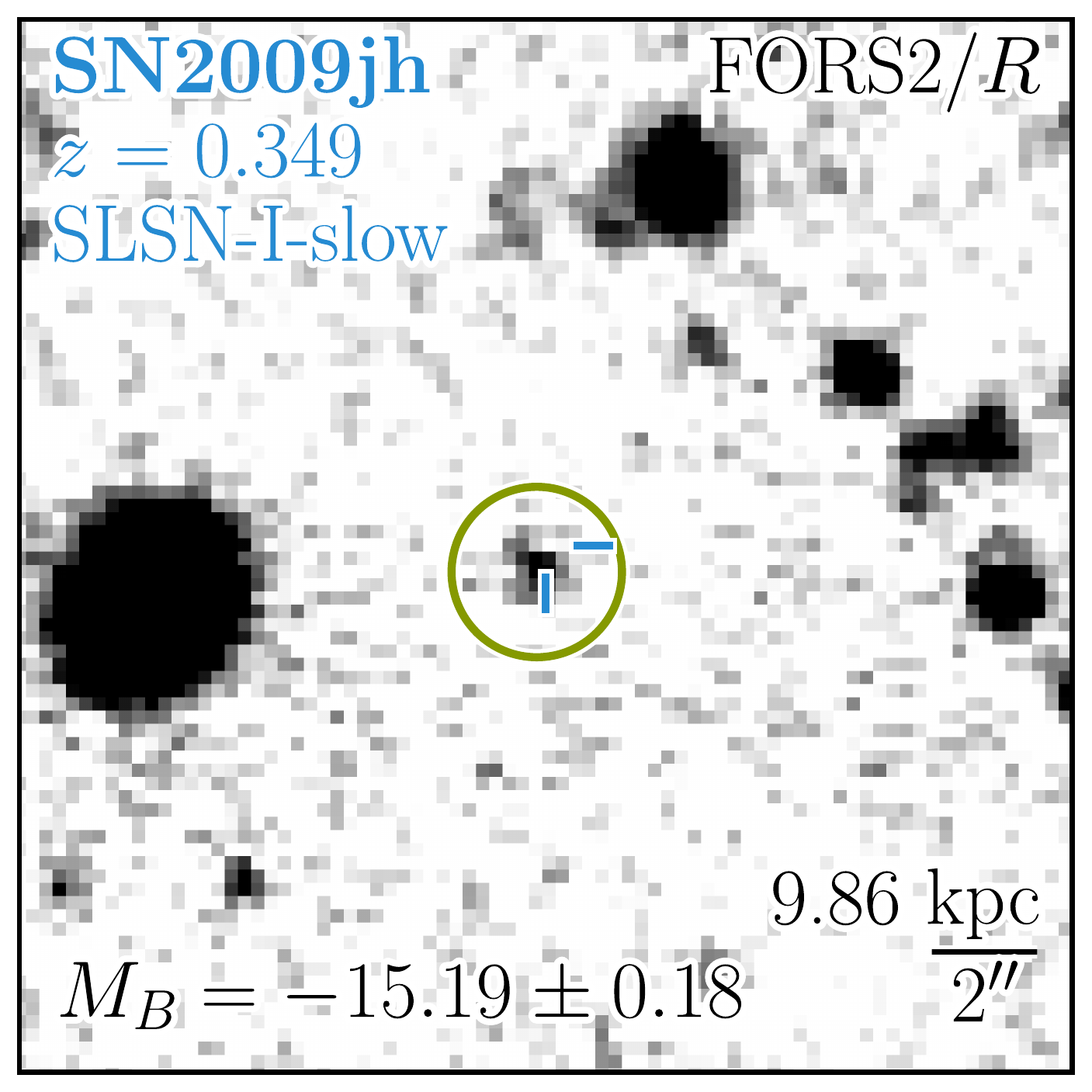}
\includegraphics[width=0.19\textwidth]{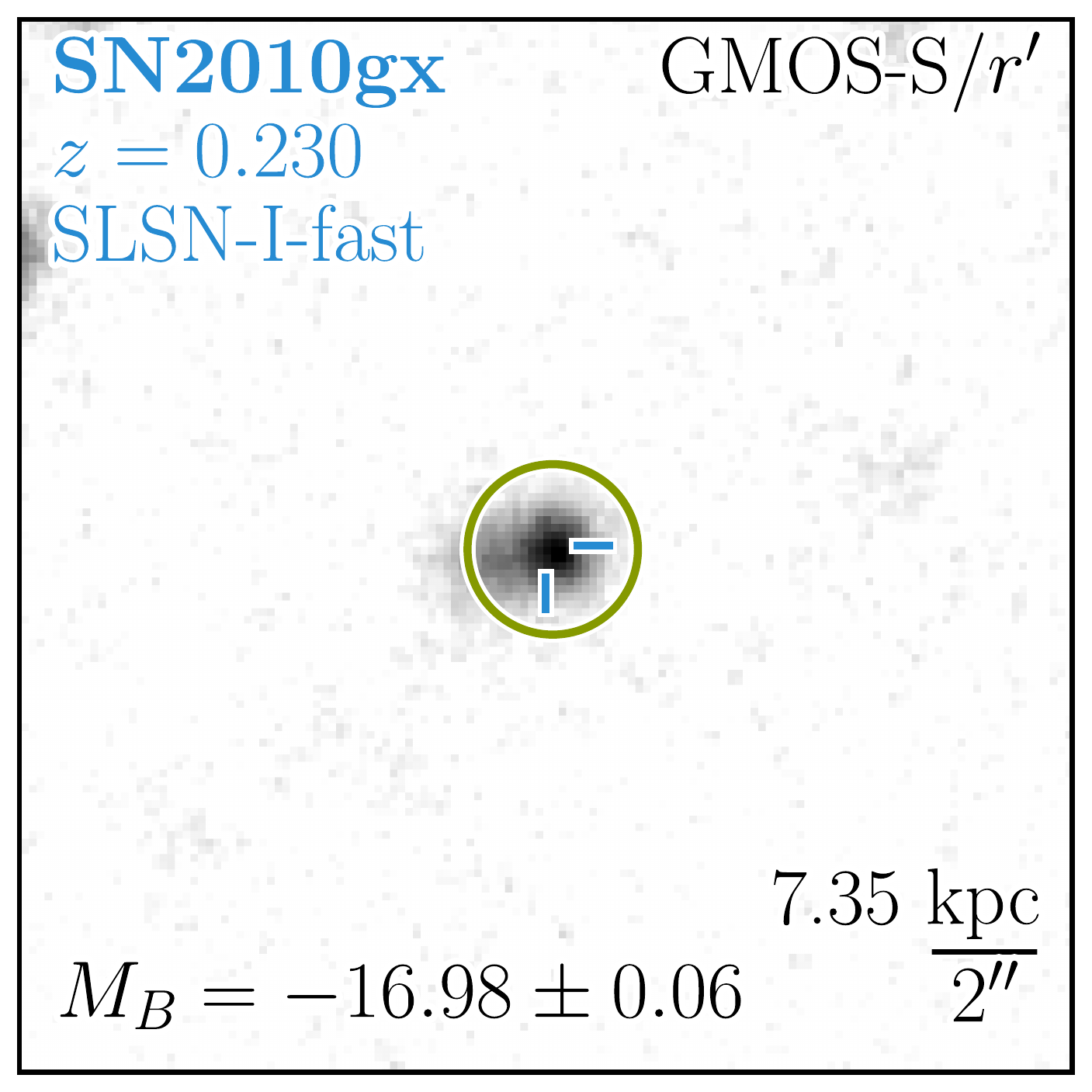}
\includegraphics[width=0.19\textwidth]{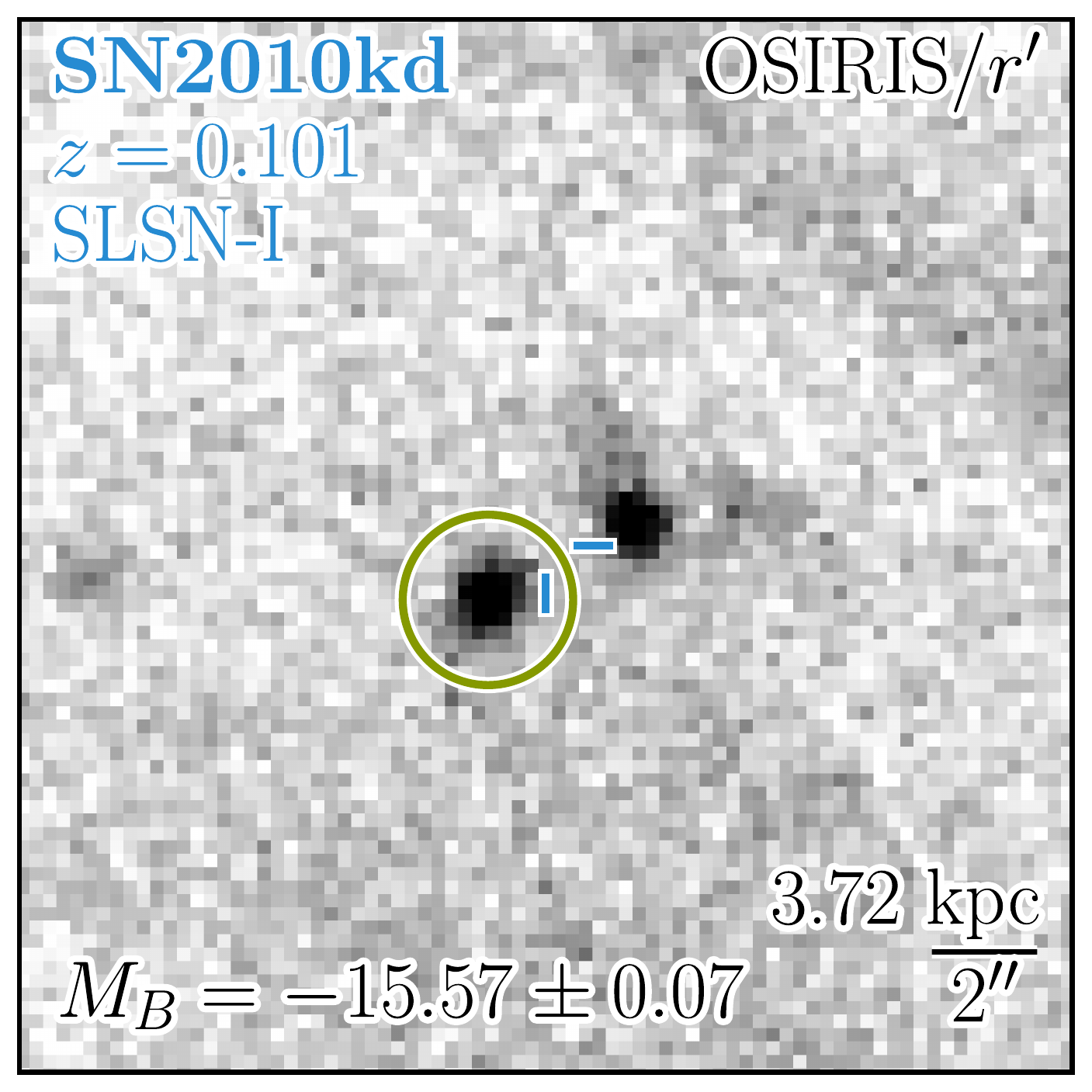}
\includegraphics[width=0.19\textwidth]{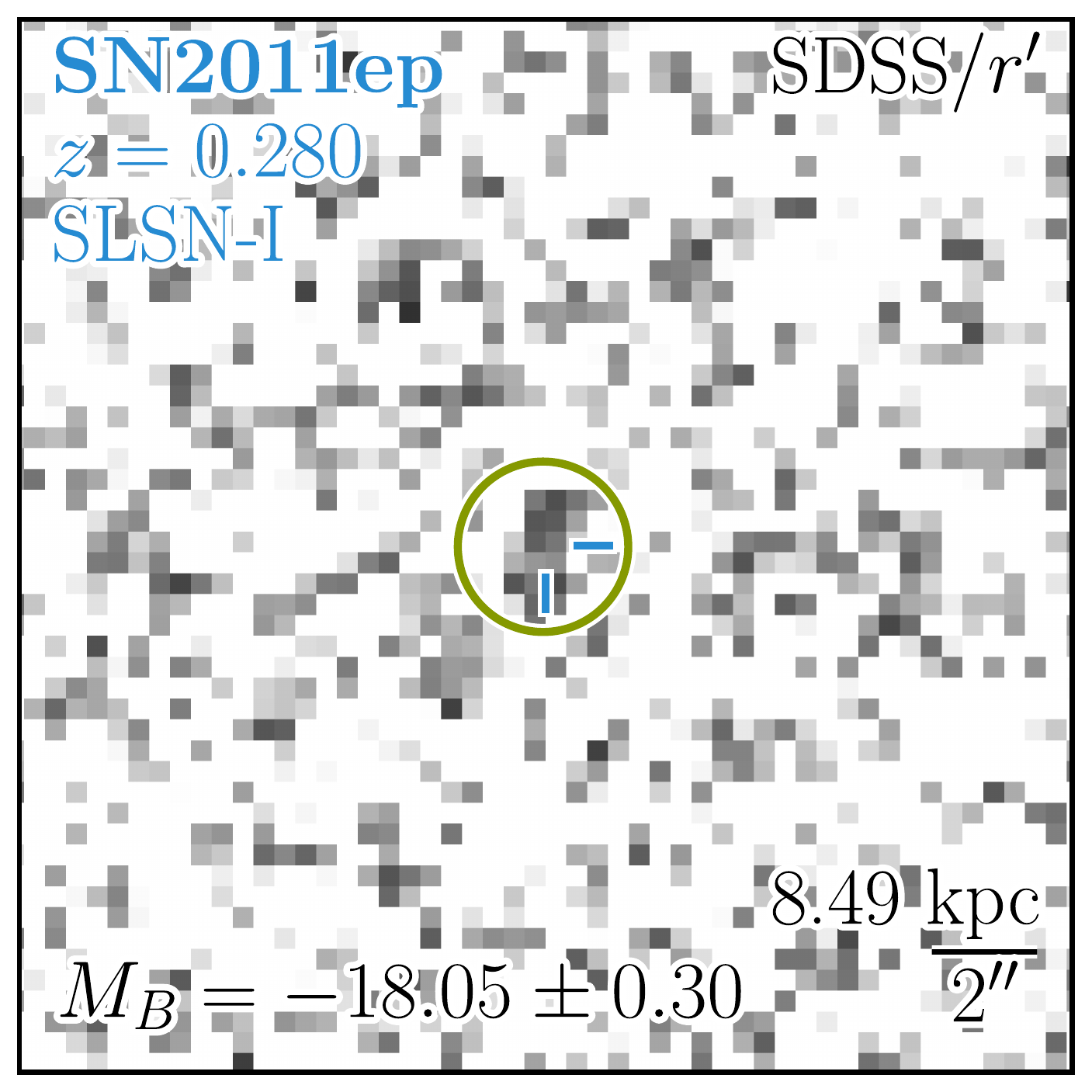}
\includegraphics[width=0.19\textwidth]{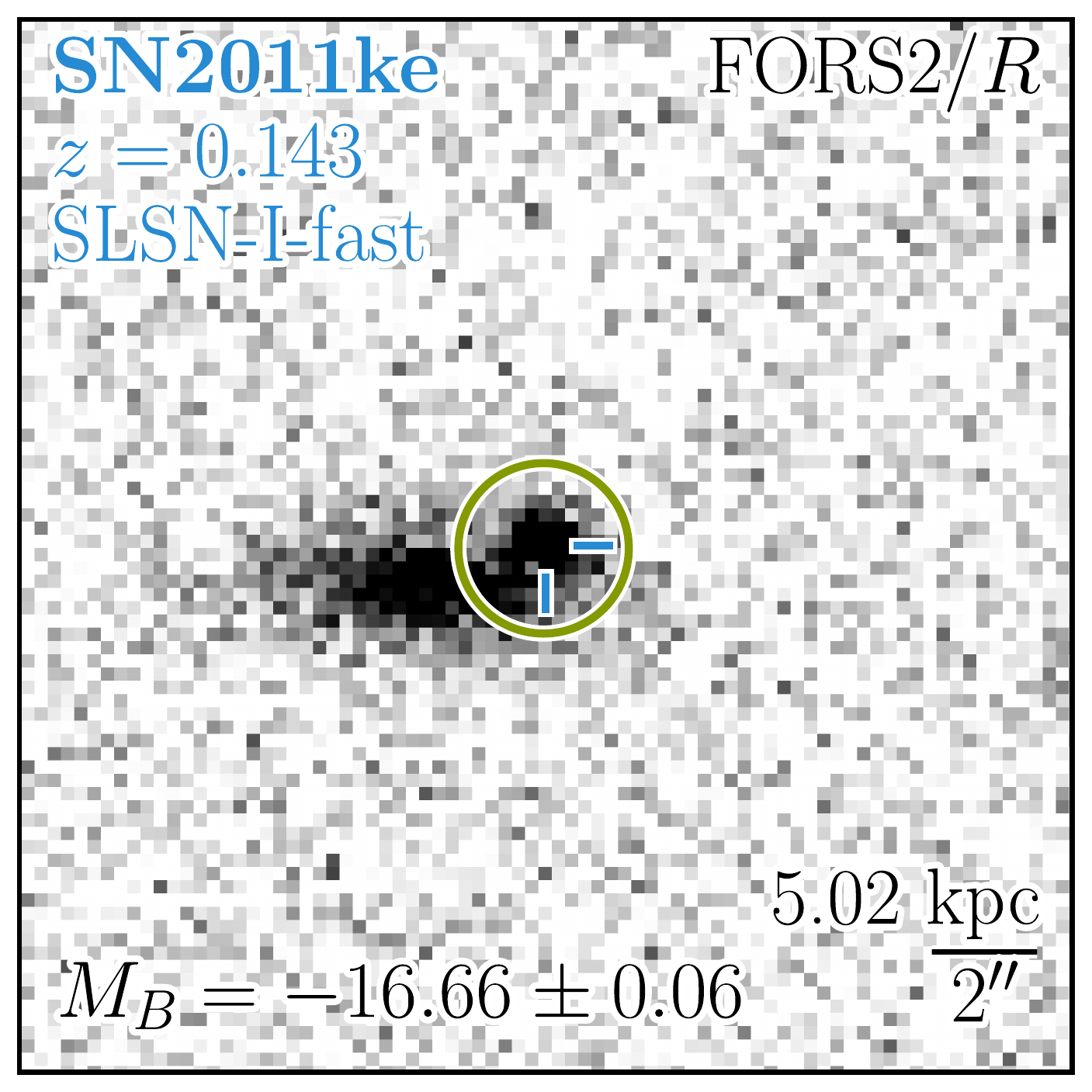}
\includegraphics[width=0.19\textwidth]{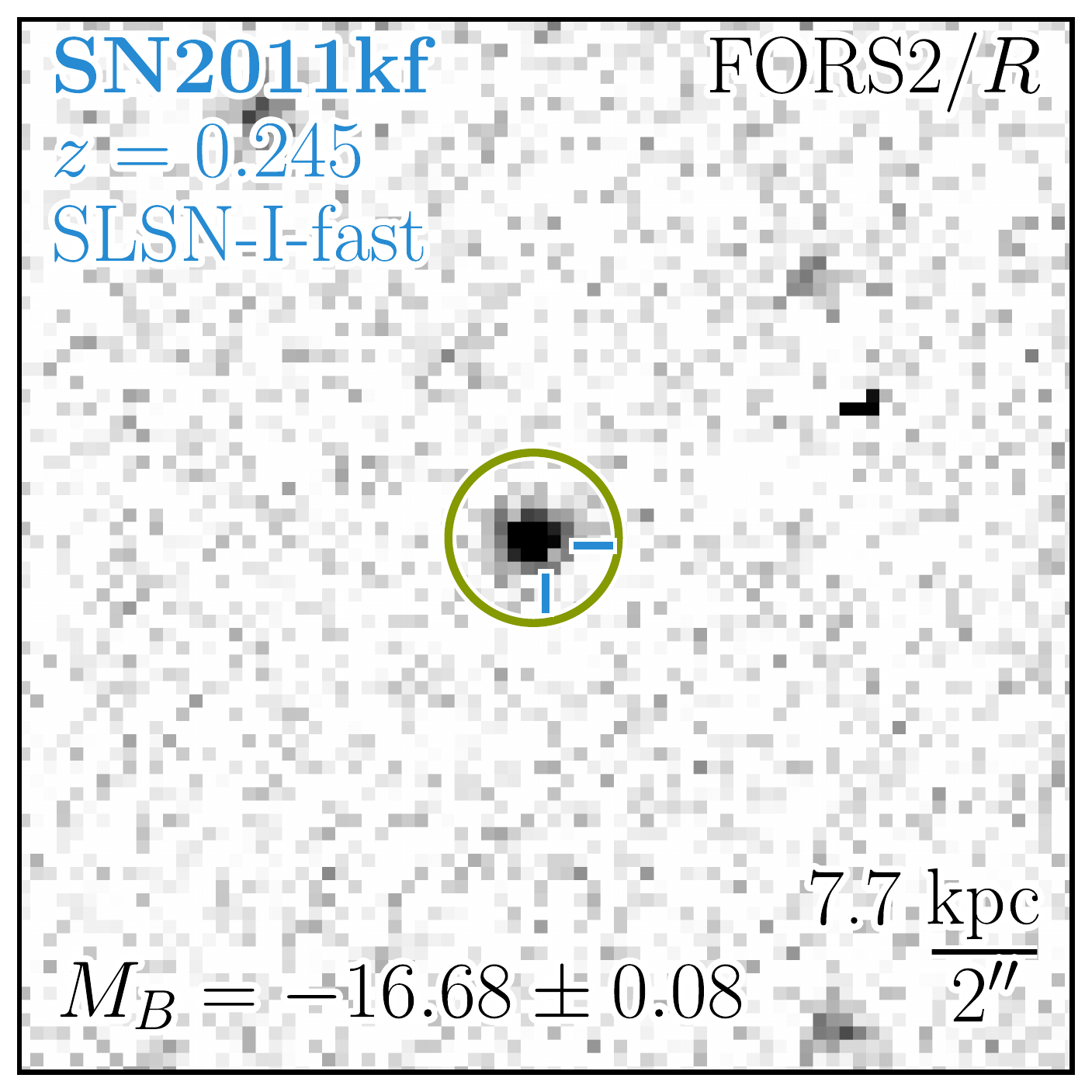}
\includegraphics[width=0.19\textwidth]{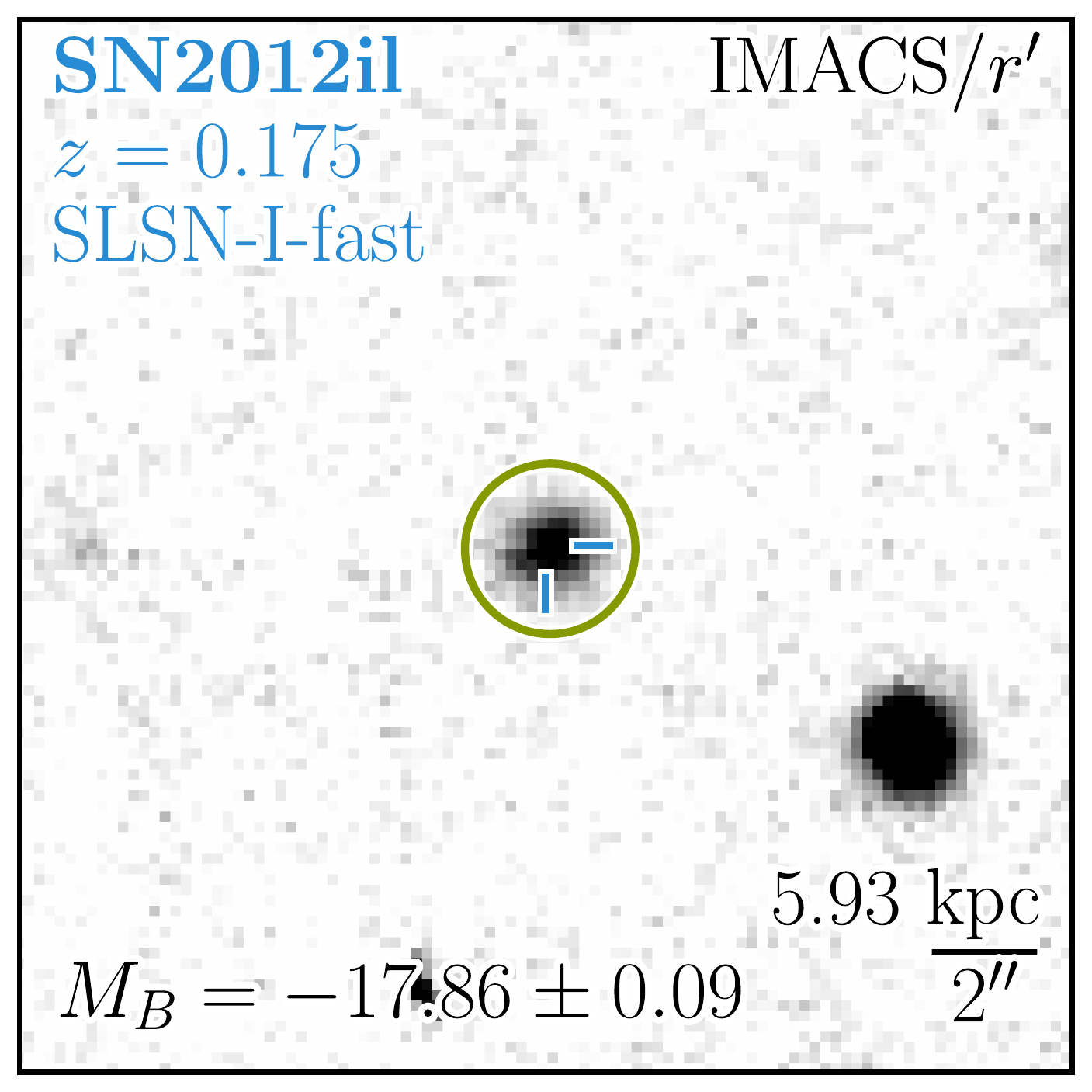}
\includegraphics[width=0.19\textwidth]{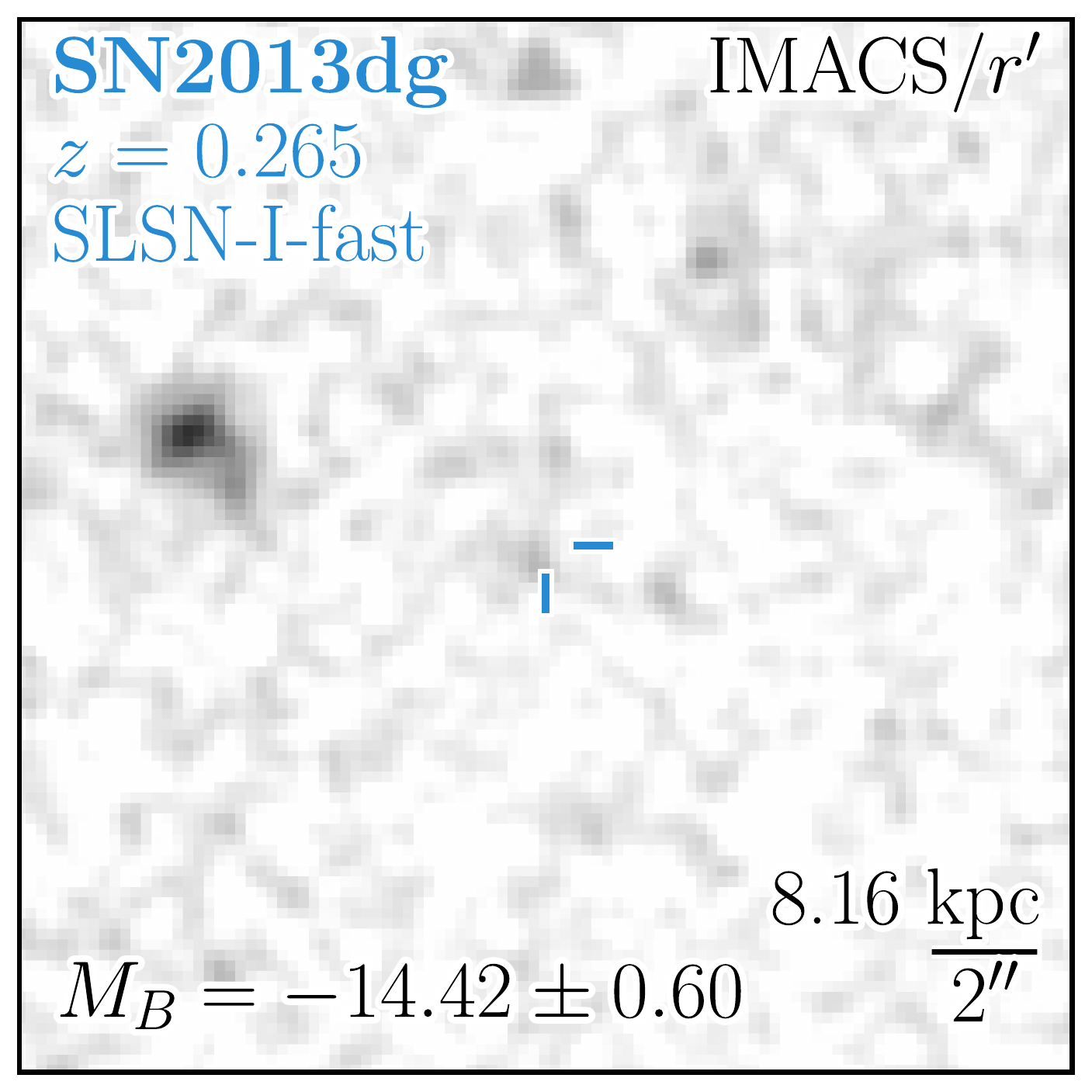}
\includegraphics[width=0.19\textwidth]{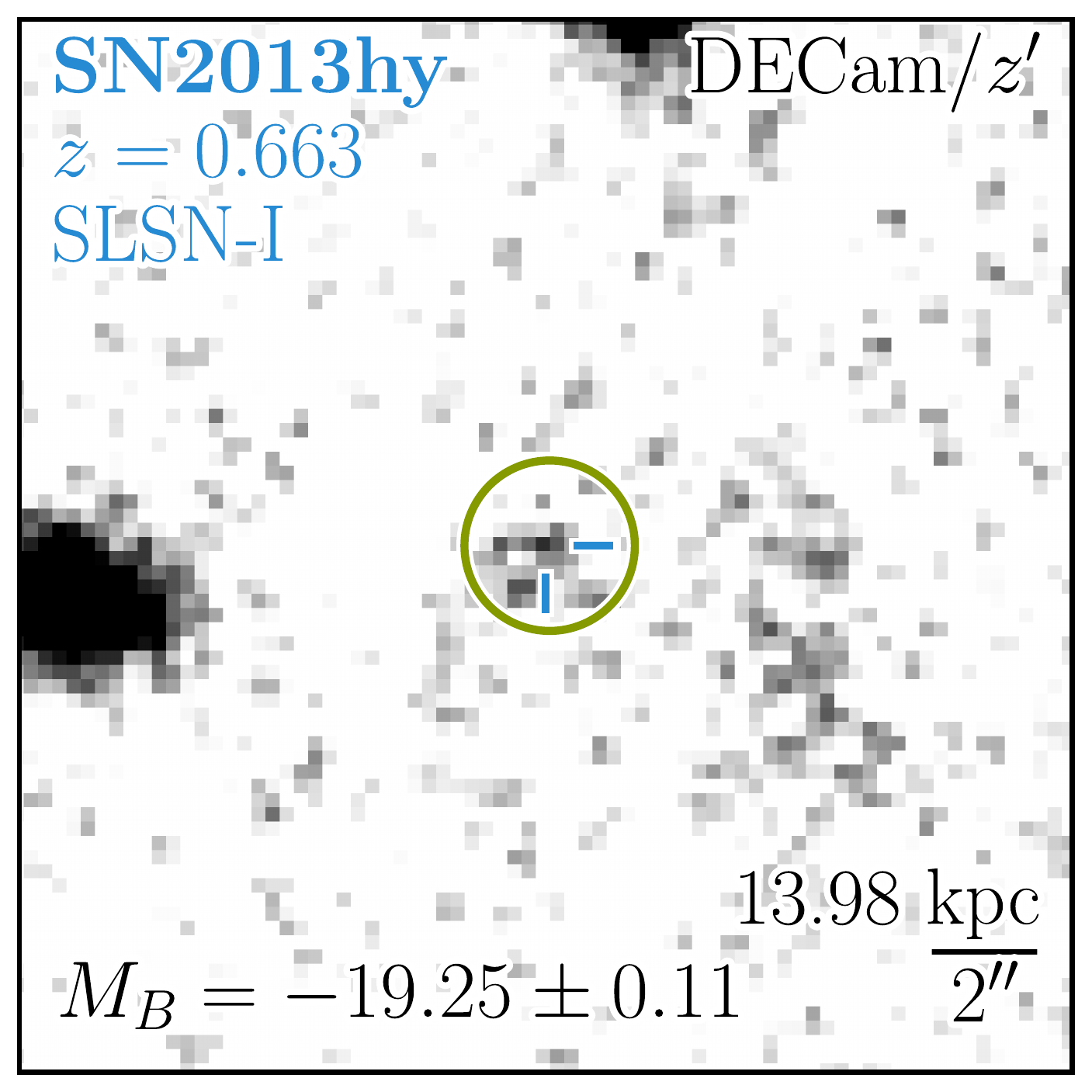}
\includegraphics[width=0.19\textwidth]{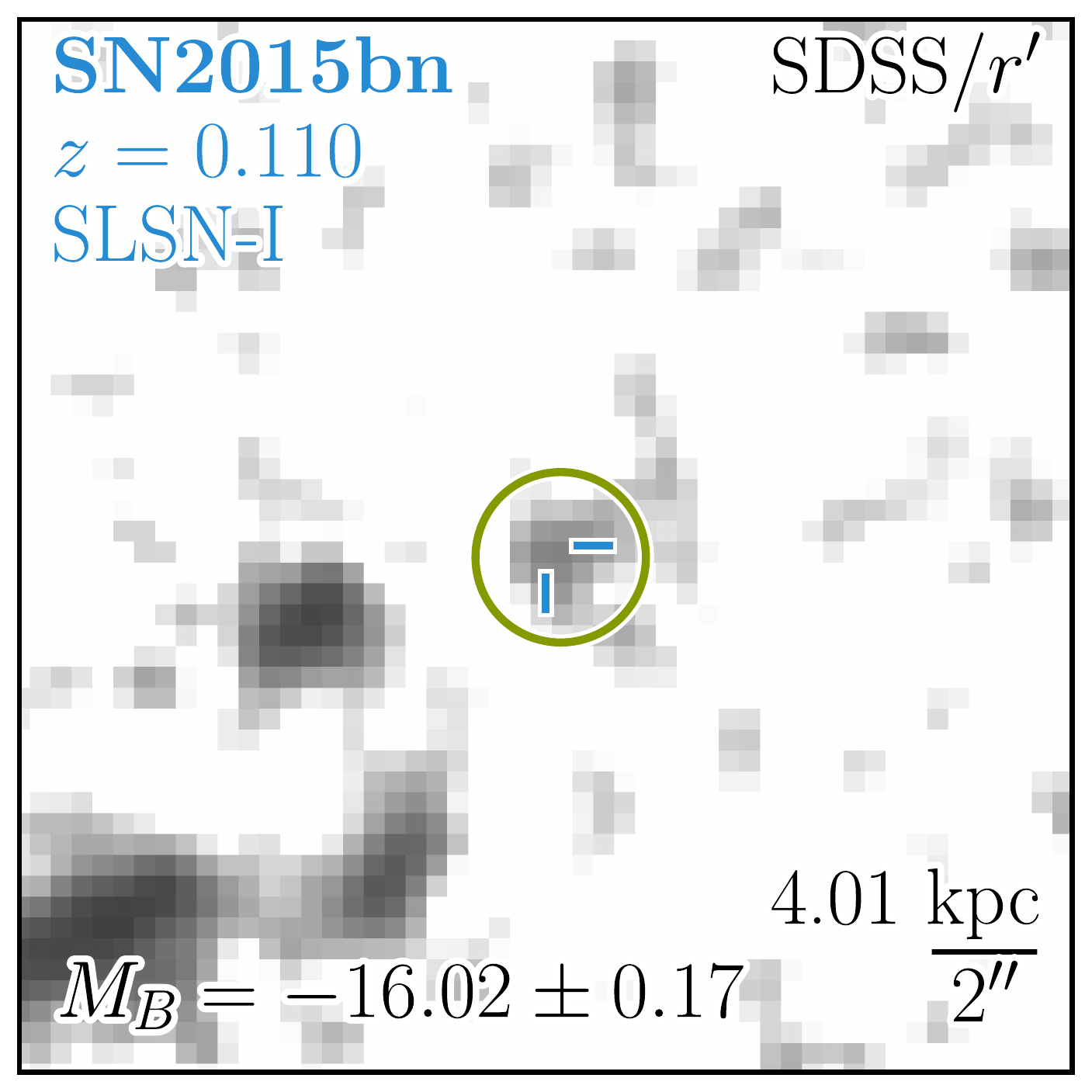}
\includegraphics[width=0.19\textwidth]{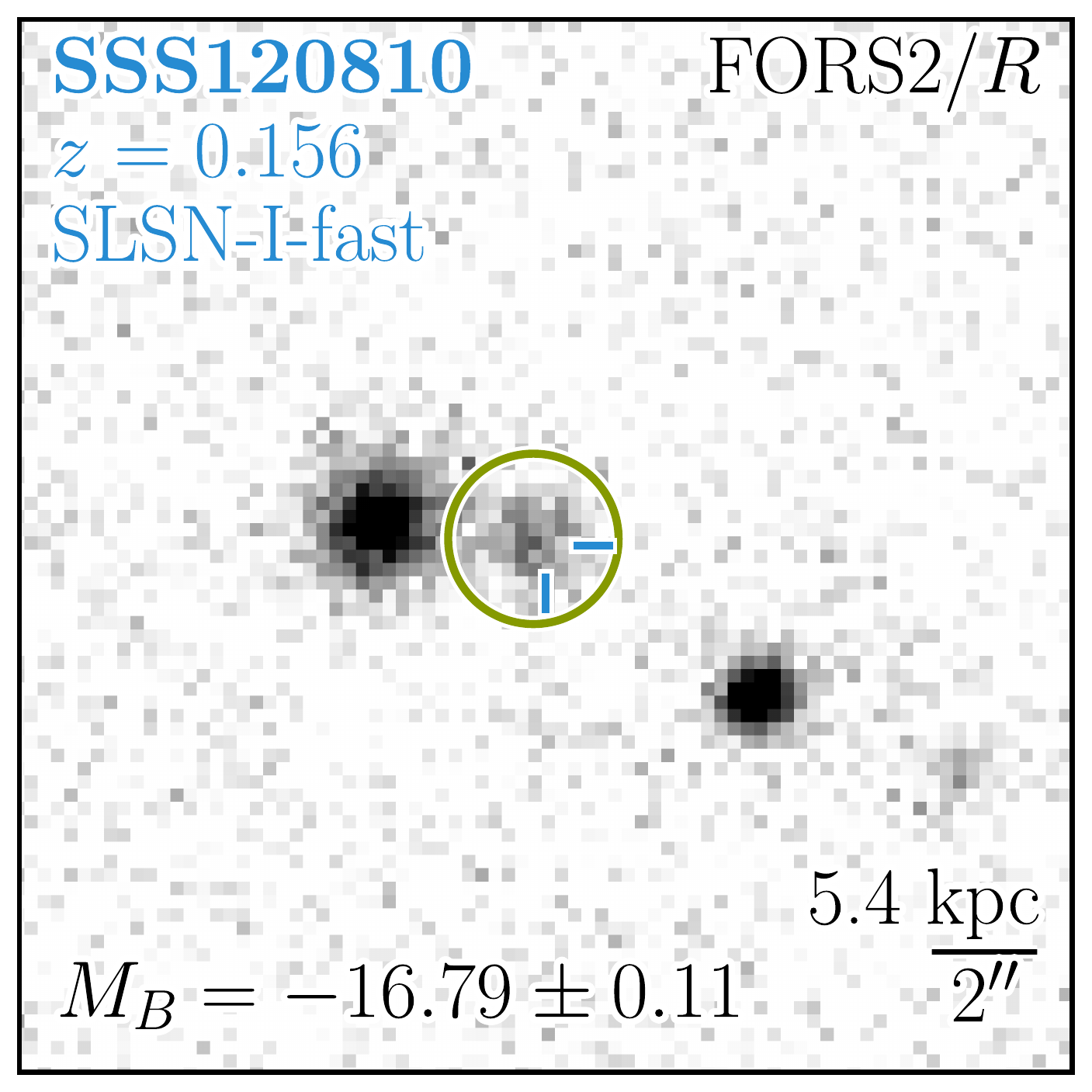}
\includegraphics[width=0.19\textwidth]{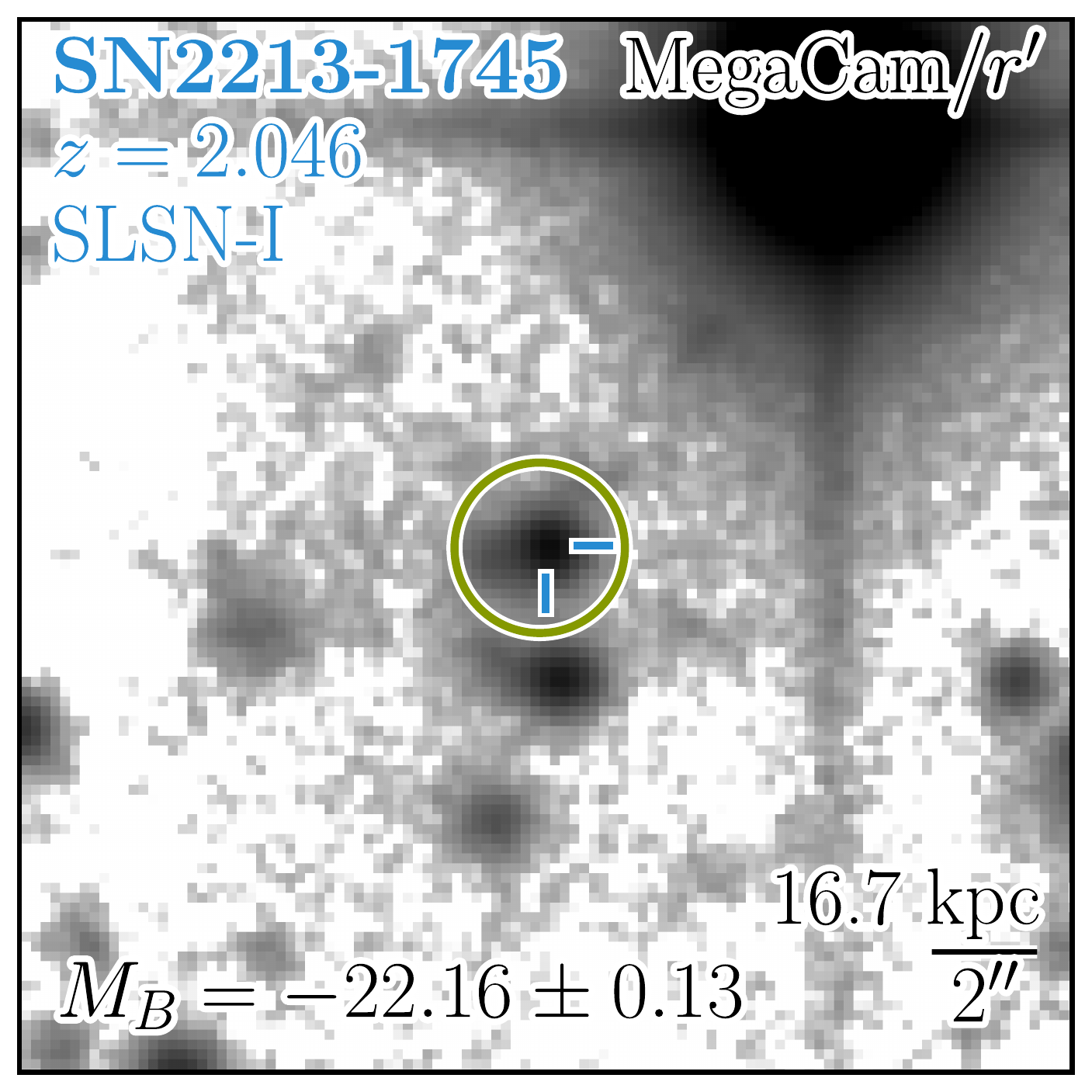}
\includegraphics[width=0.19\textwidth]{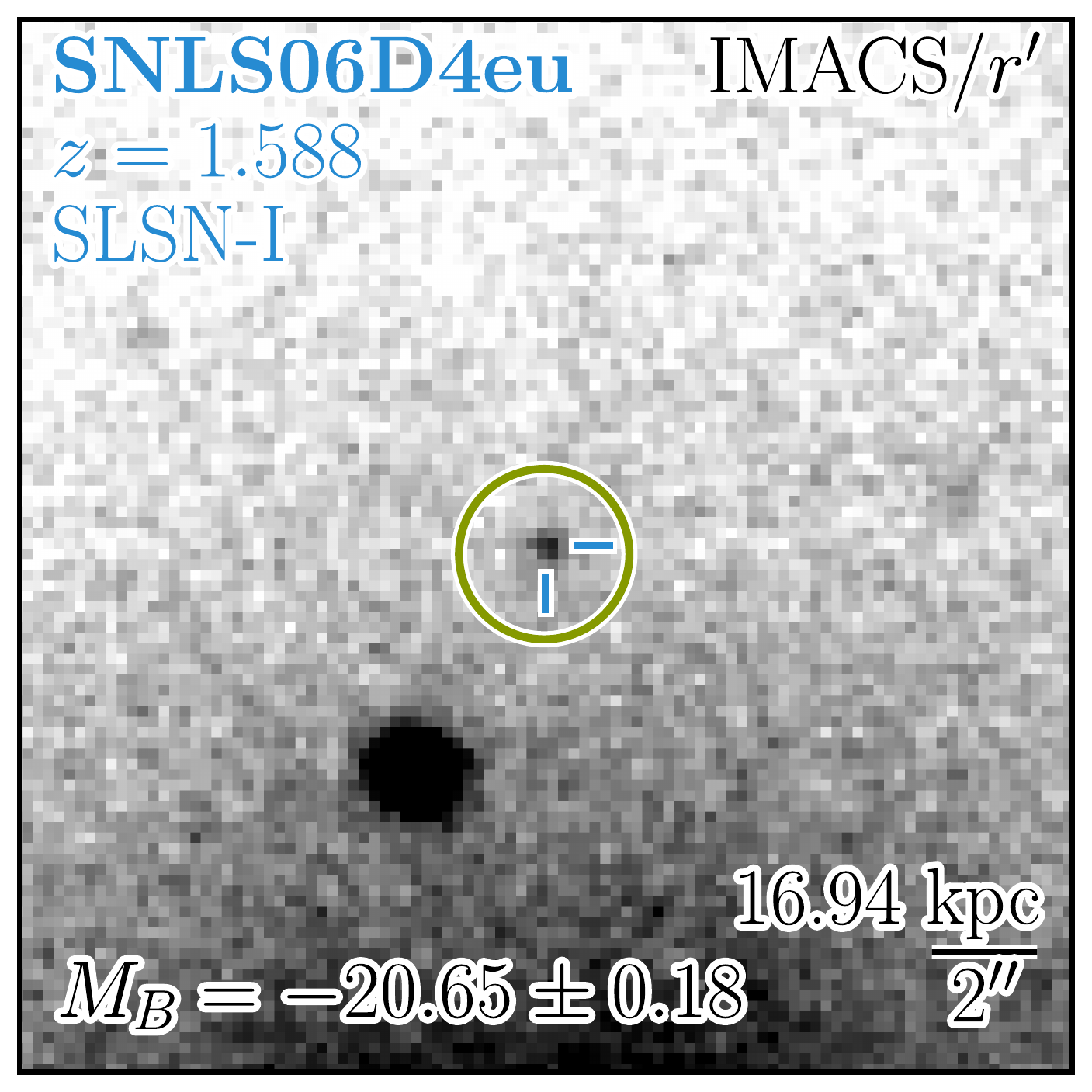}
\caption{
(Continued)
}
\end{figure*}
\clearpage

\begin{figure*}
\includegraphics[width=0.19\textwidth]{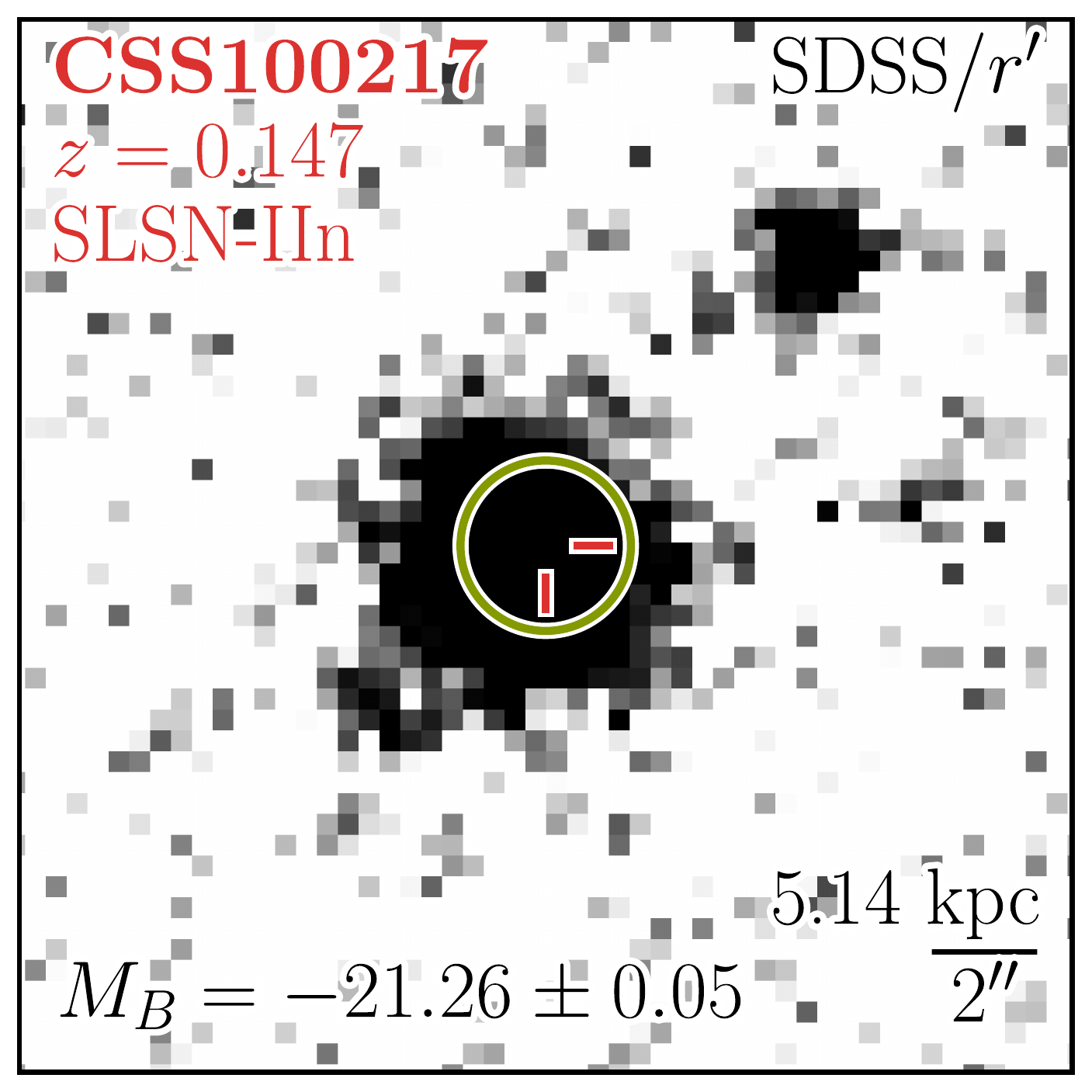}
\includegraphics[width=0.19\textwidth]{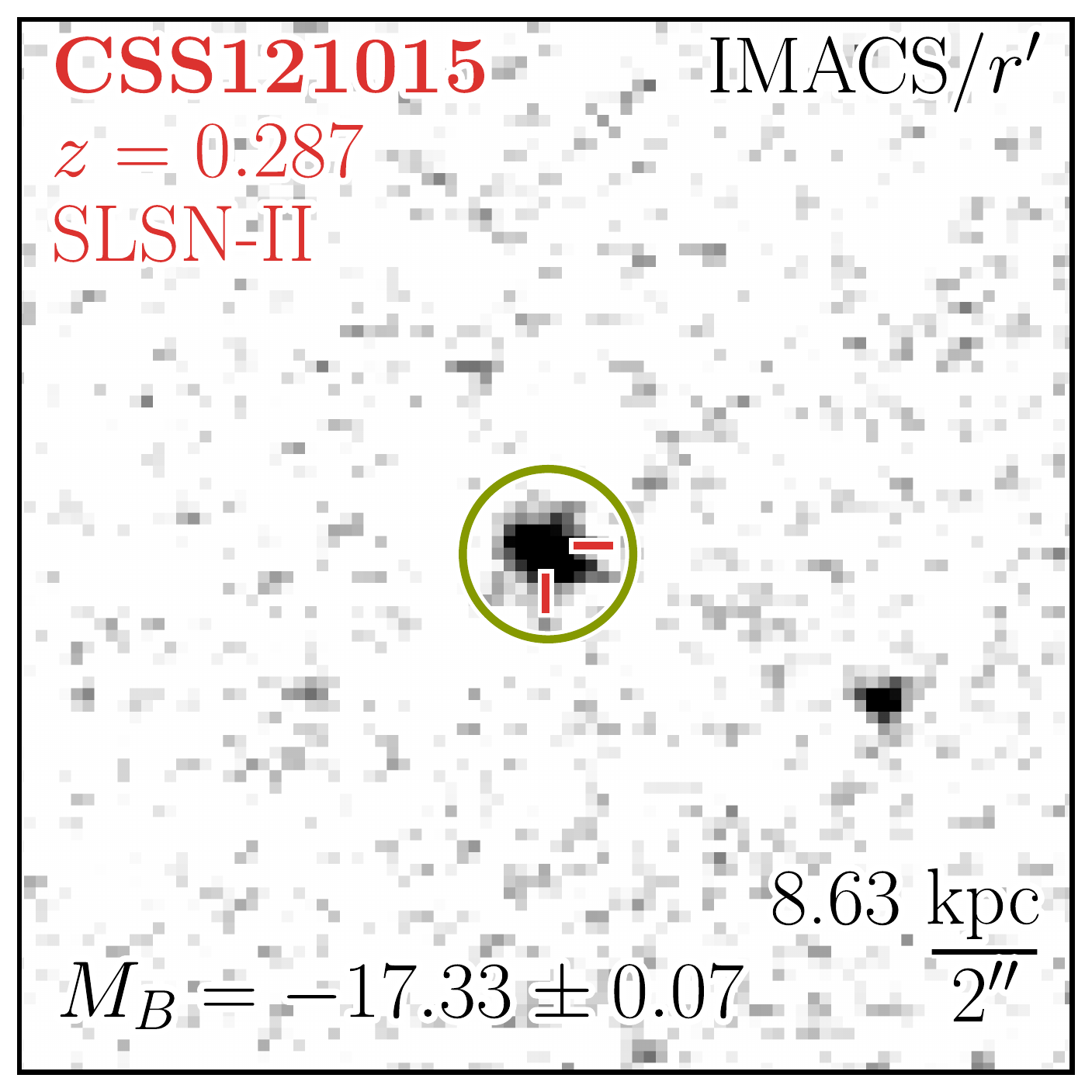}
\includegraphics[width=0.19\textwidth]{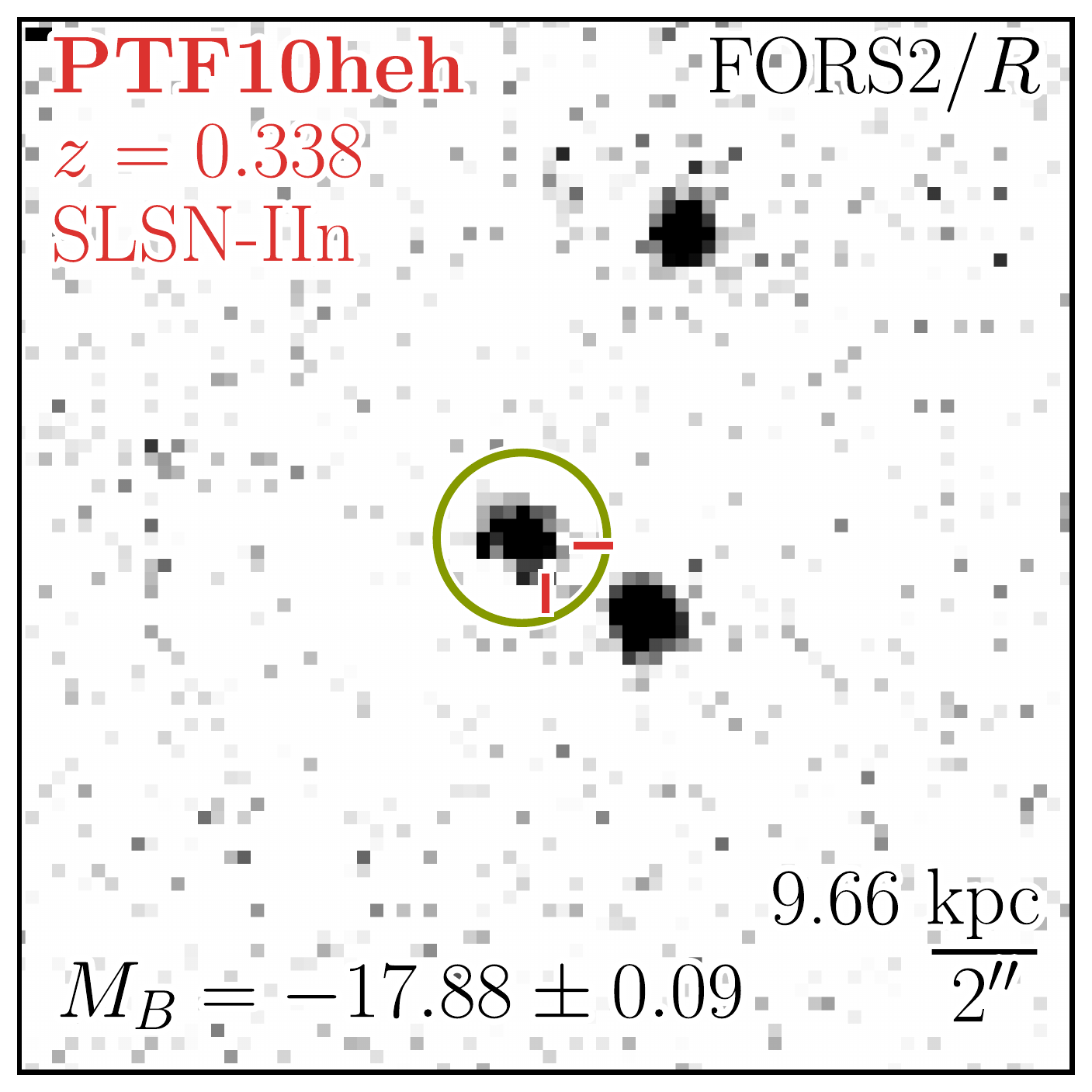}
\includegraphics[width=0.19\textwidth]{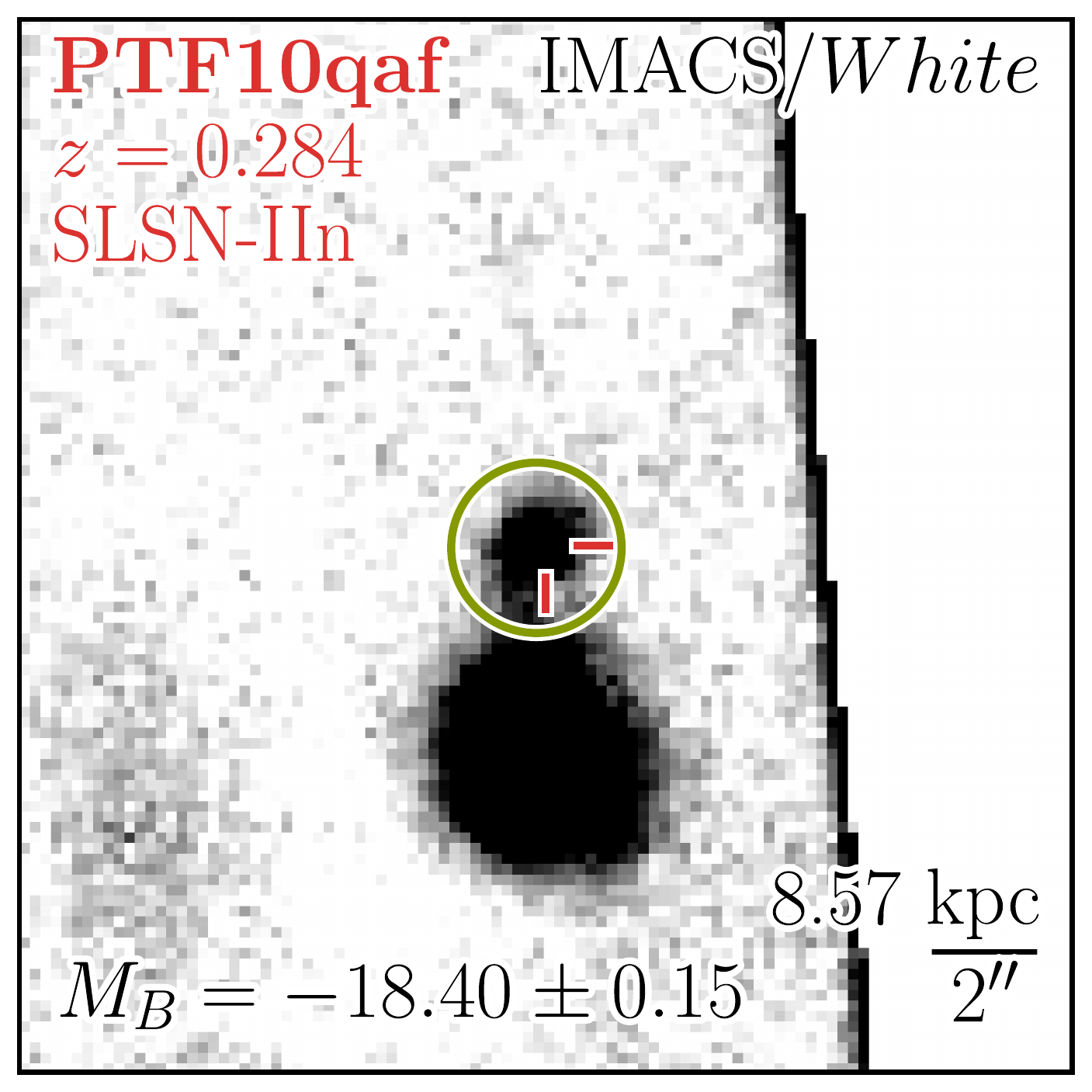}
\includegraphics[width=0.19\textwidth]{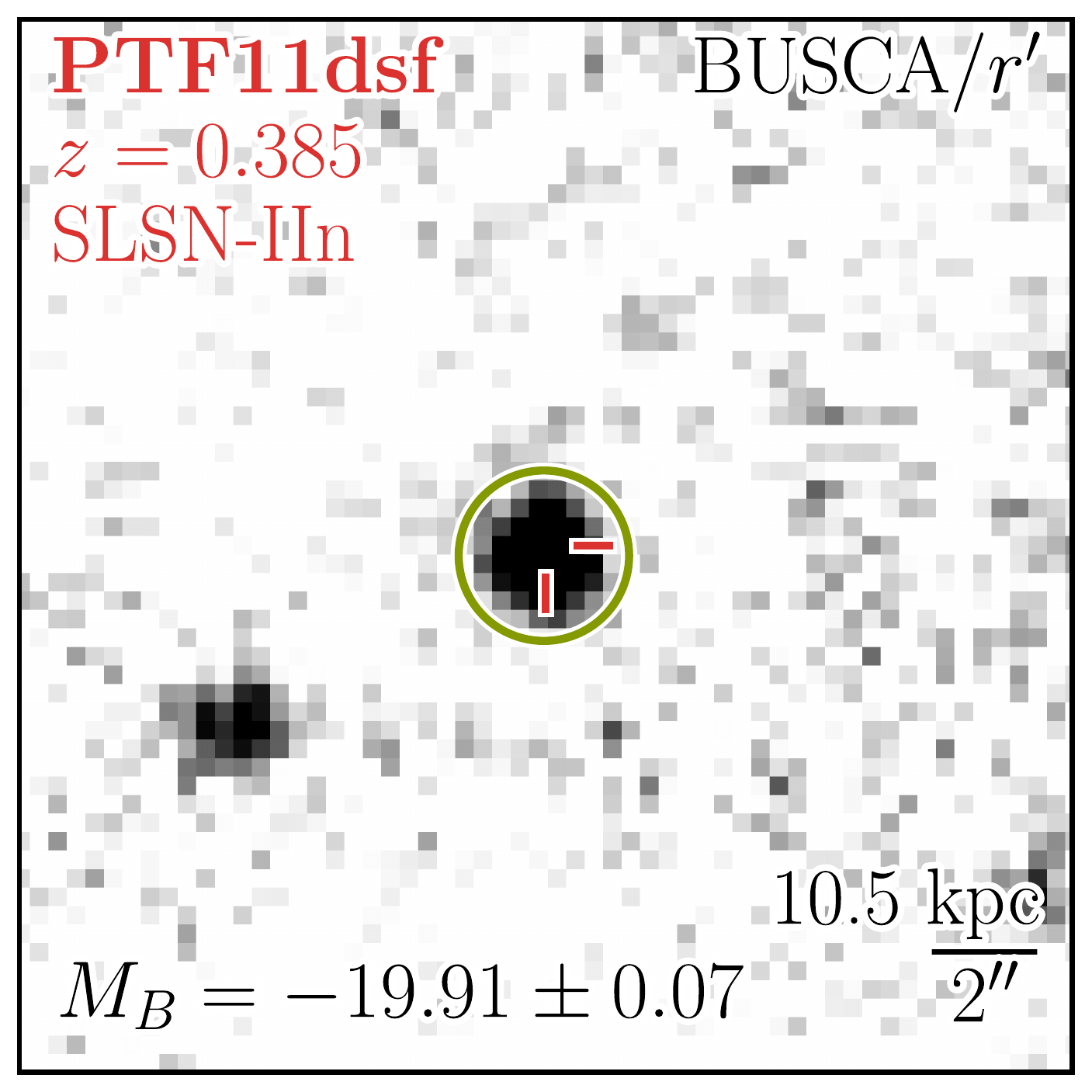}
\includegraphics[width=0.19\textwidth]{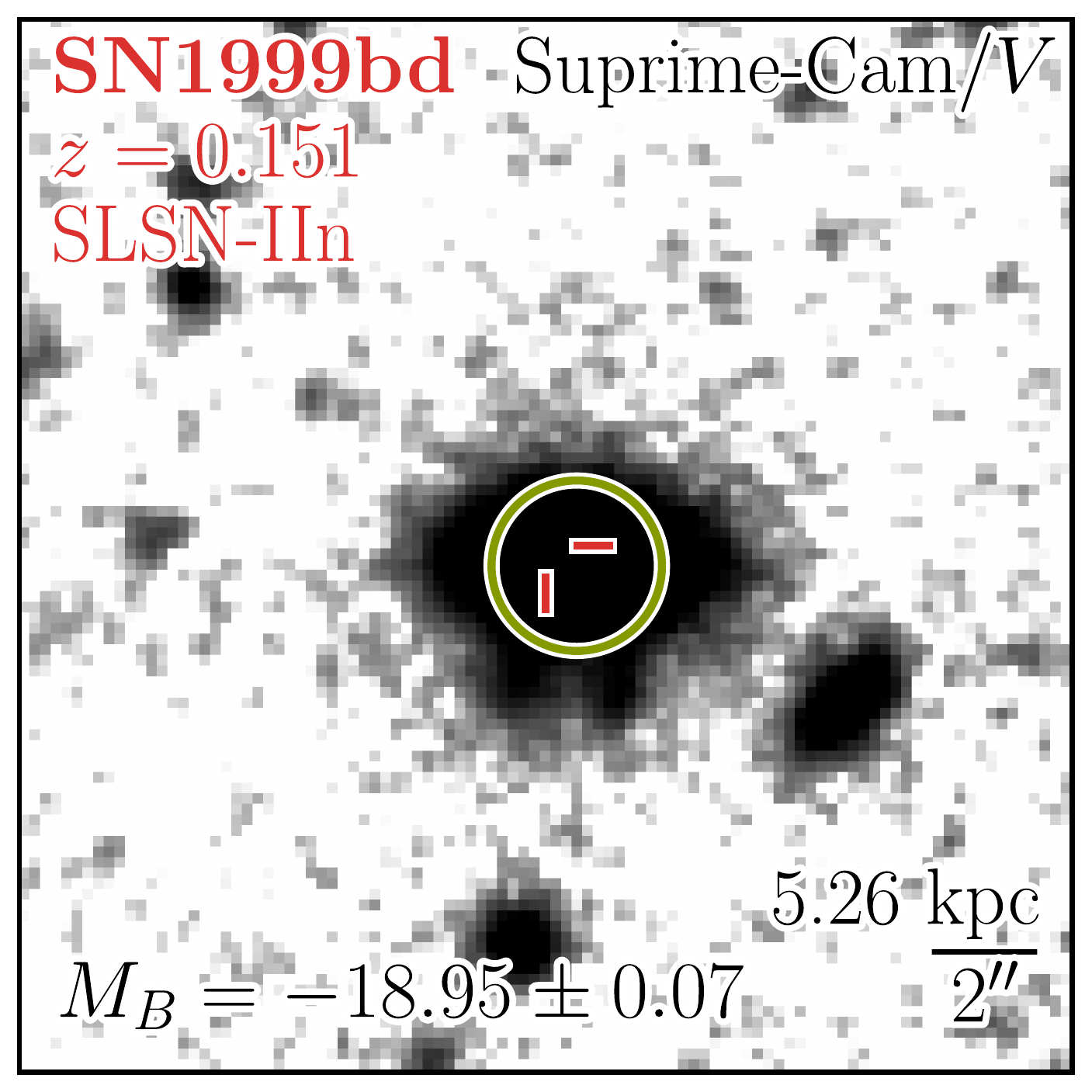}
\includegraphics[width=0.19\textwidth]{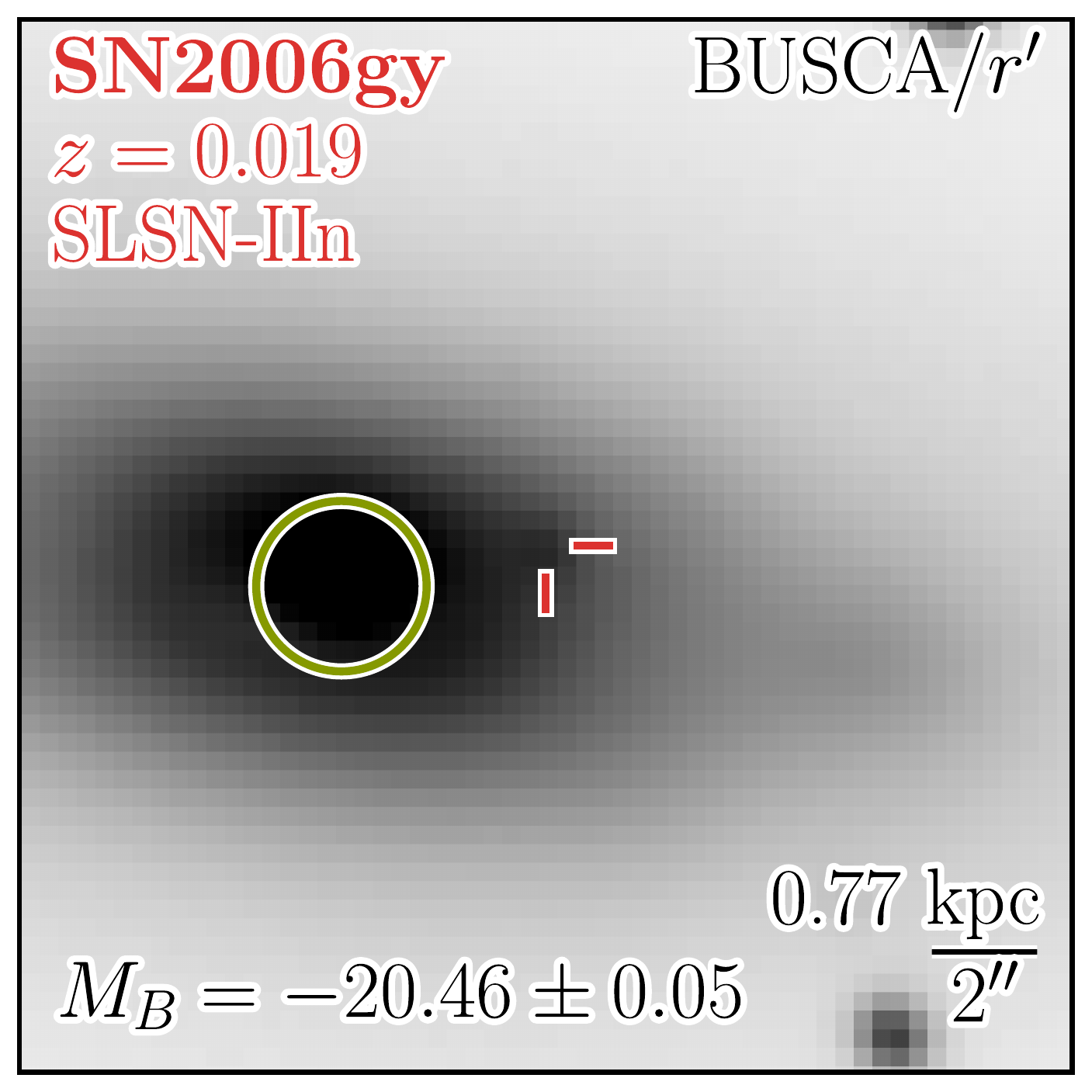}
\includegraphics[width=0.19\textwidth]{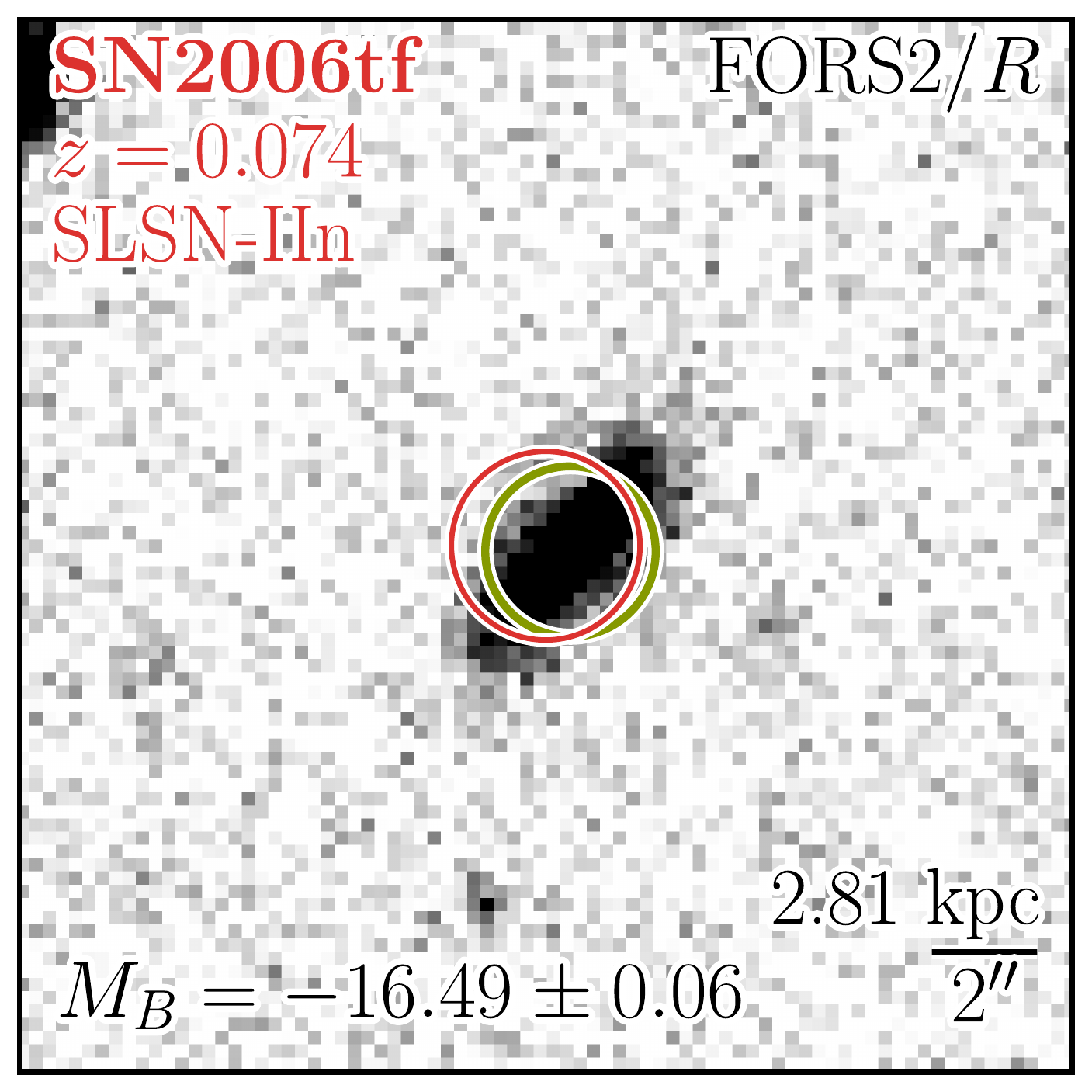}
\includegraphics[width=0.19\textwidth]{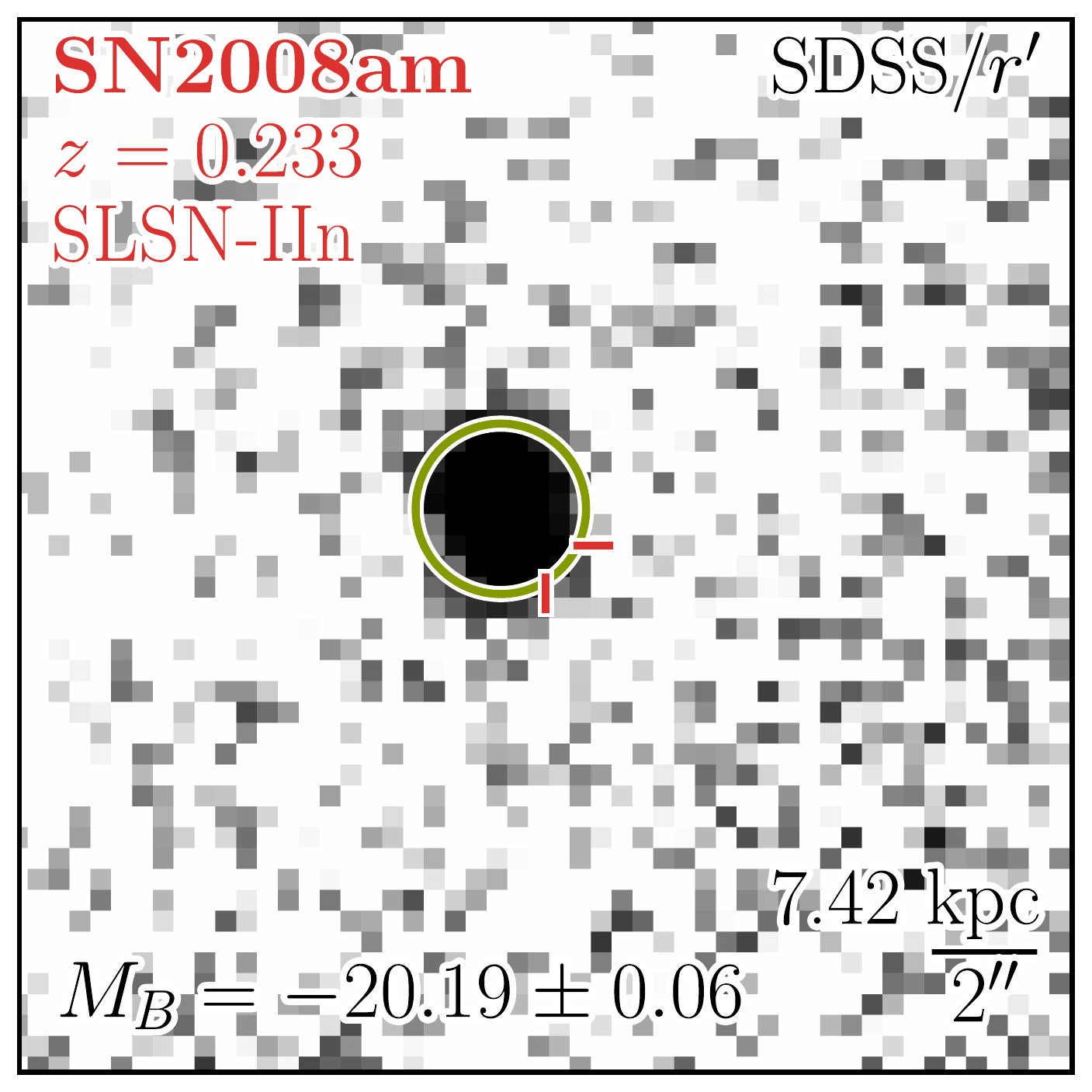}
\includegraphics[width=0.19\textwidth]{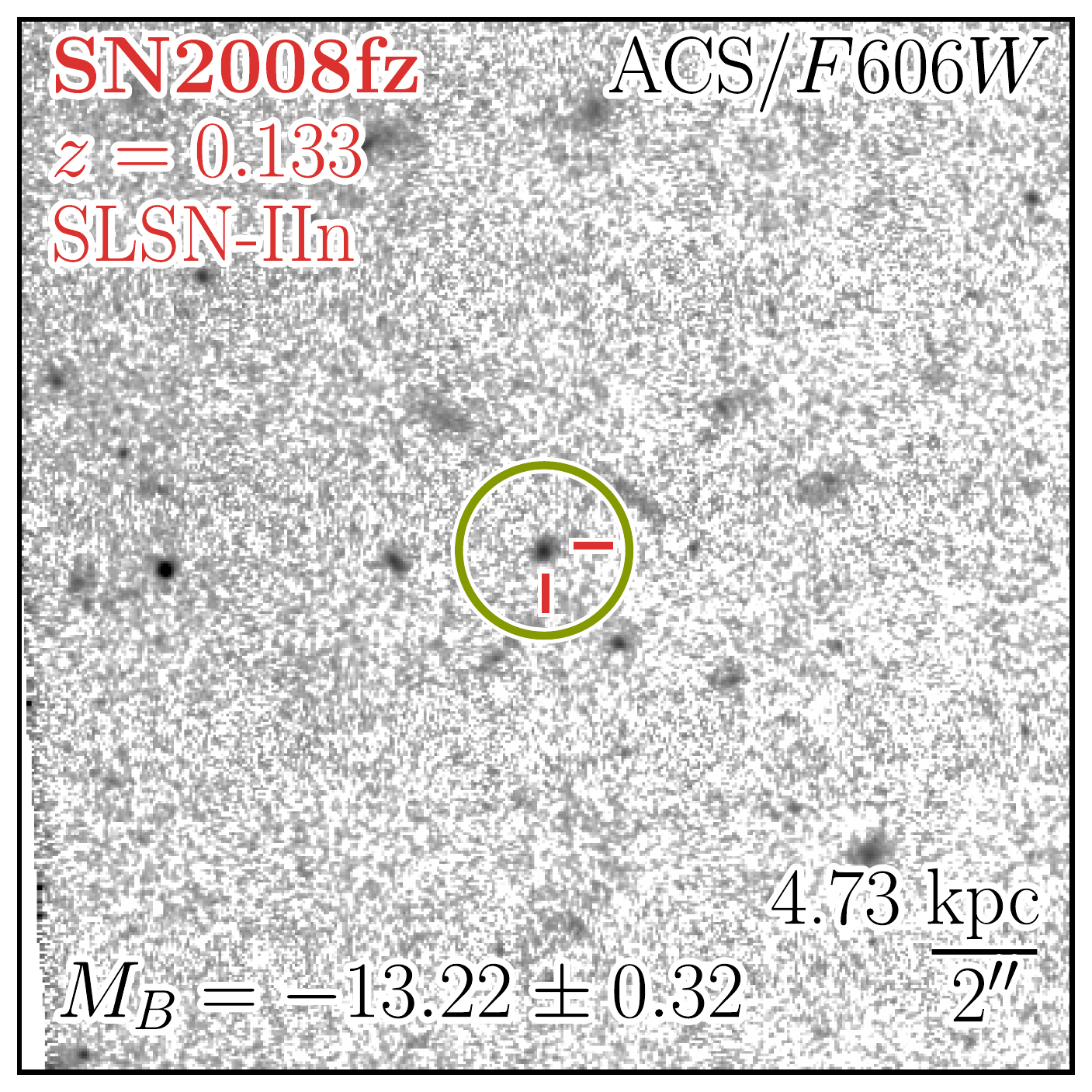}
\includegraphics[width=0.19\textwidth]{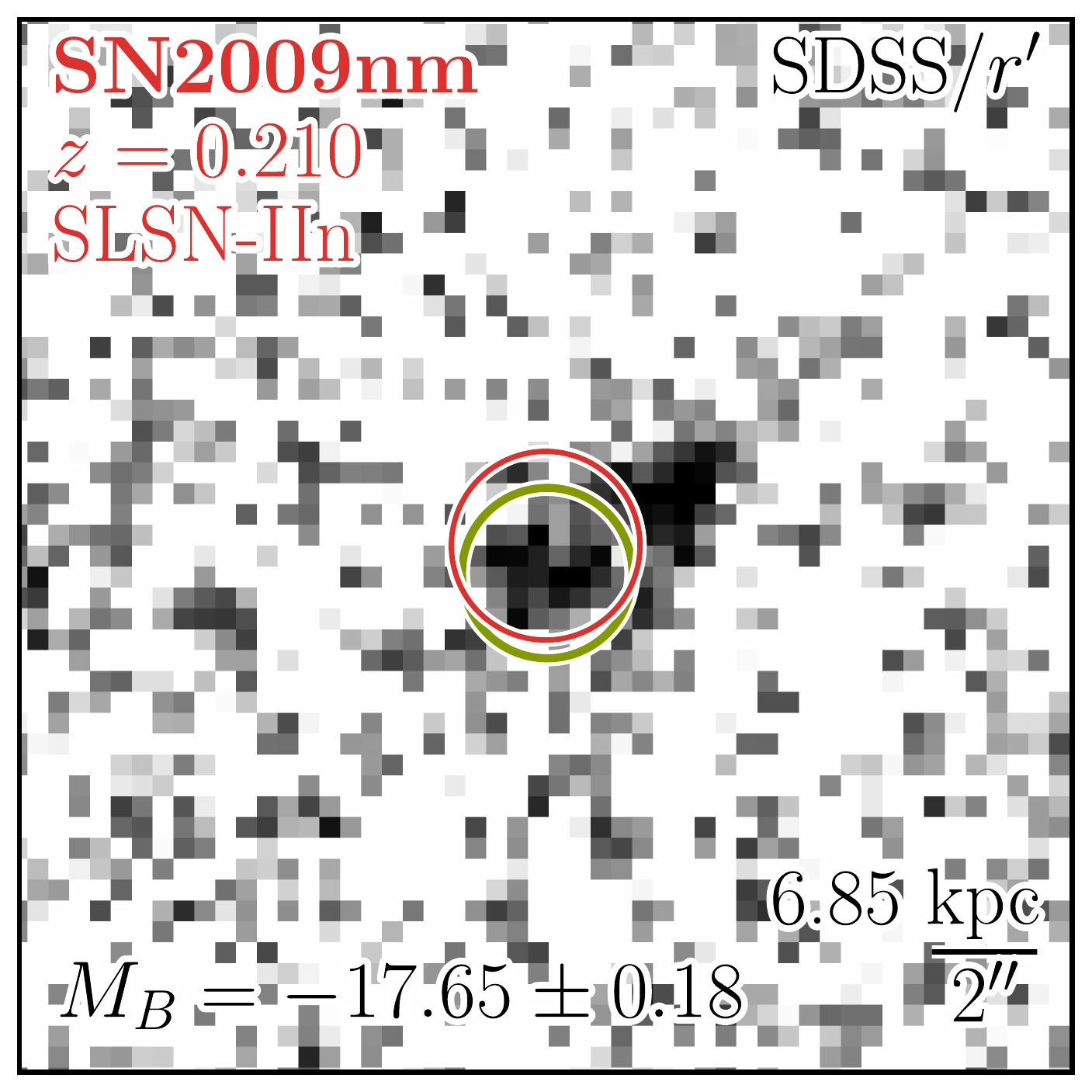}
\includegraphics[width=0.19\textwidth]{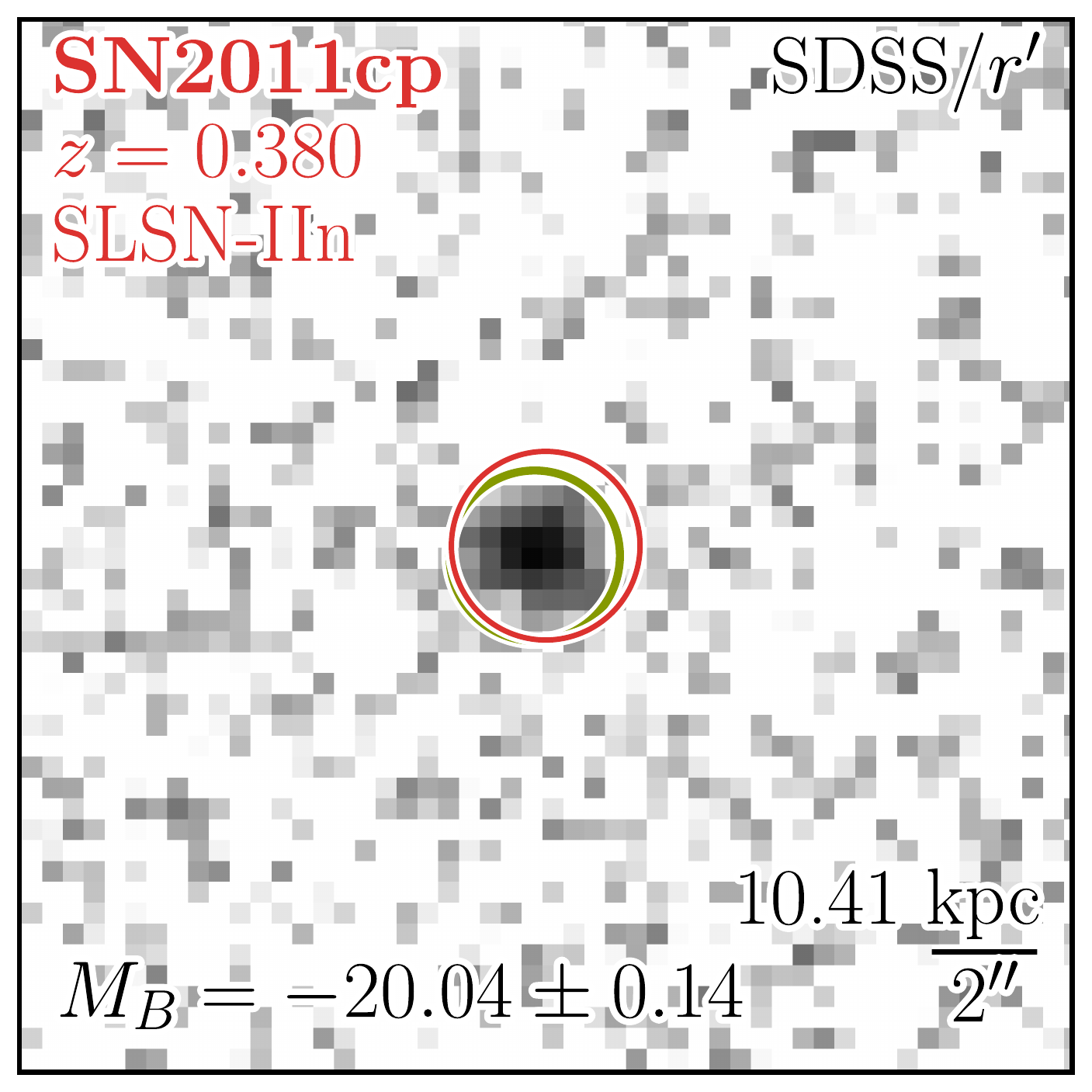}
\includegraphics[width=0.19\textwidth]{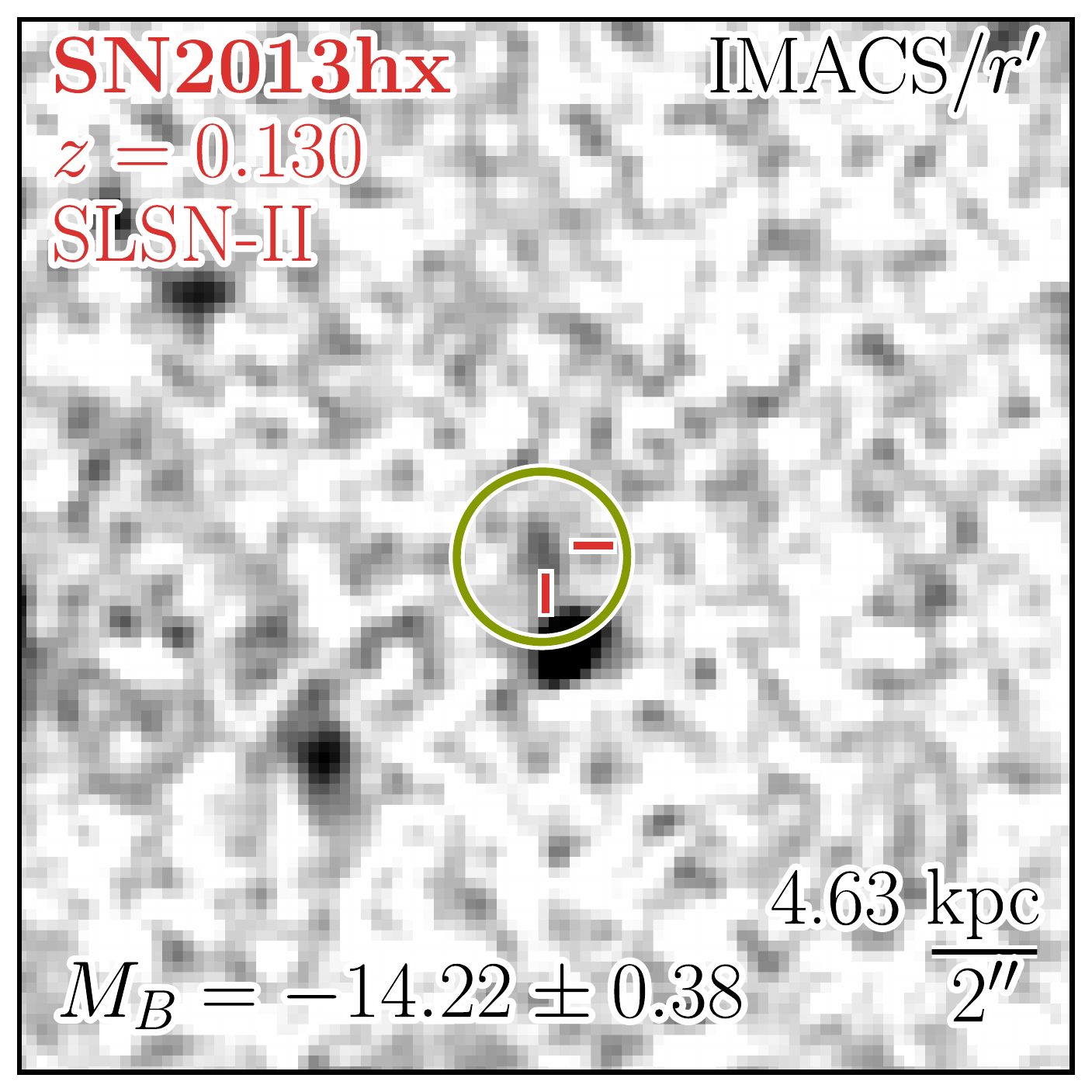}
\caption{Similar as Fig. \ref{fig:poststamp} but for H-rich SLSNe. The red crosshair marks
the position of the SNe after aligning a SN and a host image. If no SN image was available,
the red circle (arbitrary radius) indicates the SN position reported in the literature.
The image of SN2013hx was smoothed with a Gaussian kernel (width of 1 px) to improve the visibility of the host.
}
\label{fig:poststamp_app_2}
\end{figure*}

\clearpage

\section{Statistical properties of the comparison samples}

\begin{table*}
\caption{Statistical properties of GRB and SN host galaxies, EELGs and star-forming galaxies from the UltraVISTA survey per redshift bin.}
\centering
\begin{tabular}{lcccccccccccccc}
\toprule
\multirow{2}*{Sample}			& \multirow{2}*{Number}	& Mean		& $m_R$		& $(R-K_{\rm s})$		& $M_B$	& \multirow{2}*{$\log M/M_\odot$}	& $\log \rm SFR$ 		&  $\log \rm sSFR$\\
			& 			& Redshift	& (mag)		& (mag)			& (mag)	& 					& $(M_\odot\,\rm{yr}^{-1})	$& $(\rm{yr}^{-1})$\\
\midrule
\multicolumn{9}{c}{\textbf{$\mathbf{z\leq0.5}$}}\\
\midrule
\multirow{2}*{\textbf{GRB}}		&\multirow{2}*{14}	&\multirow{2}*{0.32}	&$ 22.03\pm0.68$ (7)	&$ 0.79\pm0.35	$ (5)	 &$-18.33\pm0.41	$&$8.83\pm0.20			$&$-0.19\pm0.13			$&$-9.15\pm0.11			$\\
\vspace{2mm}				&			&			&$ 1.67^{+0.61}_{-0.44}$&$ 0.62^{+0.32}_{-0.21}	$&$1.50^{+0.34}_{-0.28}	$&$0.65^{+0.17}_{-0.13}		$&$0.07^{+0.22}_{-0.05}		$&$0.07^{+0.14}_{-0.04}		$\\

\textbf{SN}\\
\multirow{2}*{\textit{GOODS}}		&\multirow{2}*{12}	&\multirow{2}*{0.41}	&\multirow{2}*{\nodata}	&\multirow{2}*{\nodata}	&$-18.82\pm0.52		$&$ 9.07\pm0.27			$&$ 0.22\pm0.19			$&$ -8.85\pm0.16		$\\
\vspace{1mm}				&			&			&			&			&$1.71^{+0.44}_{-0.35}	$&$ 0.88^{+0.23}_{-0.18}	$&$ 0.66^{+0.16}_{-0.13}	$&$ 0.52^{+0.12}_{-0.10}	$\\
\multirow{2}*{\textit{L11, S12, S13}}	&\multirow{2}*{105}&\multirow{2}*{0.05}	&\multirow{2}*{\nodata}	&\multirow{2}*{\nodata}	&$-19.51\pm0.20			$&$ 9.52\pm0.09			$&$ 0.22\pm0.11			$&$ -9.47\pm0.11		$\\
\vspace{2mm}				&			&			&			&			&$2.01^{+0.14}_{-0.13}	$&$ 0.90^{+0.07}_{-0.06}	$&$ 0.94^{+0.09}_{-0.08}	$&$ 0.78^{+0.09}_{-0.08}	$\\

\textbf{EELG}\\
\multirow{2}*{\textit{VUDS}}		&\multirow{2}*{9}	&\multirow{2}*{0.34}	&$ 24.92\pm0.18$ (11)  	&$ 1.05\pm0.07	$ (11)	&$ -15.65\pm0.31	$&$ 7.38\pm0.16			$&$ -1.14\pm0.17		$&$ -8.50\pm0.22		$\\
\vspace{1mm}				&			&			&$ 0.53^{+0.15}_{-0.12}$&$ 0.03^{+0.05}_{-0.02}$&$ 0.89^{+0.27}_{-0.20}	$&$ 0.37^{+0.14}_{-0.10}	$&$ 0.45^{+0.15}_{-0.11}	$&$ 0.42^{+0.31}_{-0.18}	$\\
\multirow{2}*{\textit{zCOSMOS}}		&\multirow{2}*{86}	&\multirow{2}*{0.30}	&$ 21.96\pm0.07$ (89)	&$ 0.53\pm0.05	$ (89)	&$ -18.47\pm0.13	$&$ 8.36\pm0.06			$&$ 0.11\pm0.06			$&$ -8.21\pm0.04		$\\
\vspace{2mm}				&			&			&$ 0.69\pm0.05		$&$ 0.48\pm0.04		$&$ 1.21\pm0.09		$&$ 0.49^{+0.05}_{-0.04}	$&$ 0.50\pm0.04			$&$ 0.27\pm0.03			$\\

\multirow{2}*{\textbf{UltraVISTA}}	&\multirow{2}*{26706}&\multirow{2}*{0.33}	&$ 22.53\pm0.01		$&$1.10\pm0.01		$&\multirow{2}*{\nodata}&$8.99\pm0.01			$&$-0.81\pm0.01			$&$-9.80\pm0.01			$\\
\vspace{2mm}				&			&			&$ 1.47			$&$0.58			$&			&$0.85				$&$0.86				$&$0.87				$\\

\midrule
\multicolumn{9}{c}{\textbf{$\mathbf{0.5<z\leq1.0}$}}\\
\midrule
\multirow{2}*{\textbf{GRB}}		&\multirow{2}*{38}	&\multirow{2}*{0.76}	&$ 24.00\pm0.39$ (12)	 &$ 1.03\pm0.15	$ (7)	 &$ -19.36\pm0.25	$&$ 9.03\pm0.12			$&$ 0.43\pm0.12			$&$-8.43\pm0.10			$\\
\vspace{2mm}				&			&			&$ 1.29^{+0.33}_{-0.26}	$&$ 0.24^{+0.28}_{-0.11}$&$ 1.54^{+0.19}_{-0.17}$&$ 0.66^{+0.10}_{-0.08}	$&$ 0.63^{+0.10}_{-0.09}	$&$0.37^{+0.09}_{-0.08}		$\\

\textbf{SN}\\
\multirow{2}*{\textit{GOODS}}		&\multirow{2}*{41}	&\multirow{2}*{0.72}	&\multirow{2}*{\nodata}	&\multirow{2}*{\nodata}	&$-19.51\pm0.30		$&$ 9.22\pm0.15			$&$ 0.25\pm0.13			$&$ -8.96\pm0.04		$\\
\vspace{2mm}				&			&			&			&			&$1.94^{+0.24}_{-0.21}	$&$ 0.93^{+0.11}_{-0.10}	$&$ 0.84^{+0.10}_{-0.09}	$&$ 0.26\pm0.03			$\\

\textbf{EELG}\\
\multirow{2}*{\textit{VUDS}}		&\multirow{2}*{21}	&\multirow{2}*{0.65}	&$ 24.73\pm0.14$ (19)	&$ 1.15\pm0.12	$ (19)	&$ -17.51\pm0.14	$&$ 8.00\pm0.07			$&$ -0.32\pm0.08		$&$ -8.34\pm0.12		$\\
\vspace{1mm}				&			&			&$ 0.61^{+0.11}_{-0.09}$&$ 0.46^{+0.10}_{-0.08}$&$ 0.65^{+0.11}_{-0.10}	$&$ 0.29^{+0.06}_{-0.05}	$&$ 0.32^{+0.06}_{-0.05}	$&$ 0.47^{+0.10}_{-0.08}	$\\
\multirow{2}*{\textit{zCOSMOS}}		&\multirow{2}*{78}	&\multirow{2}*{0.70}	&$ 22.44\pm0.05$ (91)	&$ 0.90\pm0.06	$ (91)	&$ -20.38\pm0.08	$&$ 9.13\pm0.04			$&$ 0.96\pm0.04			$&$ -8.14\pm0.04		$\\
\vspace{1mm}				&			&			&$ 0.52\pm0.04		$&$ 0.58^{+0.05}_{-0.04}$&$ 0.66^{+0.06}_{-0.05}$&$ 0.31\pm0.03			$&$ 0.30\pm0.03			$&$ 0.28\pm0.03			$\\
\multirow{2}*{\textbf{UltraVISTA}}	&\multirow{2}*{52689}&\multirow{2}*{0.77}	&$ 23.69 \pm 0.01	$&$ 1.91\pm0.01		$&\multirow{2}*{\nodata} &$ 9.60\pm0.01			$&$ -0.42\pm0.01		$&$-10.02\pm0.01		$\\
\vspace{2mm}				&			&			&$ 1.02			$&$ 0.82		$&			&$ 0.64				$&$ 0.93			$&$1.06				$\\

\midrule
\multicolumn{9}{c}{\textbf{$\mathbf{1.0<z\leq4.0}$}}\\
\midrule
\textbf{SN}				&			&			&			&			&			&				&				&				\\
\multirow{2}*{\textit{GOODS}}		&\multirow{2}*{5}	&\multirow{2}*{1.16}	&\multirow{2}*{\nodata}	&\multirow{2}*{\nodata}	&$-21.12\pm0.73		$&$ 10.02\pm0.49		$&$ 1.28\pm0.40			$&$ -8.76\pm0.30		$\\
\vspace{2mm}				&			&			&			&			&$1.44^{+0.69}_{-0.47}	$&$ 0.95^{+0.45}_{-0.30}	$&$ 0.79^{+0.37}_{-0.25}	$&$ 0.55^{+0.27}_{-0.18}	$\\

\textbf{EELG}\\
\multirow{2}*{\textit{3D-HST}}		&\multirow{2}*{22}	&\multirow{2}*{1.75}	&\multirow{2}*{\nodata}	&\multirow{2}*{\nodata}	&\multirow{2}*{\nodata}	&$ 8.72\pm0.11			$&$ 0.83\pm0.10			$&$ -7.97\pm0.11		$\\
\vspace{1mm}				&			&			&			&			&			&$ 0.46^{+0.10}_{-0.09}		$&$ 0.41^{+0.08}_{-0.07}	$&$ 0.37^{+0.10}_{-0.08}	$\\
\multirow{2}*{\textit{WISPS}}		&\multirow{2}*{7}	&\multirow{2}*{1.62}	&\multirow{2}*{\nodata}	&\multirow{2}*{\nodata}	&\multirow{2}*{\nodata}	&$ 8.62\pm0.23			$&$ 1.28\pm0.43			$&$ -7.32\pm0.36		$\\
\vspace{1mm}				&			&			&			&			&			&$ 0.57^{+0.21}_{-0.15}		$&$ 1.06^{+0.37}_{-0.27}	$&$ 0.87^{+0.34}_{-0.24}	$\\
\multirow{2}*{\textbf{UltraVISTA}}	&\multirow{2}*{70413}	&\multirow{2}*{1.60}	&$ 24.76 \pm 0.01	$&$2.43\pm0.01		$&\multirow{2}*{\nodata}&$ 9.98\pm0.01			$&$-0.16\pm0.01			$&$-10.13\pm0.01		$\\
\vspace{2mm}				&			&			&$ 1.05			$&$1.10			$&			&$ 0.56				$&$1.19				$&$ 1.23			$\\
\bottomrule
\end{tabular}
\tablecomments{
The first row of each element shows the mean value and its error and the second row the
standard deviation of the sample. The values of the $R$-band brightness, the
$B$-band luminosity and the $R-K_{\rm s}$ colour are not corrected for host attentuation.
}
\label{tab:statresult_non_SLSN}
\end{table*}

\begin{figure*}
\includegraphics[width=0.95\textwidth]{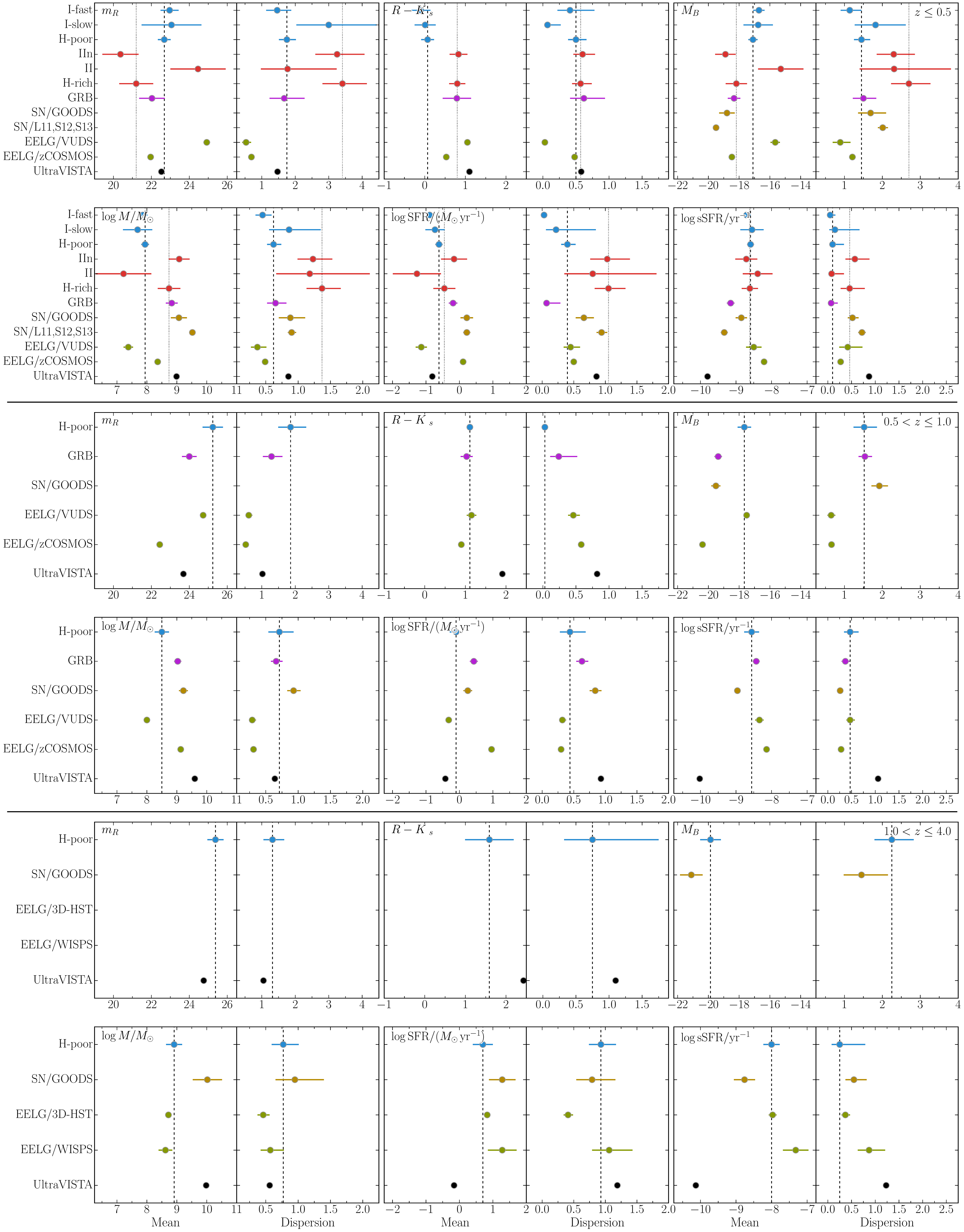}
\caption{Statistical properties of the SLSN host galaxy populations and of the comparison samples.
Top: $z\leq0.5$. Centre: $0.5 < z \leq 1.0$. Bottom: $1.0 < z \leq 4.0$. For each property,
the mean and the dispersion are displayed, as well as their uncertainties (for details see Sect. \ref{sec:statistics})
The vertical lines indicate location of the H-poor (dashed) and H-rich (dotted) SLSN host
populations in the diagnostics plots. Note, the exceptionally blue colours of
SLSN-I hosts at $z<0.5$ and huge dispersions of some SLSN-IIn host properties.
The measurement values are listed in Tables \ref{tab:statresult}~and \ref{tab:statresult_non_SLSN}.
}
\label{fig:statplot_app}
\end{figure*}
\label{lastpage}

\end{document}